\documentclass[11pt]{article}

\usepackage{tikz}
\usepackage{verbatim}
\usetikzlibrary{backgrounds,calc,shadings,shapes.arrows,shapes.symbols,shadows}
\definecolor{switch}{HTML}{006996}

\makeatletter
\pgfkeys{/pgf/.cd,
  parallelepiped offset x/.initial=2mm,
  parallelepiped offset y/.initial=2mm
}
\pgfdeclareshape{parallelepiped}
{
  \inheritsavedanchors[from=rectangle] % this is nearly a rectangle
  \inheritanchorborder[from=rectangle]
  \inheritanchor[from=rectangle]{north}
  \inheritanchor[from=rectangle]{north west}
  \inheritanchor[from=rectangle]{north east}
  \inheritanchor[from=rectangle]{center}
  \inheritanchor[from=rectangle]{west}
  \inheritanchor[from=rectangle]{east}
  \inheritanchor[from=rectangle]{mid}
  \inheritanchor[from=rectangle]{mid west}
  \inheritanchor[from=rectangle]{mid east}
  \inheritanchor[from=rectangle]{base}
  \inheritanchor[from=rectangle]{base west}
  \inheritanchor[from=rectangle]{base east}
  \inheritanchor[from=rectangle]{south}
  \inheritanchor[from=rectangle]{south west}
  \inheritanchor[from=rectangle]{south east}
  \backgroundpath{
    \southwest \pgf@xa=\pgf@x \pgf@ya=\pgf@y
    \northeast \pgf@xb=\pgf@x \pgf@yb=\pgf@y
    \pgfmathsetlength\pgfutil@tempdima{\pgfkeysvalueof{/pgf/parallelepiped
      offset x}}
    \pgfmathsetlength\pgfutil@tempdimb{\pgfkeysvalueof{/pgf/parallelepiped
      offset y}}
    \def\ppd@offset{\pgfpoint{\pgfutil@tempdima}{\pgfutil@tempdimb}}
    \pgfpathmoveto{\pgfqpoint{\pgf@xa}{\pgf@ya}}
    \pgfpathlineto{\pgfqpoint{\pgf@xb}{\pgf@ya}}
    \pgfpathlineto{\pgfqpoint{\pgf@xb}{\pgf@yb}}
    \pgfpathlineto{\pgfqpoint{\pgf@xa}{\pgf@yb}}
    \pgfpathclose
    \pgfpathmoveto{\pgfqpoint{\pgf@xb}{\pgf@ya}}
    \pgfpathlineto{\pgfpointadd{\pgfpoint{\pgf@xb}{\pgf@ya}}{\ppd@offset}}
    \pgfpathlineto{\pgfpointadd{\pgfpoint{\pgf@xb}{\pgf@yb}}{\ppd@offset}}
    \pgfpathlineto{\pgfpointadd{\pgfpoint{\pgf@xa}{\pgf@yb}}{\ppd@offset}}
    \pgfpathlineto{\pgfqpoint{\pgf@xa}{\pgf@yb}}
    \pgfpathmoveto{\pgfqpoint{\pgf@xb}{\pgf@yb}}
    \pgfpathlineto{\pgfpointadd{\pgfpoint{\pgf@xb}{\pgf@yb}}{\ppd@offset}}
  }
}
\makeatother

\tikzset{
  pmt/.style={
    fill=white,
    minimum width=1cm,
    minimum height=1cm,
    yscale=-1,
    path picture={
      \draw[color = white, draw=white]
      (path picture bounding box.south west) rectangle 
      (path picture bounding box.north east);
      \coordinate (A-center) at ($(path picture bounding box.center)!0!(path
        picture bounding box.south)$);
      \coordinate (A-west) at ([yshift=0.1cm]path picture bounding box.west);
      \coordinate (A-east) at ([yshift=0.1cm]path picture bounding box.east);
      \draw ($(A-west)!0!(A-west)$)
        arc (180:0:0.5 and 0.15);
      \draw [dashed] ($(A-west)!0!(A-west)$)
        arc (180:360:0.5 and 0.15) ;
      \draw ($(A-west)!0!(A-west)$)
        arc (180:0:0.5 and 0.4);
      \draw ($(A-west)!0!(A-west)$)
        arc (180:360:0.5 and 0.35);
      \draw ([xshift=+0.35cm,yshift=-0.35cm]A-west)
         -- ([xshift=-0.35cm,yshift=-0.35cm]A-east)
         -- ([xshift=-0.35cm,yshift=-0.59cm]A-east)
         -- ([xshift=+0.35cm,yshift=-0.59cm]A-west)
         -- ([xshift=+0.35cm,yshift=-0.35cm]A-west);
    }  
  },
  receive card/.style={
    rectangle,
    fill=green!20, draw,
    minimum width=0.57cm,
    minimum height=0.75cm,
  },
  daughter card/.style={
    rectangle,
    fill=blue!20, draw,
    minimum width=0.57cm,
    minimum height=0.75cm,
  },
  backend/.style={
    rectangle,
    fill=white, draw,
    minimum width=0.57cm,
    minimum height=0.75cm,
  },
  trigger processor/.style={
    rectangle,
    fill=red!20, draw,
    minimum width=0.57cm,
    minimum height=0.75cm,
  },
  farm/.style={
    rectangle,
    fill=red!20, draw,
    minimum width=0.57cm,
    minimum height=0.75cm,
  },
  switch/.style={
    rectangle,
    fill=blue!20, draw,
    minimum width=10cm,
    minimum height=0.75cm,
  },
}

\usetikzlibrary{calc, shadings, shadows, shapes.arrows}

\tikzset{%
  interface/.style={draw, rectangle, rounded corners, font=\LARGE\sffamily},
  ethernet/.style={interface, fill=yellow!50},% ethernet interface
  serial/.style={interface, fill=green!70},% serial interface
  speed/.style={sloped, anchor=south, font=\large\sffamily},% line speed at edge
  route/.style={draw, shape=single arrow, single arrow head extend=4mm,
    minimum height=1.7cm, minimum width=3mm, white, fill=switch!20,
    drop shadow={opacity=.8, fill=switch}, font=\tiny}% inroute/outroute arrows
}
\usepackage{amssymb}
\usepackage{amsmath}
\usepackage{graphicx}
\usepackage{epstopdf}
\usepackage[figuresright]{rotating}
\usepackage{verbatim}
\usepackage[section]{placeins}
\usepackage{multirow}
\usepackage{datetime}
\usepackage{cite}
\usepackage[affil-it]{authblk}
\usepackage{hyperref}
\usepackage{gensymb}
\usepackage{cleveref}
\usepackage{subfig}
\usepackage{bm}
\usepackage{grffile}
\usepackage{lineno}
\usepackage{todonotes}
\usepackage{xspace}
\usepackage{enumitem}
\usepackage{setspace}
\usepackage{verbatim}
\usepackage{etoolbox}
\usepackage{subfig}
\makeatletter
\preto{\@verbatim}{\topsep=0pt \partopsep=0pt }
\makeatother

\newcommand{\FHC}{\ensuremath{\nu}-mode\xspace}
\newcommand{\RHC}{\ensuremath{\overline{\nu}}-mode\xspace}

\newdateformat{mydate}{\monthname[\THEMONTH] \THEDAY, \THEYEAR}
\setlist[itemize]{noitemsep, topsep=0pt}

\begin{document}

%\vspace*{-1.5cm}
%\begin{flushright}
%LAPPD/MCP\\
%\today
%\end{flushright}

%\vspace{0.1in}
\title{TITUS: the Tokai Intermediate Tank for the Unoscillated Spectrum}

%\date{\mydate\today}
\date{}

%Authors

\author[15]{C.\ Andreopoulos~\footnote{Also at STFC/RAL.}}

\author[3]{F.C.T.\ Barbato}
\author[17]{G.\ Barker}
\author[9]{G.\ Barr}
\author[13]{P.\ Beltrame}
\author[2]{V.\ Berardi}
\author[11]{T.\ Berry}
\author[14]{A.\ Blondel}
\author[17]{S.\ Boyd}
\author[14]{A.\ Bravar}

\author[2]{F.S.\ Cafagna}
\author[16]{S.\ Cartwright}
\author[2]{M.G.\ Catanesi} 
\author[4]{C.\ Checchia}
\author[16]{A.\ Cole}
\author[4]{G.\ Collazuol}
\author[13]{G.A.\ Cowan}

\author[12]{T.\ Davenne} 
\author[7]{T.\ Dealtry}
\author[12]{C.\ Densham}
\author[3]{G.\ De Rosa}
\author[10]{F.\ Di Lodovico}
\author[13]{E.\ Drakopoulou}
\author[1]{P.\ Dunne}

\author[7]{A.\ Finch}
\author[12]{M.\ Fitton}

\author[17]{D. Hadley}
\author[10]{K.\ Hayrapetyan}

\author[2]{R.A.\ Intonti}

\author[1]{P.\ Jonsson}

\author[11]{A.\ Kaboth$^*$}
\author[10]{T.\ Katori}
\author[7]{L.\ Kormos} 
\author[6]{Y.\ Kudenko}

\author[8]{J.\ Lagoda}
\author[10]{J.\ Lasorak}
\author[4]{M.\ Laveder}
\author[7]{M.\ Lawe}
\author[1]{P.\ Litchfield}
\author[4]{A.\ Longhin} 
\author[5]{L.\ Ludovici}

\author[1]{W.\ Ma}
\author[2]{L.\ Magaletti}
\author[16]{M.\ Malek} 
\author[15]{N.\ McCauley}
\author[4]{M.\ Mezzetto}
\author[11]{J.\ Monroe} 

\author[12]{T.\ Nicholls}
\author[13]{M.\ Needham}
\author[14]{E.\ Noah}
\author[12]{F.\ Nova}

\author[7]{H.M.\ O'Keeffe} 
\author[10]{A.\ Owen}

\author[3]{V.\ Palladino} 
\author[15]{D.\ Payne}
\author[16]{J.\ Perkin} 
\author[13]{S.\ Playfer} 
\author[15]{A.\ Pritchard}
\author[10]{N.\ Prouse}

\author[2]{E.\ Radicioni}
\author[14]{M.\ Rayner}
\author[3]{C.\ Riccio}
\author[10]{B.\ Richards} 
\author[15]{J.\ Rose}
\author[3]{A.C.\ Ruggeri}

\author[9]{R.\ Shah}
\author[11]{Y.\ Shitov~\footnote{Also at Imperial College London.}}
\author[9]{C.\ Simpson~\footnote{Also at University of Tokyo/IPMU.}}
\author[13]{G.\ Sidiropoulos}
\author[12]{T.\ Stewart} 

\author[10]{R.\ Terri}
\author[16]{L.\ Thompson}
\author[12]{M.\ Thorpe}

\author[1]{Y.\ Uchida} 

\author[9]{D.\ Wark$^*$}
\author[1]{M.O.\ Wascko} 
\author[12]{A.\ Weber~\footnote{Also at Oxford University.}}
\author[10]{J.R.\ Wilson}

%Affiliations
\affil[1]{\small Imperial College London, Department of Physics, London, United Kingdom} 
\affil[2]{INFN Sezione di Bari and Universit\`{a} e Politecnico di Bari, Dipartimento Interuniversitario di Fisica, Bari, Italy}
\affil[3]{INFN Sezione di Napoli and Universit\`{a} di Napoli, Dipartimento di Fisica, Napoli, Italy}
\affil[4]{INFN Sezione di Padova and Universit\`{a} di Padova, Dipartimento di Fisica, Padova, Italy}
\affil[5]{INFN Sezione di Roma and Universit\`{a} di Roma “La Sapienza”, Roma, Italy}
\affil[6]{Institute for Nuclear Research of the Russian Academy of Sciences, Moscow, Russia}
\affil[7]{Lancaster University, Physics Department, Lancaster, United Kingdom}
\affil[8]{National Centre for Nuclear Research, Warsaw, Poland}
\affil[9]{Oxford University, Department of Physics, Oxford, United Kingdom}
\affil[10]{Queen Mary University of London, School of Physics and Astronomy, London, United Kingdom}
\affil[11]{Royal Holloway University of London, Department of Physics, London, United Kingdom}
\affil[12]{STFC, Rutherford Appleton Laboratory, Harwell Oxford, and Daresbury Laboratory, Warrington, United Kingdom}
\affil[13]{University of Edinburgh, School of Physics and Astronomy, Edinburgh, United Kingdom} 
\affil[14]{University of Geneva,  Section de Physique, DPNC, Geneva, Switzerland}
\affil[15]{University of Liverpool, Department of Physics, Liverpool, United Kingdom}
\affil[16]{University of Sheffield, Department of Physics and Astronomy, Sheffield, United Kingdom}
\affil[17]{University of Warwick, Department of Physics, Coventry, United Kingdom}

\maketitle
%\end{center}
\vspace{-0.25in}
\noindent
%\pagebreak
%\linenumbers
%\setstretch{1.65}
\begin{abstract}

The TITUS, Tokai Intermediate Tank for Unoscillated Spectrum,
detector, is a proposed Gd-doped Water Cherenkov tank with a
magnetised muon range detector downstream. It is located at J-PARC
atabout 2\,km from the neutrino target and it is proposed as a
potential near detector for the Hyper-Kamiokande experiment.  Assuming
a beam power of 1.3\,MW and 27.05$\times 10^{21}$ protons-on-target
the sensitivity to CP and mixing parameters achieved by
Hyper-Kamiokande with TITUS as a near detector is presented.  Also,
the potential of the detector for cross sections and Standard Model
parameter determination, supernova neutrino and dark matter are
shown.
\end{abstract}

%\newpage
\tableofcontents
\setcounter{tocdepth}{3}
\newpage

\section{Experimental overview}
\label{sec-expoverview}
\begin{figure}[htb]
\centering
\includegraphics[width=0.9\linewidth]{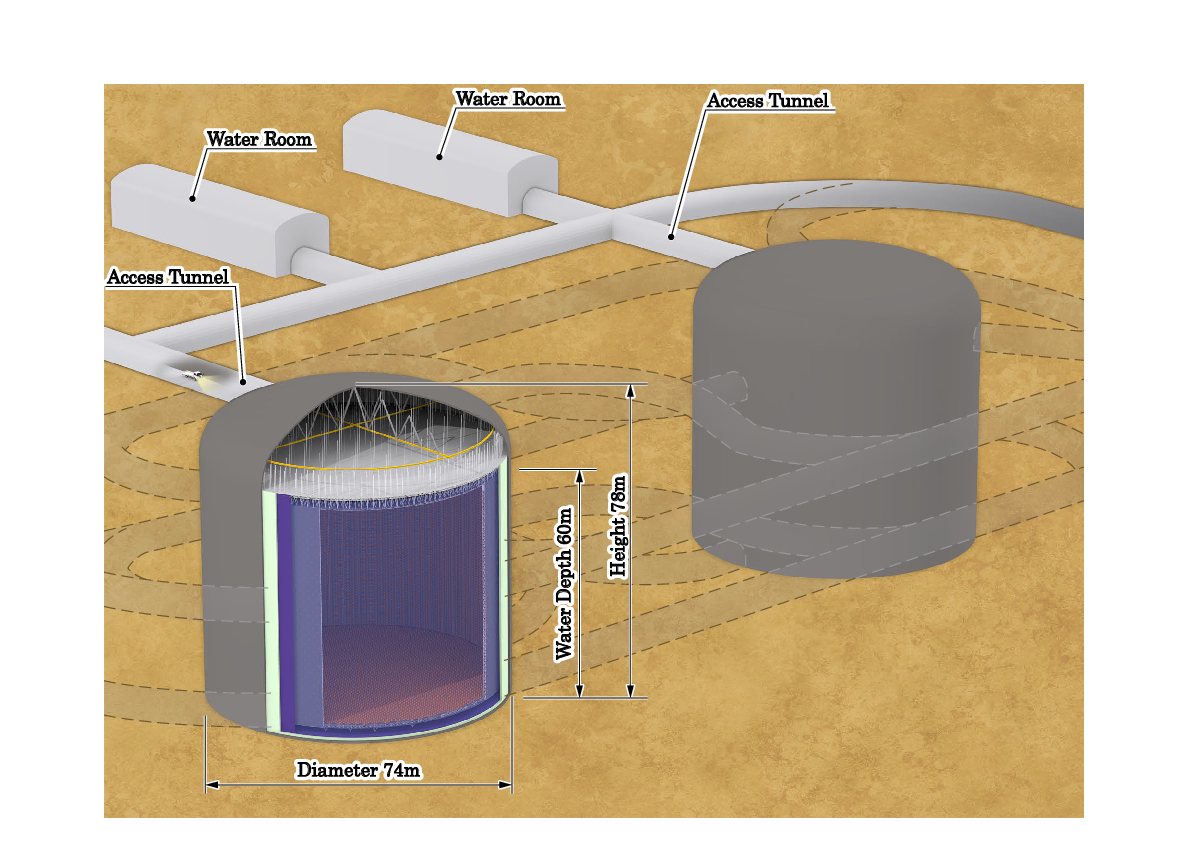}
\caption{Drawing of the Hyper-Kamiokande two-tank configuration.}
\label{fig:hyperk}
\end{figure}

The proposed Hyper-Kamiokande (Hyper-K, HK) detector~\cite{Abe:2015zbg} is
a half Mton water Cherenkov (WC) detector with a two-tank
configuration, as shown in Figure~\ref{fig:hyperk}, where the first
tank (a cylinder with diameter 74\,m and height 60\,m) is scheduled to
start operation around 2026 and the second identical tank starts six
years later.

%Hyper-Kamiokande is the successor to the Super-Kamiokande (Super-K)
%detector. It is a multipurpose experiment and it also acts as the far
%detector of a long baseline neutrino experiment with a beam produced at
%J-PARC. The planned beam power is 1.3\,MW. In this document we assume a total
%running time of 10 years.

Hyper-K will act as the far detector for a long-baseline neutrino
experiment using 0.6\,GeV neutrinos produced by a 1.3\,MW proton beam
at J-PARC; in this document we assume a total running time of 10 years
and a total exposure of 27.05$\times 10^{21}$ protons-on-target (POT).
In addition, it is a multipurpose non-accelerator experiment whose
large fiducial mass will allow it to address topics such as
atmospheric neutrinos, the search for proton decay and astrophysical
neutrinos.

The accelerator neutrino event rate observed at Hyper-K depends on the
oscillation probability, neutrino flux, neutrino interaction
cross-section, detection efficiency, and detector fiducial mass of
Hyper-K.  The neutrino flux and cross-section models can be
constrained by data collected at the near detector, ND280, situated
close enough to the neutrino production point such that oscillation
effects are negligible. The T2K collaboration has successfully applied
a method of fitting the near detector data with parameterised models
of the neutrino flux and interaction
cross-sections~\cite{Abe:2015awa}.  However, there are several
limitations to the T2K approach that we aim to overcome with the
current proposal.

The main goal of the detector proposed in this paper is to measure the
neutrino beam spectrum before oscillating andbeing detected at the far
detector.  Many of the uncertainties on the modelling of neutrino
interactions arise from uncertainties in nuclear effects, implying
that an ideal near detector should then include the same nuclear
targets as in the far detector. The performance of a WC detector can
be enhanced using gadolinium doping that permits tagging of the final
state neutrons thanks to a very high cross section for neutron capture
on Gd. In the case of charged-current quasi-elastic interactions
(CCQE), which are the principal target for oscillation and
CP-violation studies, the outgoing nucleon is a proton for neutrino
interactions and a neutron for antineutrino interactions. Thus, the
gadolinium doping, similar to that proposed by J.~Beacom and
M.~Vagins~\cite{FSneutrons}, will allow us to distinguish between
neutrinos and antineutrinos, a capability usually restricted to
magnetised detectors. However, whereas a magnetised detector
distinguishes neutrinos and antineutrinos by measuring the charge of
the produced lepton, the Gd-doped WC detector will do so by means of
the final state nucleons.
%
%Gadolinium salts dissolved in water have high neutron capture
%cross-sections and produce $\sim$8 MeV in gammas, several tens of
%microseconds after the initial event. This delayed 8 MeV signal is
%much easier to detect than the 2 MeV gammas from neutron capture in
%pure water.
The Gd-doped WC detector will be complemented by a magnetised Muon
Range Detector (MRD), which will detect and measure the charge of
muons that exit from the WC tank into the MRD (approximately 20\% of
the total muon yield). Whilst it is important to detect the high
energy muons, thus the high energy tail of the neutrino spectrum, this
can also serve as a direct calibration method for the gadolinium.  A
correction to the susceptibility caused by the paramagnetic nature of
the Gd$_2$(SO$_4$)$_3$ will be applied.

\begin{figure}[htb]
\centering
\includegraphics[width=0.9\linewidth]{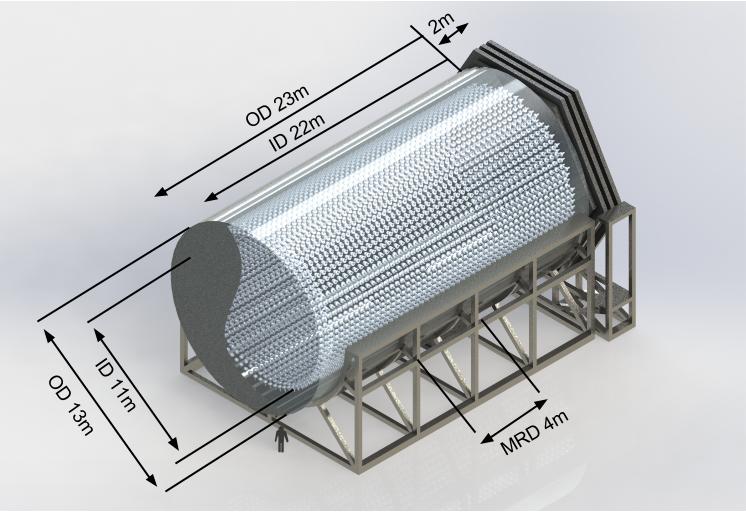}
\caption{Schematic diagram of the TITUS detector with
dimensions of the inner (ID) and outer (OD) detectors shown.}
\label{fig:detector}
\end{figure}

The proposed Gd-doped WC detector with a magnetised downstream MRD is
called the Tokai Intermediate Tank for Unoscillated Spectrum
(TITUS). Its total WC volume is 2.1\,kton and it is planned to be
situated approximately 2\,km from the neutrino beam production target. As such it is
also referred to as an intermediate detector.
Figure~\ref{fig:detector} shows a schematic of the TITUS detector.

%Since the T2K experiment should continue running up to 2025 (T2K
%phase-2), with an increasing beam up to 1.3\,MW, the proposed
%detector can also serve as near detector for the T2K experiment.
In recent years, much theoretical work has been done to calculate
contributions to the CCQE reconstructed final states, identified by a
muon and no pions in the final state, from non-CCQE processes such as
two body currents or final state interactions that can absorb a pion.
The short and long range correlated nucleon pairs contribute to the
neutrino interaction differently.  Although such correlations have
been known in nuclear physics for many years, the importance of these
was only recently realised~\cite{Alvarez-Ruso:2014bla,Agashe:2014kda}
by high energy physics community.  These nuclear effects often lead to
the ejection of multiple nucleons in the final state and are referred
to here as multi-nucleon processes.  Among them, n-particle n-hole
interactions (npnh) may account for as much as 30\% of the total
cross-section in 1-10 GeV region.  Identifying a nucleon in the final
state will help address this issue.  CCQE and non-CCQE neutrino
interactions typically produce different numbers of neutrons;
therefore the ability to tag neutrons in the final state can provide
discrimination between signal and background.
%The oscillation probability depends on the neutrino energy, while we
%estimate the neutrino energy from the observed four momentum of the
%final state charged lepton. Correctly modeling the relationship
%between neutrino energy and final state lepton kinematics is essential
%to correctly applying the oscillation probability, even when there is
%a constraint on the event rate from the near detector data. 
%The ability to be able to distinguish the main neutrino interaction
%CCQE with respect to the other final states, and the ability to
%measure the intrinsic electron neutrino background will help to
%overcome the different beam’s energy dependence and flavor content at
%the near and far detectors.
The neutron tagging techniques will also be useful to a broader
program of physics beyond oscillation physics. For example, the
neutron tagging can help in separating signal from background in
proton decay final states.  Moreover, in the detection of diffuse
supernova neutrino background, neutron tagging can be used to separate
genuine neutrinos from various radiogenic and spallation
backgrounds. In the event of a core collapse supernova, the detection
of neutrons can be used to help discriminate among different
interactions in the water such as inverse beta decay and
neutrino-oxygen scattering.
%Finally, the main disadvantage of the WC detector is the inability to
%separate positively and negatively charged leptons, and hence
%antineutrino and neutrino interactions. This ability is especially
%important for a CP violation measurement where the wrong sign
%contribution to the neutrino flux should be well understood, and this
%can be addressed using gadolinium.

\section{Physics goals}
\label{sec-physics}
The novel design of the TITUS detector will permit significant
improvement in the determination of the oscillation parameters and
neutrino interaction measurements. The main characteristics for
oscillation physics are:
\begin{itemize} 
\item the same target as the far detector;
\item the ability to distinguish neutrinos and antineutrinos;
\item distinguishing between neutrino nucleon interaction modes based on the neutron multiplicity;
\item full containment of the neutrino spectrum including the high energy tail, reducing the error on the kaon component of the beam;
\item measurement of the intrinsic electron neutrino contamination of the beam;
\item measurement of the charged and neutral current differential
  cross sections.
\end{itemize}
The beam observed by a detector positioned at the same angle off-axis,
approximately 2\,km from the beam target, is very similar to that seen
by the far detector. This minimises the need to re-weight the near
detector beam spectrum to match the far detector, thereby reducing
systematic errors and the dependence on external measurements and
simulations.

Moreover, the physics studies that can be performed by TITUS also
include rare and exotic final states:
\begin{itemize} 
\item cross section determination;
\item Standard Model measurements;
\item supernova neutrinos;
\item non-standard physics and dark matter searches.
\end{itemize}

In the following sections, we will first address the optimisation of
TITUS, starting from the detector baseline in section~\ref{sec-beam},
neutron capture in section~\ref{subsec-neutron}, and external
backgrounds in section~\ref{sec-background} before moving on to the
tank in section~\ref{sec-detector}, which includes discussions of the
data acquisition and calibration. Thereafter, two sensitivity studies
will be presented.  The first will discuss a basic study (see
section~\ref{sec-basicstrategy}) and the second will use a full
software and reconstruction chain (see section~\ref{sec-software}).
Finally, section~\ref{sec-other} will discuss other physics studies,
including neutrino cross section measurements, Standard Model-related
measurements, supernova bursts, and dark matter measurements before
the conclusions are discussed in section~\ref{sec-conclusions}.

\section{Baseline and beam}
\label{sec-beam}
There are two main factors in determining the baseline for TITUS.  The
first is to have a flux similar to the one at the far detector to
directly measure and constrain the electron neutrino charged-current
($\nu_{e}$CC) and neutral current neutral pion (NC1$\pi^{0}$) spectra,
the main backgrounds for $\nu_e$ appearance measurements. The second
is to be close enough to the production target to have approximately
one event per beam spill.  Practical considerations regarding
available land for excavation are then explored to provide possible
locations for the TITUS detector.

\subsection{Baselines}
In 2001 six sites, with baselines ranging from 1.5 to 2.5\,km, were
investigated for a possible future intermediate detector to complement
the existing near detector, ND280.  These locations are along the
direction that connects J-PARC's neutrino target station to the
Tochibora mine, the site for the Hyper-K detector.

\begin{figure}[hbt]
  \centering \includegraphics[width=1.\textwidth]{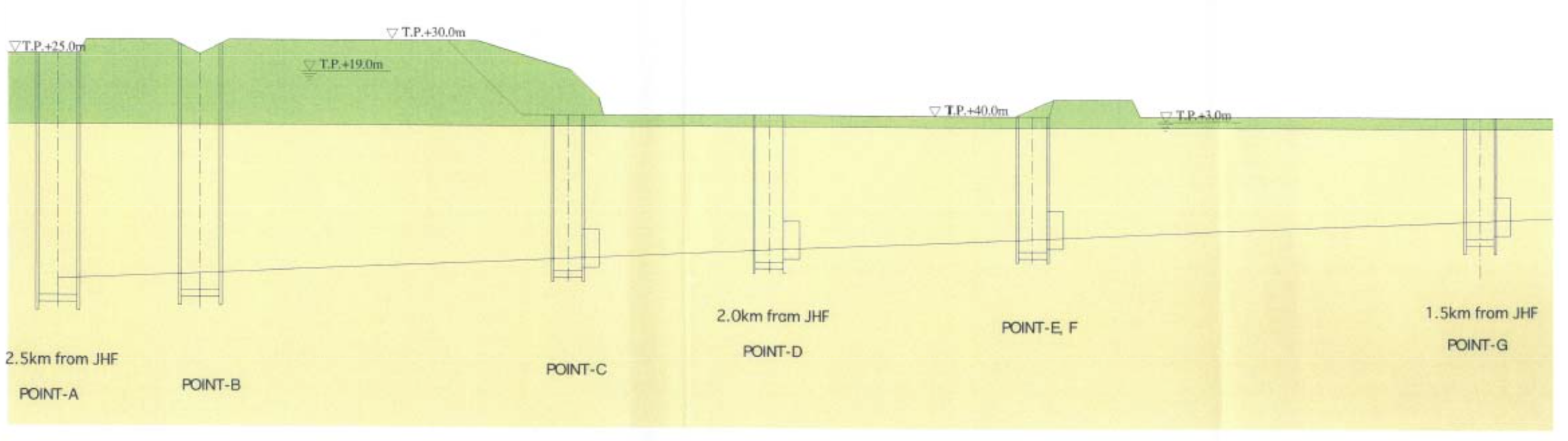} \caption{\label{fig:land}Schematic
  view of the six candidate near detector sites, with distance from
  the J-PARC neutrino target, cavity depth and ground elevation.  }
\end{figure}

The lie of the land, shown in Figure~\ref{fig:land} with the candidate
sites (points A to G), has an important impact on the civil
engineering works required to build the underground cavity.

The longest baseline considered corresponds to Point A in
Figure~\ref{fig:land} at about 2.5\,km. Due to a ground elevation of
about 25\,m, it requires a cavity about 90\,m deep, with a diameter of
30\,m. This has a severe impact on the total civil engineering cost;
it was estimated that point A is about four time as expensive as point
D, which has a 2\,km baseline, a 50\,m cavity depth, and 18\,m cavity
diameter. Because of such high costs locations A--C will not be taken
into account in the presented studies. Before excavation a boring
survey will need to be performed at the candidate site.
%At present there are no reliable estimates of the total civil engineering
%costs. As they strongly depend on the actual quality of the soil at the 
%candidate site, they will have to be evaluated after a boring survey has
%been performed.
%
%The civil engineering costs for a near detector site around point D
%were evaluated in 2004
%, based on the studies from 2001 when the
%construction of a 2~km T2K near detector was considered. 
%The costing includes the excavation of the underground facility and
%support surface buildings with, in addition, the costs of installation
%and material for cranes and elevators necessary to transport people
%and equipment in and out of the underground hall. Although these
%estimates provide initial figures to work on, they will need to be
%reevaluated, in particular taking into account the actual quality of
%the soil, that will be determined with a boring survey.
%
%Table~\ref{tab:cavitycosts} summarizes these costs, based on detailed engineering designs and quotation.
%
%\begin{table}[tbp] \centering
%\begin{tabular}{c| r}\hline\hline
%ITEM/SYSTEM & COST (M\$)\\
%\hline
%Hall excavation/construction & 9.14 \\
%Surface buildings & 0.72 \\
%Air Conditioning, water and services & 0.51 \\
%Power facilities &  0.55\\
%Cranes & 0.09\\
%Elevator & 0.24\\\hline
%TOTAL & 11.25 \\
%\hline\hline
%\end{tabular}
%\caption{Estimate of civil engineering costs to build a near detector hall at a 2~km distance from the J-PARCneutrino target.}\label{tab:cavitycosts}
%\end{table}

\subsection{Neutrino beam flux considerations}
\label{sec:fluxcon}

Neutrino fluxes are generated for three baselines using the same
simulation program~\cite{T2Kflux} as is used to simulate the
T2K neutrino flux. The baselines considered are 1000\,m, 1838\,m,
and 2036\,m from the point where the proton beam collides with the
target, with an $8\times8$\,m$^{2}$ plane for the 1000\,m baseline and
$12\times12$\,m$^{2}$ for the others, as well as 295\,km away assuming
the Hyper-K detector is located at the current position of Super-K.
For each baseline, the horn currents are assumed to be +320\,kA for a
$\nu_{\mu}$-enhanced beam and $-$320\,kA for a
$\overline{\nu}_{\mu}$-enhanced beam.  The neutrino flavours generated
for each horn current in the simulation are $\nu_{\mu}$,
$\overline{\nu}_{\mu}$, $\nu_{e}$, and $\overline{\nu}_{e}$.

\begin{figure}[htpb]
\begin{center}
\includegraphics[width=0.45\linewidth]{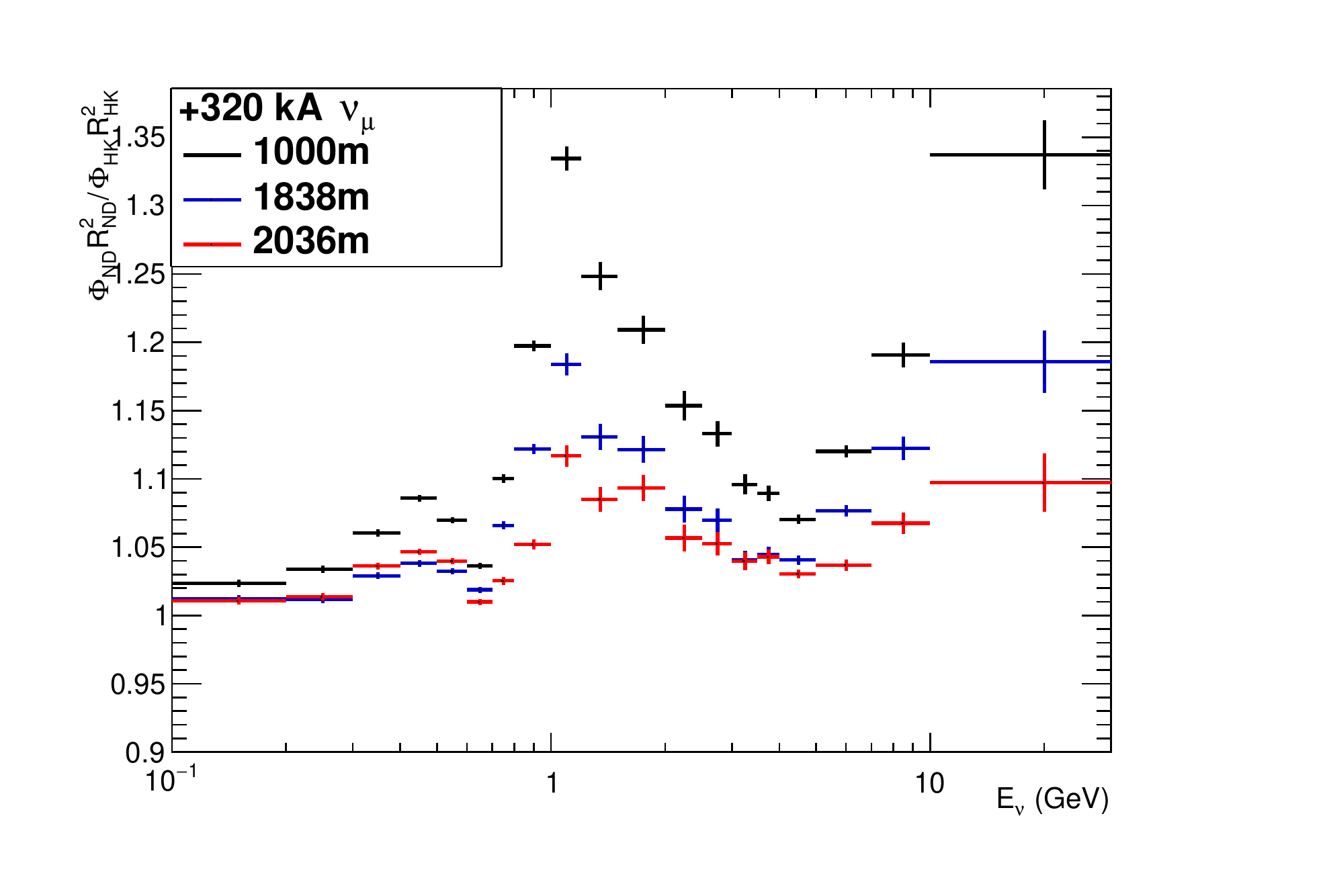}
\includegraphics[width=0.45\linewidth]{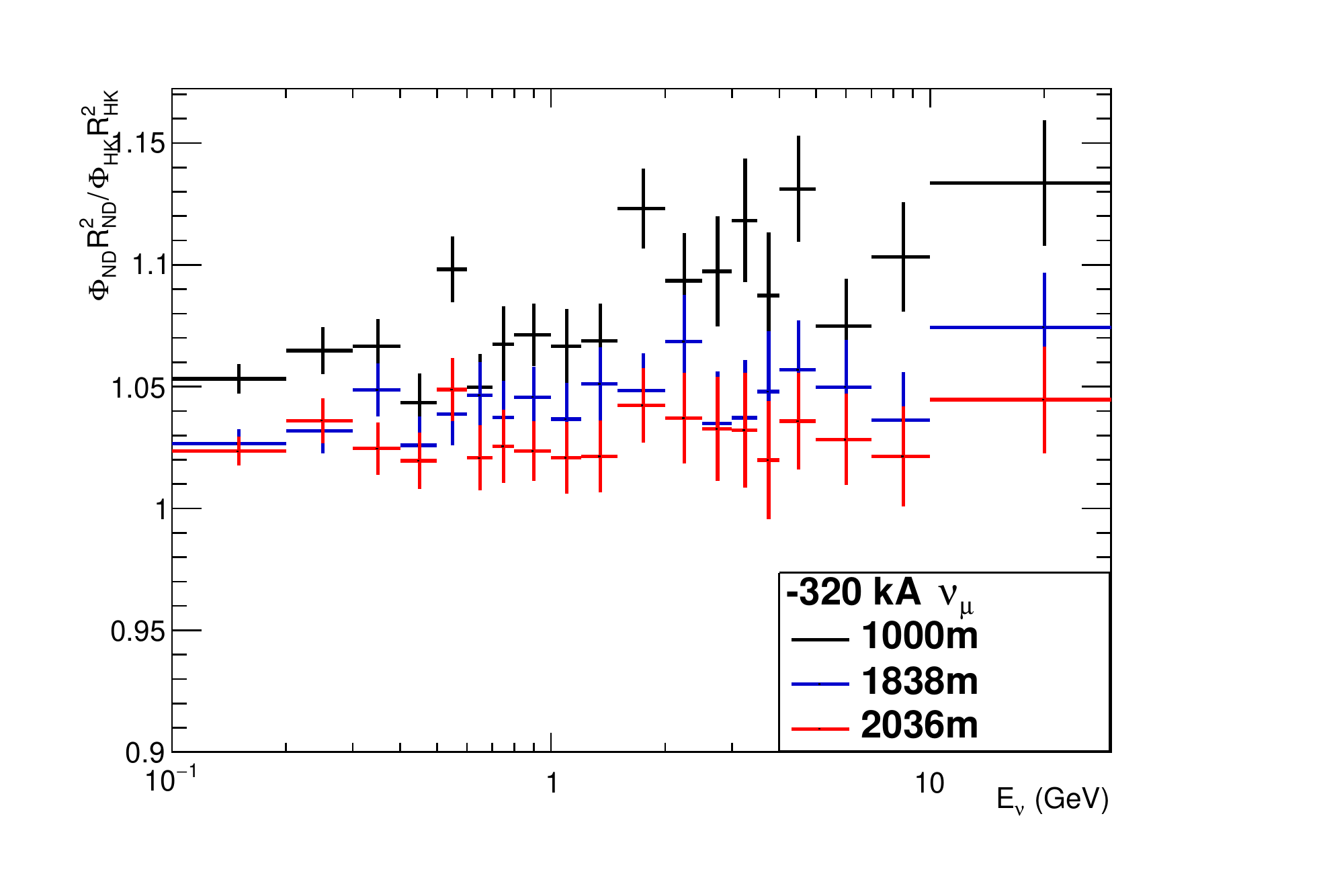}
\includegraphics[width=0.45\linewidth]{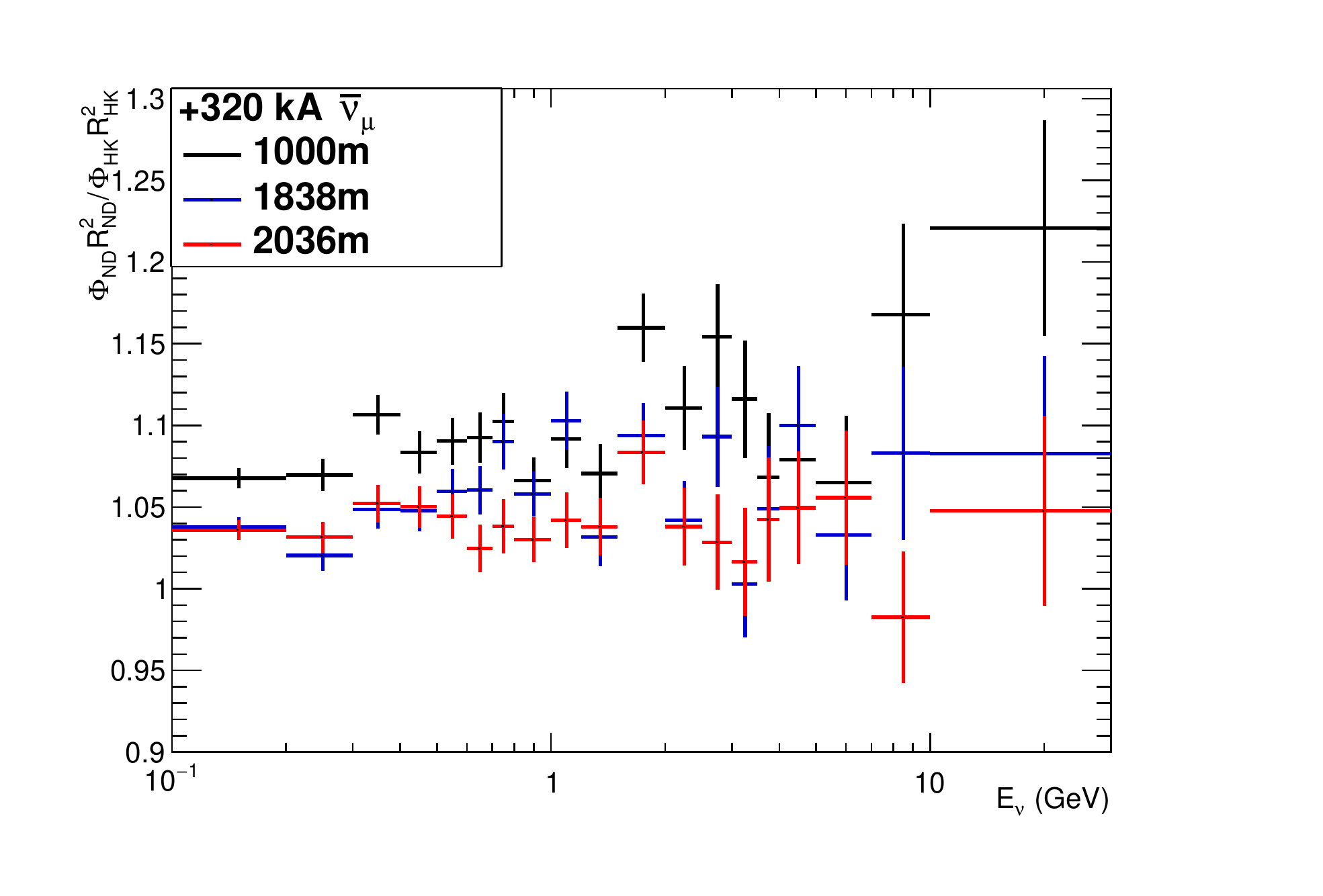}
\includegraphics[width=0.45\linewidth]{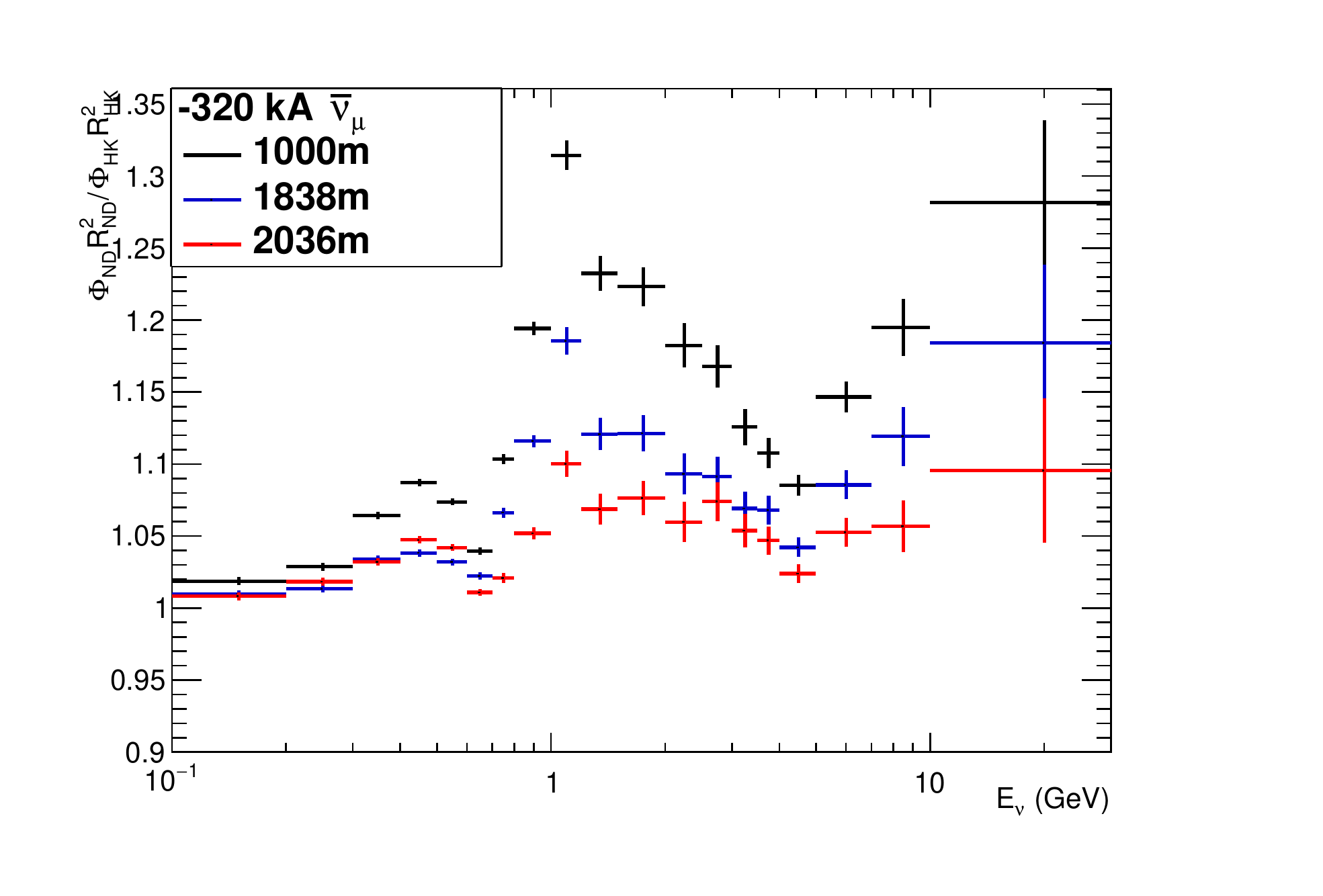}
\includegraphics[width=0.45\linewidth]{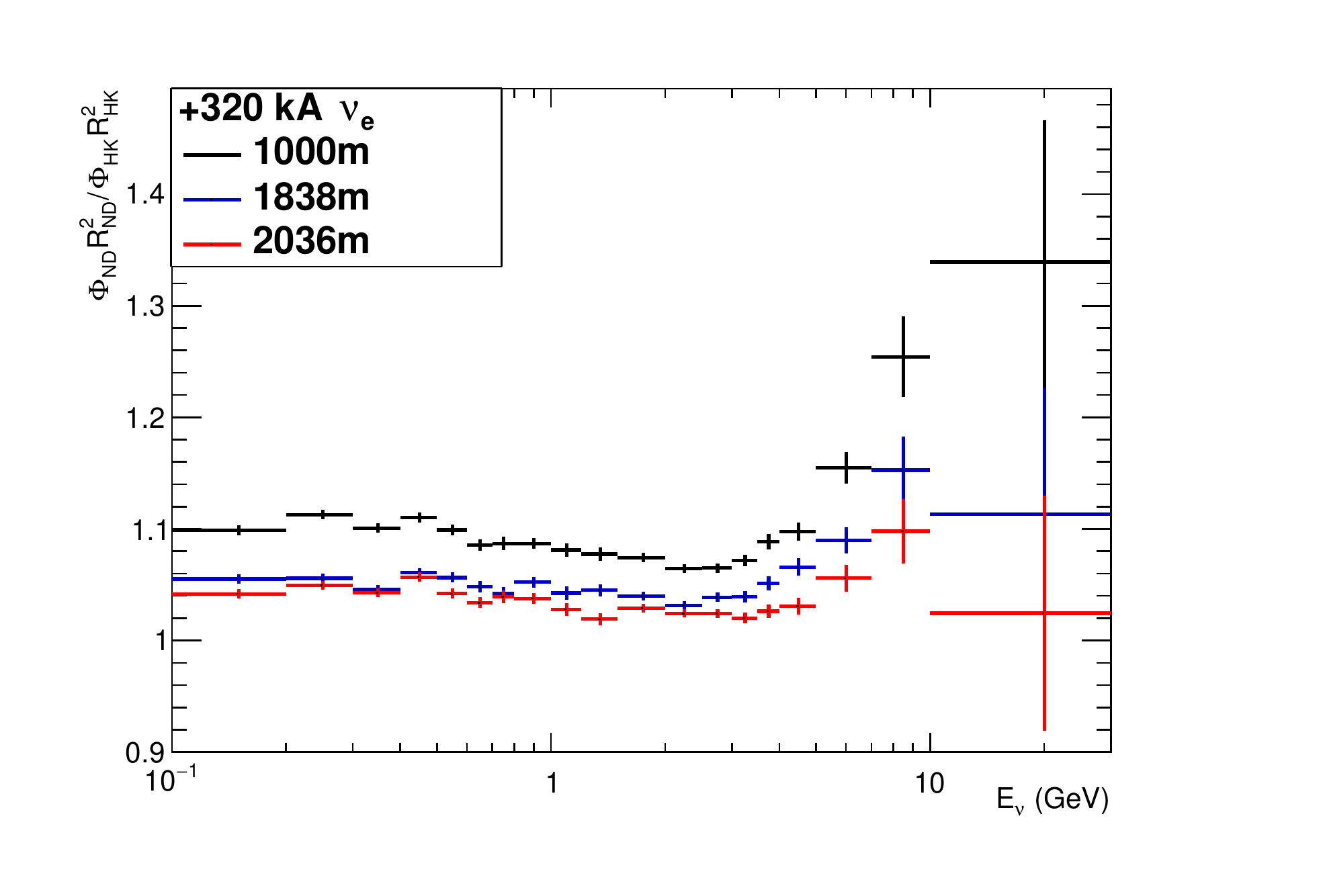}
\includegraphics[width=0.45\linewidth]{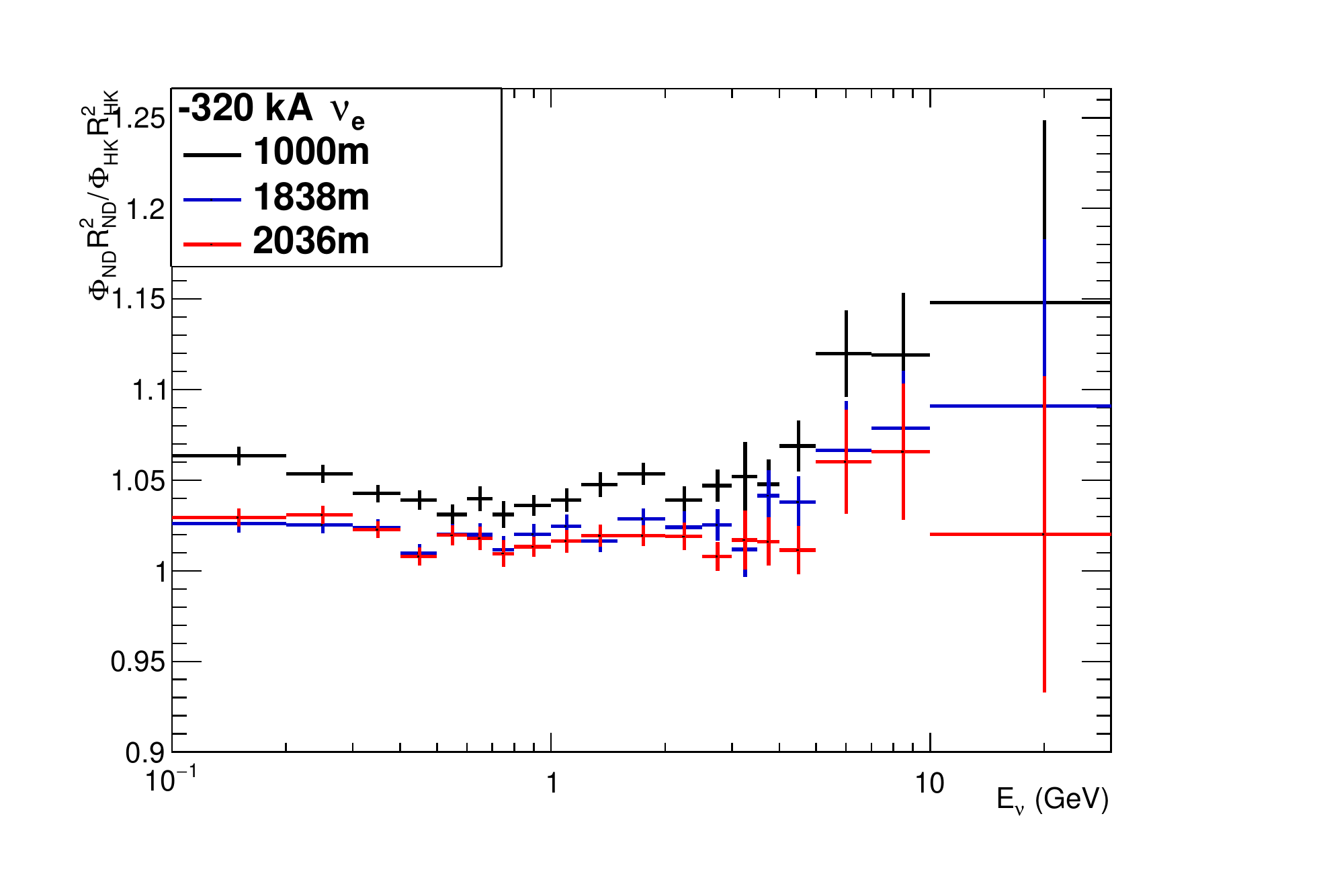}
\includegraphics[width=0.45\linewidth]{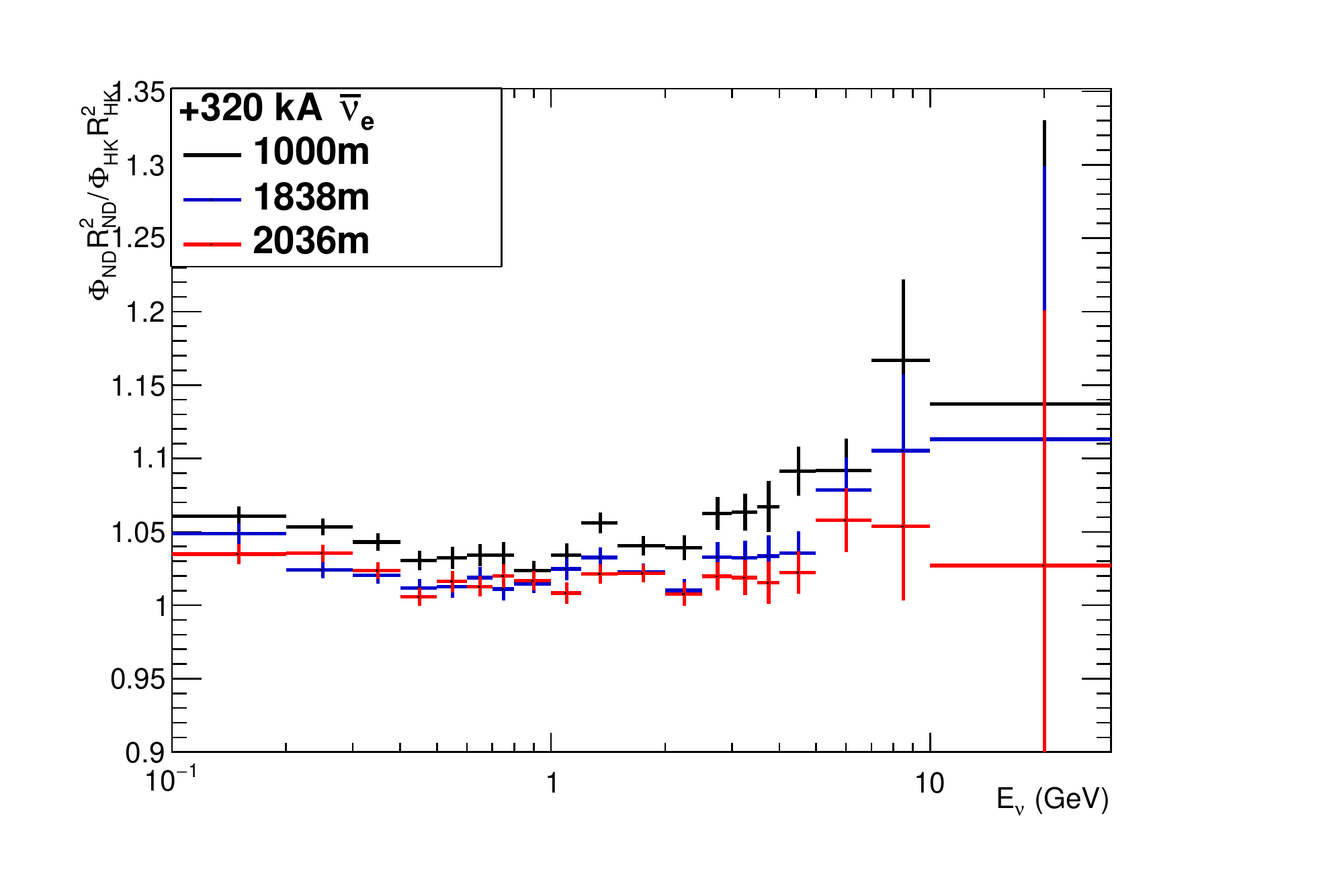}
\includegraphics[width=0.45\linewidth]{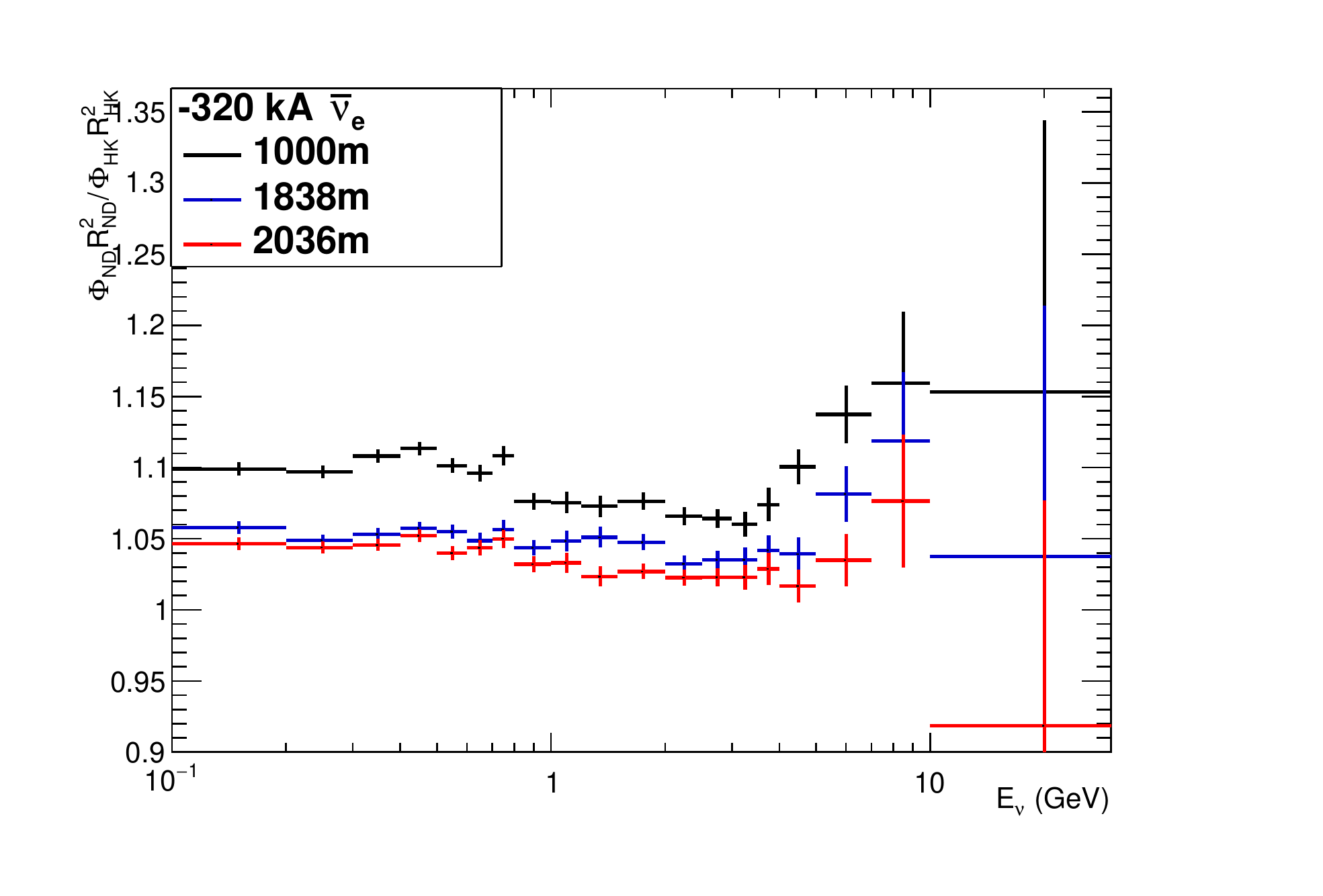}
\caption{
Flux ratios to the far detector normalised to the baselines for the
$\nu_\mu$-enhanced beam (left) and $\overline{\nu}_\mu$-enhanced beam
(right) for the different neutrino types. Errors are statistical
only. The peaks, as seen in the top left and second-from-the-top right
plots, are from the change in the ratio from the pion contribution to
the flux. This is related to both the horn current and the solid angle
subtended by the detectors with respect to the neutrino production
points and hence is more prominent at a closer baseline.}
%have to do with effects from the phase space of the meson decay points
%which allow for a neutrino to pass through the detector;
\label{fig:beambase}
\end{center}
\end{figure}

The J-PARC accelerator delivers beam in discrete ``spills'', each of
which consists of a number of narrow ``bunches''. Here we assume that
the time between spills is 1.3\,s and that each spill has a window of
1.3\,$\mu$s, contains 8 bunches, and delivers $3.8 \times10^{14}$
protons-on-target, equivalent to a 1.3\,MW beam. In addition, we
assume that each bunch has a 1$\sigma$ width of 25\,ns and that events
occur within $\pm 50$\,ns of the bunch~\cite{IshidaHKbeam}. The
normalisation used for the neutrino fluxes for baseline and tank
optimisation studies will be reported on a per-spill and per-bunch
basis.
%The
%number of protons-on-target (POT) and the timing of a spill for
%possible upgrades to the beamline at J-PARC are taken
%from~\cite{IshidaHKbeam}.

%For a spill, we use $2.2\times 10^{14}$\,POT, where a spill is assumed
%to have a window of 1.3\,$\mu$s and a repetition rate of 1.3\,s with 8
%bunches in a spill.  Additionally, we assume each bunch has a
%1$\sigma$ width of 25\,ns and that an event occurs within $\pm$50\,ns
%of the bunch.
The ratios of the flux at each baseline compared to Hyper-K for each
flavour, with the baselines taken into account, are shown in
Figure~\ref{fig:beambase}.  The baseline which has the least variation
in the flux relative to the Hyper-K far detector for all neutrino
flavours is at 2036\,m, followed closely by 1838\,m.  The neutrino
flux at 1000\,m shows greater variation due to the fact that the beam
is observed more as a line source than a point source as it would be
at the far detector.  This is independent of the neutrino flavour or
the horn current. From Figure~\ref{fig:beambase}, it is apparent that
a longer baseline is preferred for the physics goals outlined in
section~\ref{sec-physics}.

\begin{table}[tbp] \centering
\begin{tabular}{c c c}\hline\hline
Baseline (m) & $\nu$-int./kT/spill & $\nu$-int./kT/bunch\\
\hline
%1000 & 1.48 & 0.18 \\
%1838 & 0.42 & 0.05 \\
%2036 & 0.33 & 0.04 \\
1000 & 2.56 & 0.31 \\
1838 & 0.73 & 0.09 \\
2036 & 0.57 & 0.07 \\
\hline\hline
\end{tabular}
\caption{
Number of beam neutrino interactions on water per kiloton per spill or
per-bunch for the $\nu_{\mu}$-enhanced beam configuration at different
baselines.  The calculation uses NEUT 5.3.3 for the cross section
model.}\label{tab:evtperkT}
\end{table}

In addition, the number of beam neutrino interactions on water
per-kiloton per-spill was calculated using the fluxes described above,
and NEUT~\cite{Hayato:2009} 5.3.3.  These are shown in
Table~\ref{tab:evtperkT} for the $\nu_{\mu}$-enhanced, or forward horn
current (FHC), beam configuration, which has a larger total event rate
than the $\overline{\nu}_{\mu}$-enhanced, or reverse horn current
(RHC), beam configuration. Table~\ref{tab:evtperkT} shows that
baselines at a distance of roughly 2\,km have a lower, but sufficient,
event rate of roughly one event per spill, which also implies lower probabilities of pileup.

\section{Neutron capture with gadolinium}
\label{subsec-neutron}
As discussed in section~\ref{sec-expoverview} gadolinium doping is
used in TITUS to enhance the efficiency of neutron capture.

Even moderately energetic neutrons, with kinetic energies from tens to
hundreds of MeV, will quickly lose energy by collisions with free
protons and oxygen nuclei in water.  The cross sections for these
capture reactions are 0.33\,barns and 0.19\,millibarns, respectively,
so to first approximation every thermal neutron is captured on a free
proton via the reaction $n + p \rightarrow d + \gamma$.  The resulting
gamma has an energy of 2.2\,MeV~\cite{Ncapturewater} and makes very
little detectable light since any Compton scattered electron is close
to the Cherenkov threshold.  The entire sequence from liberation to
capture takes around 200\,$\mu$s, with only a very small dependence
(plus or minus a few $\mu$s) on initial neutron energy.

The situation is much improved by adding a water-soluble gadolinium
compound, gadolinium chloride, GdCl$_3$, or the less reactive though
also less soluble gadolinium sulphate, Gd$_2$(SO$_4$)$_3$, to the
water. Naturally occurring gadolinium has a neutron capture cross
section of 49700\,barns, and these captures produce an $\sim$8.0\,MeV
gamma cascade.  The visible energy will be around 4--5\,MeV in a WC
detector.  Due to the larger cross section of gadolinium, adding 0.2\%
by weight (about 0.1\% Gd) of one of these compounds is sufficient to
cause 90\% of the neutrons to capture visibly on gadolinium.
%instead of invisibly on hydrogen.
Following the addition of gadolinium, the time between neutron
liberation and capture is reduced by an order of magnitude to around
20~$\mu$s, greatly suppressing accidental backgrounds.

The gadolinium neutron capture is an established technique for low
energy physics such as reactor oscillation
experiments~\cite{Cao:2013tsa}. The plan with TITUS is to extend this
technique to physics around 1\,GeV.
%Past, and current neutrino oscillation experiments focussed on
%understanding on the lepton kinematics.  However, precise neutrino
%energy reconstruction require an understanding of the hadronic
%system~\cite{Martini_erec1,Martini_erec2,Nieves_erec,Mosel_erec}.
The main motivation is that the nucleon multiplicity provides
information about the primary interaction.  This can
be seen from the following reactions:

$\bullet$ $\nu_{\mu}$CCQE,     ~$\sim$    0 neutron,~$\nu_\mu+n\rightarrow \mu^-+p$\\
$\bullet$ $\nu_{\mu}$CC-npnh,    ~$\sim$ 0.3 neutron,~$\nu_{\mu}+(n+p)\rightarrow \mu^-+p+p~,~\nu_{\mu}+(n+n)\rightarrow \mu^-+n+p$\\
%$\bullet$ $\nu_{\mu}$NC,       ~$\sim$  0.5 neutron,~$\nu_{\mu}+n\rightarrow \nu_{\mu}+n~,~\nu_{\mu}+p\rightarrow \nu_{\mu}+p$\\
$\bullet$ $\bar\nu_{\mu}$CCQE, ~$\sim$    1 neutron,~$\bar\nu_{\mu}+p\rightarrow \mu^++n$\\
$\bullet$ $\bar\nu_{\mu}$CC-npnh,~$\sim$ 1.7 neutron,~$\bar\nu_{\mu}+(n+p)\rightarrow \mu^++n+n~,~\bar\nu_{\mu}+(p+p)\rightarrow \mu^++p+n$

where the average number of neutrons is obtained using NEUT 5.3.3.  We
assume that the correlated nucleon pairs for the npnh interactions are
dominated by neutron-proton pairs.  Na\"{i}vely, we expect a different
number of neutrons from different primary interactions, either
neutrino or antineutrino, hitting single nucleon or correlated
nucleons. This could be modified event-by-event by nuclear effects
such as re-scattering, charge exchange, and absorption in the nuclear
medium. Nonetheless, it is expected that the primary interaction
information is statistically conserved\cite{lalakulich2012}.

\begin{figure}[htb]
\centering
\begin{tabular}{cc}
\includegraphics[scale=0.4]{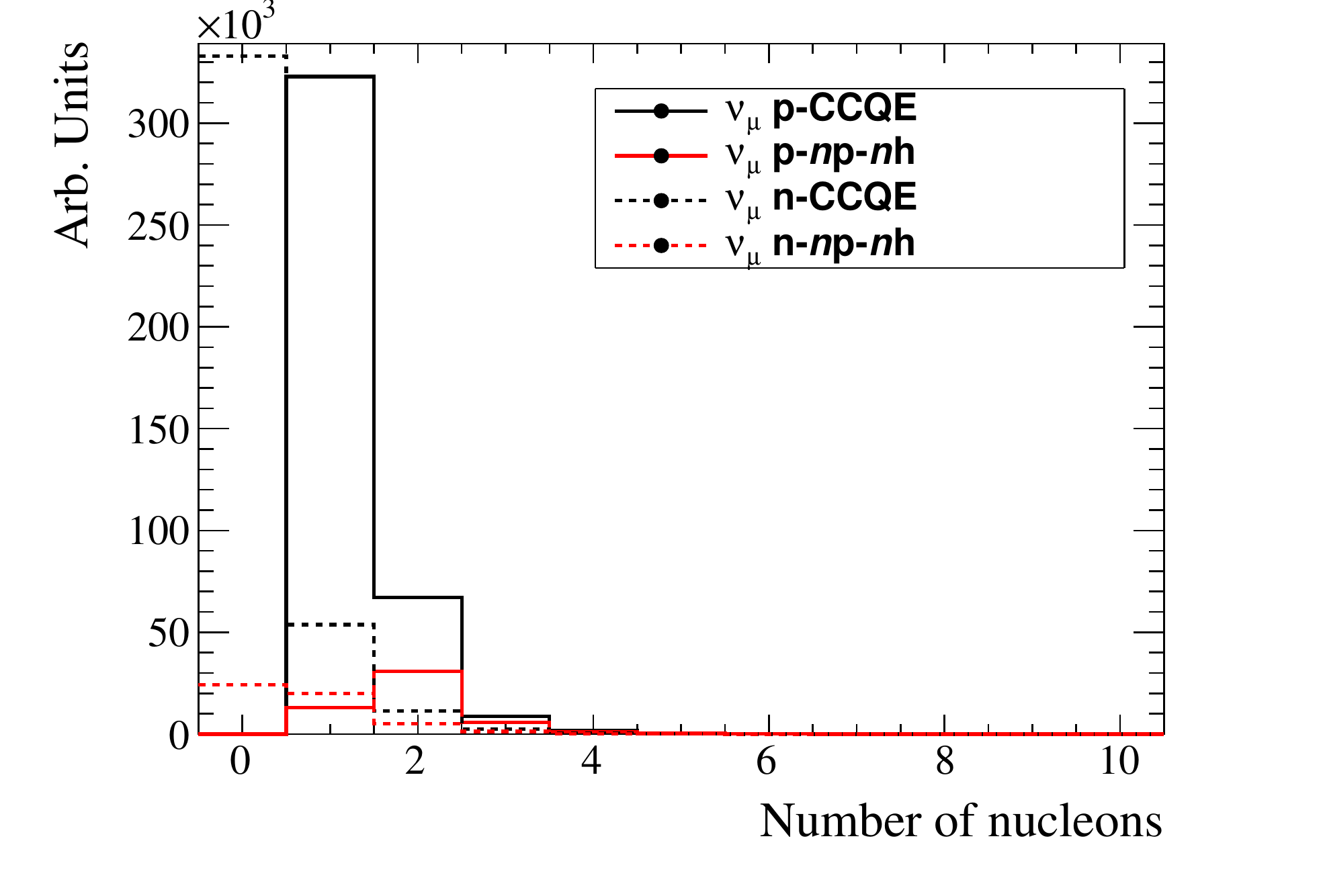}
\includegraphics[scale=0.4]{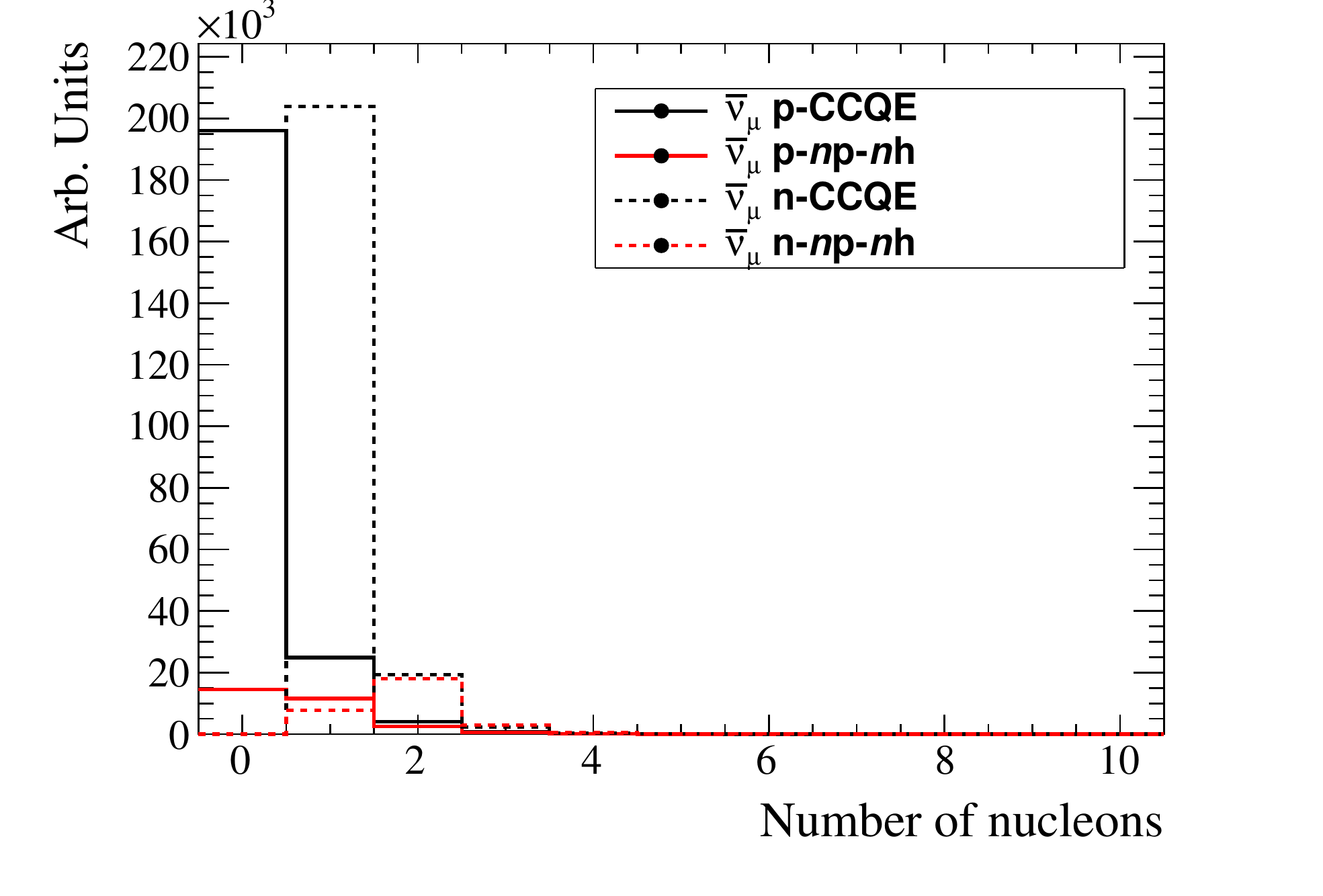}\\
\includegraphics[scale=0.4]{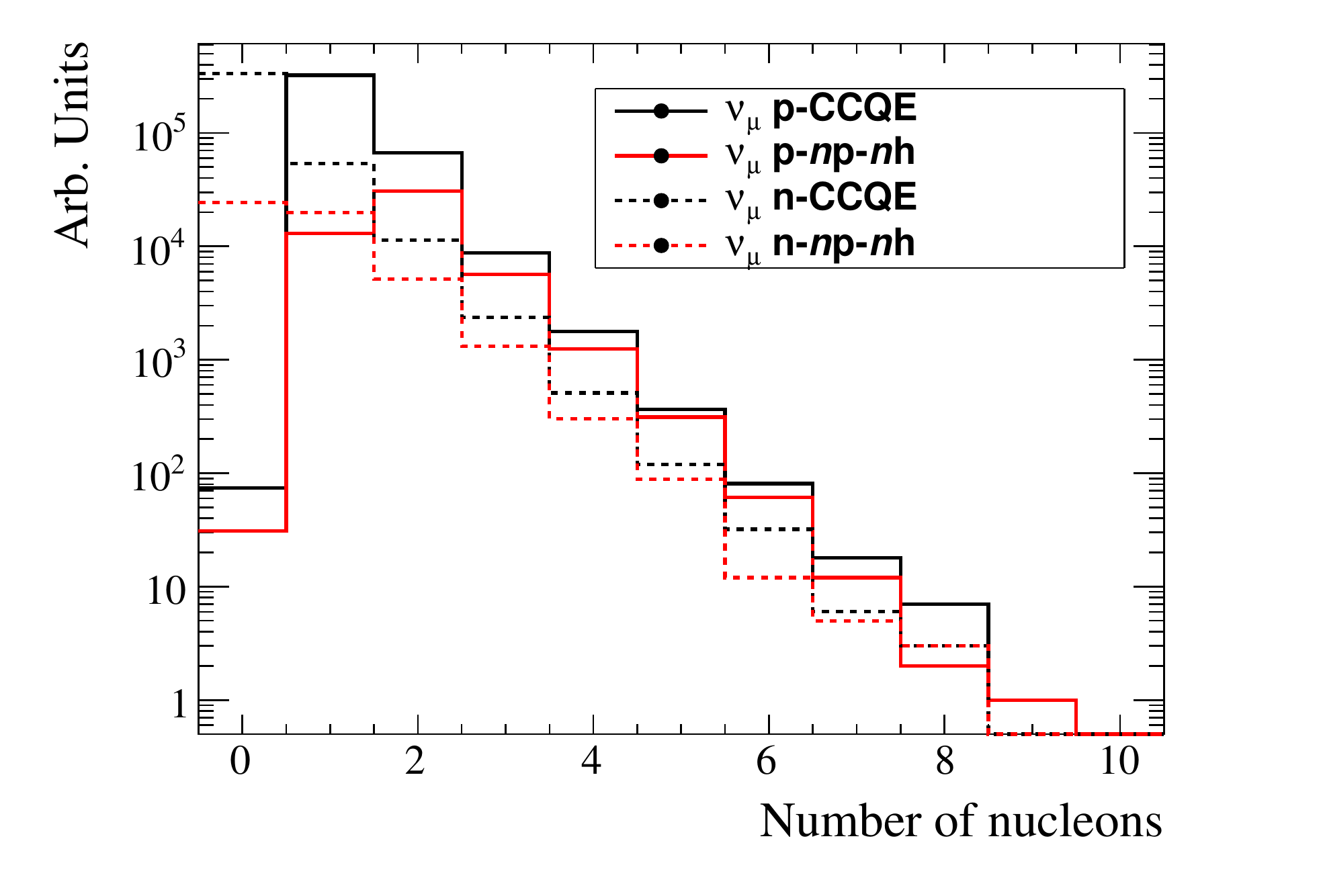}
\includegraphics[scale=0.4]{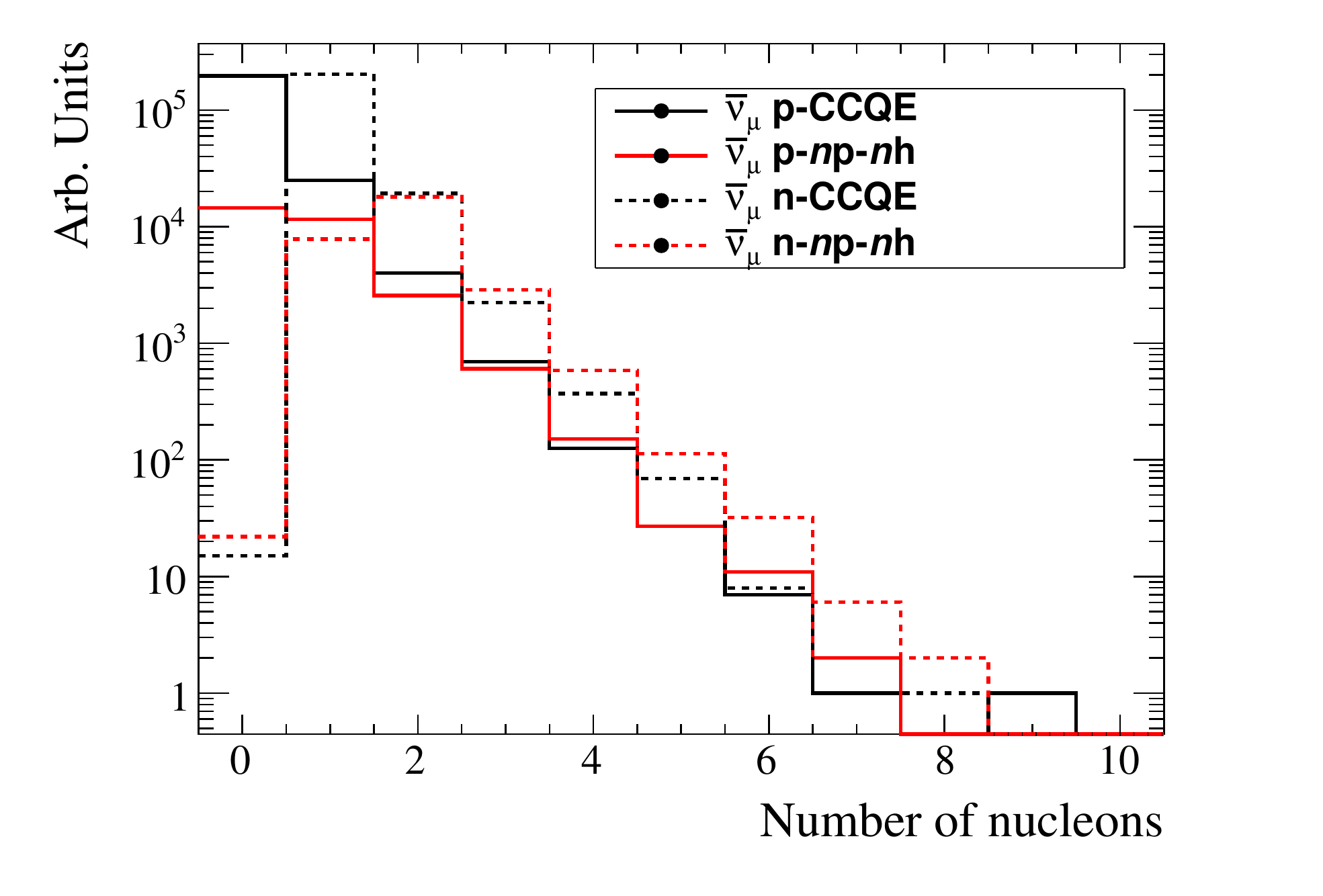}\\
\end{tabular}
\caption{\label{fig:ncap} Number of protons and neutrons escaping
  from the target nuclei (water) with the CCQE or CC-npnh
  interactions after final state interactions.  
  The top left plot is for muon neutrino interactions and
  the top right plot is for muon antineutrino interactions.  
  The same information is shown on the bottom plots, but on a log scale.
  }
\end{figure}

Figure~\ref{fig:ncap} shows the number of outgoing neutrons for
different interaction channels that we expect to dominate for a
particular beam horn current configuration.  We used NEUT 5.3.3 to
simulate the neutrino/antineutrino interactions with a water target.
NEUT simulates the interactions with correlated nucleon pairs, on top
of genuine CCQE interactions.  In the figure, the different final
states for $\nu$CCQE, $\bar\nu$CCQE, $\nu$CC-npnh, and
$\bar\nu$CC-npnh display different neutron multiplicity spectra,
indicating that this could be a new tool to identify the primary
interaction in cases where standard water Cherenkov detectors cannot.
%This is the power of neutron m%multiplicity measurement.
%
%Finally, the Gd$_2$(SO$_4$)$_3$ is paramagnetic. Due to the presence
%of the magnetised MRD, a correction due to the non-zero susceptibility
%of the magnetic field is applied of the order a less then 1\% needs to
%be applied.

A comparable technology, the proton multiplicity measurement, is also
under development by LArTPC~\cite{ArgoNeuT_Hammer}.  These two
nucleon-counting techniques are complementary, and new information
about the outgoing nucleons will improve the performance of neutrino
energy reconstruction, usually focused on the lepton kinematics,
necessary to measure the CP violating Dirac phase in the lepton
sector.

\section{External background and pile-up}
\label{sec-background}
The external background consists of particles originating from sources
other than neutrino interactions in the detector. They can be
misidentified as a signal due to a reconstruction error or
re-interaction inside the detector volume. We consider two possible
sources of background: the interactions of beam neutrinos in the
surroundings of the detector, which coincide with the beam window, and
accidental cosmic rays. This section presents the methods used to
estimate the background rate, leading to the optimisation of the
detector design.

\subsection{Sand interactions}
Neutrinos from the beam interact not only in the detector, but also in
the surrounding sand and pit structures. The particles coming from
these interactions that enter the detector cannot be removed by
cutting on the bunch time, because they produce signal in the same
time as the interactions in the detector.  These interactions will be
referred to as ``sand interactions'' below.

Neutral particles coming from the sand interactions can re-interact
inside the detector producing a false signal in the fiducial volume,
whilst charged particles may be mis-reconstructed as starting inside
the detector. Additionally, the sand events can pile up with the
interactions in the detector. Particles from these interactions
therefore lower the selection efficiency and purity.

A dedicated simulation allows us to predict the rate of these
particles entering the detector. The simulation is performed in two
steps.

First, the neutrino interactions are simulated with the NEUT generator
(version 5.3.3). The neutrinos from the beam are allowed to
interact in a rectangular volume filled with sand and positioned at
the distance of 2036\,m with respect to the target.  The size of the
volume is 100\,m (L) $\times$ 40\,m (W) $\times$ 40\,m~(H). There are
no measurements concerning the chemical composition and density of the
sand at the candidate TITUS sites, so it is assumed that the sand is
pure SiO$_2$ with a density equal to 2.15\,g\,cm$^{-3}$.

The next step is the propagation of particles produced in the neutrino
interactions using the Geant4 package (version
v9r4p04n00)~\cite{GEANT4}. The particles which reached the surface of
a box big enough to encapsulate the water tank and the proposed MRD
(23\,m (L) $\times$ 13.86\,m (W) $\times$ 12\,m~(H), see section~\ref{sec:mrddesign}) are saved for
further propagation through the detector setup. The box is placed
centrally inside the sand volume.

The primary particles are tracked through the sand until they enter
the detector box, stop, decay or exit the geometry setup. Secondary
particles produced in re-interactions are tracked as well, in the same
way. However, not all the primary or secondary particles are
propagated: low-energy particles produced at the distance of about
10\,m from the detector have no chance of entering it and are
therefore skipped to reduce the CPU time needed for the
simulation. The cuts were tuned using a smaller sample to ensure that
the final number of particles entering the detector box is not
affected.

The numbers of particles entering the box are summarised in
Table~\ref{table:sand} for a generated sample of $2.5 \times
10^{20}$~POT. Not all of them will enter the tank or MRD and produce a
signal because they will not produce enough Cherenkov light to be
detected or their direction is too close to the detector axis.  Note
that the total contribution of these particles will be given in the
detector optimisation studies in section~\ref{subsec-tank}.
\begin{table}[htb]
\begin{center}
\begin{tabular}{l|c|c}
\hline\hline
 & Number of particles & Rate per spill\\
\hline
muons    & 375 615   & 0.33 \\
% muons - & 355 947
% muons + & 19 668
neutrons & 9 179 373 & 8.08 \\
% +2 antineutrons
protons  & 45 885    & 0.04 \\
% +1 antiproton
charged pions & 31 444 & 0.03 \\
% pi + (211) 21 225
% pi - (-211) 10 219
photons  & 3 046 799 & 2.68 \\
electrons and positrons & 214 878 & 0.19 \\
% e- & 162 810
% e+ & 52 068
other    & 2 233 & 0.002 \\
\hline\hline
\end{tabular}
\caption{Number of particles of various types produced in the sand
  interactions and entering the detector box surface (see text for
  explanation). The second and third columns show numbers for the
  whole generated sample of $2.5 \times 10^{20}$~POT, and the rate per
  spill of $2.2 \times 10^{14}$~POT, respectively.}
\label{table:sand}
\end{center}
\end{table}

\begin{figure}[htb]
\begin{center}
\includegraphics[width=0.49\linewidth]{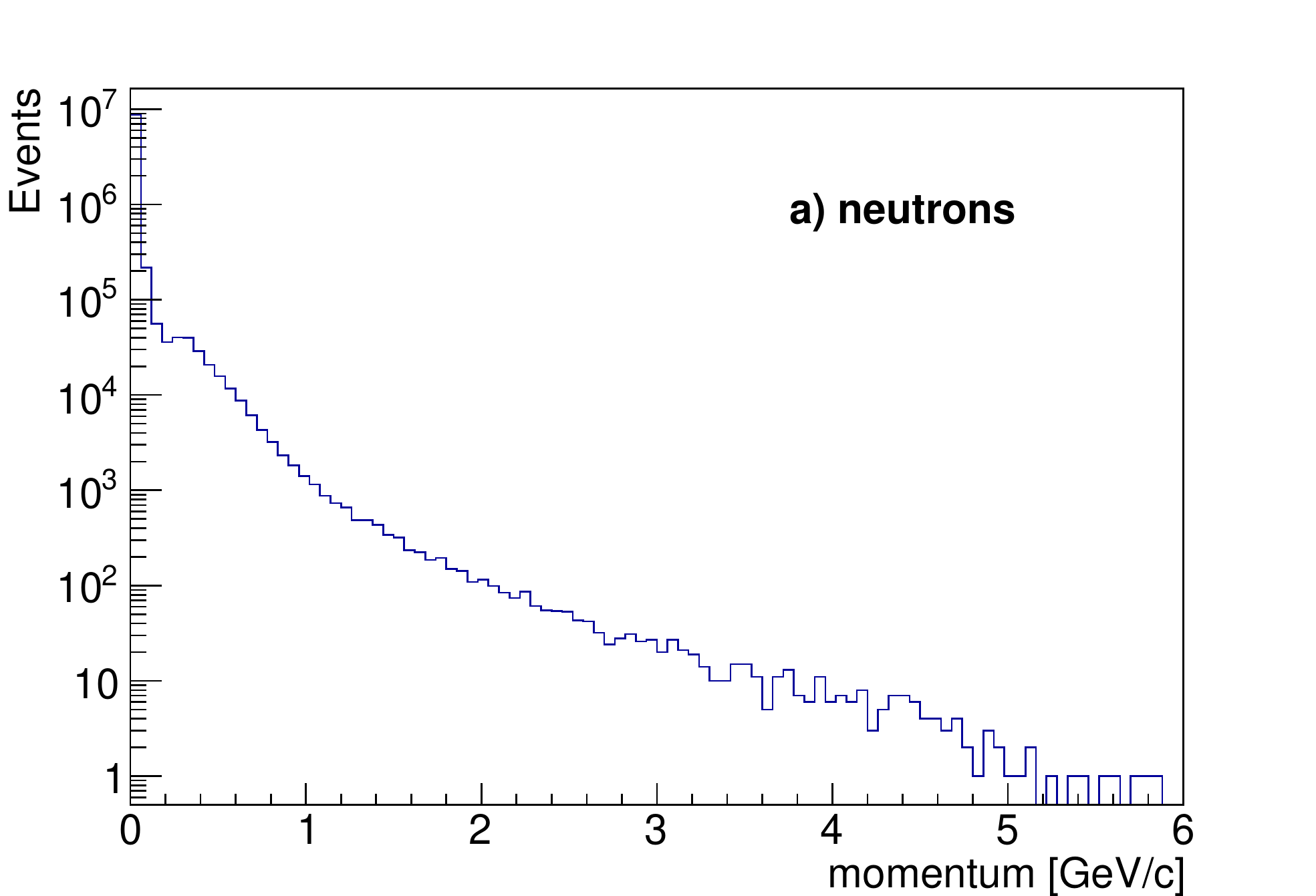}
\includegraphics[width=0.49\linewidth]{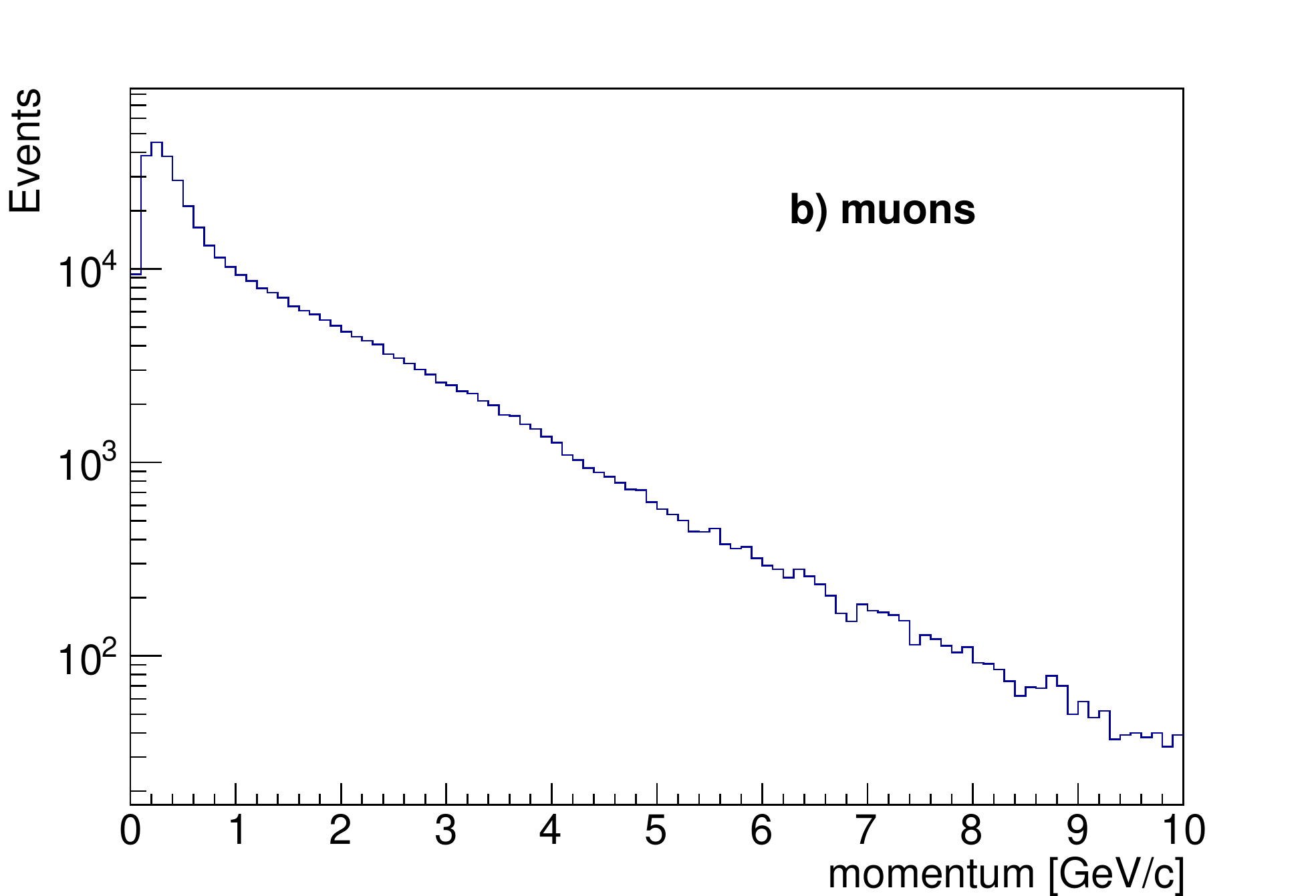}
\includegraphics[width=0.49\linewidth]{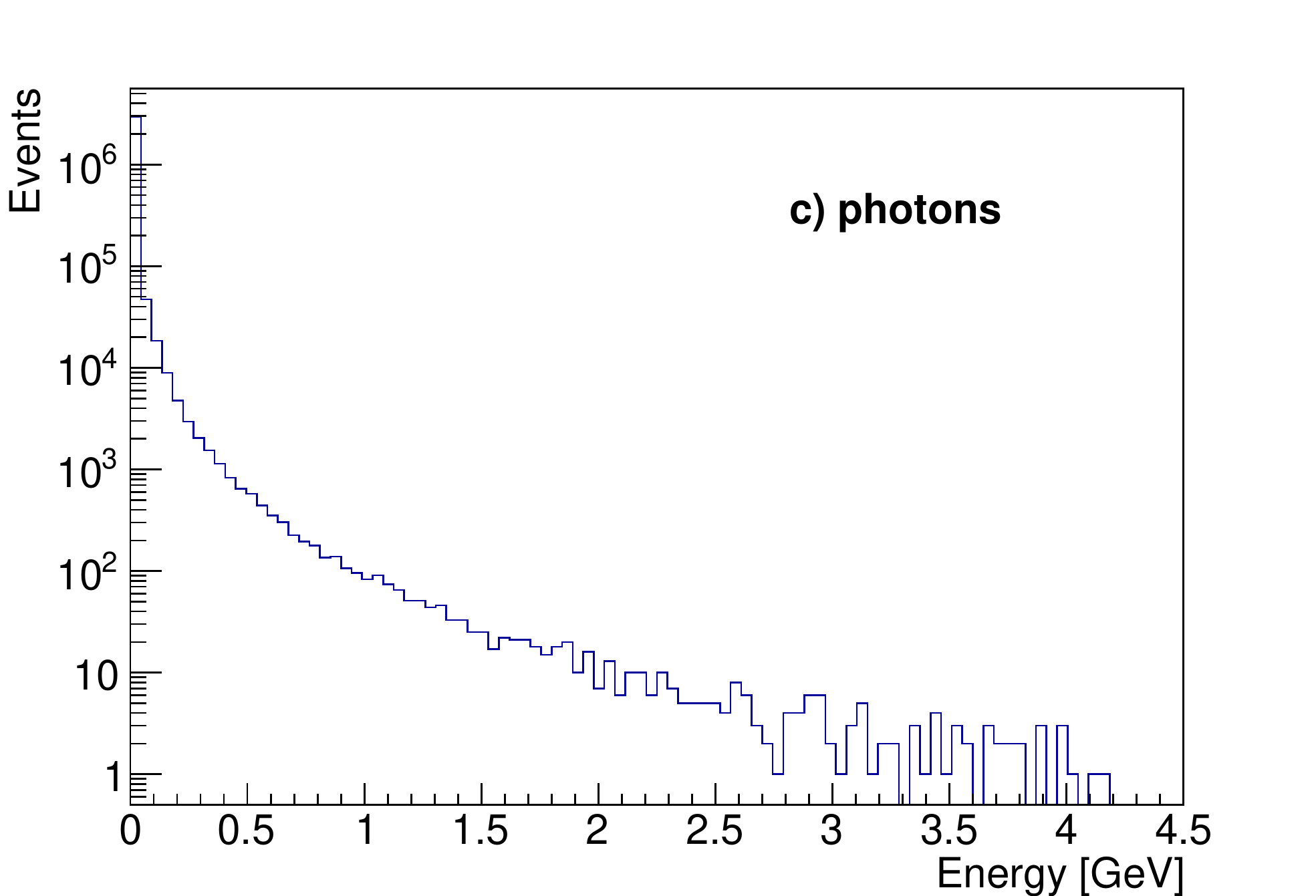}
\includegraphics[width=0.49\linewidth]{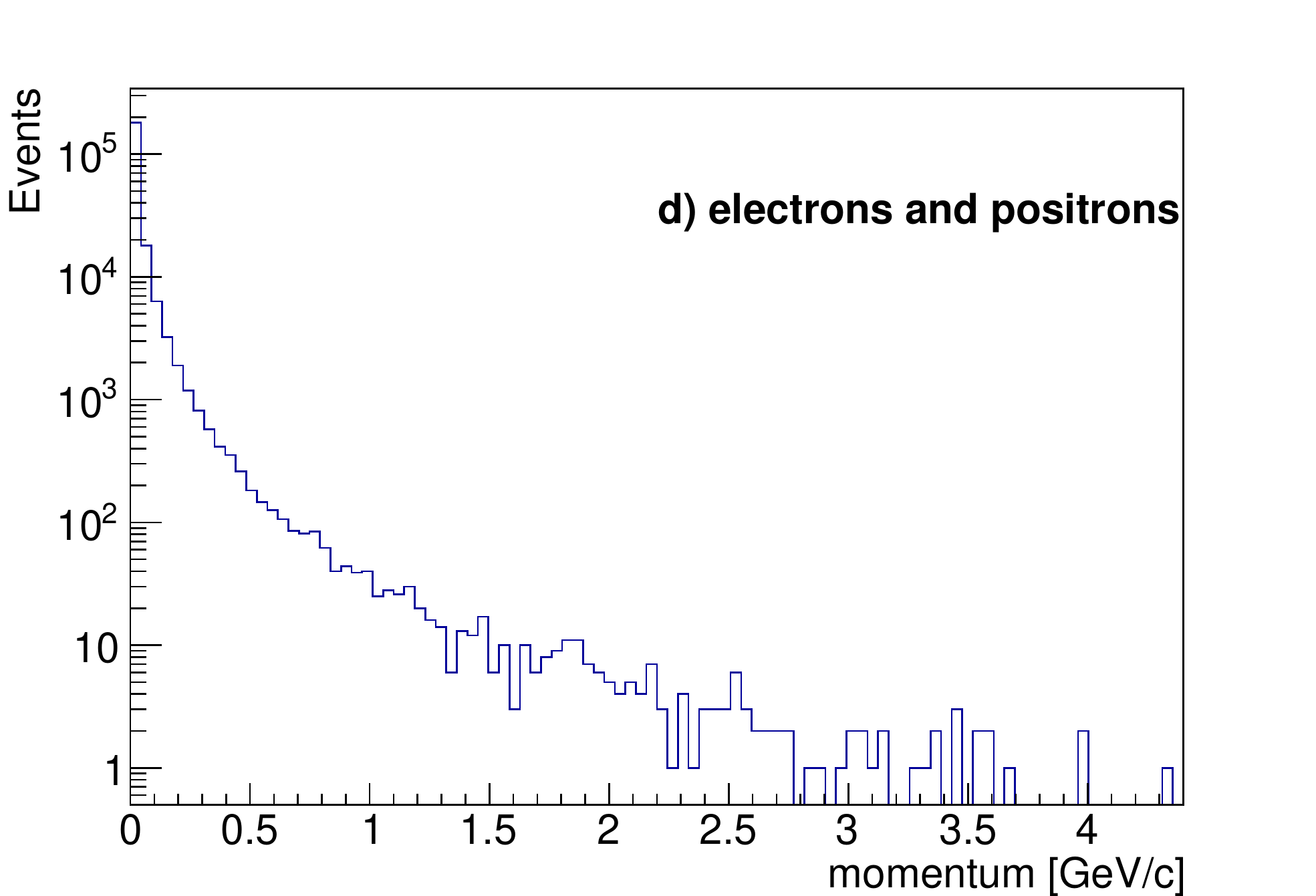}
\caption{Momentum distributions for the a) neutron, b) muons, c)
  photons and d) electrons and positrons created in the neutrino
  interactions in the sand and entering the detector box for a whole
  generated sample of $2.5 \times 10^{20}$~POT.}
\label{fig:sandmomentum}
\end{center}
\end{figure}

The most numerous particles are neutrons, which are mostly slow (about
75\% have a momentum below 20\,MeV/c). The momentum distribution for
neutrons is shown in Figure~\ref{fig:sandmomentum}~a).
% <0.1GeV:  8 862 194 = 96.5%
% <0.01GeV: 2 187 311 = 23.8%
% <0.02GeV: 6 816 116 = 74.3%

Muons are essentially all produced directly in the neutrino
interactions and as a result their energy is quite high. The momentum
distribution, as shown in Figure~\ref{fig:sandmomentum}~b), is peaked
at about 200\,MeV/c and 97\% of muons have a momentum higher than that
value.

Photons, electrons and positrons arriving to the detector are mostly
(about 98\%) produced in electromagnetic cascades produced by
particles crossing the sand. There is also a small fraction of primary
photons emitted from the target nuclei and electrons produced by
interactions of electron neutrinos. The energy distribution of photons
and the momentum distribution of electrons and positrons are shown in
Figure~\ref{fig:sandmomentum}~c) and d), respectively. The
distributions are dominated by low energy particles: 90\% of photons
and 50\% of $e^\pm$ have a~momentum below 10\,MeV/c.

% gamma <0.2GeV: 3 031 626 = 99.5%
%       <0.1GeV: 3 007 286 = 98.7%
%       <0.01GeV: 2 748 826 = 90%
% e+e- <0.2GeV: 209 120 = 97%
%      <0.1GeV: 200 698 = 93.4%
%      <0.02GeV: 145 627 = 67.8%

The particles that reach the surface of the box can then be further
tracked by the detector simulation tool. The results are used in the
tank optimisation and veto studies, described in
section~\ref{subsec-tank}.

\subsection{Cosmic background sources}

Cosmic-ray muons and muon-induced spallation neutrons constitute
additional sources of background. Both may cause events that coincide
with the beam window, producing a non-beam- induced background. For
the tank optimisation studies, we use the results from C.~Galbiati and
J.F.~Beacom~\cite{Galbiati} at sea level, which are reported in
Table~\ref{tab:cosmics}.  The cosmic ray muon background is reported
in Table~\ref{tab:cosmics} as a flux (events per square meter per
hour), whereas the neutron background scales with the detector water
mass in kilotons (kT). Once these backgrounds are scaled to the
1.3\,$\mu$s window of the neutrino beam, the cosmic-induced background
contributes a negligible amount to the total event rate regardless of
detector orientation or shape.
\begin{table}[htb] \centering
\begin{tabular}{c c c}\hline\hline
Depth (m) & $\Phi_{\mu}$ & \# of neutrons \\
 &  ($\mu $m$^{-2}$day$^{-1}$) & (events kT$^{-1}$ day$^{-1}$) \\
\hline
0 & $1.44\times10^{6}$ & $7.2\times10^{6}$ \\
\hline\hline
\end{tabular}
\caption{ Cosmic muon flux, $\Phi_{\mu}$, and number of spallation
  neutron background events at sea level. Taken
  from~\cite{Galbiati}.}\label{tab:cosmics}
\end{table}

\section{Detector design}
\label{sec-detector}
In this section we describe the different components of the
detector. A detailed optimisation has been performed for the tank and
the MRD. Realistic solutions for a WC detector with gadolinium doping
are proposed for the electronics, DAQ, photosensors and calibration.

\subsection{Tank}
\label{subsec-tank}
Detailed studies have been performed to optimise the size of the inner
detector (ID), the tank orientation, the baseline, and the addition of
an outer detector (OD), while taking into account the physics goals of
the detector.

To optimise the tank size, we study both $\nu_{\mu}$-enhanced and
$\overline{\nu}_{\mu}$-enhanced beam CC interactions at 2036\,m, the
preferred detector baseline, since these interactions provide the
signal sample for the oscillation studies at Hyper-K. For each
generated CC interaction, a vertex is randomly thrown in the
tank %five times for a given radius and length to determine if the
muon from the interaction would generally be contained in the ID.
%This was done due to constraints of computing time for the event generation and to be able to better test the high momentum tail of the sample generated.  
From each point thrown in the detector, the distance to the tank wall along
the direction of travel for the outgoing lepton is calculated as well
as the energy loss assuming a constant loss of 1.981\,MeV/cm~\cite{Agashe:2014kda}.  The
muon is considered contained in the ID if it has a non-positive
kinetic energy when propagated to the tank wall.

\begin{figure}[htb]
\begin{center}
\includegraphics[width=0.45\linewidth]{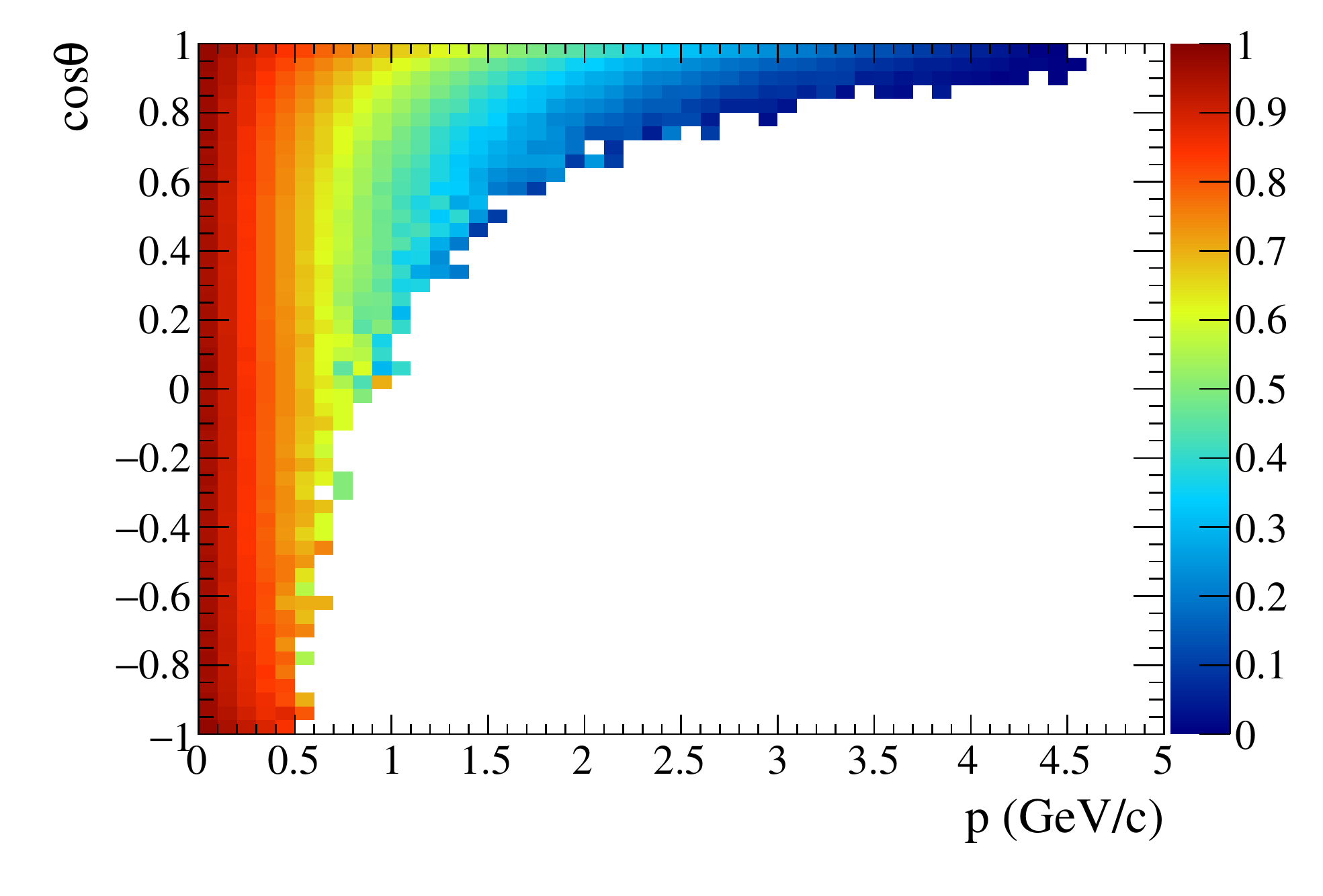}
\includegraphics[width=0.45\linewidth]{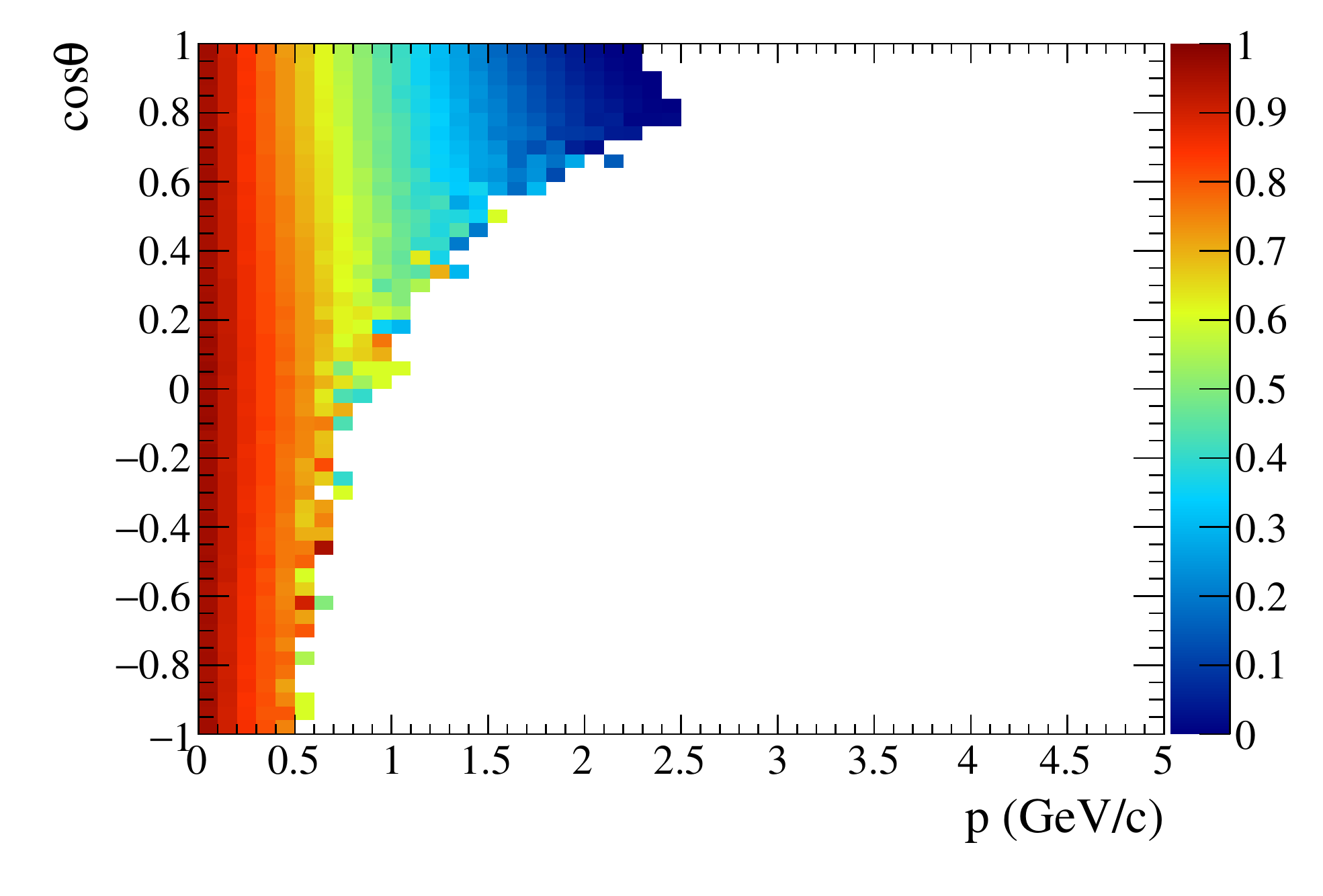}
\includegraphics[width=0.45\linewidth]{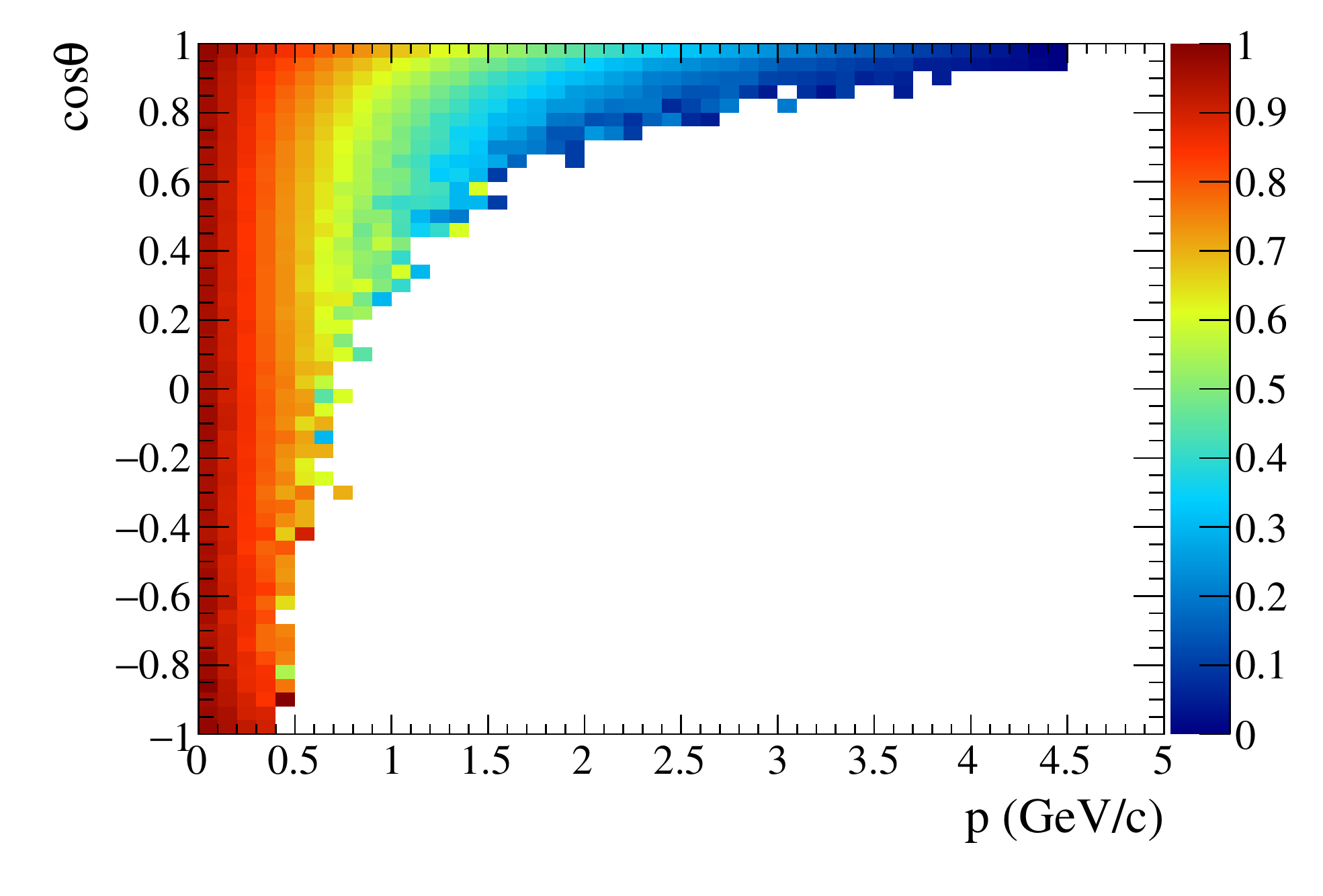}
\includegraphics[width=0.45\linewidth]{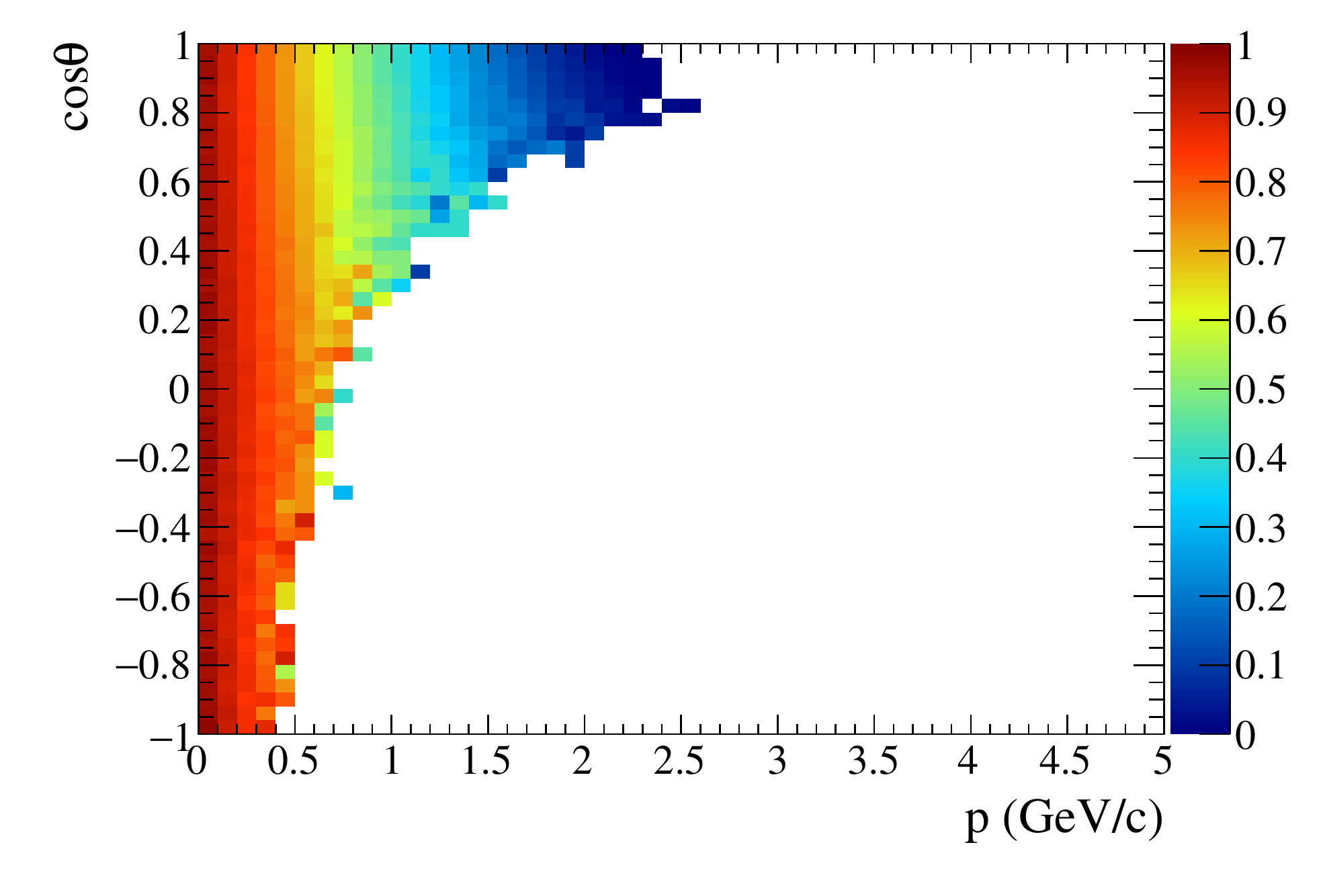}
\caption{
Fraction of muons that are fully contained in the nominal TITUS ID as
a function of the muon momentum and angle with respect to the beam.
The top(bottom) row shows $\nu_{\mu}$($\overline{\nu}_{\mu}$)
interactions.  The left column has the z-axis of the cylindrical tank
along the beam direction while the right has a vertical z-axis.  }
\label{fig:murange}
\end{center}
\end{figure}

Using the total event rate calculated from
Table~\ref{tab:baselineevtrate}, tanks of different radius and length
have been studied, and it is concluded that an ID of radius of 5.5\,m
and length of 22\,m, corresponding to a water mass of 2.1\,kT, has the
desired performance.  For an ID of this size, the overall fraction of
muons contained as a function of their momentum and angle with respect
to the beam that are contained is greater for a tank oriented along
the beam than perpendicular to it, as shown in Table~\ref{tab:murange}
and Figure~\ref{fig:murange}.
%Combined with the total event rate calculated from
%Table~\ref{tab:baselineevtrate} of a tank at 2036\,m for different
%radii and lengths, it is concluded that an ID of radius of 5.5\,m and
%length of 22\,m, which has a volume of 2\,kT, with the z-axis along
%the beam direction had the desired performance and would allow for
%better sampling of the high energy tail of the neutrino beam, which
%can be useful for higher energy atmospheric neutrino interaction
%constraints as well.  For an ID of this size, the overall fraction of
%muons contained as a function of their momentum and angle with respect
%to the beam that are contained is greater for a tank oriented along
%the beam than perpendicular to it, as shown in Table~\ref{tab:murange}
%and Figure~\ref{fig:murange}.  In addition, the probability of having
%more than one neutrino interaction per spill occurring is lower at a
%baseline of roughly 2\,km for any horn current; this aids our ability
%to match any neutron captures observed in the ID with the neutrino
%interaction that produced it.
The 2.1\,kT ID gives a similar number of events at 2036\,m and
1838\,m, as seen in the last column in Table~\ref{tab:h2onuint}.
This shows that at a baseline of $\sim$2\,km, the size of the TITUS ID
does not need to undergo a re-optimisation.  For a baseline of 1\,km,
due to the higher probability of event pileup, the TITUS ID volume
should be reduced by roughly a factor of four, unless we introduce the
Outer Detector, as we will see later in this section.

%In order to be able to match a neutrino interaction and a
%captured neutron from the same interaction in a Gd-doped detector, it
%is necessary to have roughly one interaction per spill occurring within the
%TITUS ID with minimal penetration of external particles. This would
%lead to fewer mis-identified neutron tagged interactions. With this
%criterion in mind, the longer baselines studied provide a more useful
%baseline with which to achieve the overall physics goals of the
%detector. There is a small difference in the number of interactions
%per bunch in the FV between the two longer baselines. 

\begin{table}[tbp] \centering
\begin{tabular}{c c c}\hline\hline
Baseline (m) & FHC & RHC \\
\hline
%1000 & 1.48 & 0.50\\
%1838 & 0.42 & 0.14\\
%2036 & 0.33 & 0.11\\
1000 & 2.56 & 0.87\\
1838 & 0.73 & 0.24\\
2036 & 0.57 & 0.19\\
\hline\hline
\end{tabular}
\caption{
The number of beam neutrino interactions per kT per spill for a
$\nu_{\mu}$-enhanced beam (FHC) and $\overline{\nu}_{\mu}$-enhanced
beam (RHC).}\label{tab:baselineevtrate}
\end{table}

\begin{table}[!htb] \centering
\begin{tabular}{c c c}\hline\hline
Beam Type  & \% Contained Oriented Up & \% Contained Oriented Along Beam \\
\hline
FHC & 62.6 & 67.8\\
RHC & 56.6 & 66.2\\
\hline\hline
\end{tabular}
\caption{
Percentage of contained muons for the inner detector z-axis oriented
up and along the neutrino beam direction for a $\nu_{\mu}$-enhanced
beam (FHC) and $\overline{\nu}_{\mu}$-enhanced beam
(RHC).}\label{tab:murange}
\end{table}

%\subsubsection{Full Tank Optimization}
We then look at the expected rate from the beam and external
background for the chosen tank (radius of 5.5\,m and length of 22\,m)
with and without MRD and OD.

For the tank configurations and baselines we have considered only
events that are in-time with the beam, which includes beam-induced
interactions on water and, where applicable, iron, sand interactions,
and cosmic sources.  The event rates for water are taken from
Table~\ref{tab:baselineevtrate}, the sand muon simulation is used to
track particles to where they enter the tank and cosmic sources are
calculated from Table~\ref{tab:cosmics}.  In each case, all water in
the tank is assumed to be Gd-doped, so the neutron capture background
can be considered for neutrons that will have a capture time nearly
in-time with the beam.  A fiducial volume cut is also applied to the
event rates.  The assumption is that the reconstruction can reasonably
select events that have a reconstructed vertex at least 1\,m away from
the ID tank wall, giving a fiducial volume of 1.27\,kT.  This cut will
be optimised as reconstruction improves, and is used here for
illustrative purposes.

We study three possible configurations:
\begin{enumerate}[nolistsep] 
\item only a tank, where an outer detector (OD) of water surrounds the ID (similar to
Super-K);
\item a tank and a downstream muon range detector (MRD), with an OD surrounding the rest of
the tank;
\item a tank and an MRD that covers both the downstream face of the
detector and 75\% of the barrel, with an upstream OD of water.
\end{enumerate}
Where the design includes an MRD, particles from iron interactions are tracked to see if any make it into the TITUS ID.  The total number of beam neutrino
interactions per spill in water for each of these studies and the
TITUS ID is given in Table~\ref{tab:h2onuint}.

\begin{table}[tbp] \centering
\begin{tabular}{c c c c c}\hline\hline
Baseline (m) & Study 1 (ev/spill) &  Study 2 (ev/spill) &  Study 3 (ev/spill) & ID (ev/spill) \\
\hline
%1000 & 4.69 & 4.50 & 3.23 & 3.08 \\
%1838 & 1.33 & 1.27 & 0.91 & 0.87 \\
%2036 & 1.06 & 1.01 & 0.73 & 0.69 \\
1000 & 8.13 & 7.80 & 5.60 & 5.34 \\
1838 & 2.31 & 2.20 & 1.58 & 1.51 \\
2036 & 1.84 & 1.75 & 1.27 & 1.20 \\
\hline\hline
\end{tabular}
\caption{
The number of beam neutrino interactions per spill for a FHC beam
configuration in the TITUS tank and ID for detector designs with no
MRD (1), downstream MRD only (2), and downstream plus barrel MRD (3);
see text for further details. The final column are the number of beam
neutrino interactions per spill just in the ID, which is a subset of
the events in each of the first three columns.}\label{tab:h2onuint}
\end{table}

\begin{table}[tbp] \centering
%\begin{tabular}{c c c c c c c}\hline\hline
\begin{tabular}{c p{1.8cm} p{1.8cm} p{1.8cm} p{1.8cm} p{1.8cm} p{1.8cm}}\hline\hline
Baseline (m) & Tank ev/spill & Tank ev/bunch & ID\hspace{0.5cm}
ev/spill & ID ev/bunch & FV ev/spill & FV ev/bunch\\
\hline
%1000 & 39.17 & 4.87 & 3.92 & 0.47 & 2.41 & 0.28 \\
%1838 & 11.19 & 1.37 & 1.25 & 0.14 & 0.79 & 0.08 \\
%2036 &  8.91 & 1.09 & 1.04 & 0.11 & 0.66 & 0.07 \\
1000 & 67.89 & 8.44 & 6.79 & 0.81 & 4.17 & 0.49 \\
1838 & 19.40 & 2.37 & 2.17 & 0.24 & 1.37 & 0.14 \\
2036 & 15.44 & 1.89 & 1.80 & 0.19 & 1.20 & 0.12 \\
\hline\hline
\end{tabular}
\caption{The number of interactions per spill and bunch, assuming a SK-style
OD, in the whole tank, the ID, and a possible fiducial region
(FV).}\label{tab:tankSK}
\end{table}

In the case where there is no MRD as part of the TITUS complex, the ID
is surrounded by an additional layer of water, similar to the Super-K
design. The OD depth is 1\,m giving the tank dimensions of 6.5\,m in
radius and 24\,m in length corresponding to a total water mass of
3.18\,kT.
%The OD design will be discussed more in section~\ref{subsec-veto}. 
The event rates per spill and per bunch assuming a $\nu_\mu$-mode beam
for the whole tank and the ID are given in Table~\ref{tab:tankSK}.  It
is assumed that all Cherenkov particles in the OD are vetoed. 
Neutrons with a kinetic energy $T_{n}<10$\,MeV are considered
captured in the OD, since they have a range less than 1\,m, and all
neutrons with $T_{n}<20$\,MeV are captured before entering the
fiducial region.

\begin{table}[tbp] \centering
%\begin{tabular}{c c c c c c c}\hline\hline
\begin{tabular}{c p{1.8cm} p{1.8cm} p{1.8cm} p{1.8cm} p{1.8cm} p{1.8cm}}\hline\hline
Baseline (m) & Tank ev/spill & Tank ev/bunch & ID \hspace{0.5cm}ev/spill & ID ev/bunch & FV ev/spill & FV ev/bunch\\
\hline
%1000 & 39.03 & 4.82 & 4.12 & 0.47 & 2.54 & 0.29 \\
%1838 & 11.14 & 1.36 & 1.31 & 0.14 & 0.83 & 0.08 \\
%2036 &  8.87 & 1.08 & 1.08 & 0.11 & 0.69 & 0.07 \\
1000 & 67.65 & 8.35 & 7.14 & 0.81 & 4.40 & 0.50 \\
1838 & 19.31 & 2.36 & 2.17 & 0.24 & 1.44 & 0.14 \\
2036 & 15.37 & 1.87 & 1.87 & 0.19 & 1.20 & 0.12 \\
\hline\hline
\end{tabular}
\caption{The number of interactions per spill and bunch assuming a downstream
MRD and the rest of the tank surrounded by a SK-style OD in the whole
tank, the ID, and a possible fiducial region.}\label{tab:tankDS}
\end{table}

For the case where there is only a downstream MRD, we assume that
there is the equivalent of 0.5\,m of iron used in the detector to
track the possible particles in the tank. The MRD itself is assumed to
be circular in this study, though more details will be given in
section~\ref{subsec-mrd}.  The rest of the ID is again assumed to be
surrounded by 1\,m of water, giving a total water mass of 3.05\,kT. It
is also assumed that none of the sand interactions enter from the
region covered by the MRD.  The event rates are given in
Table~\ref{tab:tankDS}.

\begin{table}[tbp] \centering
%\begin{tabular}{c c c c c c c}\hline\hline
\begin{tabular}{c p{1.8cm} p{1.8cm} p{1.8cm} p{1.8cm} p{1.8cm} p{1.8cm}}\hline\hline
Baseline (m) & Tank ev/spill & Tank ev/bunch & ID \hspace{0.5cm}ev/spill & ID ev/bunch & FV ev/spill & FV ev/bunch\\
\hline
%1000 & 9.86 & 1.04 & 4.97 & 0.43 & 3.26 & 0.27 \\
%1838 & 2.92 & 0.29 & 1.55 & 0.13 & 1.03 & 0.08 \\
%2036 & 2.36 & 0.23 & 1.28 & 0.10 & 0.85 & 0.06 \\
1000 & 17.09 & 1.80 & 8.61 & 0.74 & 5.65 & 0.46 \\
1838 &  5.06 & 0.50 & 2.69 & 0.23 & 1.78 & 0.14 \\
2036 &  4.09 & 0.40 & 2.21 & 0.17 & 1.47 & 0.10 \\
\hline\hline
\end{tabular}
\caption{The number of interactions per spill and bunch assuming a downstream
and barrel MRD and upstream of the tank surrounded by a SK-style OD in
the whole tank, the ID, and a possible fiducial
region.}\label{tab:tankMRD}
\end{table}

Finally, the event rates for the case with a MRD covering both 75\% of
the TITUS barrel and the downstream region and a 1\,m OD upstream of
the TITUS ID are computed.  The remaining region around the barrel is
assumed to have a neutron absorbing material, e.g. boron-doped
polystyrene, to further reduce the possible number of neutrons
entering the tank.  An additional veto is assumed to be placed between
the polystyrene and the TITUS tank to veto particles created in the
material that may enter the tank.  The event rates are given in
Table~\ref{tab:tankMRD}. The earlier conclusion on detector location
based on pile-up rates is independent of the detector configuration.

%whether a MRD surrounds the tank,
%assuming a veto exists between the TITUS ID and the two MRD designs
%investigated.  The proposed MRD design in the next subsection falls in
%between these two scenarios and thus will not affect the detector
%optimization.

%The size of the ID and OD can also afford to shrink by as much as
%25\,cm each to take into account a support structure for the PMTs and
%electronics, but this has not been taken into account since the whole
%structure will depend on what is ultimately chosen for the overall
%detector design.  What is known is that the OD will be optically
%separated from the ID while the water circulation system will not.
%Based on the detector optimization considerations given in
%section~\ref{subsec-tank}, 

Based on the above studies, the area around the ID not covered by the
MRD needs to have 1\,m of water between the external wall of the tank
and the optical separation between the OD and ID.  This is to act as a
neutron shield for the ID due to incoming neutrons from interactions
in the surrounding material, veto charged particles from interactions
in the OD or outside the detector, and to determine if particles from
interactions in the ID deposited all their energy in the ID, or if
they exited.

The MRD design in section~\ref{sec:mrddesign} complicates a Super-K
style OD in that some space needs to be cut out in the middle of the
barrel to accommodate a small MRD.  The overall tank design must take
this feature into account.  For both the downstream and barrel
components, the area of the MRD that faces the tank must have at least
one additional layer of scintillating strips to identify charged
particles entering or exiting from the MRD into the tank.

\begin{comment}
FRANCESCA to RYAN: The text below belongs to the PMT section.
\\
The regions of the OD with the water veto will have a sparser coverage
of PMTs than the ID, though the exact PMT choice has not been made.
Options include 10-inch PMTs with a wavelength shifting (WLS) plate,
similar to what Super-K is currently using for its OD, in order to
catch as much light as possible; PMTs without the WLS plate but with a
higher photocoverage; or a hemispherical version of the optical
modules used by the ANTARES experiment~\cite{ANTARES}, see
section~\ref{OD-ANTARES}.

An example of the number of 10-inch PMTs needed for the veto in the
case of only a downstream MRD is given as an estimate of the total
possibly needed.  For the upstream-facing endcap, 240 PMTs covering
12.8\% of the total area are needed.  A similar coverage for the
barrel gives 1920, for a total of 2160 PMTs in the veto region.  With
the proposed MRD, this number is an overestimate.  This number can
also be reduced if a lower photocoverage gives a similar physics
performance.

In all cases, the OD will need to be able to collect enough light,
which can be enhanced with having the remaining area covered in white
plastic sheets, from the available PMTs to be able to determine if a
charged particle above Cherenkov threshold entered the OD without
being swamped by the captured neutron background.
\end{comment}

\subsection{Magnetised muon range detector}
\label{subsec-mrd}
\label{sec:mrddesign}
Outside the TITUS tank a magnetised iron tracking detector with a
1.5\,T field is proposed. This serves to range out muons and measure
their momentum and charge.  It complements the water Cherenkov
detector by both increasing the sample size and directly constraining
the ``wrong-sign" components in the $\nu_\mu$ and $\bar{\nu}_\mu$
beams, which are an important source of uncertainty in super-beam
measurements of CP violation.
However, a correction to the susceptibility caused by the paramagnetic
nature of the Gd$_2$(SO$_4$)$_3$ will be applied.  The correction
depends on the temperature. At about 12\degree it is around 0,4\%.

\begin{figure}[hbt]
  \centering \includegraphics[width=.7\textwidth]{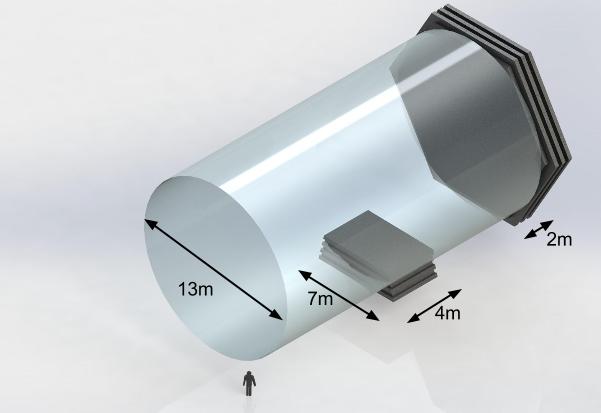} \caption{\label{fig:MRD}
  The proposed configuration of the magnetised muon range detectors of
  the TITUS experiment.  A large magnetised MRD with radius 13\,m to
  match the Cherenkov tank and thickness 2\,m, is placed downstream of
  the tank, to measure the momentum of forward-going muons.  A smaller
  magnetised side-MRD is placed on the side of the tank to allow
  low-background measurements of the antineutrino cross-section in
  this high-angle region of phase-space.  }
\end{figure}

The design, which is illustrated in Figure~\ref{fig:MRD}, includes a
full downstream magnetised muon range detector (MRD) and a small
magnetised side-MRD.  The diameter of the downstream magnetised MRD
matches that of the tank (13\,m including both ID and OD) and has a
thickness of 2\,m, allowing the forward scattered muons which escape
the tank to be included in the oscillation fit up to a momentum of
2\,GeV/c.  The smaller side-MRD has dimensions of 4\,m$\times$7\,m
with a thickness of 2\,m, and is also magnetised, allowing a
measurement of the less well understood high-$Q^2$ region of phase
space.  This region may be useful for testing and discriminating
between neutrino interaction models, and should help with the wrong
sign measurement in antineutrino mode, since the muons have different
angular distributions for $\nu$ and $\bar{\nu}$ interactions.

Tracking muons in magnetised iron is a well-established technique
which has been successfully employed in several experiments
(e.g.~\cite{SMRDref}) where charge reconstruction efficiencies are
typically in the range 95\% -- 98\%.  The design of the TITUS
magnetised MRD features 6\,cm thick planes of iron interleaved with
orthogonally arranged pairs of scintillator planes which sample the
position of particle tracks.  The magnetised MRD will measure the
momentum and total energy of muons leaving the water volume, up to a
range-out momentum of 2\,GeV/c.  The high position resolution of the
scintillator plates (1\,cm) will enable precise measurements of the
curvature and a strong, well-understood particle identification (PID)
via the direction of curvature.

%---------------------------------------------------------------
%\subsection{Scintillator detectors}
%---------------------------------------------------------------
To provide a 1\,cm spatial resolution highly segmented scintillator
detectors consisting of scintillator bars with wavelength-shifting
(WLS) fibers and Multi-pixel Geiger mode avalanche photodiodes as
photosensors are considered as a realistic option for active elements
of the MRD.  The real challenge lies in the required fine granularity
and size of the scintillator detectors for the MRD.  Each individual
element should have good characteristics to detect minimum ionising
particles (MIPs) with high efficiency in such a large detector system.
A detector option, successfully realised in ND280, is based on
7-10\,mm thick extruded polystyrene scintillator bars with WLS fibers
embedded with an optical glue. The length of the bars covered by a
chemical reflector by etching the scintillator surface in a chemical
agent can be up to 8\,meters.  For the readout, 1\,mm Kuraray Y11 WLS
fibers can be applied.  The results of the tests long extruded bars
are presented in Ref.~\cite{mineev}. It is important to stress that
even 7\,mm thick very long detectors (length of 16\,meters) provide a
light yield of 16\,photoelectrons per MIP and more than 99\%
efficiency for detection of MIP's.  It should be noted that the time
resolution of about 2\,ns can be reached for these detectors.  Shorter
scintillator detectors will be needed for a smaller magnetised
side-MRD placed on the side of the tank. In this case, the light yield
of 60\,photoelectors/MIP and time resolution of $\sigma = 0.7$ ns was
obtained for detectors of about 3\,m long~\cite{likhacheva}. This good
performance allows us to instrument both the MRD's with such
detectors.

Higher energy muons ($p_\mu \gtrsim 1$\,GeV/c) will travel through
many iron planes and their charge can be measured with very high
efficiency by reconstructing their curved trajectories in the 1.5\,T
magnetic field inside the iron.  This sample is particularly
interesting with regard to the validation of the complementary
gadolinium charge reconstruction technique.  This novel technique,
described in section~\ref{subsec-neutron}, is powerful as it can be
applied to all events, but the MRD can be used to both calibrate the
neutron capture and provide a more precise measurement of the charge
separation.

A measurement of the efficiency of the gadolinium technique using the
high energy MRD sample will allow us to exploit its full potential.
The mean charge reconstruction efficiency for all events in the
downstream MRD is estimated to be 95\%.  Furthermore, the precise
calibration of the performance of the gadolinium charge reconstruction
in TITUS will be greatly advantageous if it is used in the far
detector, Hyper-K, or indeed in other neutrino experiments, as it will
help in minimising the neutrino interaction modelling systematic
error.

The most interesting sample of muons is those resulting from neutrino
events near the oscillation maximum at $E_\nu\sim 0.6$\,GeV.  The muon
charge is more difficult to reconstruct using the traditional method
as the tracks may traverse only a handful of planes.  The design has
therefore been optimised for the lower energy muon spectrum of Hyper-K
using a novel arrangement of the first three iron planes by the
introduction of double scintillator planes and 10\,cm air gaps, both
of which increase charge reconstruction efficiency for short tracks.
In this case one does not fit tracks, but rather measures the angle of
the particle trajectory before and after each iron plane, and observes
the direction of curvature.  This technique is ultimately limited by
multiple scattering; however, several such measurements will allow an
efficiency of 90\% for events at the oscillation peak, a figure which
is comparable with the efficiency expected from the independent
gadolinium measurement.  When these two methods are combined it will
be possible to obtain $\sim96$\% pure $\nu_\mu$ and $\bar{\nu}_\mu$
samples from events in the oscillation peak in TITUS.

%\begin{figure}[hbt]
%  \centering
%  \includegraphics[width=.5\textwidth]{figs/BabyMind.png}
%  \caption{\label{fig:babymind}Temporary Baby-MIND figure. Mark will make a better figure.}
%\end{figure}

The principle for this detector is being optimised at the University
of Geneva through the Baby-MIND detector~\cite{BabyMind} which is
conceived as a proof of principle of low energy charge reconstruction
using magnetised iron.  It has been shown that the detector can be
magnetised using aluminium coils which do not reduce the
reconstruction efficiency and require minimal power to operate.  This
design is discussed in detail in \cite{BabyMind}. It has been accepted
as part of the WAGASCI experiment~\cite{Wagasci}.

% The mechanical design of the
%Baby-MIND is at an advanced stage.  Construction will begin later in
%2015, and data taking will begin in a test beam at CERN in 2016.
%Shortly afterwards it will be placed in the off-axis ND280 near
%detector complex of the T2K experiment as part of the Wagasci
%experiment \cite{Wagasci}.  
This will be a valuable proof of principle for the TITUS MRD.  The
design of the side-MRD is the same as that of the Baby-MIND, but with
increased transverse dimensions of 4\,m\,$\times$\,7\,m, allowing the
coil arrangement to be the same.

In addition to constraining the wrong-sign component, the MRD will
measure the momenta of muons leaving the water tank by ranging them
out in the iron in the traditional fashion.  In the nominal TITUS
design, 18\% of the muons exit the tank, with a preference for the
forward direction.  This number is much greater than in the far
detector due to the necessarily limited size of any intermediate
detector.  Without an MRD these events would be excluded from the
oscillation analysis.  The MRD therefore also gives a statistical
benefit of over 10\%, allowing muons which escape the tank to be
included in the analysis.

\subsection{Photosensors}
\label{subsec-pmt}
TITUS will contain over 3000 photodetectors and their 
performance is critical for the success of the experiment. Excellent temporal and spatial resolution is
required in order to precisely determine the vertex position of each
CCQE interaction. The default configuration for TITUS, assumed in all
studies unless otherwise stated, uses 12" photomultiplier tubes (PMTs)
with the same quantum efficiency as the 20" Hamamatsu Type R3600 PMTs
currently used by Super-K, but with flat rather than
hemispherical morphology of the glass. However, several alternative photosensor
technologies may offer improved performance and are currently under
study. In this section we present status reports on hybrid PMTs that
have the advantage of lower cost,
multi-PMTs and Large Area Picosecond PhotoDetectors (LAPPDs).

\subsubsection{Hybrid photosensors}
We have investigated the latest 8'' hybrid photomultiplier tube
(HPD) developed by Hammamatsu, the R12112. 
%(see Figure~\ref{detector}). 
These devices use an avalanche photodiode
instead of dynodes to achieve high single photon sensitivity and good
time resolution.  The 20'' equivalent of this PMT is being considered as
a possible candidate photosensor for the Hyper-K far detector. The
R12112 has a bi-alkali photocathode and is specified to have a quantum
efficiency of 27.2\% at 380\,nm. At 10\,kV the gain is $10^5$ with a dark
current of 20\,nA.  

\begin{figure}[t]
\begin{center}
\resizebox{0.49\textwidth}{!}{
\includegraphics{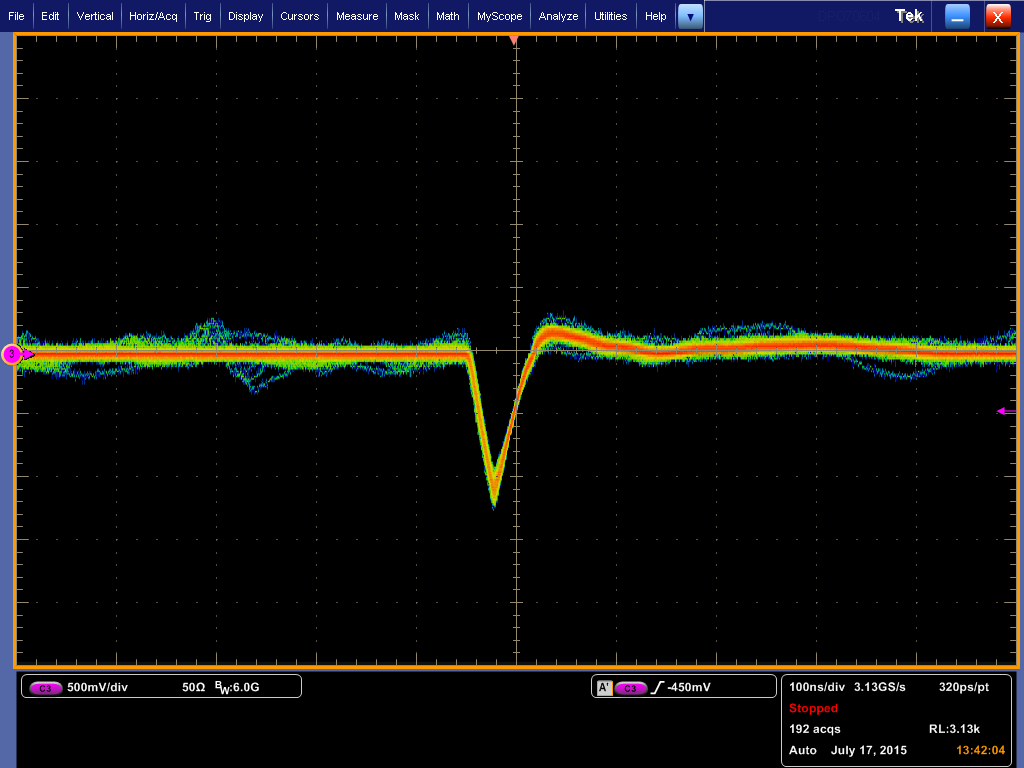}}
\resizebox{0.49\textwidth}{!}{
\includegraphics{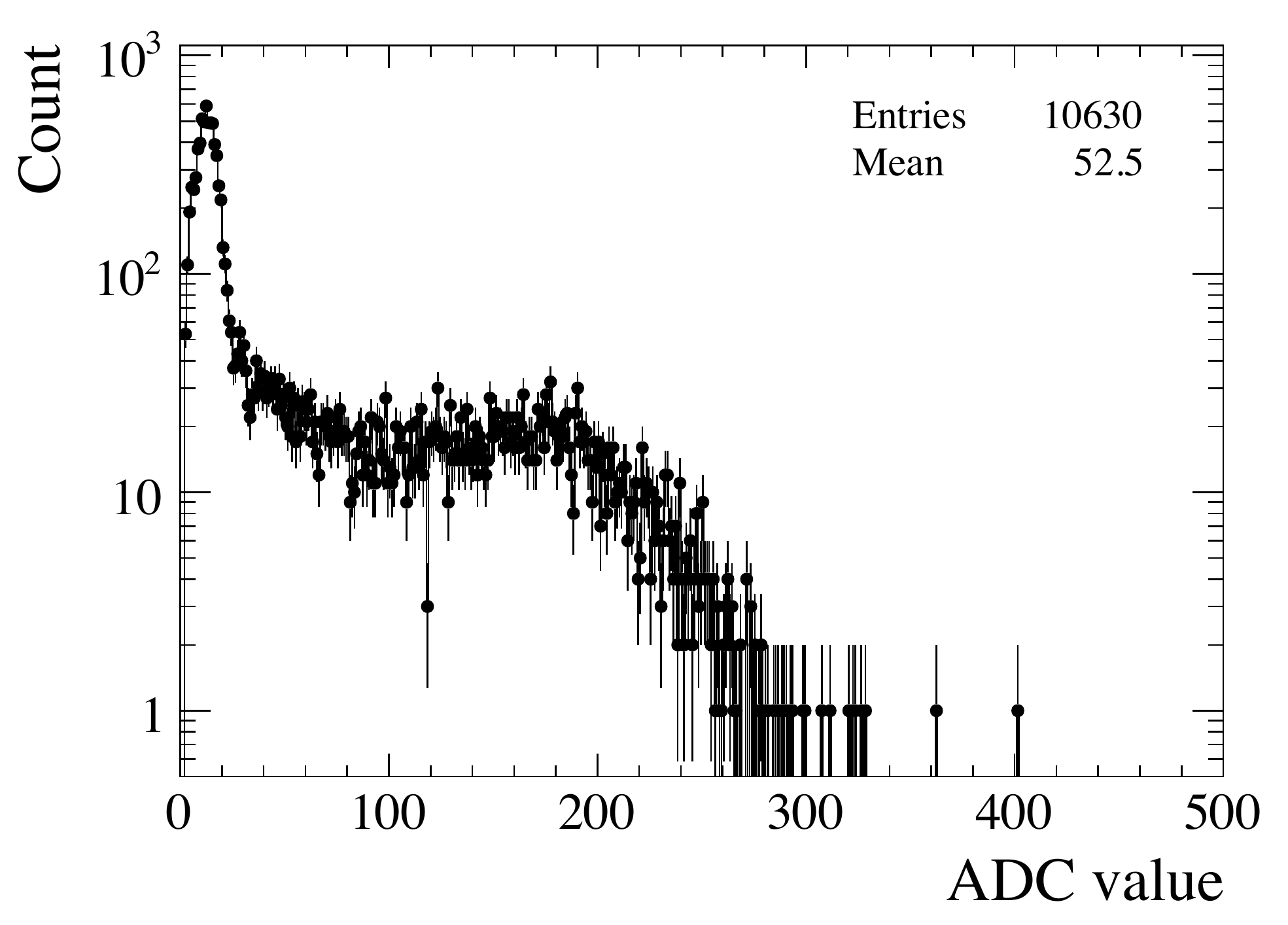}}
\end{center}
\caption{
(Left) Oscilloscope trace of R12112 photodetector signal in response
to pulsed LED light source inside the testing station. HV and LV
control voltages were set to 2.5\,V and 1.2\,V, respectively.  (Right)
Spectrum of ADC channels from a test HPD when triggering on cosmic
muons.}
\label{detector_signal}
\end{figure}

A photosensor test system is currently being
installed in the laboratory at the University of Edinburgh. Single
photon signals can be generated either by cosmic muons passing through
a 10\,mm quartz plate mounted directly above the HPD or by a pulsed
LED source.
% (see
%Figure~\ref{detector}) to verify these performance numbers and further
%characterise these devices for use in TITUS. Particular attention is
%needed to understand the performance of the HPDs in the presence of a
%magnetic field, as they may be operated in the fringe field of the
%TITUS MRD, should it be magnetised.
%
%The testing facility can be configured to use either a pulsed LED to
%generate single photons or detect Cherenkov light generated by cosmic
%muons passing through a 10\,cm quartz plate mounted directly above the
%HPD. 
%In the cosmic muon configuration a cosmic muon trigger system is used
%for data acquisition (the event rate is $\sim 10$ Hz).
Figure~\ref{detector_signal} (left) shows the amplified signal from
the HPD when using the pulsed LED source while
Figure~\ref{detector_signal} (right) shows the spectrum of analog-to-digital converter (ADC)
channels recorded by the data acquisition system when triggering on
cosmic muons. The single-photon peak is clearly visible above ADC
channel 100. Work is ongoing to further optimise and calibrate the HPD
testing facility. Further measurements will be carried out, including a
test of the effect of magnetic fields on these devices.
%
%\begin{figure}[t]
%\begin{center}
%\resizebox{0.4\textwidth}{!}{
%\includegraphics{figs/HPD_station.jpg}}
%\resizebox{0.4\textwidth}{!}{
%\includegraphics{figs/HPD_station_interior.jpg}}
%\end{center}
%\caption{(Left) Photosensor testing station. (Right) R12112 photodetector
%mounted in the testing station.  The scintillator paddles mounted
%above the HPD are used to trigger on cosmic muons.}
%\label{detector}
%\end{figure}
%

\subsubsection{Multi-PMTs}
An interesting option for the photodetector system is based on the
multi-PMT digital optical modules (DOMs) used by the Km3NeT
experiment~\cite{Km3NetTRD}.

\begin{figure}[htb]
\centering
\begin{tabular}{cc}
\includegraphics[width=0.4\textwidth]{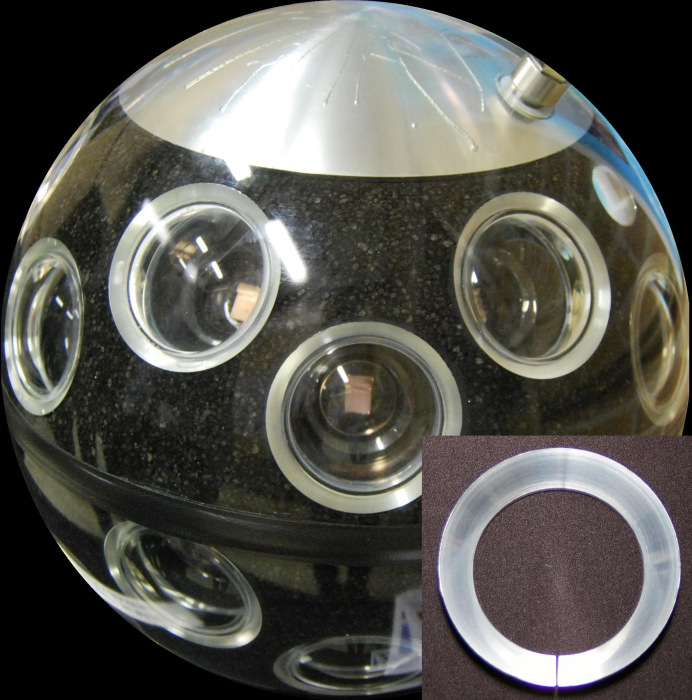}&
\includegraphics[width=0.4\textwidth]{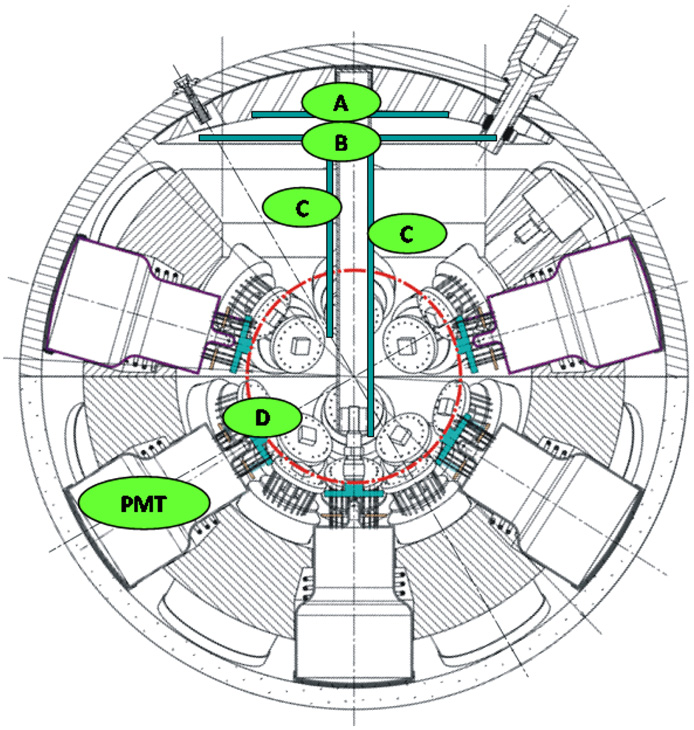}\\
\end{tabular}
\caption{(Left) A mechanical reference model of a multi-PMT DOM with reflectors
surrounding 3'' PMTs. (Right) Schematic drawing of a multi-PMT DOM;
D: PMTs including their bases; A: aluminium cooling structure; B and
C: front-end electronic components~\cite{Km3Net-mpmt,Km3Net-mpmt2}.}
\label{km3net-mpmt}
\end{figure}

The detection element of the KM3NeT deep-sea Cherenkov detector is a
pressure resistant glass sphere that contains photomultipliers, with
dedicated electronics, embedded in a transparent silicone gel to
ensure mechanical and optical coupling~\cite{Km3NetTRD}. A photograph and schematic diagram of
the KM3NeT multi-PMT DOM is shown in Figure~\ref{km3net-mpmt}.

This design has various advantages compared to traditional optical
modules with a single large PMT~\cite{ANTARES, NEMO}. In particular,
while the total photocathode area is the same or even larger than in
the case of a single large PMT, there is an increase in segmentation
%it houses from three to four times the photocathode area in a single
%glass sphere and has an almost uniform angular coverage. 
of the photocathode area that will help in distinguishing single-photon from multi-photon hits, resulting in an efficient background rejection
with a low background rate even at the single DOM level. Indeed,
two-photon hits can be unambiguously recognised if the two photons hit
separate tubes, which should occur in 85\% of all cases. In addition,
adjacent tubes can be selected to enhance the signal
coming from a single direction, whereas the background is mostly
randomly distributed~\cite{Km3Net-OM}. Thus, the arrival of more than
one photon at the DOM is identified with a high efficiency and purity
and provides a sensitivity to the direction of the detected light.

The reliability of the multi-PMT DOM is high, since the failure of a
single PMT should minimally degrade the performance of the
photodetector system.

%In TITUS, we envisage the use of a hemispherical DOM containing 3'' PMTs.
%PMTs is considered as detector system for the OD.  

%The design for a TITUS multi-PMT DOM is under investigation, as regards both the type and number of PMTs and for the vessel to be
%used.

The preliminary design assumes a hemispherical multi-PMT DOM with seven
3'' PMTs, one on the top of the hemisphere and the others arranged on the vertices of a
hexagon. 
The choice of small PMTs is based on the possibility of having a quantum
efficiency above 30\%, a small transit time spread, no need to shield
from the Earth's magnetic field and lower cost. The baseline option
is the same as the 3'' PMT used in KM3NeT~\cite{Km3Net-OM}, but 
an innovative small surface area hybrid photodetector can also be considered
(see for example~\cite{VSiPMT}).  %{\bf All the PMTs face the OD region. (?)}

The vessel is important to protect the photomultipliers and associated
equipment against the hydrostatic pressure and water. In KM3NeT commercially available
borosilicate glass spheres are used for the
vessels. Earlier studies from the KM3NeT Collaboration indicated that
sources of noise in the optical module include light produced
either by the scintillation or Cherenkov effect or by radioactive
contamination ($^{40}$K) in the glass material
itself~\cite{Km3NetVessel}.  This background can be reduced in TITUS
by using a custom-built acrylic pressure vessel. Several studies are
ongoing to choose the best shape and material.

Detailed simulation will be needed to determine the optimal design
and the best PMT choice for TITUS detector.

\subsubsection{LAPPDs}
\label{sec:lappds}
\begin{figure}[hbt]
  \centering
  \includegraphics[width=.6\textwidth]{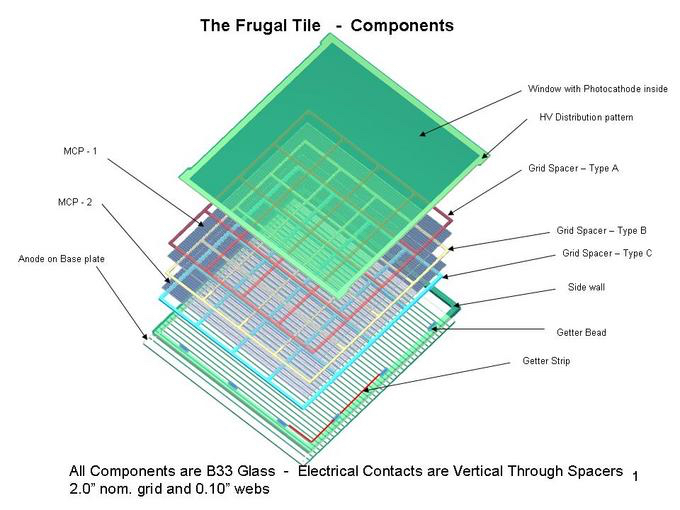}
  \includegraphics[width=.38\textwidth]{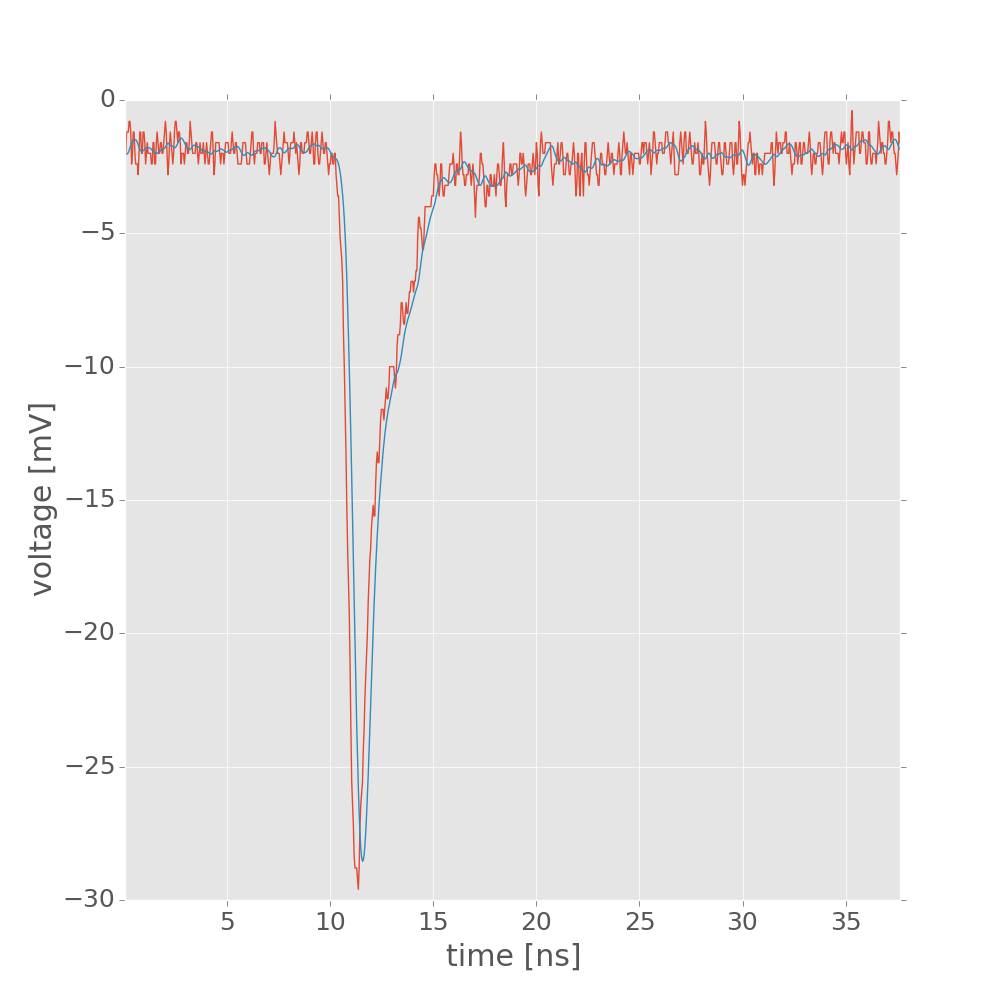}
  \caption{\label{fig:LAPPDlayout}(Left) Schematic view of the LAPPD tile.
  (Right) Oscilloscope trace of a single photon single from a MCP
    tile in the Edinburgh test facility. The red line is the
    raw signal, the blue is after digital filtering.
  }
\end{figure}

An attractive alternative photosensor, though not yet 
commercially available, is the Large Area Picosecond PhotoDetector (LAPPD)~\cite{lappdref}.
The LAPPD is an imaging detector based on micro-channel plate (MCP)
technology, with a $20\times 20$\,cm tile basic layout, shown in
Figure~\ref{fig:LAPPDlayout} (left).
It is able to resolve the position and time of single
incident photons within an individual sensor. This maximises the use of
the fiducial volume as it permits the reconstruction of events very close
to the wall of the detector, where the light can only spread over a
small area.  Preliminary Monte Carlo studies indicate that the
measurement of Cherenkov photon arrival space-time points with
resolutions of 1\,cm and 100\,ps will allow the detector to function as
a tracking detector, with track and vertex reconstruction approaching
size scales of just a few centimetres~\cite{anghel}. Imaging detectors
would enable photon counting by separating the space and time
coordinates of the individual hits, rather than simply using the total
charge. This means truly digital photon counting and would translate
directly into better energy resolution and better discrimination
between dark noise and photons from neutron capture with a time resolution better than 100\,ps and a
spatial resolution of the order of 3\,mm for single photons. The
design of the LAPPD is based on low-cost materials, well-established industrial
techniques and advances in material science. 

This technology is still in development and the number of available
tiles is limited. The ANNIE (Accelerator Neutrino Neutron Interaction
Experiment) experiment at Fermilab~\cite{Anghel:2015xxt} aims to be
the first neutrino experiment to test the tiles when available.  This
will allow the benefits of imaging photosensors to be
understood. Small $6\times6$\,cm prototype MCP tiles using a similar
technology have been obtained by the Edinburgh and Sheffield groups
from Argonne Lab~\cite{Wang:2016xnu} and tests are being made in the
laboratory to understand the performance of these devices, in
particular in a magnetic field. Figure~\ref{fig:LAPPDlayout} (right)
shows an example oscilloscope trace when the test LAPPD is illuminated
with an LED source.

\begin{comment}

The design hinges on two ``next generation'' MCPs~\cite{minot2014},
each consisting of millions of conductive glass capillaries (4–25
$\mu$m in diameter) fused together and sliced into a thin plate. Each
capillary works as an independent secondary-electron
multiplier. Single electrons that hit a pore on one side of the plate
convert into large bunches of electrons that cascade from the other
side, with typical amplification from a pair of plates of $10^7$.  The
two MCPs are enclosed between a glass window with a photocathode
applied to its inside surface, and a bottom anode plate is segmented
into striplines.
%LAPPD tiles have demonstrated excellent gain uniformity. 

In terms of LAPPD availability, testing and simulation, there is
important synergy between TITUS and ANNIE ~\cite{Anghel:2015xxt}.  LAPPDs
will be transitioning, by the end of 2015, to pilot industrial
production at Incom, Inc. If LAPPDs are a viable technology for TITUS,
a fast readout system and DAQ will be need to be developed.  In parallel,
detailed simulation will be performed to determine the ideal PMT and
LAPPD coverage to discriminate between different event categories.
\end{comment}

%---------------------------------------------------------------
\subsubsection{Photosensors for MRD}
%---------------------------------------------------------------
 
The MRD detector will consist of a large number of scintillator
detectors and therefore will have a large number of readout channels
that require the usage of very compact, insensitive to magnetic field
photosensors with a high sensitivity to the green light emitted from
WLS fibres. Multi-pixel Geiger mode avalanche photodiodes are recettly
used as photosensors in such detectors. Detailed information about
these photodiodes and their basic principle of operation can be found
in Ref.~\cite{renker}.  The first application of Geiger mode avalance
photodiodes in a large-scale experiment has been done in the near
neutrino detector ND280~\cite{nd280} of the T2K experiment where
approximately 56000 Hamamatsu Multipixel Photon Counters
(MPPCs)~\cite{mppc} are used.
    
Manufacturers have advanced in developing new generations of
multi-pixel Geiger photodiodes referred as SiPMs in recent years.
Various new SiPM types were developed by different companies
(Hamamatsu, KETEK, SensL, AdvanSiD).  The performance of Hamamatsu
MPPC's was greatly improved since 2009 when the ND280 was
commissioned. The dark noise rate per mm$^2$ of active area was
reduced by more than 10 times. The significant decrease of the dark
noise enables us to operate the new MPPCs with higher over-voltage to
achieve higher gain and this regime is relatively immune to
temperature changes. The optical crosstalk and after pulses were
significantly suppressed from the level of 10-20\% to less than 1\%
that allowed to operate with lower level signals in scintillator
detectors.  The photon detection efficiency of new devices was also
increased by about 2 times. The detailed study of new Hamamatsu MPPC's
and their parameters can be found in Ref.~\cite{hosomi}. All these
improvements are very important for applications of MPPC's in large
size detectors like the TITUS MRD.

\subsection{Electronics and readout}
\label{subsec-elect}
To read out the 3000+ PMTs of the TITUS Cherenkov detector, we propose
a trigger-less readout system based on well-established waveform
digitising technology (WFD).  The signals from the PMTs will be
continuously sampled using flash ADCs operating in the GHz range with
12 to 14 bit resolution.  The WFD technology combines in a single
device traditional operations like constant fraction discrimination
for timing, peak sensing and charge integration for energy
measurements, etc.  The advantage of this technology is that critical
decisions on whether to read out events can be delayed until the whole
waveform can be used to select interesting events PMT by PMT.  Another
advantage of this technology compared to traditional constant fraction
discriminators and Time to Digital Converters (TDCs) is that pile-up
can be effectively recognised and corrected for, with double hit
resolution of the order of the rise-time of input PMT signals,
permitting in principle, an individual measurement of each Cherenkov
photon reaching the PMT.

\begin{figure}[htbp]
\begin{center}
\includegraphics[width=0.8\textwidth]{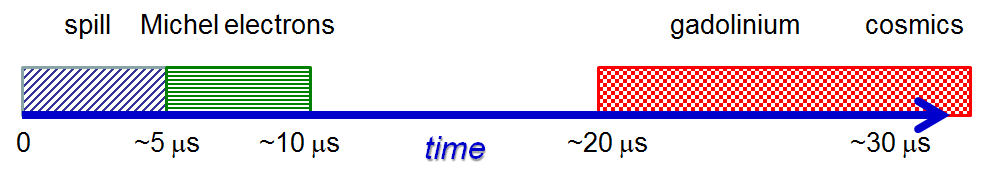} \\
\end{center}
\caption{Diagram showing approximate arrival times of Cherenkov light from
muons, Michel electrons and neutron capture, relative to the spill
window.}
\label{fig:timing}
\end{figure}

The electronics design for TITUS needs to accommodate fast sampling of
the Cherenkov light from the muon tracks ($\sim100$\,ps resolution
over 100\,ns duration) over the whole 5\,$\mu$s spill, the Michel
electrons, and the delayed capture of the thermal neutrons, which
occurs approximately 30\,$\mu$s later, as illustrated in
Figure~\ref{fig:timing}.

%%%%%%%%%%
%%%%%%%%%%

The digitising electronics can be located close to the PMTs or several
10\,m away.  In both cases we consider using a differential
transmission line using cat VI network cables.  The single ended
output from the PMT will be converted to a differential signal using
high frequency RF transformers.  The advantage of adopting RF
transformers is that the ground of the PMT (including the HV base and
possible preamplifiers) will be fully decoupled from the readout
electronics ground.  With PMT gains around $10^7$ there is no need to
further amplify the PMT outputs, whereas if HPDs are used, a
preamplifier will be required.  On the other end of the transmission
line, the differential PMT signal will be received by a large
bandwidth amplifier which will also condition the signal before
feeding it into the flash ADC.

All the readout chain is based on commercially available components
used in the telecommunication industry. There are several commercially
available flash ADC operating with 1\,GHz sampling frequencies and 12
or 14 bit resolution.  The flash ADC will be coupled to a high
performance Field Programmable Gate Array (FPGA), also operating in
the GHz range.  Several flash ADCs can be driven by the same FPGA.
The whole system will be synchronised using external clocks, operating
in the MHz range, to which on board clock frequency multipliers will
be phase loop locked (PLL).
 
Different waveform processing algorithms can be run in the FPGA.  In
the first stage, these algorithms will be used for baseline
subtraction and zero suppression, etc.  Next, different constant
fraction discriminator algorithms can be used to extract the timing
information from the waveforms.  With a sampling frequency of 1\,GHz a
timing resolution better than 50\,ps can be expected using fine-tuned
algorithms.  Further, charge-integrating algorithms can be used to
determine the charge of the input pulse.  Given that the signal
duration is of the order of several tens of ns, the signal is sampled
several times, so that the final ADC resolution will be higher than
the resolution of the flash ADC (roughly $
1/\sqrt{N_{\mbox{\scriptsize{samples}}}}$).

The drawbacks of this design are the cost and the considerable power
consumption (3 to 5\,W) per channel due to the high sampling frequency
of ADCs, the high performance FPGAs operating in the GHz range and
large bandwidth amplifiers.  While the cost is not a serious issue (10
to 15\% of the cost of a 20 inch PMT), especially for a near detector
with a small number of PMTs, the power consumption can become an issue
for large systems comprising up to 100,000 PMTs.  The electronics
industry is constantly developing new components (better performance,
cheaper, lower power consumption) and the latest products will be used
for the construction of the readout electronics.

The trigger and readout electronics will be developed jointly by the
University of Geneva (UniGE), Queen Mary University of London (QMUL),
et al.  UniGE has significant prior experience with data acquisition
systems and the development and implementation of fast WFD systems
(NA61/SHINE experiment at CERN, MICE at RAL).  QMUL also has
significant expertise from work in the T2K and SNO+ experiments.

To reduce the power dissipation (and cost) one could consider using
flash ADCs with lower sampling frequency coupled to lower performance
FPGAs.  For instance, one could consider splitting and phase shifting
the input signal into a 4 ch.\ 250\,MHz flash ADC, for an effective
1\,GHz sampling frequency of the input signal, without any loss in
performance but lower overall power dissipation.

An interesting option, particularly as regards lowering power
consumption is to use circular switched capacitor arrays (SCA), which
sample and store the waveforms into a capacitor array with a several
GHz sampling frequency.  The stored charges in the capacitor array are
then digitised at a much lower frequency, around 50\,MHz.  UniGE has
significant experience in developing and using digitising electronics
using the DRS4 SCA developed at PSI~\cite{psi_drs}.  The potential
disadvantage of this solution based on the DRS4 SCA is the time
required to read out the capacitor array (at least 15\,$\mu$s),
i.e. the dead time during which the SCA is not able to sample input
signals, and the finite depth of the capacitor array ($\sim
10$\,$\mu$s at 1\,GHz sampling frequency).  To solve this readout /
deadtime issue, which could be of a problem for observing the delayed
neutron capture, we will run several SCAs in parallel for the same
input signal.  As soon as one SCA is full and being readout, the
sampling of the signal continues on the following SCA.  In this way,
there is virtually no deadtime in the system.  Indeed, in a special
operation mode, the DRS4 chip can also sample the input with one
channel while reading out another channel at the same time, with
slightly increased noise.  This mode can be used to build a system
with up to eight parallel analog buffers and with virtually zero dead
time for event rates below the typical readout time (up to 500\,kHz).

A new version of the capacitor array, the DRS5, is under development
and expected for Summer 2016.  Instead of storing the sampled waveform
in a linear array, the waveform will be stored in a two-dimensional
array (matrix) consisting of short waveform segments.  The design of
the chip is such that each segment can be read out individually,
instead of the full array as for the DRS4.  Instead of digitising the
whole waveform, only interesting parts of the waveform would be
digitised, making it an almost deadtimeless system and overcoming the
limitations of the DRS4 ASIC.  Four to eight such segments would
typically be required to digitise an entire 100\,ns PMT pulse,
allowing a continuous acquisition rate of up to 500\,kHz.
%%The question, however, remains how to identify efficiently
%%(i.e. trigger on) the interesting segments.
As soon as this new ASIC is available, UniGE will start testing its
performance.

\subsection{DAQ}
\label{subsec-daq}
The Data Acquisition (DAQ) systems for the TITUS detector will collect
raw (digitised) data output from the electronics and write formatted
data to storage for offline analysis.

The data rate that the DAQ must support depends on the trigger
strategy, which will be designed to store all interesting physics
processes, discard non-physics events and provide reliable data
storage onsite. Interactions in the TITUS detector will have energies
ranging from a few MeV to several GeV. Beam events may be selected by
an external trigger, if that is available. Alternatively, a simple
N$_{\rm hits}$ threshold can be applied to select interactions that
should be stored to disk. N$_{\rm hits}$ is defined as the number of
WFD-identified peaks from PMT waveforms in a given time period; if the
number of firing PMTs is greater than a pre-determined number
threshold, the event will be read out. Cosmic ray interactions may be
selected by a similar N$_{\rm hits}$ trigger or even with a random
trigger given the high rate of cosmic muons in the detector (4~kHz
from Table~\ref{tab:cosmics}).

The acquisition window should be long enough to catch neutron
captures; because these happen with an average delay of 30~$\mu$s
since the primary vertex, a window of at least 300~$\mu$s should be
chosen. Assuming for instance one cosmic ray trigger is issued in
between every pair of beam triggers leads to the event rates collected
in Table~\ref{tab:event rates}. Thus the DAQ system must be capable of
supporting instantaneous event rates of order 50 GB/s and average
rates of few MB/s.

\begin{table}
\centering
\begin{tabular}{lrrr}
\hline\hline
\multicolumn{4}{c}{Event rates}\\
\hline
event type & $n$ hits in one DAQ window  & instantaneous rate & average rate \\
\hline
beam neutrino & 3982 & 0.77\,\text{GB}/\text{s} & 0.18\,\text{MB}/\text{s} \\
cosmic muon & 17669 & 3.4\,\text{GB}/\text{s} & 2.7\,\text{MB}/\text{s} \\
PMT dark noise & 15120 & 2.9\,\text{GB}/\text{s} & 1.3\,\text{MB}/\text{s} \\
$\mu$ sand events & 1314 & 0.25\,\text{GB}/\text{s} & 0.059\,\text{MB}/\text{s} \\
$n$ sand events & 420 & 0.081\,\text{GB}/\text{s} & 0.019\,\text{MB}/\text{s} \\
${}^{214}$Bi in PMT & 875 & 0.17\,\text{GB}/\text{s} & 0.078\,\text{MB}/\text{s} \\
${}^{208}$Tl in PMT & 131 & 0.025\,\text{GB}/\text{s} & 0.012\,\text{MB}/\text{s} \\
\hline
total & 39512 & 7.6\,\text{GB}/\text{s} & 4.4\,\text{MB}/\text{s} \\
\hline\hline
\end{tabular}
\caption{Instantaneous and average data rates for different event
categories. The numbers are obtained assuming that: the detector is
instrumented with 6000 PMTs; each waveform is digitised in 30 samples,
resulting in 58 bytes per hit; each PMT has a dark noise rate of
8.4~kHz; the glass of each PMT is contaminated with 27\,Bq of
${}^{214}$Bi and 2.8\,Bq of ${}^{208}$Tl as was measured for the SK
PMTs, and each of those decays generates respectively 18 and 26 hits
on average; and finally beam and cosmic ray triggers are taken in
alternation.  }\label{tab:event rates}
\end{table}

In the event of a supernova, it will be beneficial to read out the
whole detector without zero-suppression for a number of seconds. For a
supernova at a distance of 10\,kpc, $\mathcal{O}$(300) neutrino
interactions are anticipated in a $\sim$10 s time frame (see
Table~\ref{table:sn_rate}), leading to an average raw data rate of
$\sim$0.3\,MB/s. This is however a tiny perturbation of the total
detector rate and cannot be triggered upon with the N$_{\rm hits}$
technique; the supernova could only be recognised by identifying the
double presence of positron hits and delayed neutron capture hits or
by an external trigger from SNEWS if the the DAQ has a buffer.

The DAQ system will use a number of custom front-end applications
interfaced with an open source or custom-built framework, which will
allow user operation via a web interface.  It is anticipated that the
same framework will be used by the Hyper-K and TITUS
detectors. The DAQ will operate on commercially available computing
hardware.

Monte Carlo simulations are currently being used to assess the
performance of potential trigger algorithms, to calculate event rates,
to determine the impact of system latency and to verify whether full
digitization of pulse shapes is required.  Results from such studies
are being used to help define whether the trigger should be
implemented in hardware or in a software farm.

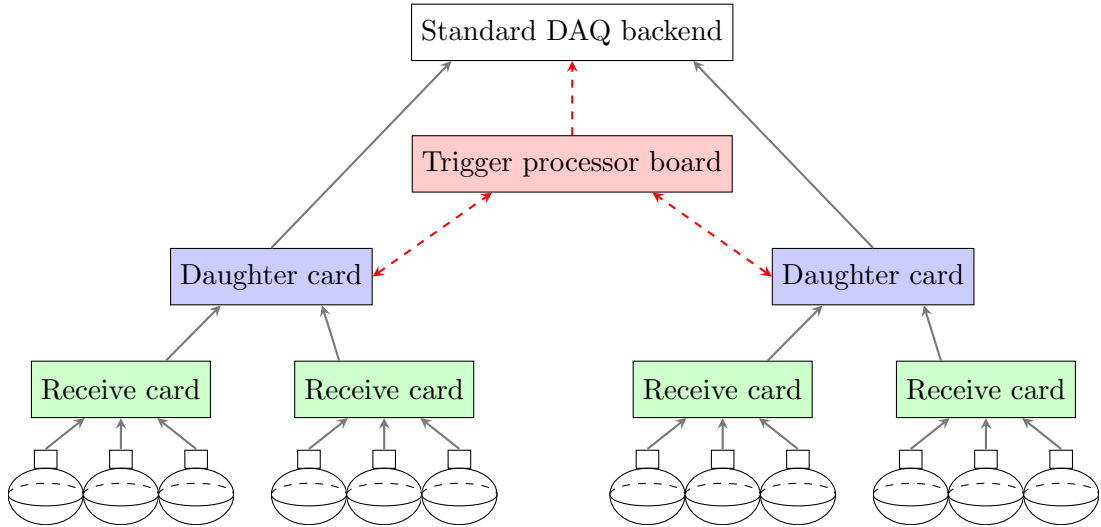
\begin{figure}[tpb]
\centering
   \tikzstyle{thick arrow oneway} = [thick,->,>=stealth]
\tikzstyle{thick arrow twoway} = [thick,<->,>=stealth]

\def\PMTsperboard{64}
\def\receivecardsperboard{4}
\def\daughtercardsperboard{39}

\begin{tikzpicture}

\node[pmt](pmt 1){}; %1};
\node[pmt, right of= pmt 1](pmt 2){}; %...};
\node[pmt, right of= pmt 2](pmt 3){}; %\PMTsperboard};

\node[receive card, above of=pmt 2,xshift=0.0cm,yshift=0.3cm]
  (receive card 1){Receive card}; %1};

\draw[thick arrow oneway,darkgray!10!gray] (pmt 1.south)--(receive card 1);
\draw[thick arrow oneway,darkgray!10!gray] (pmt 2.south)--(receive card 1);
\draw[thick arrow oneway,darkgray!10!gray] (pmt 3.south)--(receive card 1);

\begin{scope}[xshift=3.5cm]
  \node[pmt](pmt 4){}; % 1};
  \node[pmt, right of= pmt 4](pmt 5){}; %...};
  \node[pmt, right of= pmt 5](pmt 6){}; %\PMTsperboard};

  \node[receive card, above of=pmt 5,xshift=0.0cm,yshift=0.3cm]
  (receive card 2){Receive card}; %\receivecardsperboard};

  \draw[thick arrow oneway,darkgray!10!gray] (pmt 4.south)--(receive card 2);
  \draw[thick arrow oneway,darkgray!10!gray] (pmt 5.south)--(receive card 2);
  \draw[thick arrow oneway,darkgray!10!gray] (pmt 6.south)--(receive card 2);
\end{scope}

\begin{scope}[xshift=8cm]
  \node[pmt](pmt 7){}; % 1};
  \node[pmt, right of= pmt 7](pmt 8){}; %...};
  \node[pmt, right of= pmt 8](pmt 9){}; %\PMTsperboard};

  \node[receive card, above of=pmt 8,xshift=0.0cm,yshift=0.3cm]
  (receive card 3){Receive card}; %1};

  \draw[thick arrow oneway,darkgray!10!gray] (pmt 7.south)--(receive card 3);
  \draw[thick arrow oneway,darkgray!10!gray] (pmt 8.south)--(receive card 3);
  \draw[thick arrow oneway,darkgray!10!gray] (pmt 9.south)--(receive card 3);
\end{scope}

\begin{scope}[xshift=11.5cm]
  \node[pmt](pmt 10){}; % 1};
  \node[pmt, right of= pmt 10](pmt 11){}; %...};
  \node[pmt, right of= pmt 11](pmt 12){}; %\PMTsperboard};

  \node[receive card, above of=pmt 11,xshift=0.0cm,yshift=0.3cm]
  (receive card 4){Receive card}; %\receivecardsperboard};

  \draw[thick arrow oneway,darkgray!10!gray] (pmt 10.south)--(receive card 4);
  \draw[thick arrow oneway,darkgray!10!gray] (pmt 11.south)--(receive card 4);
  \draw[thick arrow oneway,darkgray!10!gray] (pmt 12.south)--(receive card 4);
\end{scope}

\node[daughter card, above of =receive card 1, xshift=2cm,yshift=0.5cm]
  (daughter card 1){Daughter card}; %1};
\node[daughter card, above of =receive card 3, xshift=2cm,yshift=0.5cm]
  (daughter card 2){Daughter card}; %\daughtercardsperboard};

\begin{scope}[thick arrow oneway,darkgray!10!gray]
  \draw ($(receive card 1.north)!0.5!(receive card 1.north east)$)--
   ($(daughter card 1.south)!0.5!(daughter card 1.south west)$);

  \draw ($(receive card 2.north)!0.5!(receive card 2.north west)$)--
   ($(daughter card 1.south)!0.5!(daughter card 1.south east)$);  

  \draw ($(receive card 3.north)!0.5!(receive card 3.north east)$)--
   ($(daughter card 2.south)!0.5!(daughter card 2.south west)$);

  \draw ($(receive card 4.north)!0.5!(receive card 4.north west)$)--
   ($(daughter card 2.south)!0.5!(daughter card 2.south east)$); 

  % \draw ($(daughter card 2.north west)!0.25!(daughter card 2.south west)$)--
  % ($(daughter card 1.north east)!0.25!(daughter card 1.south east)$);

  % \draw ($(daughter card 2.north west)!0.75!(daughter card 2.south west)$)--
  % ($(daughter card 1.north east)!0.75!(daughter card 1.south east)$);

\end{scope} 

\node[trigger processor, above of =daughter card 1, xshift=4cm,yshift=0.5cm](trigger processor 1){Trigger processor board}; 

\node[backend, above of =trigger processor 1, xshift=0cm,yshift=0.75cm](backend 1){Standard DAQ backend}; 

%\node[backend, above of =daughter card 1, xshift=3.5cm,yshift=0.75cm](backend 1){}; 
%\node[trigger processor, below of =backend 1, xshift=0cm,yshift=-0.0cm](trigger processor 1){}; 

% = = = = = = = = = = = = = = = =
% Labels
% = = = = = = = = = = = = = = = =

% \node[xshift=-3.05cm,yshift=0.2cm,left of = pmt 3,align=left](lev1)
%   {PMTs};

% \node[xshift=0.9cm,yshift=0.3cm,above of = lev1,align=left](lev2)
%   {Receive cards};

% \node[xshift=1.6cm,yshift=0.4cm,above of = lev2,align=left](lev3)
%   {Daughter cards};

% \node[xshift=2cm,yshift=0.75cm,above of = lev3,align=right](lev4)
%   {Trigger processing unit};

% \node[xshift=1.55cm,yshift=0.75cm,above of = lev4,align=right](lev5)
%   {Standard DAQ backend};

% % = = = = = = = = = = = = = = = =
% % Background rectangle - removed
% % = = = = = = = = = = = = = = = =

% \path ($(pmt 3.south west)!0.9!(lev1.south east)-(0,0.4cm)$) coordinate (A)
%   --([yshift=0.86cm]A |- lev4.north east)coordinate (B)--
%   ($(B)+(11.2cm,0)$)coordinate (C);

% = = = = = = = = = = = = = = = =
% Backend connections
% = = = = = = = = = = = = = = = =

% interconnections of backend 1
\begin{scope}[thick arrow oneway,darkgray!10!gray]
  \draw (daughter card 1.north)--
  ($(backend 1.south)!0.75!(backend 1.south west)$);

  \draw (daughter card 2.north)--
  ($(backend 1.south)!0.75!(backend 1.south east)$);
\end{scope}

% interconnections of trigger processor 1
\begin{scope}[thick arrow twoway,dashed,red]
  \draw (daughter card 1.east)--
  ($(trigger processor 1.south)!0.5!(trigger processor 1.south west)$);

  \draw (daughter card 2.west)--
  ($(trigger processor 1.south)!0.5!(trigger processor 1.south east)$);

  % \draw (daughter card 1.east)--
  % (trigger processor 1.west);

  % \draw (daughter card 2.west)--
  % (trigger processor 1.east);
\end{scope}

\begin{scope}[thick arrow oneway,dashed,red]
  \draw(trigger processor 1.north)--(backend 1.south);
\end{scope}

% = = = = = = = = = = = = = = = =
% Data path
% = = = = = = = = = = = = = = = =

% \draw[transform canvas={xshift=-0.1cm,yshift=0.1cm}, dashed, stealth-, very thick, red!80!black,shorten <=0.2cm, shorten >=0.0cm]
%   ($(backend 1.south)!0.5!(backend 1.south west)$)--
%   (daughter card 1.north);

% \draw[transform canvas={xshift=-0.1cm,yshift=0.1cm}, dashed, stealth-, very thick, red!80!black,shorten <=0.2cm, shorten >=0.0cm]
%   ($(daughter card 1.south)!0.5!(daughter card 1.south west)$)--
%   ($(receive card 1.north)!0.5!(receive card 1.north east)$);

% \draw[transform canvas={xshift=-0.1cm,yshift=0.1cm}, dashed, stealth-, very thick, red!80!black,shorten <=0.2cm, shorten >=0.0cm]
%   (receive card 1)--
%   ($(pmt 1.north)!0.5!(pmt 1.north)$);

\end{tikzpicture}\caption{ \label{fig:DAQ:hardware} Block
    diagram of a DAQ system with a hardware-based trigger.  The paths
    of data and triggers are shown with the solid grey arrows and
    dashed red arrows respectively.  N$_{\rm hits}$ triggers are
    created on the trigger processor board, and sub-N$_{\rm hits}$
    triggers are created on the daughter cards; both are distributed
    by the trigger processor board.  The numbers of daughter cards,
    receive cards, and PMTs have been reduced for clarity.}
\end{figure}

A hardware-based trigger, as shown in Figure~\ref{fig:DAQ:hardware},
would merge information from a group of PMTs on a ``receive card'' and
data from several receive cards would be merged on a daughter
card. Each daughter card would send N$_{\rm hits}$ information onto a
trigger processor board; if the number of PMTs firing in an event
exceeded a given N$_{\rm hits}$ cut, the event would be saved.  If the
number of hits was not sufficient, a sub-N$_{\rm hits}$ trigger would
be activated in the electronics firmware of each daughter card.  One
way this could be implemented is to use the FPGA logic to divide the
detector into smaller cells e.g. 500 cells containing a given number
of PMTs.  Each cell would have a look-up table of fixed time offsets
for each PMT within it. The FPGA would look for coincidences in
offset-corrected PMT times in a specific time window, in a given cell
location in the compartment. A local N$_{\rm hits}$ threshold would be
applied to each cell and, if at least one cell passes this
requirement, the whole event would be written to disk.

\begin{figure}[tpb]
\centering
   \tikzstyle{thick arrow oneway} = [thick,->,>=stealth]
\tikzstyle{thick arrow twoway} = [thick,<->,>=stealth]

\def\PMTsperboard{64}
\def\receivecardsperboard{4}
\def\daughtercardsperboard{39}

\begin{tikzpicture}

\node[pmt](pmt 1){}; %1};
\node[pmt, right of= pmt 1](pmt 2){}; %...};
\node[pmt, right of= pmt 2](pmt 3){}; %\PMTsperboard};

\node[receive card, above of=pmt 2,xshift=0.0cm,yshift=0.3cm]
  (receive card 1){Receive card}; %1};

\draw[thick arrow oneway,darkgray!10!gray] (pmt 1.south)--(receive card 1);
\draw[thick arrow oneway,darkgray!10!gray] (pmt 2.south)--(receive card 1);
\draw[thick arrow oneway,darkgray!10!gray] (pmt 3.south)--(receive card 1);

\begin{scope}[xshift=3.5cm]
  \node[pmt](pmt 4){}; % 1};
  \node[pmt, right of= pmt 4](pmt 5){}; %...};
  \node[pmt, right of= pmt 5](pmt 6){}; %\PMTsperboard};

  \node[receive card, above of=pmt 5,xshift=0.0cm,yshift=0.3cm]
  (receive card 2){Receive card}; %\receivecardsperboard};

  \draw[thick arrow oneway,darkgray!10!gray] (pmt 4.south)--(receive card 2);
  \draw[thick arrow oneway,darkgray!10!gray] (pmt 5.south)--(receive card 2);
  \draw[thick arrow oneway,darkgray!10!gray] (pmt 6.south)--(receive card 2);
\end{scope}

\begin{scope}[xshift=8cm]
  \node[pmt](pmt 7){}; % 1};
  \node[pmt, right of= pmt 7](pmt 8){}; %...};
  \node[pmt, right of= pmt 8](pmt 9){}; %\PMTsperboard};

  \node[receive card, above of=pmt 8,xshift=0.0cm,yshift=0.3cm]
  (receive card 3){Receive card}; %1};

  \draw[thick arrow oneway,darkgray!10!gray] (pmt 7.south)--(receive card 3);
  \draw[thick arrow oneway,darkgray!10!gray] (pmt 8.south)--(receive card 3);
  \draw[thick arrow oneway,darkgray!10!gray] (pmt 9.south)--(receive card 3);
\end{scope}

\begin{scope}[xshift=11.5cm]
  \node[pmt](pmt 10){}; % 1};
  \node[pmt, right of= pmt 10](pmt 11){}; %...};
  \node[pmt, right of= pmt 11](pmt 12){}; %\PMTsperboard};

  \node[receive card, above of=pmt 11,xshift=0.0cm,yshift=0.3cm]
  (receive card 4){Receive card}; %\receivecardsperboard};

  \draw[thick arrow oneway,darkgray!10!gray] (pmt 10.south)--(receive card 4);
  \draw[thick arrow oneway,darkgray!10!gray] (pmt 11.south)--(receive card 4);
  \draw[thick arrow oneway,darkgray!10!gray] (pmt 12.south)--(receive card 4);
\end{scope}

\node[switch, above of =receive card 1, xshift=5.75cm, yshift=0.5cm]
  (switch){Ethernet switch};

\begin{scope}[thick arrow oneway,darkgray!10!gray]
  \draw ($(receive card 1.north)!0.0!(receive card 1.north east)$)--
   ($(switch.south)!0.75!(switch.south west)$);

  \draw ($(receive card 2.north)!0.0!(receive card 2.north west)$)--
   ($(switch.south)!0.4!(switch.south west)$);  

  \draw ($(receive card 3.north)!0.0!(receive card 3.north east)$)--
   ($(switch.south)!0.4!(switch.south east)$);

  \draw ($(receive card 4.north)!0.0!(receive card 4.north west)$)--
   ($(switch.south)!0.75!(switch.south east)$); 
\end{scope} 

\node[farm, above of =switch, xshift=-3cm,yshift=0.5cm](farm 1){Processing farm node}; 
\node[farm, above of =switch, xshift=+3cm,yshift=0.5cm](farm 2){Processing farm node}; 

\node[backend, above of =farm 1, xshift=3cm,yshift=0.75cm](backend 1){Standard DAQ backend}; 

% = = = = = = = = = = = = = = = =
% Backend connections
% = = = = = = = = = = = = = = = =

% interconnections of backend 1
\begin{scope}[thick arrow oneway,darkgray!10!gray]
  \draw (farm 1.north)--
  ($(backend 1.south)!0.75!(backend 1.south west)$);

  \draw (farm 2.north)--
  ($(backend 1.south)!0.75!(backend 1.south east)$);
\end{scope}

% interconnections of farm 1
\begin{scope}[thick arrow oneway,darkgray!10!gray]
  \draw ($(switch.north)!0.60!(switch.north west)$)--
  ($(farm 1.south)!0.0!(farm 1.south west)$);

  \draw ($(switch.north)!0.60!(switch.north east)$)--
  ($(farm 2.south)!0.0!(farm 2.south east)$);
\end{scope}

\end{tikzpicture}\caption{ \label{fig:DAQ:software} Block
    diagram of a DAQ system with a software-based trigger.  The path
    of data is shown with the solid grey arrows.  Triggers are created
    on a processing farm node, and read out to the standard DAQ
    backend.  The numbers of processing farm nodes, receive cards, and
    PMTs have been reduced for clarity.}
\end{figure}
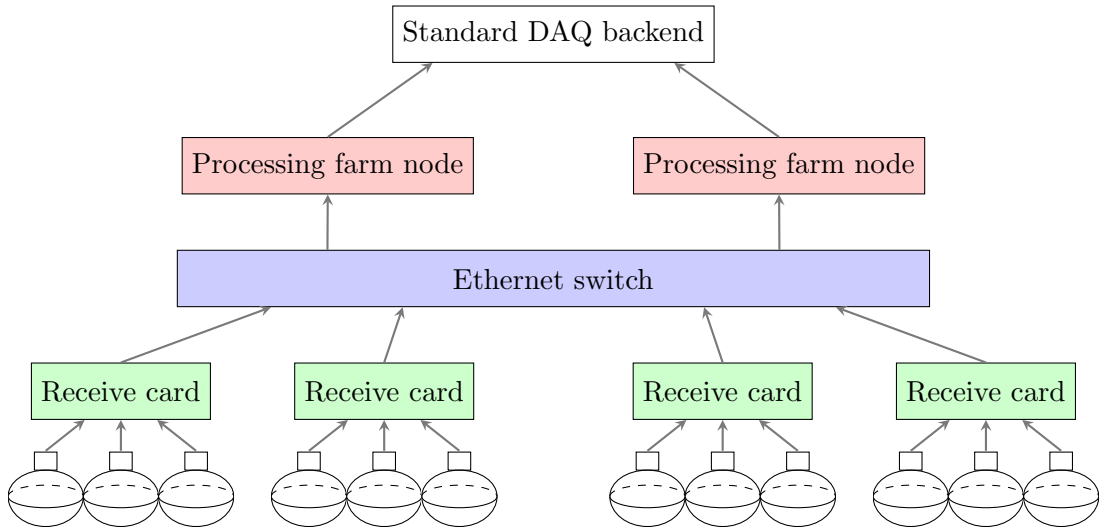

In a software trigger system, as shown in
Figure~\ref{fig:DAQ:software}, all hits would be transmitted from the
electronics to the DAQ and decisions on which time windows to keep
would made in a farm.  Data from a group of PMTs would be collected by
a receive card, which would be connected to a large ethernet switch.
The outputs of the switch would be connected to a processing farm of
Linux machines or graphical processing units or similar.  These
processing nodes would see data from the entire detector, divided into
time windows.  Fast algorithms would be implemented on these
processing nodes and decisions regarding whether to write the event to
storage would be made at this point.  Algorithms using spatial and
time information are currently being developed.

Distributed cluster technology will be used for the DAQ and trigger
farm, such that if one node fails, its processes automatically run on
another node.  This would allow faulty computer hardware to be
exchanged with minimal impact on data taking.  The computer cluster
design is currently flexible until the final DAQ design and hardware
choices are made.

\subsection{Calibration}
\label{subsec-calib}
In order to reduce the systematic uncertainties of TITUS to the level
required by Hyper-K, it will be essential to thoroughly calibrate the
TITUS detector. This will be performed using two calibration systems:
one integrated light source and a deployment system for a variety of
radioactive calibration sources.

A full calibration of the TITUS detector involves several individual
sub-tasks: calibration of the PMT array, measurement of detector
response parameters, and determination of the neutron detection
efficiency. To calibrate the PMT array, the absolute time of each PMT
with respect to the rest of the array must be measured, the dependence
of the time on the deposited PMT charge must be determined for both
single and multiple photon scenarios, and the charge response of the
PMT to photons must be determined. The detector response parameters
include the extinction and scattering of light in the water as a
function of wavelength, any dependence of the PMT response on angle,
and the overall detector gain. Finally, as TITUS uses Gd doping to
increase sensitivity to neutrons, the neutron detection efficiency
must be determined as a function of position in the tank. These needs
can be met by the two calibration systems described below.

\subsubsection{Integrated light source system}

TITUS will include an integrated light source system for all optical
and PMT calibrations. This system consists of a number of light
injection points connected via optical fibres to light pulsers in the
electronics. Light pulses of 1--2\,ns can be produced relatively
inexpensively using LEDs or similar solid state optical devices
allowing multiple optical sources to be deployed around the edge of
the detector. This system consists of an LED coupled to an optical
fibre, which is then connected to an optical diffuser on the PMT
support of TITUS. The optical diffuser is used to shape the light
inside TITUS and a number of designs are possible, providing different
calibration pulses for different needs. The key challenges are the
coupling of the LEDs to the optical fibre, minimising dispersion in
the fibre to maintain short optical pulses and achieving the required
dynamic range. There are two points at which the light produced can be
monitored, and TITUS will use both. First, the optical coupling
between the LED and the fibre can be monitored by a solid-state device
such as an MPPC or photodiode built into the LED-fibre coupling
system. Second, a return fibre from the optical diffuser can also be
implemented. These two systems will enable both pulse-by-pulse and
long-term monitoring of the light entering the detector, allowing PMT
and optical calibration data to be taken without the
manpower-intensive calibration source deployment previously used in
water Cherenkov detectors. These data can be collected either in
dedicated high-rate calibration runs or interspersed during normal
data-taking.

The calibration of the PMT timing requires a short duration light
pulse of known origin and time. The integrated calibration system,
from any given fibre, provides this but clearly cannot illuminate the
entire PMT array at once.  To minimise the number of fibres required
the optical diffuser for the PMT calibration must provide a wide
opening angle, of order 30 degrees, to illuminate a significant number
of PMTs on the far side of the detector. The diffuser must be
carefully designed to ensure that there is no time dependence as a
function of angle. To achieve the overall calibration of global time
offset of the array, PMTs must be illuminated by at least two fibres
to allow the fibre times to be cross calibrated. For TITUS, four-fold
degeneracy of the PMT calibration fibre points is the target to permit
improved cross calibration and redundancy against single point
failures in the fibres. This system will enable the calibration of PMT
timing, the dependence of time on charge and the PMT time response.

The integrated calibration system can also be used to measure optical
scattering, extinction and the PMT response. While the basic elements
of the system are the same as that used for PMT calibration, a number
of changes are required meaning that fibres and diffusers used for
these calibrations are different. These properties must be measured as
a function of wavelength, thus several LED types will be used to
provide light at 6 different wavelengths between 330\,nm and
500\,nm. To measure scattering a narrow beam is required from the
optical diffuser; the scattering length is measured by monitoring the
light level of PMTs outside the narrow beam as a function of the path
length of the beam through the detector. Optical extinction is
measured by monitoring the light levels at specific PMTs inside the
optical beams; unlike scattering, wide-angle beams are important for
this calibration to provide a variety of path lengths. The pulse by
pulse monitoring of the calibration system is essential for this
calibration as the light level at given PMTs is the key measurement of
the system. The measured light level at the PMTs is a combination of
extinction and PMT response as a function of angle; several light
paths and angles are needed for these to be decoupled in the analysis,
requiring a variety of diffuser points and diffuser directions within
TITUS.

\subsubsection{Calibration source deployment system}

In addition to the integrated system TITUS will also have the option
to deploy calibration sources similar to other water Cherenkov
detectors. This system will consist of a number of source deployment
points above the detector from which sources may be lowered. In
addition to the z-axis option that these points provide, optionally a
full 3D manipulator system may be developed allowing sources to be
deployed over a wider range of the detector volume. A variety of
sources can be deployed through this system including an optical
source as a backup to the integrated calibration system. This system
would be heavily utilised for the neutron calibration system.

A number of radioactive sources can be developed to provide neutrons
for calibration. These include $^{252}$Cf and AmBe sources that were
previously used to calibrate the neutron response of the SNO
detector. These sources produce neutrons at a known rate, and by
comparing the rate of measured Gd captures to this rate the neutron
detection efficiency and any possible variation across the detector
can be measured. Neutron captures on Gd also provide a source of
events inside the detector of known energy providing a further
calibration of the detector response in addition to the neutron
response of the detector.
 
Overall these two calibration systems provide the data that will be
required to characterise TITUS and reduce the systematic uncertainties
to the required level. The integrated calibration system permits
calibration and monitoring of the detector without the deployment of
specialised manpower while calibration sources can be used for more
extensive calibrations during the time when the neutrino beam is off.

\section{Basic selection and sensitivity studies}
\label{sec-basicstrategy}
\newcommand{\enu}{\ensuremath{E_{\nu}}}
\newcommand{\enutrue}{\ensuremath{E_{\nu}^{\mathrm{true}}}}
\newcommand{\enuqe}{\ensuremath{E_{\nu}^{QE}}}

\newcommand{\pipm}{\ensuremath{\pi^{\pm}}}
\newcommand{\piz}{\ensuremath{\pi^{0}}}
\newcommand{\pizero}{\ensuremath{\pi^{0}}}

\newcommand{\nubar}{\ensuremath{\overline{\nu}}}

\newcommand{\nue}{\ensuremath{\nu_e}}
\newcommand{\numu}{\ensuremath{\nu_\mu}}
\newcommand{\nuebar}{\ensuremath{\overline{\nu}_e}}
\newcommand{\numubar}{\ensuremath{\overline{\nu}_\mu}}

\newcommand{\selMu}{\ensuremath{\mathrm{1R}\mu}}
\newcommand{\selE}{\ensuremath{\mathrm{1R}e}}

\newcommand{\selMuEfficiency}{\ensuremath{79\%}}
\newcommand{\selEEfficiency}{\ensuremath{76\%}}

\newcommand{\deltaCP}{\ensuremath{\delta_{CP}}}
\newcommand{\deltacp}{\ensuremath{\delta_{CP}}}

The primary purpose of TITUS is the measurement of oscillation
parameters in combination with Hyper-K, in particular $\deltaCP$.  In
this section the ability of TITUS to select charged-current muon and
electron events is evaluated and the enhancement of samples with
neutron tagging is explored using a methodology based on Monte Carlo
truth vectors smeared and weighted according to the efficiencies and
resolutions seen in Super-K. This allows fast evaluation of the
performance of the detector before a full simulation and
reconstruction is performed as presented in section~\ref{sec-software}
but has the disadvantage that cuts cannot be optimised to reflect the
differences between TITUS and Super-K, most notably a smaller tank
size but larger sample statistics. The results in this section focus
on the performance of the TITUS detector without the MRD option; the
latter will be added in future sensitivity studies. As in
section~\ref{subsec-tank}, the tank is: 5.5\,m in radius, 22\,m in
length and surrounded by an OD. We use a 40\% photocoverage, similar
to that of Super-K, and hence we assume a similar reconstruction
performance to Super-K can be achieved~\cite{Abe:2015awa}. The
sensitivity results are shown for a 516\,kton Hyper-K detector, where
the second identical tank starts 6 years after the first, and a beam
power of 1.3\,MW. We assume that the ratio of running time in
neutrino-mode (\FHC) and antineutrino mode (\RHC) is 1 to 3. Finally,
we compare the sensitivity improvement that TITUS brings with respect
to Hyper-K and Hyper-K plus ND280 alone.

\subsection{Basic selection}
\label{subsec-basicselection}
The simulated events used in this analysis are based on Monte Carlo
truth vectors smeared and weighted according to the efficiencies and
resolutions seen in Super-K. In what follows we will refer to this as
``table-based'' reconstruction. The detector response model is
calculated as a function of the distance of the most energetic
particle from the wall.  This takes into account the smaller size of
TITUS relative to Super-K.  An implicit assumption in the usage of
these tables is that the TITUS reconstruction will be able to achieve
at least the same performance as Super-K.

MC vectors were generated with NEUT version 5.4.2. In this version
there is an updated CCQE model with n-particle, n-holes interactions
(npnh) included. Secondary hadron interactions and neutron capture on
gadolinium are simulated with a Geant4-based simulation of the TITUS
detector in WCSim~\cite{WCSim} version 1.5.0. We use the Geant4
version 4.9.6 with G4NDL4.2 and the Photon Evaporation model (i.e. the
flag $\rm G4NEUTRONHP\_$$\rm USE\_$$\rm ONLY\_$ $\rm
PHOTONEVAPORATION$ is set) so that energy is conserved; this is due to
the fact that one model conserves energy and the other emits gamma
capture photons at the correct energy.  A 90\% neutron capture
efficiency on Gd is assumed for a 0.2\% doping with a 95\% efficiency
for reconstructing neutrons
%actual number is 94.78% captured on gadolinium
as obtained from full reconstruction studies described in
section~\ref{subsec-fullselection}.

For the studies below, the signal definition is a true $\nu$CCQE
interaction for \FHC running and $\overline{\nu}$CCQE for \RHC
running.  In addition, we can use the number of neutron captures
detected to further enhance the signal definition and, at the same
time, have a background-enhanced sample to further constrain some of
the background systematics.  Further information is given in the
following subsections.

\subsection{Lepton Selection}

The Super-K single-ring muon ($\selMu$) and single-ring electron
($\selE$) selections were applied to simulated events in TITUS.  The
$\selMu$ selection requires that the ring passing the muon PID to be
fully contained in the fiducial volume (FCFV), no additional Cherenkov
rings, the reconstructed muon momentum $p_\mu > 200$\,MeV and 0 or 1
detected Michel electrons.  The $\selE$ selection requires a FCFV
single ring that passes the electron PID, the electron reconstructed
momentum $p_e>100$\,MeV, the reconstructed neutrino energy
E$_{\nu}^{\text{QE}}<1250$\,MeV, and finally Michel electron and
$\pizero$ veto cuts applied to remove the background.

To cope with backgrounds near the edge of the tank, additional
fiducial cuts are applied on top of the $\selE$ and $\selMu$
selections.  Events with vertices less than 2\,m from the wall were
rejected: close to the wall there is a large $\mu$ contamination in
the $e$ sample due to the PID algorithm not having enough information
to separate $\mu$ and $e$.  Events where the distance from the vertex
to the wall along the direction of the reconstructed lepton track is
less than 4\,m were rejected.  It should be noted that this Super-K
selection is tuned for a larger volume, which makes performance near
the walls less important.  There may be scope to re-optimise the cuts
in order to expand this fiducial volume.

The standard deviation of the difference between reconstructed
quantities and their true values is shown in Table~\ref{tab:res}.
This is evaluated for true charged-current events with no pions in the
final state reconstructed within the fiducial volume and passing the
event selection.  Table~\ref{tab:res} shows the resolutions for both
the $\selE$ and $\selMu$ event selections.  The overall effect of this
smearing on the $\enuqe$ resolution (the important reconstructed
parameter for the oscillation analysis) is 24\% (8\%) for selected
electrons (muons).

\begin{table}
\centering
\begin{tabular}{lrr}
\hline\hline
\multicolumn{3}{c}{Detector Resolutions}\\
\hline
Quantity & Electron & Muon\\
\hline
Visible energy [GeV] & 0.075 & 0.042\\
Visible energy [\%] & 9.0 & 6.0\\
Lepton Angle [degrees] & 2.4 & 1.7\\
Vertex Position [m] & 0.21 & 0.12\\
E$_{\nu}^{\text{QE}}$ [GeV] & 0.17 & 0.09\\
E$_{\nu}^{\text{QE}}$ [\%] & 24.0 & 8.0\\
\hline\hline
\end{tabular}
\caption{Detector resolution for various reconstructed quantities for true
charged current events containing no pions passing the electron and
muon event selection.  
}\label{tab:res}
\end{table}

\begin{figure}[!tb]\centering
\subfloat[$\selE$ \FHC]{\label{fig:selection:1refhc} 
\includegraphics[width=0.49\textwidth]{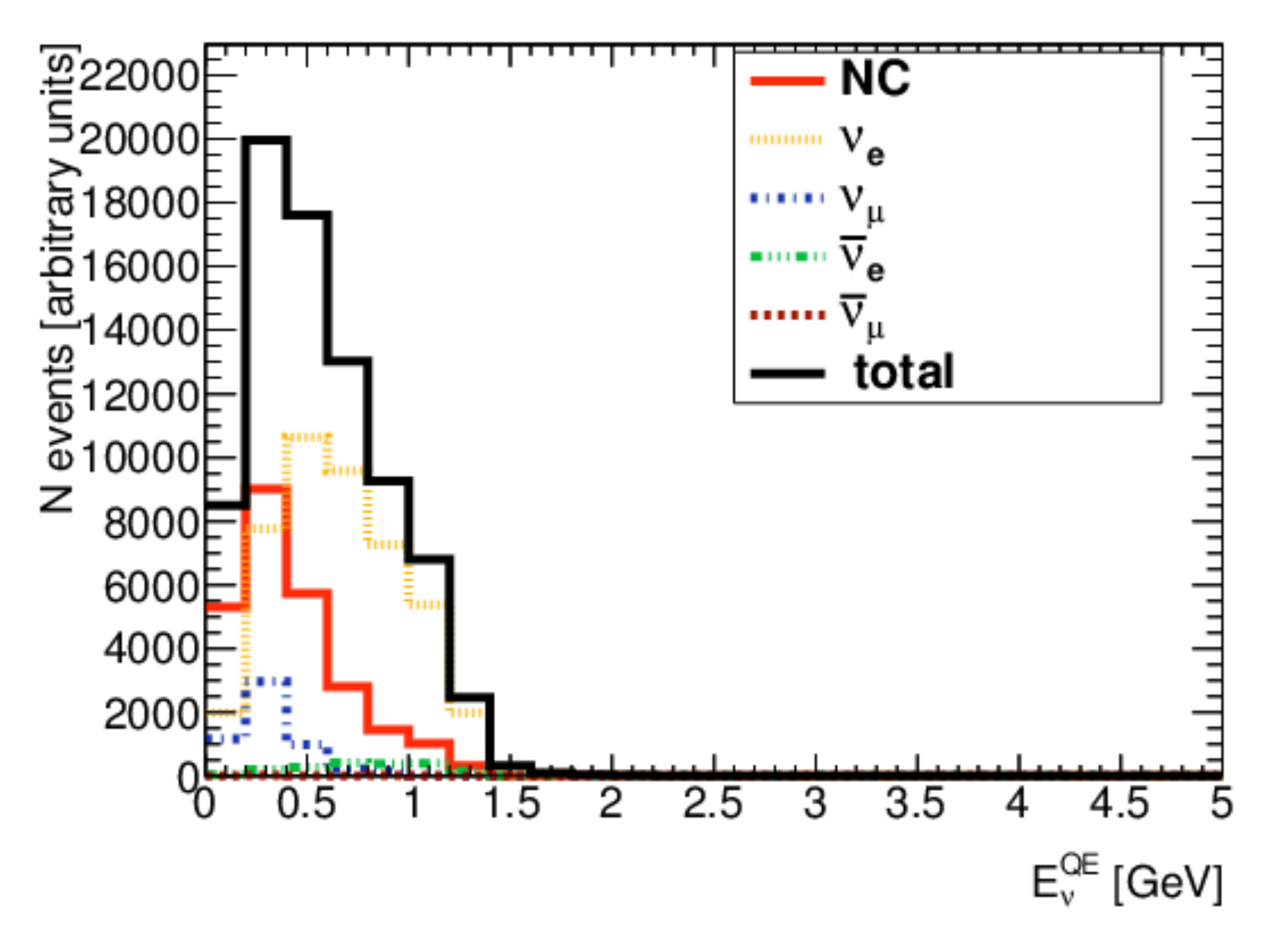}
}
\subfloat[$\selE$ \RHC]{\label{fig:selection:1rerhc} 
\includegraphics[width=0.49\textwidth]{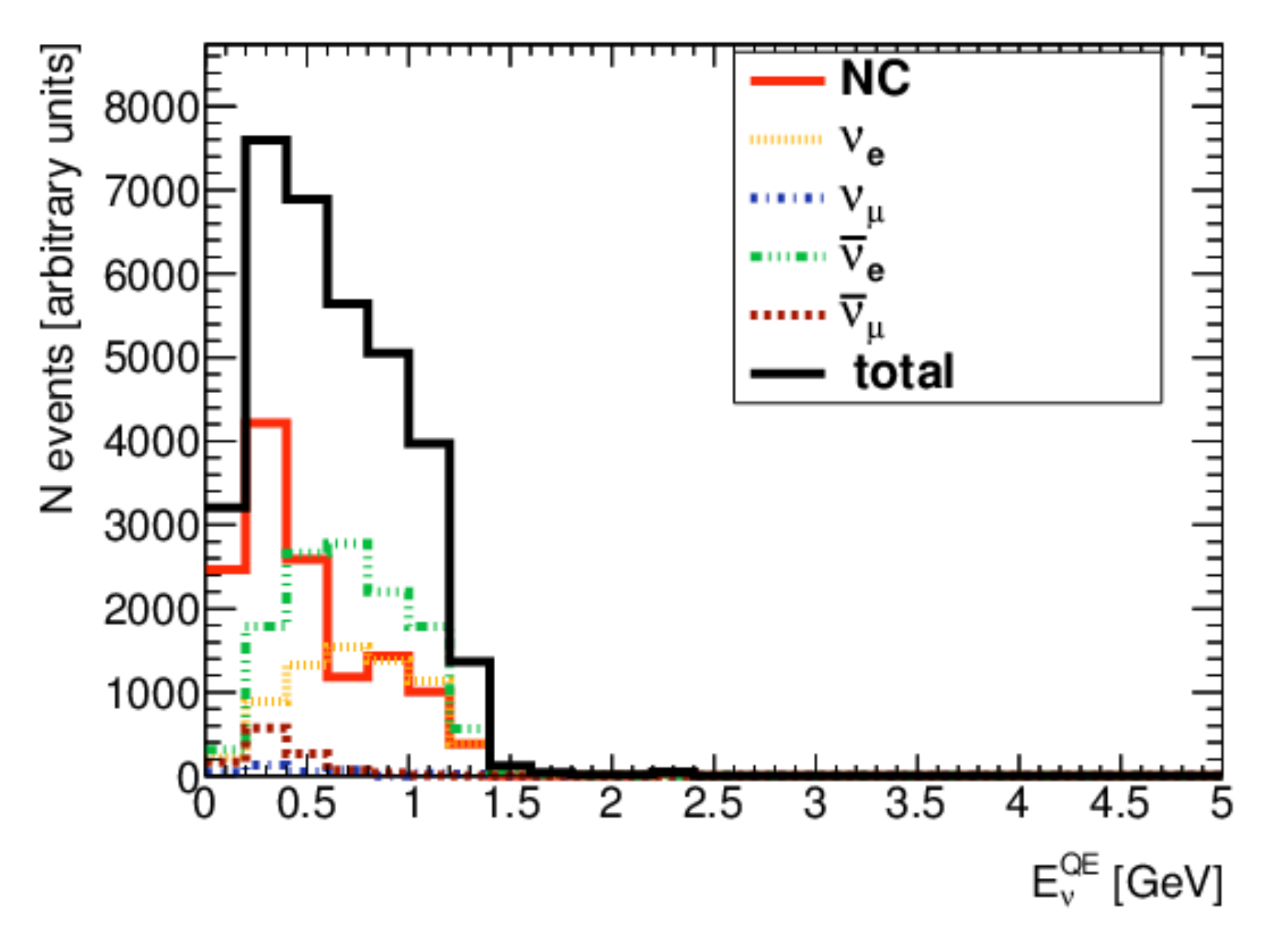}
}
 
\subfloat[$\selMu$ \FHC]{\label{fig:selection:1rmufhc} 
\includegraphics[width=0.49\textwidth]{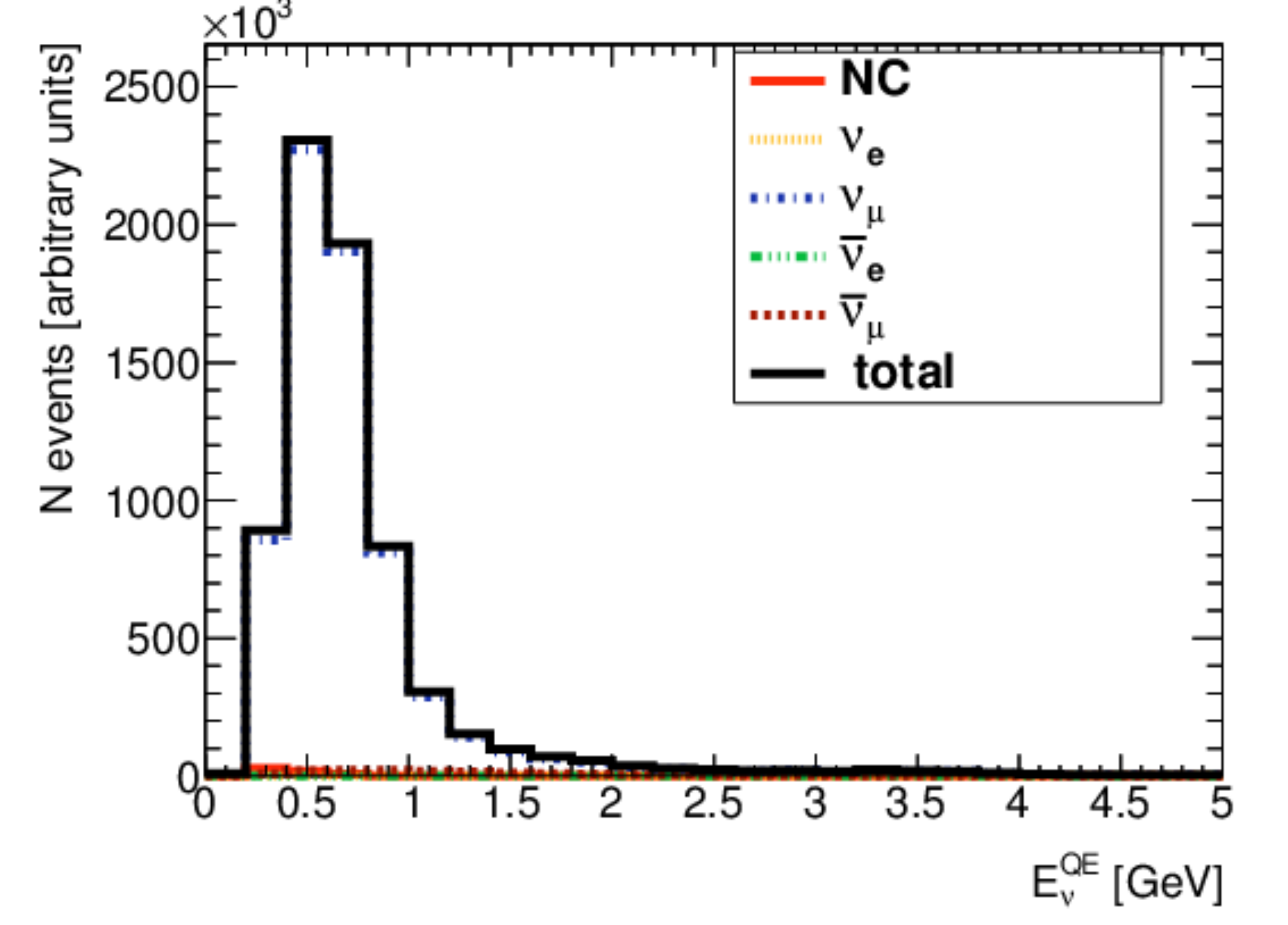}
}
\subfloat[$\selMu$ \RHC]{\label{fig:selection:1rmurhc} 
\includegraphics[width=0.49\textwidth]{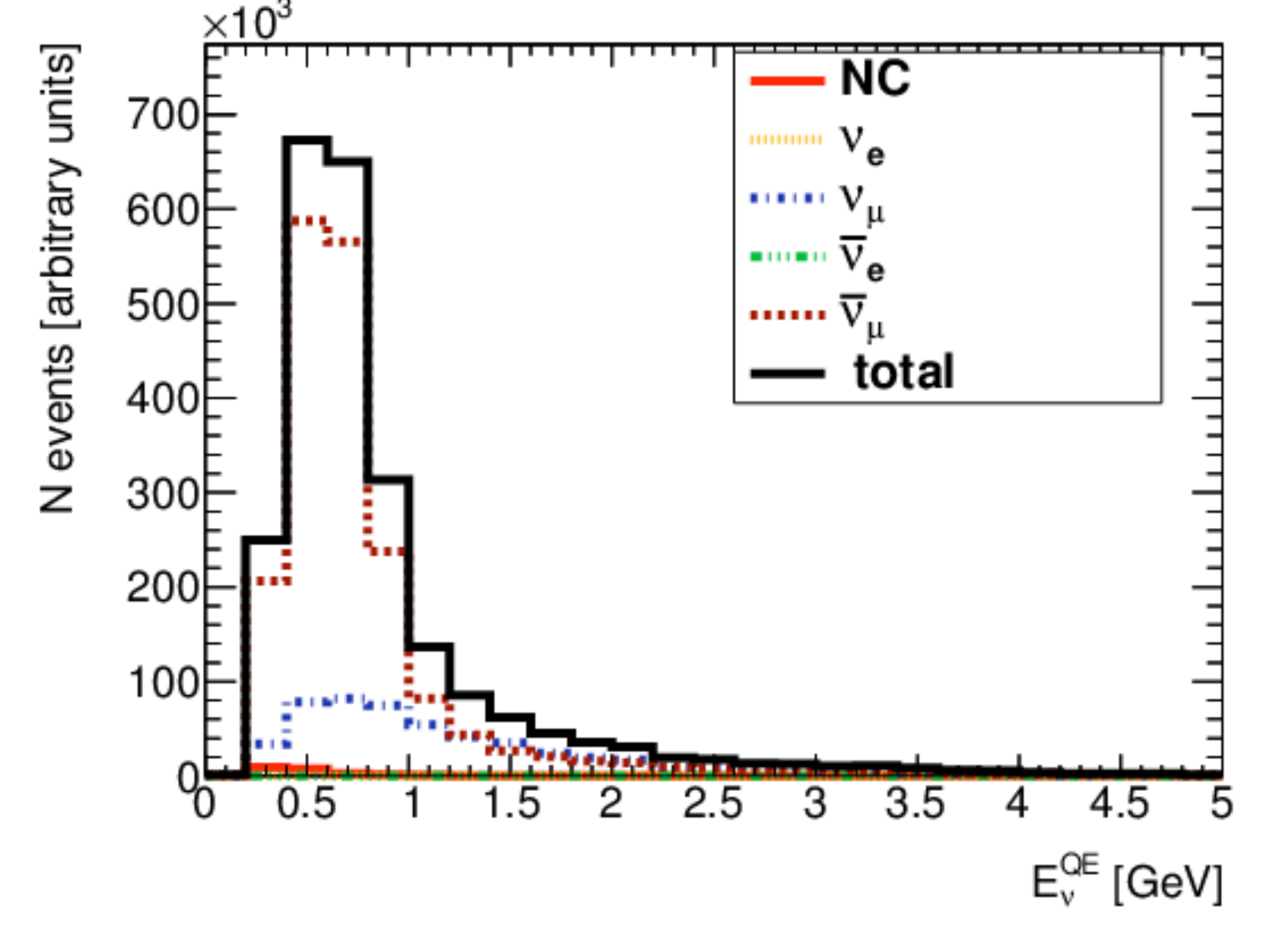}
}
 
\caption{E$_{\nu}^{\text{QE}}$ distribution of the TITUS $\selE$
  (upper plots) and $\selMu$ (lower plots) samples for the neutrino
  (left plots) and antineutrino (right plots) modes,
  respectively. \label{fig:selectiontitus} }
\end{figure}

The selected $\selE$ and $\selMu$ samples are shown in
Figure~\ref{fig:selectiontitus}.  The selections are applied to two
data samples with different beam conditions: neutrino mode (\FHC) and
antineutrino mode (\RHC).
%For comparison, the same samples in Hyper-K are shown in Figure
%\ref{fig:selectionhk}.  The composition of each sample is shown in
%Figure \ref{tab:eventtable1rmu,tab:eventtable1re}.
The muon sample is dominated by \numu CC0$\pi$ events, i.e. events
with a muon and no mesons escaping the nucleus, which are selected
with an efficiency of 79\%.  The dominant background in the muon
selection appears in the wrong-sign \numu CC0$\pi$ events that make up
18\% of the muon sample during antineutrino mode running.

The $\selE$ selection efficiency for $\nu_e$CC$0\pi$ events is
\selEEfficiency.  There is a significant NC background, 
dominated by neutral current $\pizero$ production, which is similar in
size to the $\nue$ signal.  This selection is optimised for Super-K:
optimisation for TITUS requires further study, but is likely to
involve tighter PID cuts to produce a cleaner $\nue$
sample. Significant performance improvements are expected.

\subsection{Neutron selection}
True CCQE reactions yield final-state neutrons for antineutrinos
($\overline{\nu} p \rightarrow l^+ n $) but not for neutrinos ($\nu
n \rightarrow l^- p$).  Although this can be modified by final-state
interactions, neutron tagging should still permit the selection of
signal-enhanced samples by imposing a neutron requirement or veto for
the $\overline{\nu}$ or $\nu$ mode respectively.  For the purposes of
this discussion, the “signal” is defined as a true $\nu$CCQE
interaction during $\nu$-mode running, and as a true
$\overline{\nu}$CCQE interaction in $\overline{\nu}$ mode.

\begin{figure}[htb]
\begin{minipage}{0.49\textwidth}
 \centering
\includegraphics[width=0.95\linewidth]{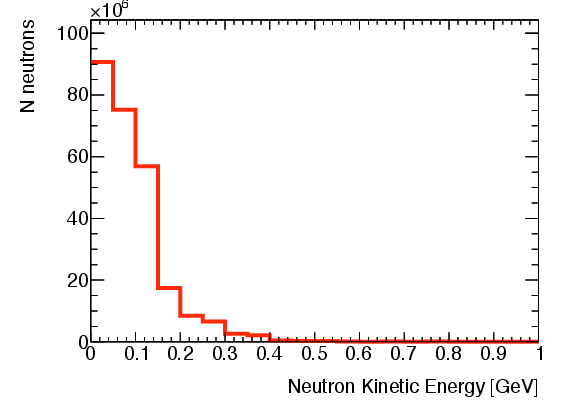}
\caption{The true kinetic energy of neutrons after final state interactions within the nucleus but before secondary interactions within the water volume.}
\label{fig:neutronenergy}
\end{minipage}
\begin{minipage}{0.49\textwidth}
  \centering
\vspace{0.5cm}
\includegraphics[width=0.95\textwidth]{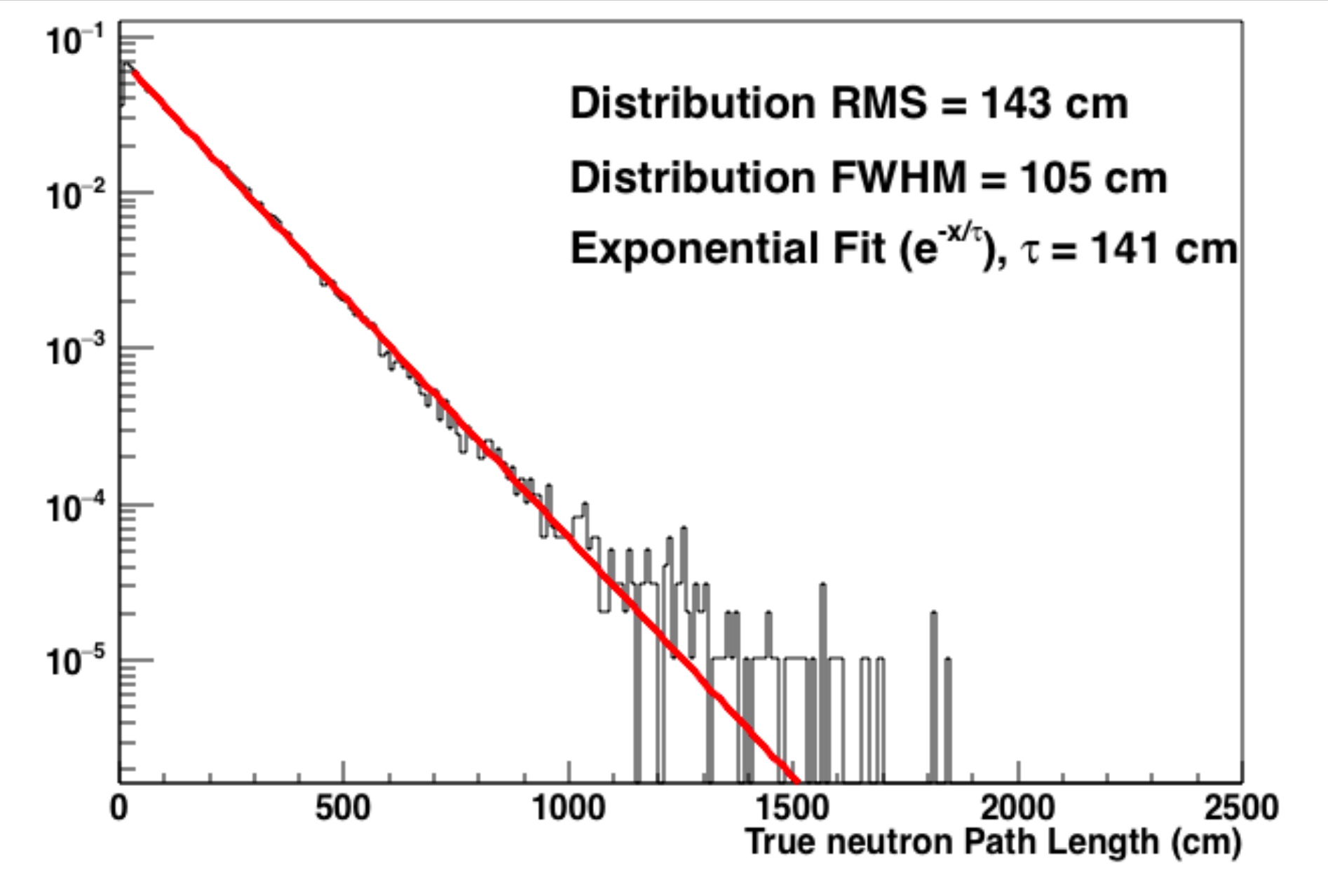}
\vspace{0.4cm}
\caption{The path length of neutrons emitted from events more than 2\,m from the walls of the TITUS detector before capture. This figure was obtained from simulations of \FHC beam events in TITUS.}
\label{fig:neutronpathlength}
\end{minipage}
\end{figure}

\begin{figure}[!tb]\centering
\subfloat[Any $N_\mathrm{neutrons}$]{\label{fig:neutrontaggingwrongsign:all} 
\includegraphics[width=0.33\textwidth]{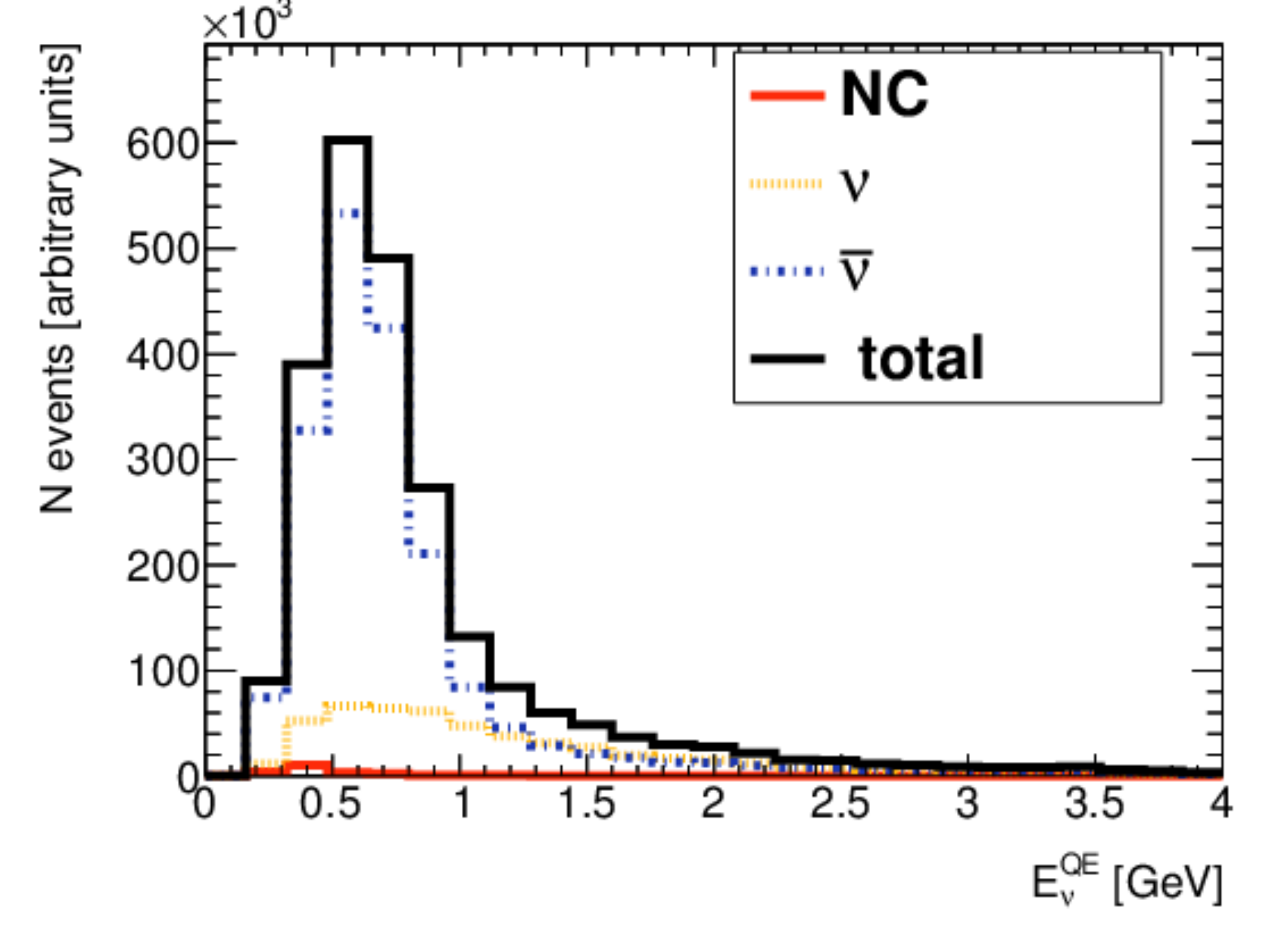}
}
\subfloat[$N_\mathrm{neutrons} = 0$]{\label{fig:neutrontaggingwrongsign:noneutron} 
\includegraphics[width=0.33\textwidth]{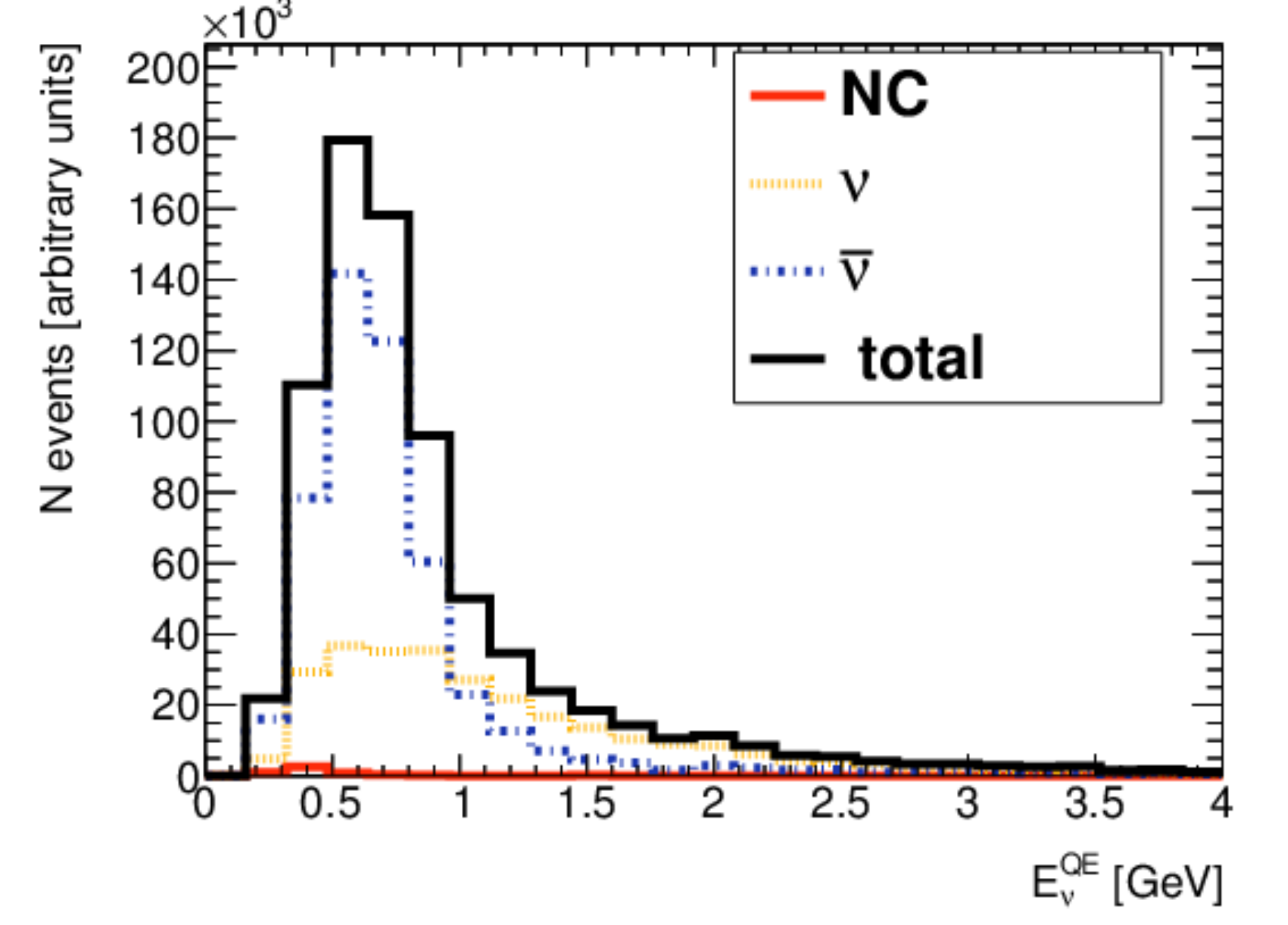}
}
\subfloat[$N_\mathrm{neutrons} \ge 1$]{\label{fig:neutrontaggingwrongsign:hasneutron} 
\includegraphics[width=0.33\textwidth]{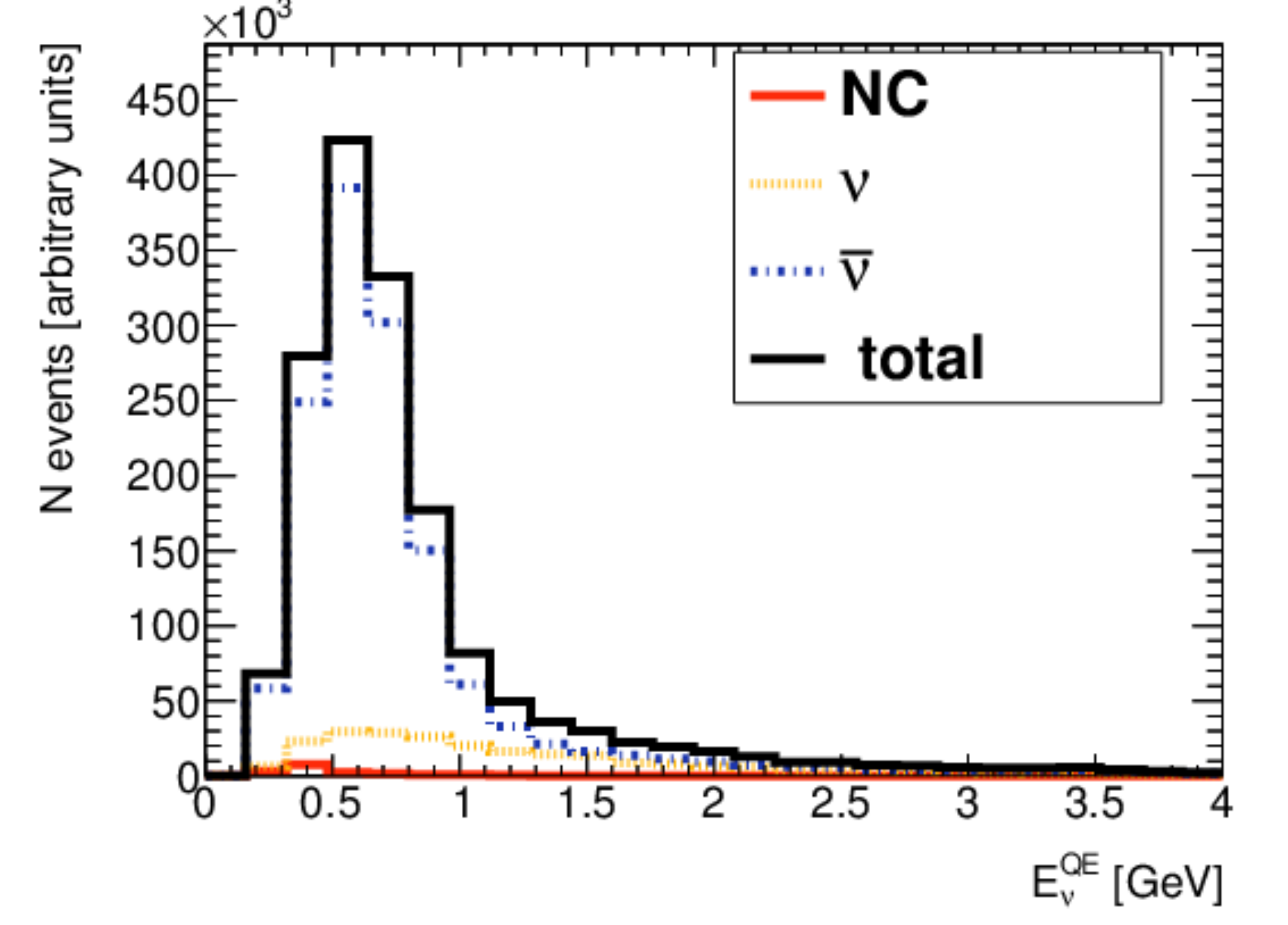}
}
\caption{The E$_{\nu}^{QE}$ distributions for the 1R$\mu$ sample during antineutrino mode running for different selections on the number of neutrons, $N_{\text{neutrons}}$: any neutrons, $N_{\text{neutrons}}$ = 0 and $N_{\text{neutrons}}\geq 1$. \label{fig:neutrontaggingwrongsign}}
\end{figure}

\begin{figure}[!tb]\centering
\subfloat[Any $N_\mathrm{neutrons}$]{\label{fig:neutrontaggingresolution:all} 
\includegraphics[width=0.33\textwidth]{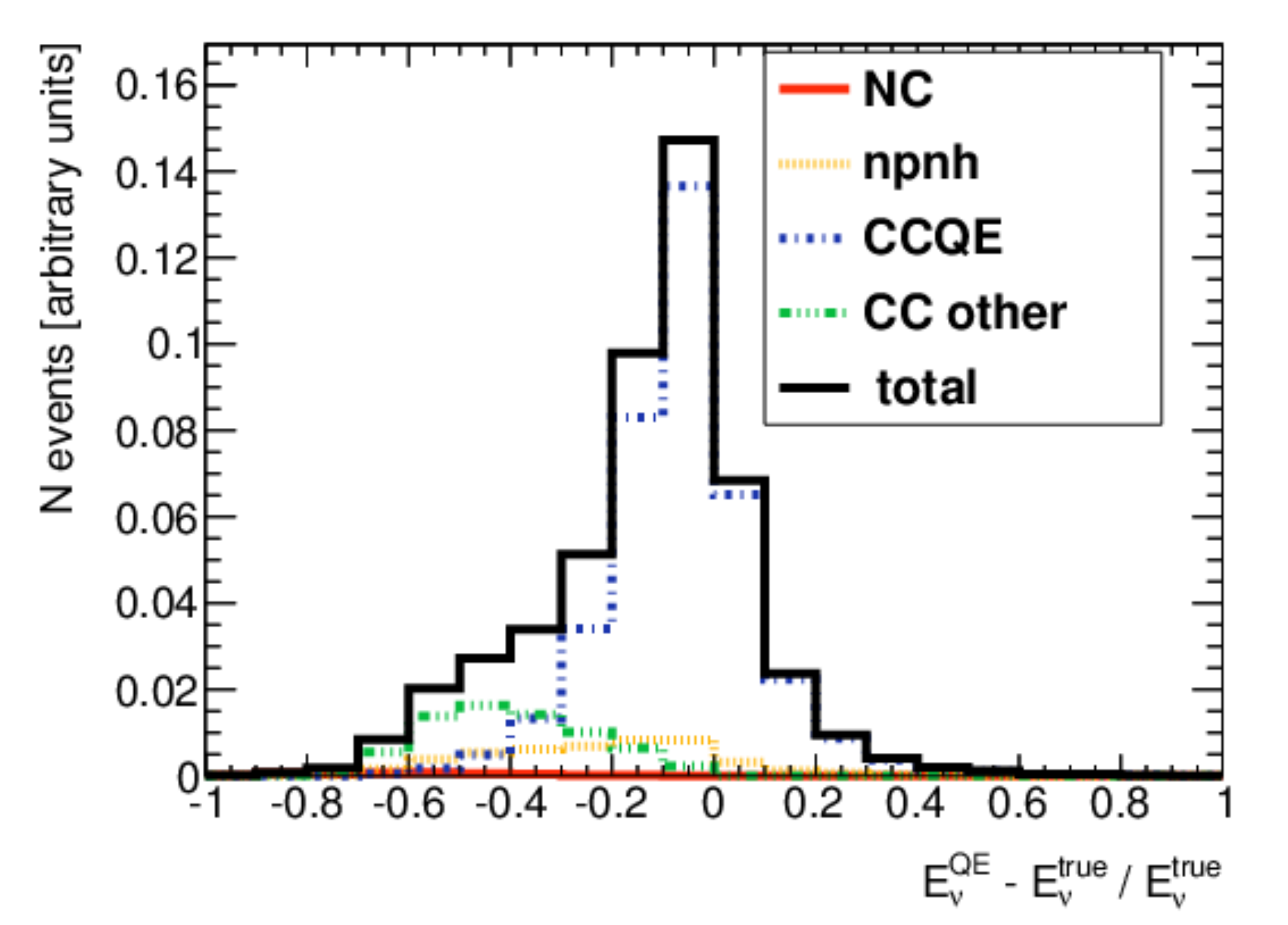}
}
\subfloat[$N_\mathrm{neutrons} = 0$]{\label{fig:neutrontaggingresolution:noneutron} 
\includegraphics[width=0.33\textwidth]{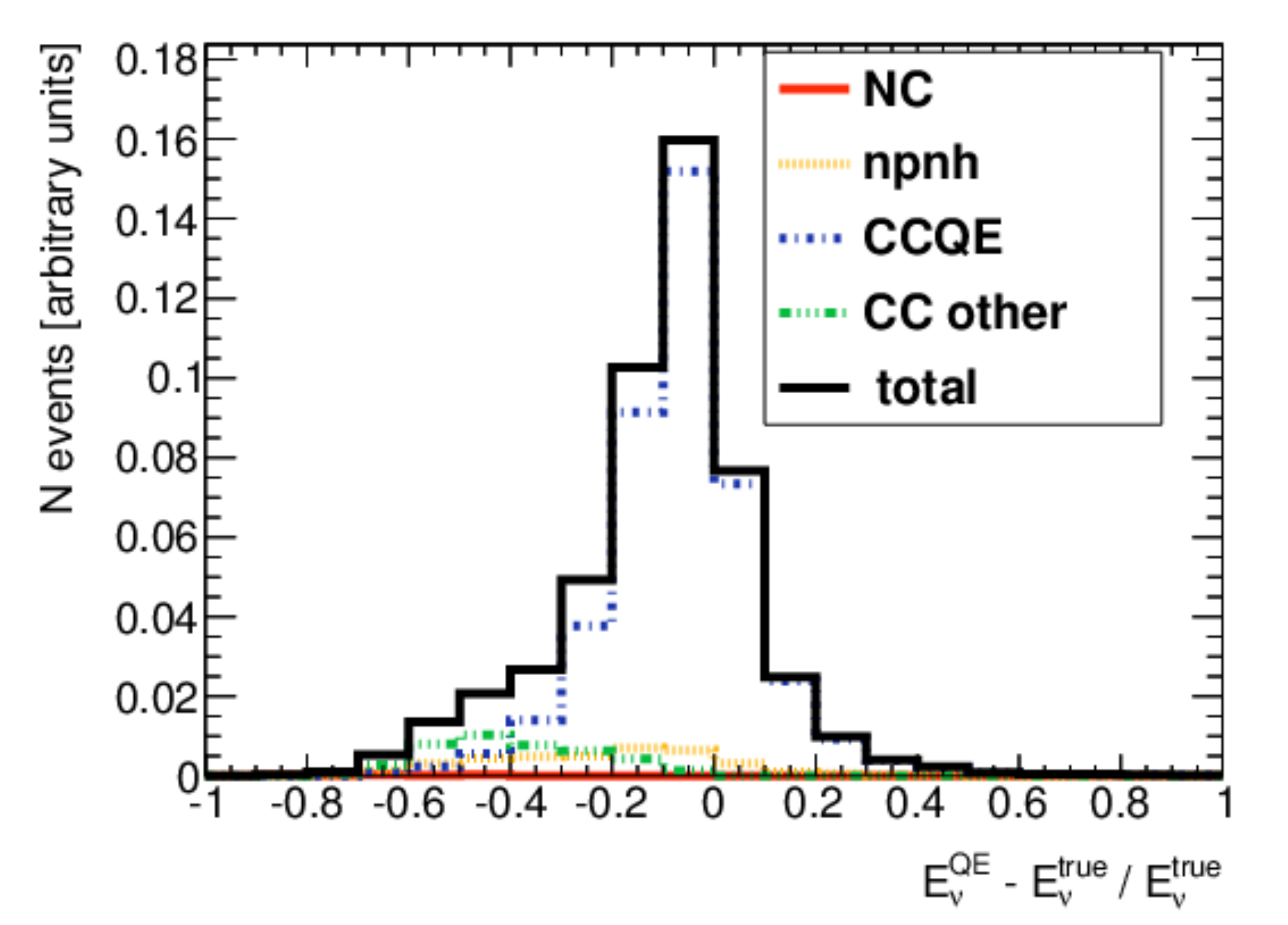}
}
\subfloat[$N_\mathrm{neutrons} \ge 1$]{\label{fig:neutrontaggingresolution:hasneutron} 
\includegraphics[width=0.33\textwidth]{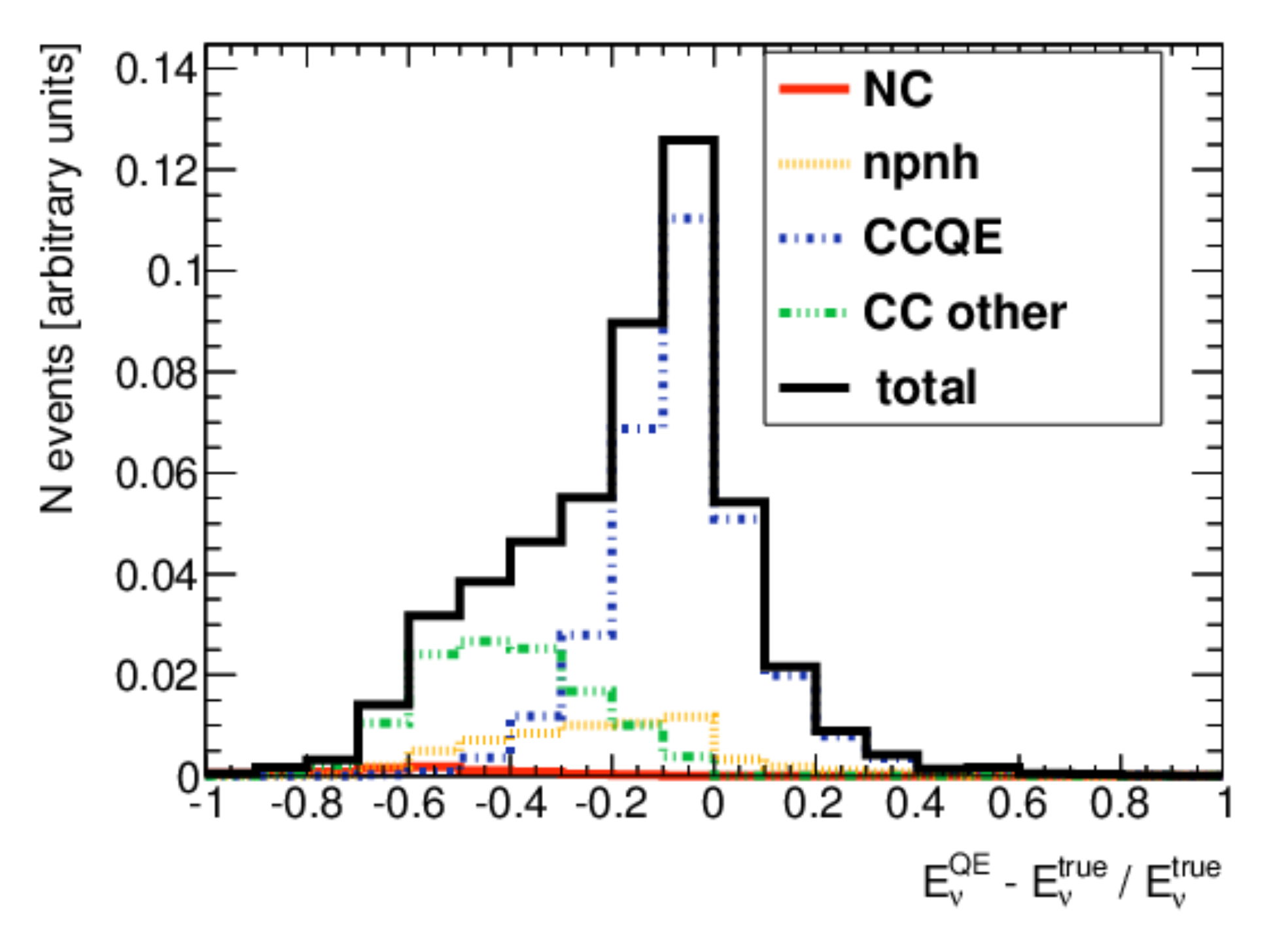}
}
\caption{The E$_{\nu}^{QE}$ distributions for the 1R$\mu$ sample during neutrino mode running for different selections on the number of neutrons, $N_{\text{neutrons}}$: any neutrons, $N_{\text{neutrons}}$ = 0 and $N_{\text{neutrons}}\geq 1$. \label{fig:neutrontaggingresolution}}
\end{figure}

The true neutron kinetic energy after FSI is shown in
Figure~\ref{fig:neutronenergy}, its path length in
Figure~\ref{fig:neutronpathlength} and the reconstructed energy
$E^{QE}_{\nu}$ and its resolution are shown in Figures
\ref{fig:neutrontaggingwrongsign} and
\ref{fig:neutrontaggingresolution} for the $\overline{\nu}$-mode and $\nu$-mode running respectively. The
breakdown of the selected sample by event topology and the
signal-to-background ratio are given in Table \ref{tab:eventtableneutronselection}. 

Imposing a neutron veto in $\nu$-mode running improves the sample
purity from $74 \, \%$ to $83 \, \%$ and the signal-to-background
ratio from 2.9 to 4.9.  The increased purity also improves the energy
resolution, as shown in Figure \ref{fig:neutrontaggingresolution}, by
reducing non-CCQE backgrounds which are incorrectly reconstructed.
Conversely, requiring a neutron in \RHC halves the wrong-sign CCQE
background from 13\% to 7\%, increasing the signal purity from 61\% to
73\% and almost doubling the signal-to-background ratio.  In addition,
reversing the neutron selection/veto provides a background-enhanced
sample which can be used for data-driven background studies.”

\begin{table}[htb]
\caption{\label{tab:eventtableneutronselection}Neutron (N) selection with gadolinium capture and the 1R$\mu$ event selection.}
\centering
\begin{tabular}{ll|rrr|rrr}
\hline\hline
& & \multicolumn{3}{c}{Fraction} & \multicolumn{3}{c}{S/B} \\
%\hline
Beam Mode & Event Topology & Any N & N$=$1 & N=0 & Any N & N$=$1 & N=0\\
\hline
\FHC & $\nu$ CCQE & 0.74 & 0.66 & 0.83 & 2.87 & 1.93 & 4.90\\
& $\nu$ CC-other & 0.23 & 0.30 & 0.16 & 0.30 & 0.43 & 0.19\\
& $\nubar$ CCQE & 0.02 & 0.03 & 0.01 & 0.02 & 0.03 & 0.01\\
& $\nubar$ CC-other & 0.01 & 0.01 & 0.00 & 0.01 & 0.01 & 0.00\\
& NC & 0.01 & 0.01 & 0.00 & 0.01 & 0.01 & 0.00\\
\hline
\RHC & $\nu$ CCQE & 0.13 & 0.07 & 0.27 & 0.15 & 0.07 & 0.37\\
& $\nu$ CC-other & 0.08 & 0.06 & 0.09 & 0.09 & 0.06 & 0.10\\
& $\nubar$ CCQE & 0.61 & 0.73 & 0.59 & 1.54 & 2.67 & 1.42\\
& $\nubar$ CC-other & 0.17 & 0.14 & 0.04 & 0.20 & 0.17 & 0.05\\
& NC & 0.01 & 0.00 & 0.01 & 0.01 & 0.00 & 0.01\\
\hline\hline
\end{tabular}
\end{table}

\subsection{Systematic Uncertainty on Selected Event Sample}
The effects of flux and cross-section systematic uncertainties on the
selected samples at TITUS and Hyper-K have been evaluated.  The flux
systematic uncertainty is based on the error model used by T2K
\cite{T2Kflux}.  Assumptions have been made about the ultimate
performance of the T2K experiment, including the use of replica target
data from the NA61/SHINE experiment.  The prior uncertainty is
estimated to be around 6\%.  There is almost 100\% correlation between
the total fluxes in each running mode between TITUS and Hyper-K
detectors, which leads to a significant cancellation of uncertainties.
There is a 60\% correlation between \FHC and \RHC running modes, which
again will lead to some cancellation of uncertainties.  

The interaction uncertainty model is based on the T2K interaction
uncertainties used as prior input to T2K oscillation analyses as in Ref.~\cite{Abe:2015awa}. This model was modified to include an
uncertainty of 50\% on the normalisation of npnh events and an
estimate of the nucleon FSI uncertainties.  A 2\% uncertainty is set
on the $\nue/\numu$ cross section ratio.  This is assumed to be
anticorrelated between $\nu$ and $\nubar$ interactions (the most
conservative estimate for the $\deltaCP$ measurement).

\begin{figure}[htb]
 \centering
\includegraphics[width=0.66\textwidth]{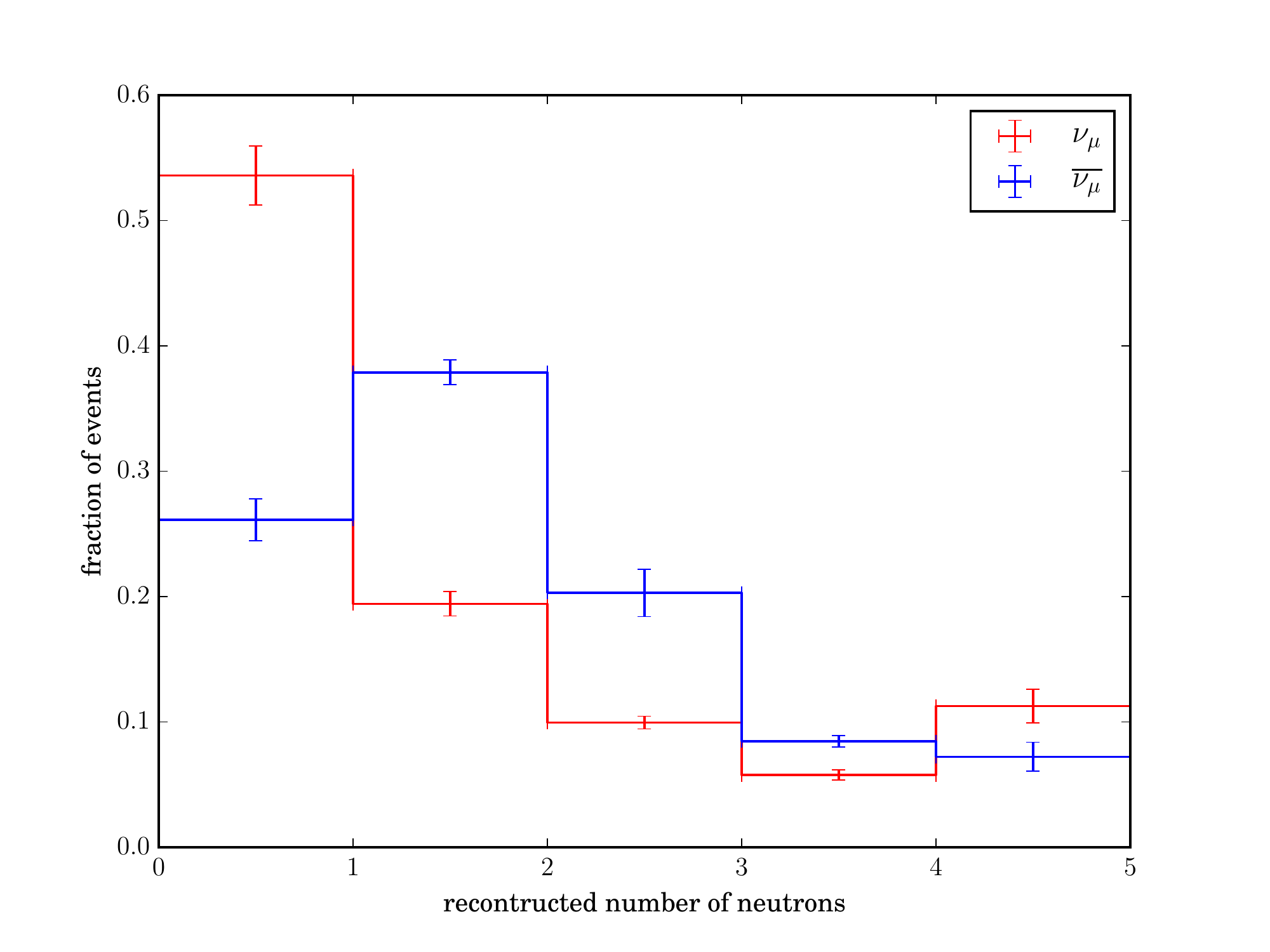}
\caption{ 
The reconstructed neutron multiplicity distribution for $\nu_{\mu}$
and $\overline{\nu}_{\mu}$ events during antineutrino beam mode
running. The error bars show the systematic uncertainty on the shape
of the reconstructed number of neutrons distribution.
}
\label{fig:neutroncountinguncertainty}
\end{figure}

A rigorous evaluation of the nucleon FSI uncertainties for the NEUT
Monte Carlo generator is an ongoing effort within the T2K
collaboration and not available at the time of this study.  The
nucleon FSI errors were evaluated with the GENIE event
generator~\cite{Andreopoulos:2009rq}, which provides reweighting tools
to vary parameters of the FSI model.  A GENIE event samples for the
TITUS \FHC and \RHC neutrino fluxes were generated.  The nucleon mean
free path and probabilities of elastic scattering, multi-nucleon
knockout, pion production and charge exchange processes were varied
within their default uncertainties as provided by GENIE.  The
variation in the GENIE neutron multiplicity due to these uncertainties
was applied to the NEUT neutron multiplicity.  The resulting
uncertainty on the neutron multiplicity distribution assumed in this
analysis is shown in
Figure~\ref{fig:neutroncountinguncertainty}. TITUS itself and other
dedicated experiments will provide a direct experimental constraints
on the neutron multiplicity for neutrino-Oxygen interactions. The
current uncertainties only refer to the NEUT generator as used in this
analysis. Differences with other generators may be larger.

%The uncertainty on the $\nu-\nubar$ cross section ratio was set to 20\%.

The effect of the systematic uncertainty on the total number of events
observed is shown in Table~\ref{tab:eventratesyst}.  In Hyper-K the
event rate in the $\selE$ sample is used and oscillation is taken into
account.  The total systematic uncertainty on the absolute event rate,
without including a near detector, is $\sim$16\%.  This is dominated
by the large neutrino-nucleus cross section uncertainty.  Including
the measurements of TITUS the total systematic error is reduced to
$\sim$3-4\%.  The TITUS statistical error is negligible.  The
statistical error (dominated by the event rate at Hyper-K) is
$\sim$2\, \%.

\begin{table}[htb]
\caption{\label{tab:eventratesyst} The effect of each systematic uncertainty
  on the Hyper-K appearance event samples.  The size of each
  systematic is shown both before and after taking into account the
  TITUS near detector information.  
}  \centering
\begin{tabular}{lrrrr}
  \hline\hline
  & \multicolumn{2}{c}{$\nu$ mode}
  & \multicolumn{2}{c}{$\overline{\nu}$ mode} \\
Systematic & No ND & With TITUS & No ND & With TITUS \\
\hline
Interaction Syst. & 13.6 & 1.4 & 11.5 & 1.0 \\
Flux Syst. & 7.5 & 0.8 & 8.0 & 1.0 \\
Detector + FSI + PN & 3.0 & 2.4 & 2.1 & 1.5 \\\hline
Total Syst. & 16.1 & 3.9 & 14.5 & 3.5 \\\hline
Statistical & 1.8 & 1.8 & 1.9 & 1.9 \\\hline\hline
Stat. + Syst. & 16.2 & 4.3 & 14.6 & 4.0 \\
\hline\hline
\end{tabular}
\end{table}
%|------------+---------------+--------------+---------------+--------------|
%| Systematic | FHC before ND | FHC after ND | RHC before ND | RHC after ND |
%|------------+---------------+--------------+---------------+--------------|
%| syst       | 0.1607        | 0.0386       | 0.1446        | 0.0354       |
%| xsec       | 0.1361        | 0.0137       | 0.1145        | 0.0101       |
%| flux       | 0.0754        | 0.0080       | 0.0796        | 0.0098       |
%| detector   | 0.0296        | 0.0235       | 0.0213        | 0.0148       |
%| stat       | 0.0183        | 0.0184       | 0.0190        | 0.0195       |
%| stat+syst  | 0.1595        | 0.0446       | 0.1408        | 0.0403       |
%|------------+---------------+--------------+---------------+--------------|

\subsection{Sensitivity studies}
\label{subsec-basicsensitivity}
\subsubsection{Fit Method}
\label{sec:basicfitmethod}
The selected event samples are binned in 2D where one axis corresponds
to the true neutrino energy and the second axis corresponds to the
observable bin.  The true $\enu$ bin edges are: (0.2, 0.4, 0.5, 0.6,
0.7, 1.0, 1.5, 2.5, 3.5, 5.0, 7.0, 30.0)\,GeV.  These are chosen to
match the input flux error matrix binning.  The observable bins are
combinations of:
\begin{itemize}
\item selected sample (2 options: $\selE$ or $\selMu$);
\item detector (2 options: TITUS or Hyper-K);
\item beam mode (2 options: \FHC or \RHC);
\item reconstructed neutrino energy (6 options with bin edges: 0.2, 0.4, 0.6, 0.7, 1.0, 1.5, 30.0\,GeV)
\end{itemize}
for a total of 48 bins.
Optionally the observable binning could include:
\begin{itemize}
\item the number of tagged neutrons, $N$, (either 2 bins for ``binary'' tagging $N = 0$, $N \ge 1$ or 3 bins for ``counting'' $N = 0$, $N = 1$, $N \ge 2$),
\end{itemize}
allowing up to 144 bins when both neutron counting and reconstructed
neutrino energy information are used.

To simulate observed event rate vectors in the far detector after
oscillation, the following weights are applied:
\begin{equation}
E^{\mathrm{selected}}_{\textrm{with osc.}}(i, {\bm\theta})
    = 
      \sum_{j}^{\enutrue\textrm{ bins}}{ 
           \sum_{k}^{\nu\textrm{ }\mathrm{type}} {
               \epsilon(i, j, k) %%%E(i, j, k) 
                                    %%%            \frac{
                                                 %     E(i, j, \nue) P(\nue \to \nu_{k} \lvert {\bm\theta})
                                                 %   + E(i, j, \numu) P(\numu \to \nu_{k} \lvert {\bm\theta})
                                                 \sum_{l}^{\nu\textrm{ }\mathrm{type}}{ E(i, j, l) P(\nu_{l} \to \nu_{k} \lvert {\bm\theta}) }
                                       %%%         }{
                                        %%%              E(i, j, k)
                                        %%%        }
           }
     },
\label{eqn:nevents}
\end{equation}

where $i$ corresponds to observable bins, $j$ corresponds to bins in
$\enutrue$, $k$ corresponds to neutrino flavour and $\bm{\theta}$ are
the six oscillation parameters.  $E(i,j,k)$ is the expected number of
events in each bin before any selection is applied.
%%%Note that the
%%%appearance of $E(i,j,k)$ in both the numerator and denominator despite
%%%the fact that it should cancel is due to practical constraints in the
%%%implementation.  
$P(\nu_l \to \nu_k | {\bm\theta})$ is the 3-flavour neutrino
oscillation probability calculated with
GLoBES~\cite{Huber:2004ka}, \cite{Huber:2007ji}.  Each bin is
multiplied by the selection efficiency, $\epsilon(i, j, k)$, to give
the distribution after selection.  The observed number of events is
calculated by summing over neutrino type and true neutrino energy.

To estimate the systematic uncertainties, Equation~\ref{eqn:nevents}
is modified with one additional parameter per observable bin, $w_{i}$,
\begin{equation}
E'(i, {\bm\theta}, w_{i}) \equiv E^{\mathrm{selected}}_{\textrm{with
osc. and syst.}}(i, {\bm\theta}, w_{i}) = w_{i}
E^{\mathrm{selected}}_{\textrm{with osc.}}(i, {\bm\theta}).
\end{equation}

The standard Poisson log-likelihood is defined as
\begin{equation}
-2\ln(\lambda({\bm\theta}, {\bm w})) = 2 \sum_{i}{
                                            E'(i, {\bm\theta}, w_{i}) - O(i) + O(i) \ln \frac{O(i)}{E'(i, {\bm\theta}, w_{i})}
                                        }.
\end{equation}
where $O(i)$ is the observed events in bin $i$.  The observed
distribution is set to the Asimov dataset~\cite{Asimov} (the most
likely data with all parameters at their nominal values).

A Bayesian method is used to calculate the measured values of the
oscillation parameters given the (fake) observed data.  Gaussian
constraints are set on the other oscillation parameters based on the
current PDG values as shown in Table~\ref{tab:oscpriors} with
$\theta_{23}$=45$^\circ$.  The prior on $\deltacp$ is set to
be uniform in $\deltacp$ space.  A flat prior is also set on
$\sin^{2}2\theta_{13}$, as the PDG value is dominated by the reactor
constraint, and it is the goal of Hyper-K to measure $\deltaCP$
independently of the constraint from the reactor experiments.  The
prior on the systematic parameters, $w_i$, is assumed to be a
multivariate Gaussian. A Monte Carlo method was used to vary the
underlying systematic parameters in Table~\ref{tab:eventratesyst} and
calculate the covariance of the values $E(i,
{\bm\theta}^{\textrm{nominal}}, 1)$.  This combined likelihood and
prior is input into a Markov Chain Monte Carlo tool to estimate the
posterior distribution on the oscillation parameters, ${\bm\theta}$.

\begin{table}[htb]
\caption{\label{tab:oscpriors}The prior uncertainties on the oscillation parameters. They are Gaussian unless otherwise stated.}
\centering
\begin{tabular}{ll}
\hline\hline
Parameter & Nominal value and Prior Uncertainty\\
\hline
$\deltacp$ & 0.0, uniform in $\deltacp$\\
$\sin^{2}2\theta_{13}$ & 0.095, uniform in $\sin^{2}2\theta_{13}$\\
$\sin^{2}2\theta_{23}$ & 1.00 $\pm$ 0.03 ($\approx$ $\sin^{2}2\theta_{23} > 0.95$ at 90\% CL)\\
$\sin^{2}2\theta_{12}$ & 0.857 $\pm$ 0.034\\
$\Delta m^{2}_{32}$ & (2.32 $\pm$ 0.10) $\times$ 10$^{\text{-3}}$ eV$^{\text{2}}$\\
$\Delta m^{2}_{21}$ & (7.5 $\pm$ 0.2) $\times$ 10$^{\text{-5}}$ eV$^{\text{2}}$\\
\hline\hline
\end{tabular}
\end{table}

The results shown are evaluated for a 1:3 POT ratio for $\nu$ and
$\nubar$ running mode.  The $\deltaCP$ precision is insensitive to the
ratio of $\nu-\nubar$ running over a wide range of running ratios.

\subsubsection{Results}

\paragraph{Uncertanties on the number of selected events in the far detector:}
%Effect of the Far Detector $\numu$ Sample on the $\deltaCP$ Measurement}
%\Cref{fig:potexposureSummary} demonstrates only a small performance
%improvement over the Hyper-K only configuration due to the ND280 near
%detector for the analysis in section~\ref{sec-basicstrategy}.  This is
%understood to be due to the power of the far detector single ring muon
%sample to constrain the single ring electron sample (which has
%sensitivity to $\delta_{CP}$).  
\Cref{tab:effectofhkmuonconstraint} shows the uncertainty 
on the far detector $\nu_{e}$ sample given constraints from different
combinations of near detector and far detector information.  
\begin{table}[htb]
\begin{center}
\begin{tabular}{l|ll|ll}
  \hline\hline
  & \multicolumn{2}{|c|}{\FHC} & \multicolumn{2}{|c}{\RHC} \\ 
Near Detector & w/o HK 1R$\mu$ & with HK 1R$\mu$ & w/o HK 1R$\mu$ & with HK 1R$\mu$\\
\hline
No near detector & 15.3\% & 6.0\% & 14.7\% & 4.7\%\\
ND280 & 7.9\% & 5.5\% & 6.6\% & 4.1\%\\
TITUS & 4.4\% & 3.8\% & 3.9\% & 3.3\%\\
\hline\hline
\end{tabular}
\caption{The expected uncertainty on the number of selected events in
  the far detector single ring electron sample.  The uncertainty is
  shown given constraints from the TITUS and ND280 near detectors and
  the Hyper-K single ring muon sample.
  \label{tab:effectofhkmuonconstraint}
}
\end{center}
\end{table}

\begin{figure}[htb]
\begin{minipage}{0.5\textwidth}
\centering
\includegraphics[width=1\linewidth]{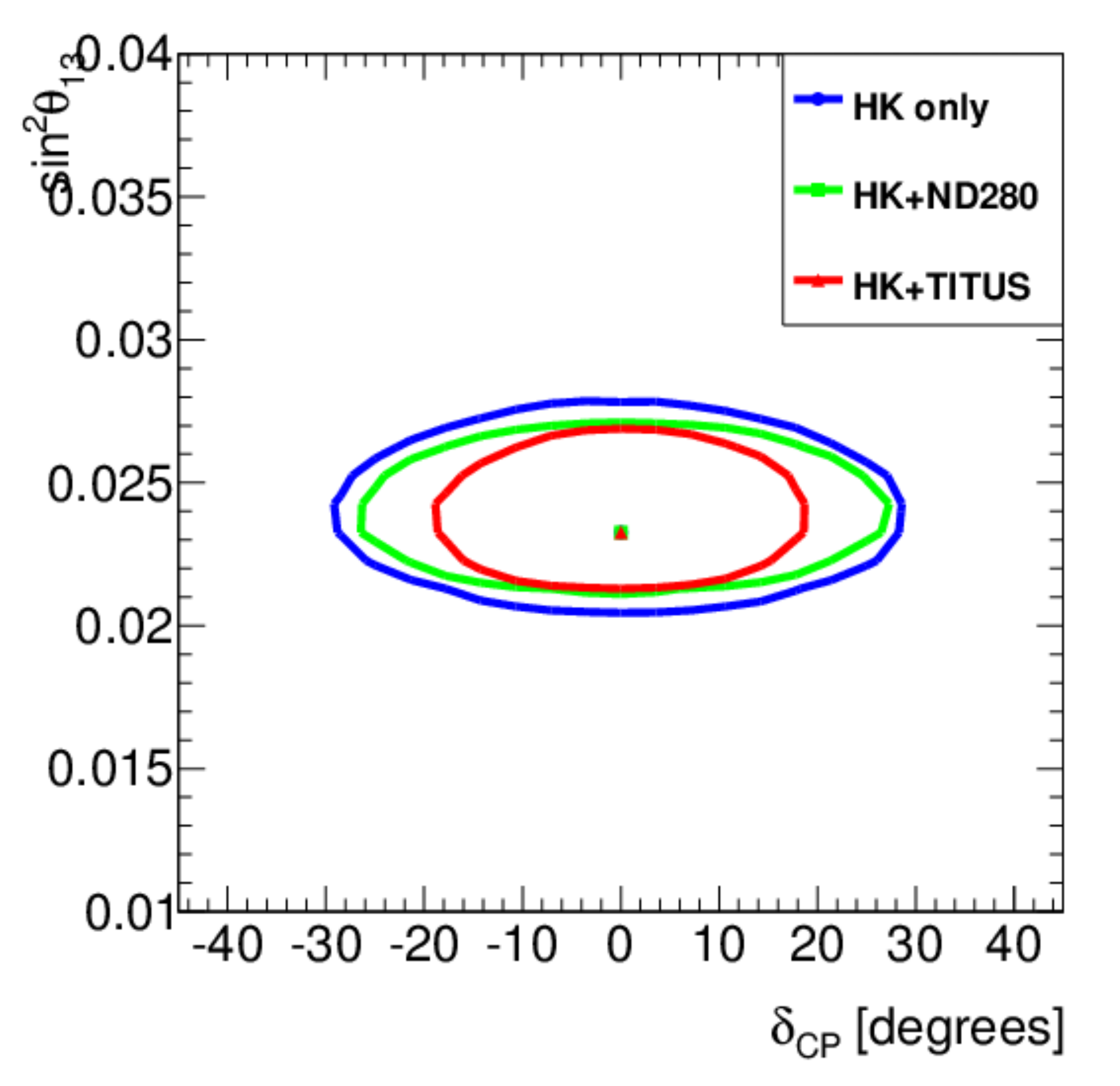}
\caption{\label{fig:contourdetcomparison}$\deltaCP$-sin$^2{2\theta_{13}}$ 90\% contour comparing the nominal TITUS+Hyper-K with ND280+HK and Hyper-K only.}
\end{minipage}
\begin{minipage}{0.5\textwidth}
\centering
\includegraphics[width=1\linewidth]{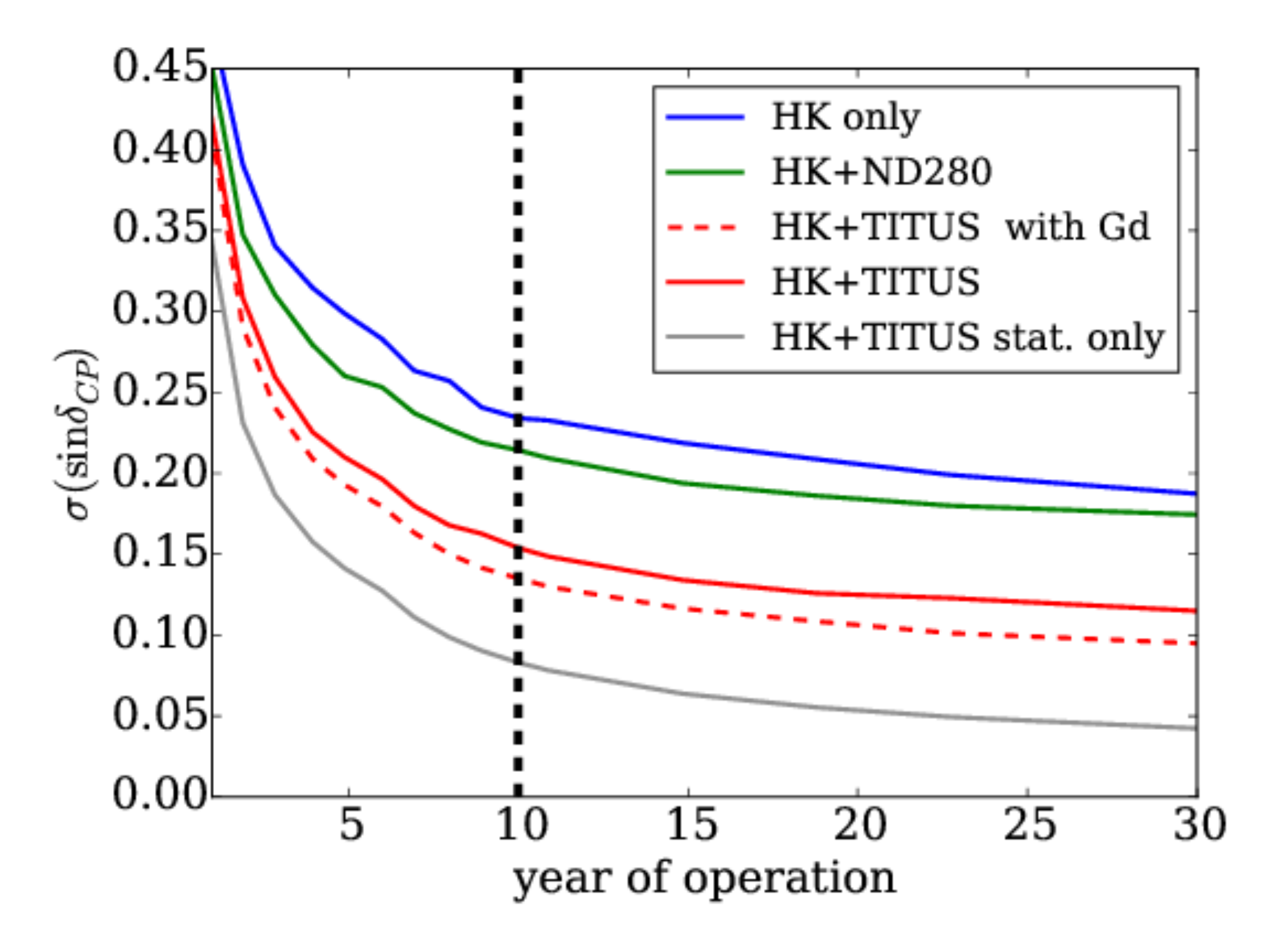}
\caption{\label{fig:potexposureSummary}$\deltaCP$ performance as a function of POT exposure for different
configurations. HK$+$TITUS with Gd means both detectors are Gd-doped.}
\end{minipage}
\end{figure}

When excluding the far detector $\nu_{\mu}$ sample from the fit, both
the TITUS and ND280 near detectors provide a significant reduction in
uncertainty in the number of expected events in the far detector
electron sample.  However, the constraint on the $\nu_{e}$ sample from
far detector $\nu_{\mu}$ sample uncertainty is very strong and changes
the balance when included in the fit. This is based on the assumptions
on the fit made in the subsection~\ref{sec:basicfitmethod}.  However,
the ND280 sample used in this analysis does not not contain a
$\nu_{e}$ sample and one should expect better performance from an
ND280 sample that includes a direct measurement of the un-oscillated
intrinsic $\nue$ spectrum.
% is at a similar level to the constraint
%provided by the ND280.  
%The ability of the far detector muon sample to
%constrain the electron sample depends on the assumptions made on the
%relative detector uncertainties for the two selections, the prior flux
%uncertainty and the correlation between the $\nue$ ($\nuebar$) and
%$\numu$ ($\numubar$) fluxes, and the uncertainty on extrapolating
%$\numu$-nucleus interaction to $\nue$-nucleus interactions.  
%Hence the
%ND280 does not provide a significant improvement in the $\deltaCP$
%sensitivity over the Hyper-K only case.  
\paragraph{Performance:}
%|-------------+-------------------+-------------|
%| mode        | sigma(sindeltacp) | dcp degrees |
%|-------------+-------------------+-------------|
%| HK only     |             0.234 |       13.53 |
%| HK+ND280    |             0.214 |       12.36 |
%| HK+TITUS    |            0.1538 |        8.85 |
%| HK+TITUS+Gd |            0.1362 |        7.83 |
%| stat only   |            0.0866 |        4.96 |
%|-------------+-------------------+-------------|
Figure~\ref{fig:contourdetcomparison} shows the expected 90\% credible
interval on $\deltaCP$ and $\sin^2{2\theta_{13}}$.  The precision on
$\deltaCP$ assuming $\deltaCP=0$ is 0.15 radians (or 8.9$\degree$).
With no systematic uncertainties, the statistical precision is $0.08$
(4.6$\degree$).
%This is comparable to the precision achieved in the Hyper-K
%physics potential studies~\cite{hkptep}.  
The precision as a function of exposure is shown in
Figure~\ref{fig:potexposureSummary}.  The Hyper-K with a ND280
analysis achieves a precision of 0.21 radians (12.4$\degree$).
Inclusion of the TITUS near detector gives an improvement of $28 \%$
compared to Hyper-K plus ND280. Further improvements may be achieved
by fitting the ND280 and TITUS data samples simultaneously.

In addition, there is a significant improvement when including neutron
tagging information.  A $\deltaCP$ precision of 0.14 radians
(7.8$\degree$) is achieved when including whether a neutron was tagged
or not in an event.  Inclusion of gadolinium in both the near and far
detector gives a 11\% improvement in the performance of the
experiment.  Adding the ability to count the multiplicity of neutrons
(the ``counting'' option) does not give any significant increase in
performance over ``binary'' neutron tagging, based on the current
resolutions.

In summary, the addition of the TITUS Water Cherenkov near detector
provides significant improvement in the $\deltaCP$ precision and
therefore the discovery reach of the Hyper-K experiment.  Using a
systematic error model based on the experience with T2K, TITUS can
significantly reduce the systematic uncertainties on the $\deltaCP$
measurement.  The addition of Gd to both near and far detectors allows
the selection of higher purity samples, the use of which improves the
$\deltaCP$ precision by a further 11\%.

\section{Full reconstruction sensitivity studies}
\label{sec-software}
The simulation and reconstruction software for TITUS is based on the
WChSandBox simulation package~\cite{Anghel:2015xxt} developed by the
ANNIE collaboration and adapted to the TITUS configuration, together
with a global reconstruction package developed for TITUS.  Both
packages are described in the following sections.  A selection is
devised for the oscillation analysis and, using the reconstructed
data, the CP sensitivity results are obtained using the VaLOR fitting
method~\cite{valor}. Simulation and reconstruction of the MRD are in
progress as discussed in Sec.~\ref{subsec-mrdsim}, and will be added
in the future. The current results are based on the WC tank only.

\subsection{WC simulation}
\label{subsec-wch}
The neutrino interactions are simulated using the both the
$\nu_{\mu}$- and $\overline{\nu}_{\mu}$-mode beam fluxes for the
2036\,m baseline (see section~\ref{sec-beam}).
%The neutrino fluxes are plotted as a function of energy for each
%flavor and horn current configuration.
The neutrino interaction generators used in the studies below are
NEUT v5.3.3 and GENIE v2.8.0.

Each generator takes a flux histogram and generates a sample according
to the event rate for that particular flavour and horn configuration,
assuming interactions are only on water, H$_{2}$O.  Only the $^{16}$O
and $^1$H are simulated.  The presumed percentage of gadolinium in the
water is negligible in terms of the overall event rate and so is
ignored in our interaction simulation.
%The generator output is then converted into a text format to be fed
%into the detector simulation.

The detector simulation is performed with a fast simulation package
called WChSandBox (or ``SandBox''). The SandBox package is
Geant4-based, with a single volume of 99.8\% water and 0.2\%
gadolinium by mass.  SandBox reads in the generator output and
propagates the primary tracks, simulating secondary interactions as it
does so. Some features of SandBox include:
\begin{itemize}
 \item a track-matching algorithm to associate the multiple Geant
 outputs that are produced when scattering occurs; 
\item separate classification of optical photons and ``particles''
 (including high energy gamma rays);
\item detailed information pertaining to neutrons, including the number of
 neutrons produced, number of neutrons captured, the time and position
 of the captures, the daughter nucleus produced and energy released
 from a neutron capture.
\end{itemize}
The SandBox outputs two files, one that formats the generator
information into a ROOT file and the other that contains the output of
the detector simulation. The simulation output files are quite large,
owing to tracking of individual optical photons. To partially mitigate
this, optical photons are prescaled by 50\% in SandBox. Even so, a
typical simulated event is about 10\,MB in size.

%Examples of SandBox simulated events can be seen in
%Figure~\ref{fig:SandBoxEvents}; the events displayed are a (a) CCQE
%single-ring $\mu$ event and a (b) NC$\pi^{0}$ two-ring event.
%
%\begin{figure}[tpb]
%\begin{center}
%\includegraphics[width=0.45\linewidth]{figs/SandBox-CCQE.png}
%\includegraphics[width=0.45\linewidth]{figs/SandBox-NCpi0.png}
%\caption{Two SandBox simulated events. On the left there is a CCQE single-ring $\mu$ event, whilst on the right there is a NC$\pi^{0}$ two-ring event.}
%\label{fig:SandBoxEvents}
%\end{center}
%\end{figure}
%
To complete the water Cherenkov part of the simulation, photosensors
are also required.
%As can be seen in the two events in
%Figure~\ref{fig:SandBoxEvents}, 
SandBox does not have this capacity enabled; instead, the sensors are
added afterward as a ``mask'' of optical photons that hit the detector
walls. This mask includes user-set details of the wavelength-dependent
quantum efficiency, as well as the size, shape, and timing resolution
of the photosensor being simulated. The photosensors for each detector
surface can be set individually, and ``hybrid'' surfaces, containing
more than one type of sensor, are also possible.
%Figure~\ref{fig:TITUS-mask} shows the TITUS detector with 20\%
%photocathode coverage.
%
%\begin{figure}[tpb]
%\begin{center}
%\includegraphics[width=0.8\linewidth]{figs/TITUS-mask.png}
%\caption{The TITUS detector shown with a photosensor mask. The mask shown has 20\% photocathode coverage with 10\,inch circular PMTs.}
%\label{fig:TITUS-mask}
%\end{center}
%\end{figure}

Once the photosensor simulation has been applied, new output ROOT
files are stored. These files are considerably smaller, as the
intermediate information (e.g., individual optical photons in a
shower) has been discarded. These ROOT files are then passed to the
reconstruction algorithms described in section~\ref{subsec-highE}.

\begin{figure}[htpb]\centering
\begin{tabular}{cc}
\includegraphics[width=0.45\textwidth]{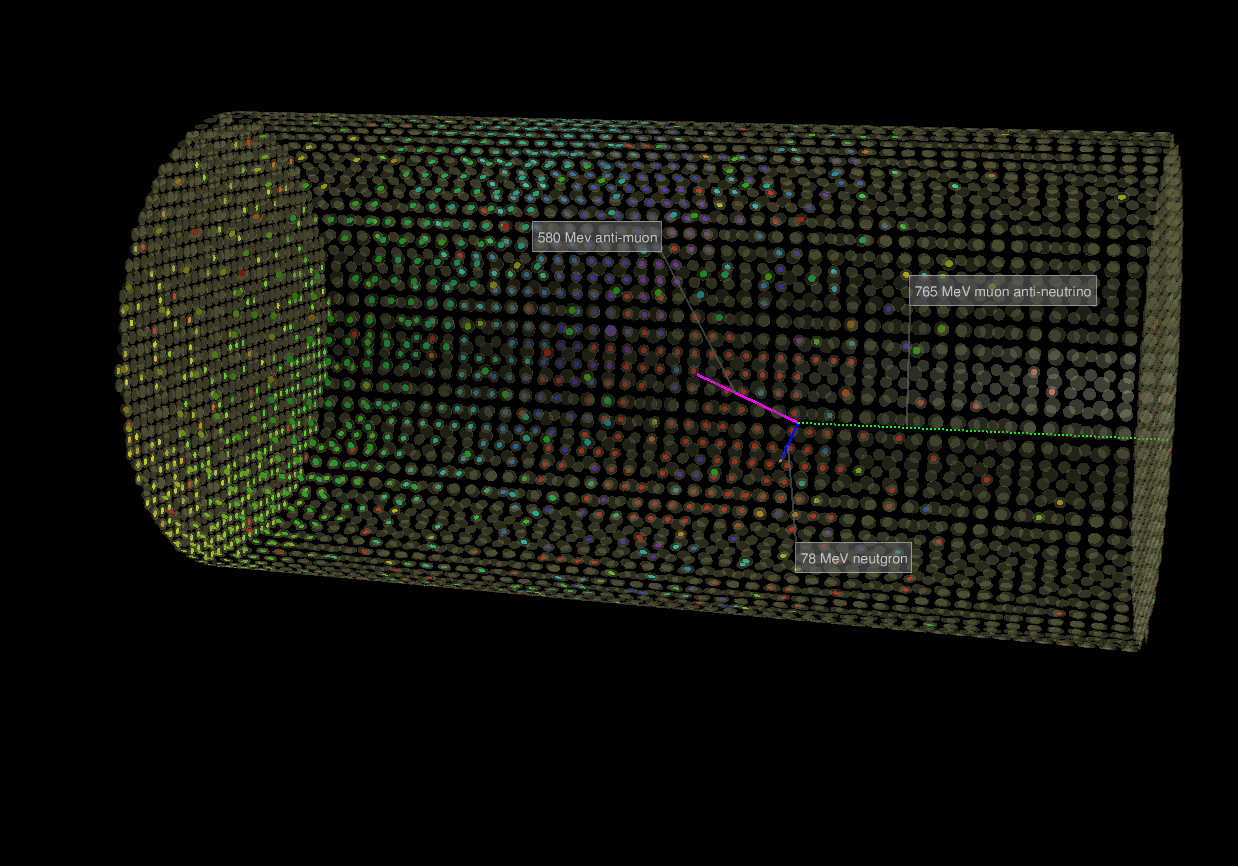}&
\includegraphics[width=0.45\textwidth]{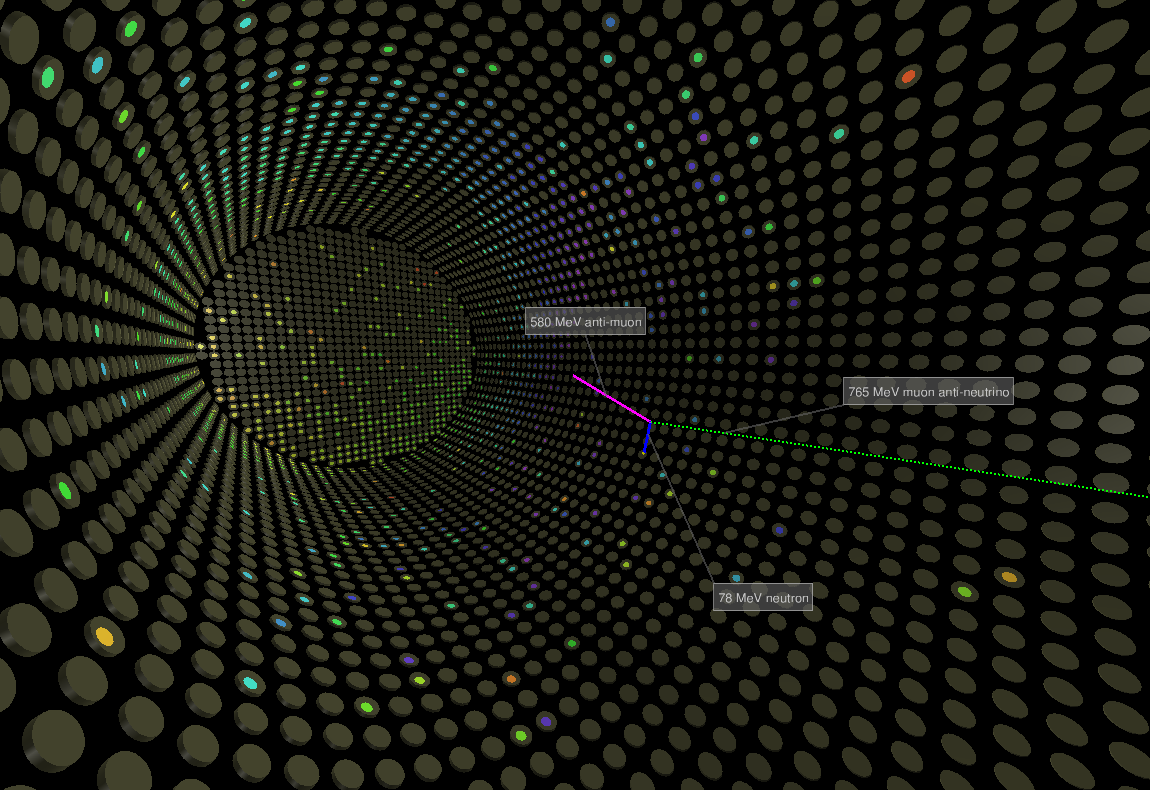}\\
\includegraphics[width=0.45\textwidth]{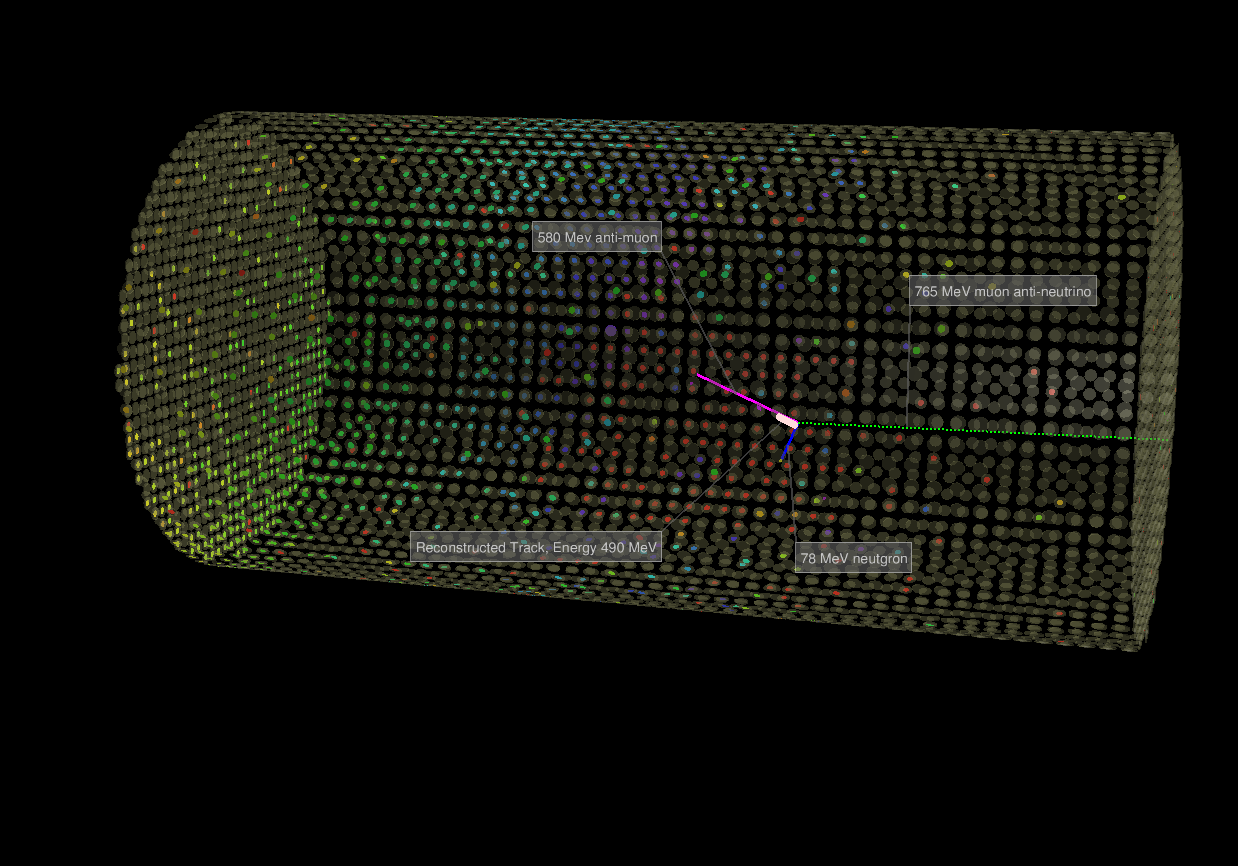}&
\includegraphics[width=0.45\textwidth]{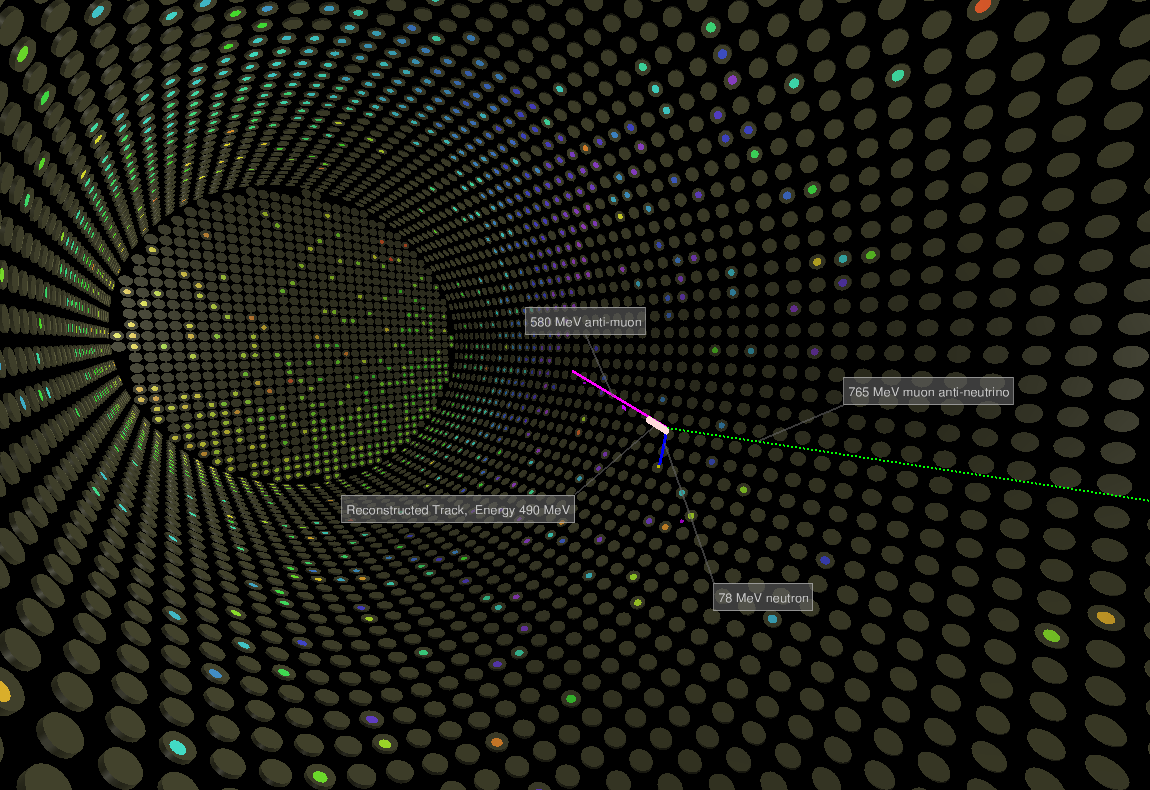}\\
\includegraphics[width=0.45\textwidth]{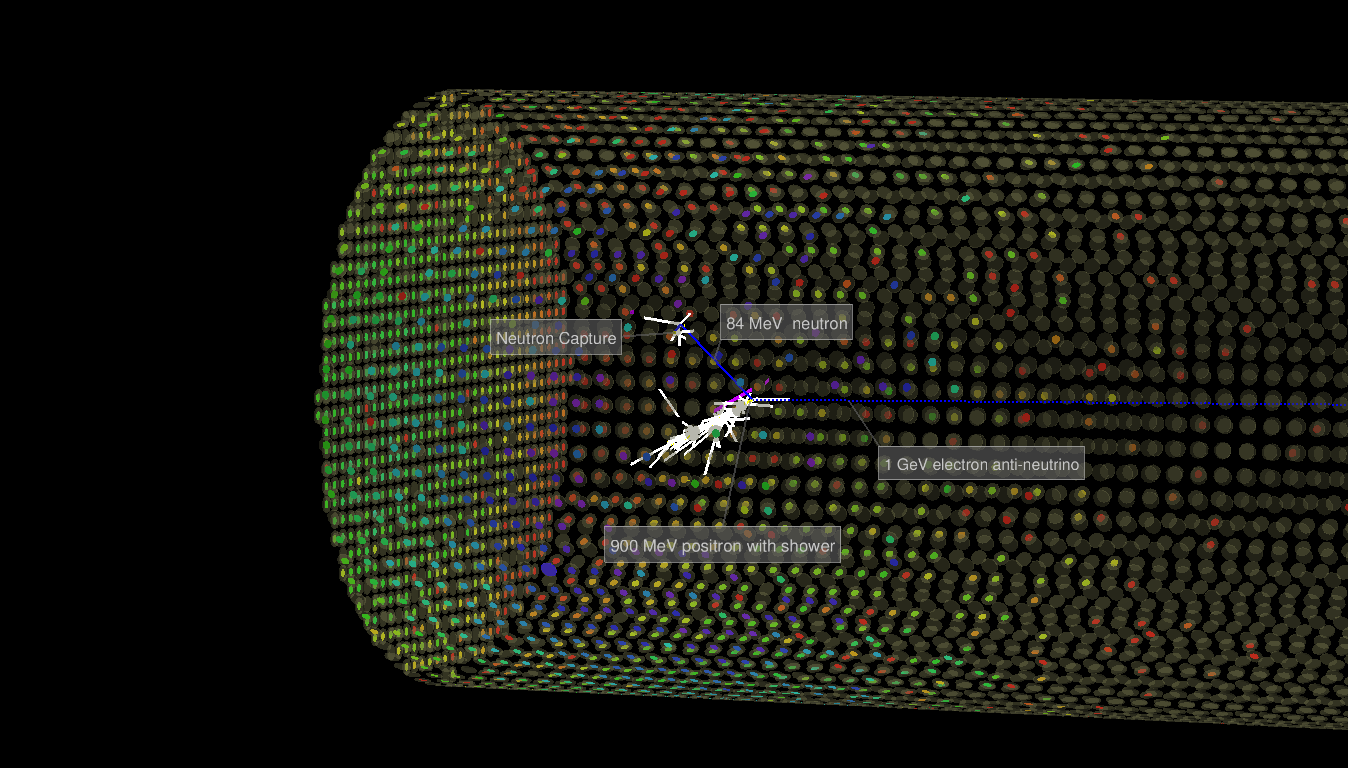}&
\includegraphics[width=0.45\textwidth]{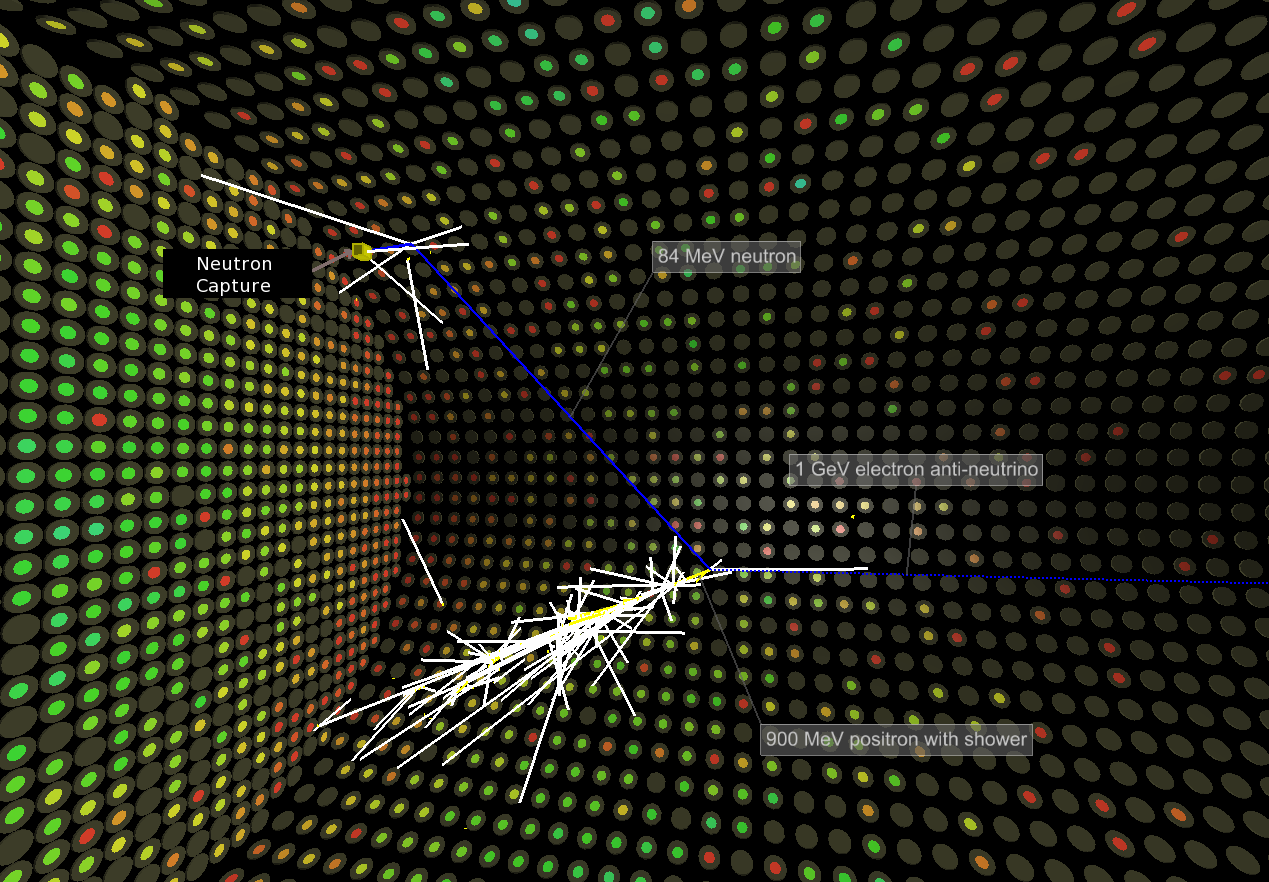}\\
\end{tabular}
\caption{
Event displays based on simulated events, shown in 3D, for
antineutrino events. Only the WC tank is shown. The first row shows an
antineutrino event in two views, the second row shows an antineutrino
event in two views with a reconstructed track, and the third row an
antineutrino event in two views with neutron capture. White lines are
high energy photons, circles represent PMTs and colours indicated
charge deposited}
\label{fig:eventDisplay2}
\end{figure}

%Figures~\ref{fig:eventDisplay1} and \ref{fig:eventDisplay2} show event
%displays for CCQE and CC1$\pi$ events in the TITUS tank. Note the
%neutron capture event in the bottom row of
%Figure~\ref{fig:eventDisplay2}.
Figure~\ref{fig:eventDisplay2} show event displays in the TITUS
tank. Note the neutron capture event in the bottom row.
%
%\begin{figure}[htpb]\centering
%\begin{tabular}{cc}
%\includegraphics[width=0.4\textwidth]{figs/eventdisplays/CCQE1.png}&
%\includegraphics[width=0.4\textwidth]{figs/eventdisplays/CCQE2_3D.png}\\
%\includegraphics[width=0.4\textwidth]{figs/eventdisplays/CCQE3_3D.png}&
%\includegraphics[width=0.4\textwidth]{figs/eventdisplays/CCQE4_3D.png}\\
%\includegraphics[width=0.4\textwidth]{figs/eventdisplays/SinglePion1_3D.png}&
%\includegraphics[width=0.4\textwidth]{figs/eventdisplays/SinglePion2_3D.png}\\
%\end{tabular}
%\caption{
%Event displays based on simulated events, shown in 3D, for neutrino
%events. Only the WC tank is shown. The first two rows show CCQE
%events, the third CC1$\pi$ events. White lines are high energy
%photons, circles represent PMTs and colours indicated charge
%deposited}
%\label{fig:eventDisplay1}
%\end{figure}

The TITUS event display program is based on the ``EVE'' package in
ROOT. It provides a three-dimensional view of events in the TITUS tank
as well as a two-dimensional ``unrolled'' view of the cylindrical
tank. The tracks of true particles are shown.  The PMTs are
represented as a set of circles on the surface of the tank.  The total
energy falling on each circle is added up and used to colour code the
circle when it is displayed.
%sections~\ref{subsec-lowE} and \ref{subsec-highE}.

\subsubsection{Event reconstruction}
\label{subsec-highE}
A set of tools has been developed for event reconstruction. All events
are first reconstructed using low-energy reconstruction tools.  Events
that then have an energy above a threshold of 60 MeV, chosen to
exclude the majority of secondary sub-events such as Michel electrons
and neutron capture signals, are then passed through the high energy
reconstruction chain.

The low-energy vertex reconstruction is similar to the ``Quad Fitter''
method used in SNO and described in section~5.3.2 of
Ref.~\cite{brice-thesis}. Combinations of four PMT hits are used to
obtain a solution for the position, direction, and time of origin of
an interaction.  The energy of the event can be approximated from the
number of hits recorded by the detector, since they are related
linearly.

While low-energy particles can be approximated as a point source, for
high-energy particles the non-zero track length must be taken into
account, along with the possibility of multiple particles producing
Cherenkov rings. The high-energy reconstruction first searches for
rings and identifies which observed photoelectrons (PE) belong to each
found ring. The timing information of each hit is then used to improve
the reconstructed vertex position and direction. Both the muon and
electron hypotheses are considered using a look-up-based method to
determine the energy and a likelihood-based method of particle
identification (PID) to determine whether the ring is electron-like or
muon-like.

\begin{figure}[ht]\centering

\includegraphics[width=0.45\textwidth]{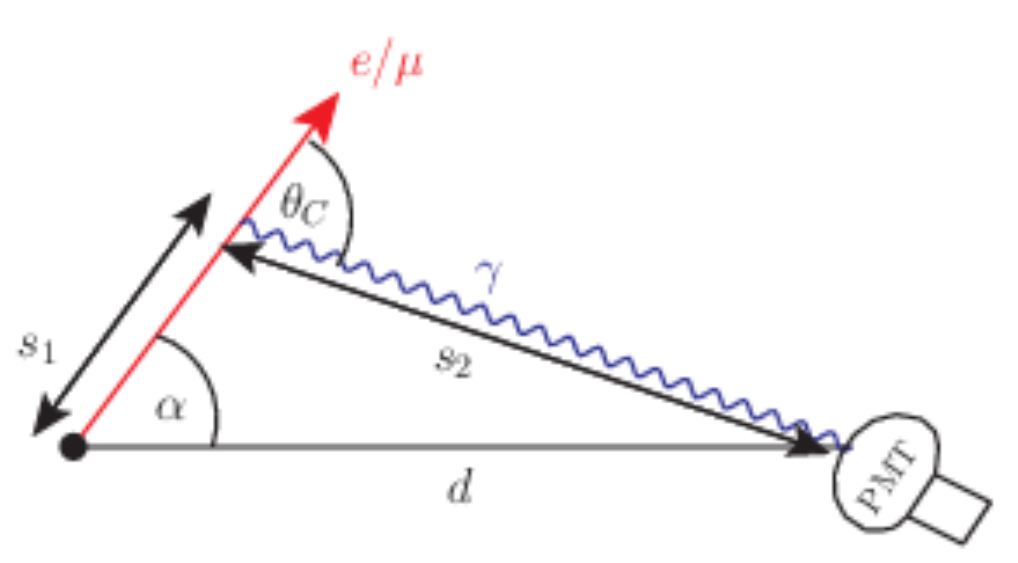}
\caption{
The Hough transform and track fit use time-of-flight information based
on the distance the charged particle travels,$s_1$, before emitting a photon at angle
$\theta_C$, which travels distance $s_2$ before arriving at a PMT
located distance $d$ from the interaction vertex at an angle $\alpha$
to the lepton track.}
\label{fig:TOF}
\end{figure}

A Hough transform, similar to that used in
Super-K~\cite{Shiozawa:1999sd}, is employed to search for rings. The
algorithm converts hits into Hough space from which the centre of the
rings can be determined.

The possible directions of the particle responsible for a given hit
are calculated using the timing information of each associated PMT
hit. The timing information yields a circle of possible directions for
each photoelectron origin and the radius of theses circles is
dependant on the angle of the particle's track and the direction from
the interaction vertex and the hit PMT, shown in
Figure~\ref{fig:TOF}. This angle, $\alpha$, is determined from an
expression of the time-of-flight (TOF) $t$, using the distance $d$
between reconstructed interaction vertex and hit PMT, and the
Cherenkov angle $\theta_C$ (which is approximated by that of a
particle travelling at $c$)

\begin{equation}\label{eq:TOF}
	t=\frac{s_1}{c}+\frac{n_\text{ref} s_2}{c}
	=\frac{d\sin(\theta_C-\alpha)}{c\sin\theta_C}+\frac{n_\text{ref} d \sin\alpha}{c\sin\theta_C},
\end{equation}
where $n_\text{ref}$ is the refractive index. The solution for the angle $\alpha$ is
\begin{equation}
\begin{split}
\tan\alpha=\frac{A-B\sqrt{A^2-B^2+1}}{B^2-A^2},\\
           A=\frac{n_\text{ref}-\cos\theta_C}{\sin\theta_C},\quad  B=\frac{ct}{d}.
   \end{split}
\end{equation}

After each ring is found the photoelectrons that are temporally
consistent (within twice the timing resolution) with originating from
this ring are removed and the algorithm re-run to find additional
rings. Any additional ring with less than 9\% of the number of hits of
the primary ring is discarded.  The remainder of the reconstruction is
based on the individual rings and the photoelectrons associated to
them.

To improve the reconstructed vertex, the PMT hit timing information is
used in a fit taking into account the track length of the
particle. The algorithm is based on one developed by
Super-K~\cite{Shiozawa:1999sd} using the goodness test value given by
\begin{equation}
G=\sum_{i\in\text{PE}} \exp\left(-\frac{(t'_i-t_0)^2}{2\sigma^2}\right).
\end{equation}

%This is the same as Equation~\ref{eq:goodness} used for the low-energy
%reconstruction, except that 
where the sum is over all observed photoelectrons,
$t'_i=t_i-t_i^\text{exp}$ is the timing residual determined by the
expected total TOF ($t_i^\text{exp}$) and PMT hit time ($t_i$) of the
$i$th observed photoelectron, $t_0$ is determined by taking the mean
value of the $t'_i$.  $\sigma$ is the time spread including PMT
transit time spread (TTS) and chromatic dispersions.

Starting from the seed vertex position given by the low-energy
reconstruction and the seed direction from the Hough transform and
maximising $G$ for the calculated TOF allows the algorithm to estimate
the vertex position and direction.

To reconstruct the energy, a lookup table is used based on the total
number of photoelectrons observed in the PMTs and the reconstructed
distance from the interaction vertex to the wall of the tank. The
lookup tables are generated from a Monte-Carlo simulation of CCQE beam
events in TITUS for electrons and muons separately.

Particle identification is based on a likelihood method similar to
those used in SNO~\cite{bonventre-thesis} and
MiniBoone~\cite{Patterson:2009ki}. The likelihoods are constructed
from two separate Monte-Carlo generated probability distributions: the
likelihood of observing the number of photoelectrons at each PMT and
the likelihood for the time that each photoelecton is observed.  These
are then used to calculate the expected number of photons received at
a given point or PMT, given its distance from the interaction vertex
and angle between the particle's track and the line from the
interaction vertex to this point. The reconstruction performs better
with relative numbers of photoelectrons at each PMT due to the
non-Poissonian nature. The log-likelihood function for the pattern of
PMT hits is given by
	
\begin{align}
\log L_\text{pattern} =& \sum_{i\in\text{PMT}}n_i\log p_i(x,E,l),\\
p_i(x,E,l)=&\frac{n_i^\text{exp}(x,E,l)}{\sum_j n_j^\text{exp}(x,E,l)},
\end{align}
where $n_i$ is the observed number of photoelectrons at the $i$th PMT,
and $p_i$ is the probability of an observed photoelectron at the $i$th
PMT and the lookup value $n_i^\text{exp}(x,E,l)$ is the expected
number of photoelectrons observed at a PMT coming from a lepton of
type $l$ with track starting at point $x$ with energy $E$.

The likelihood for the arrival time for a photoelectron at a PMT are
calculated using similar lookup tables.The log-likelihood function for
the timing of PMT hits is given by
\begin{equation} \log L_\text{time} =-\sum_{i\in\text{PE}}
\frac{(t_i-t_0-t_i^\text{exp})^2}{2\sigma^2}, \end{equation}
where $t_i$ is the arrival time of the $i$th photoelectron, $t_0$ is
the reconstructed interaction time and $t_i^\text{exp}$ is the
expected arrival time.

\begin{figure}[htb]\centering
\begin{tabular}{cc}
\includegraphics[width=0.4\textwidth]{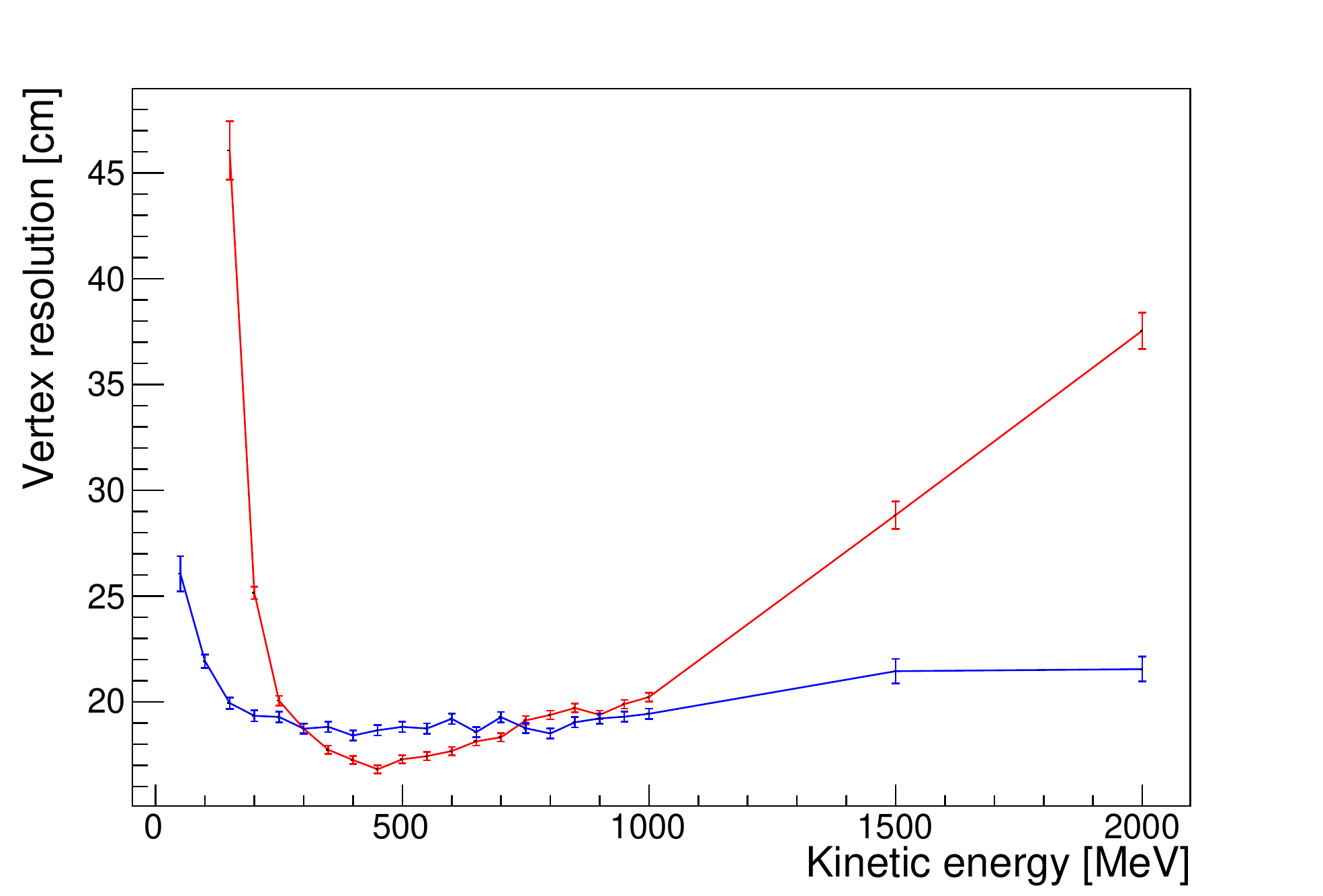} &
\includegraphics[width=0.4\textwidth]{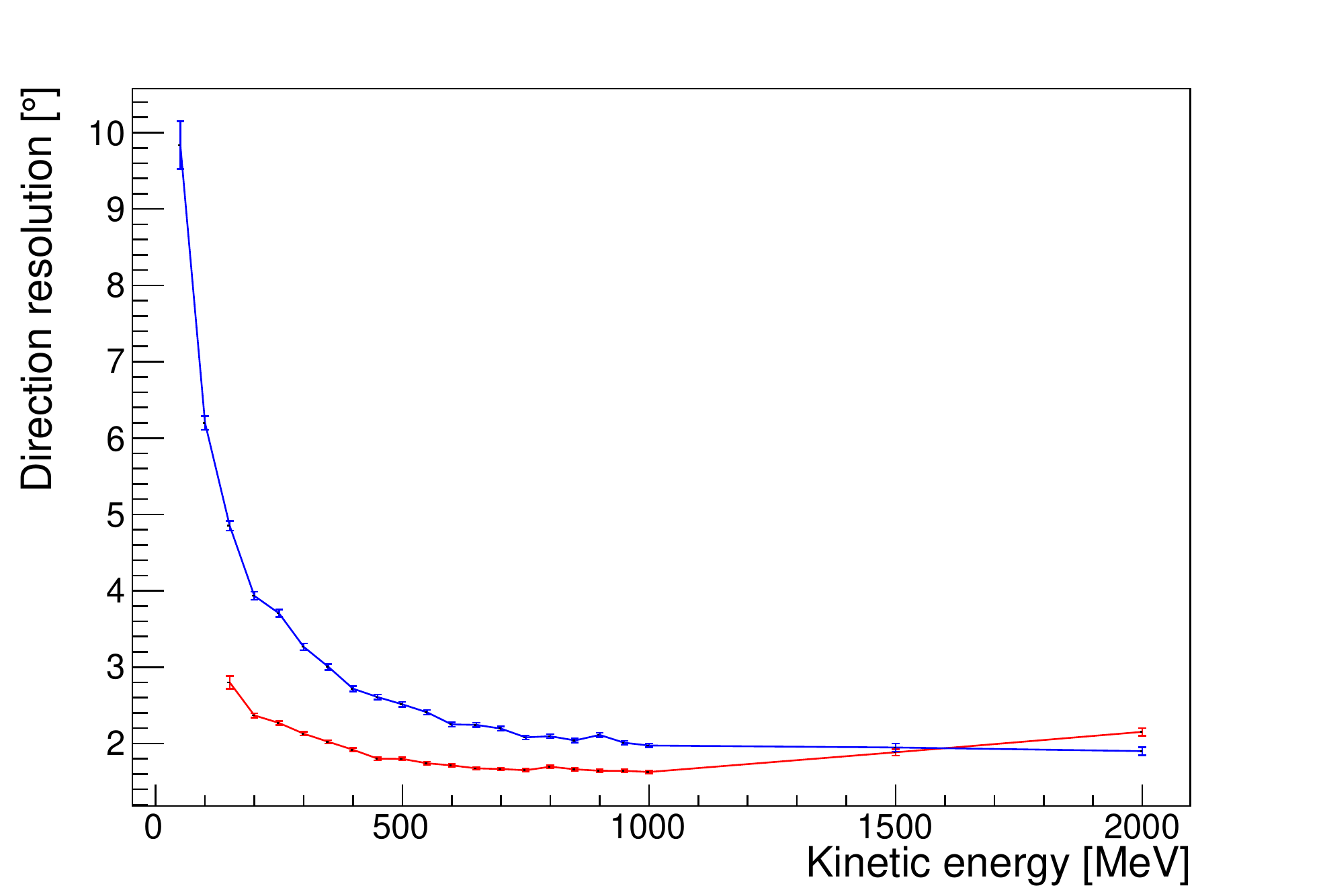}\\
\includegraphics[width=0.4\textwidth]{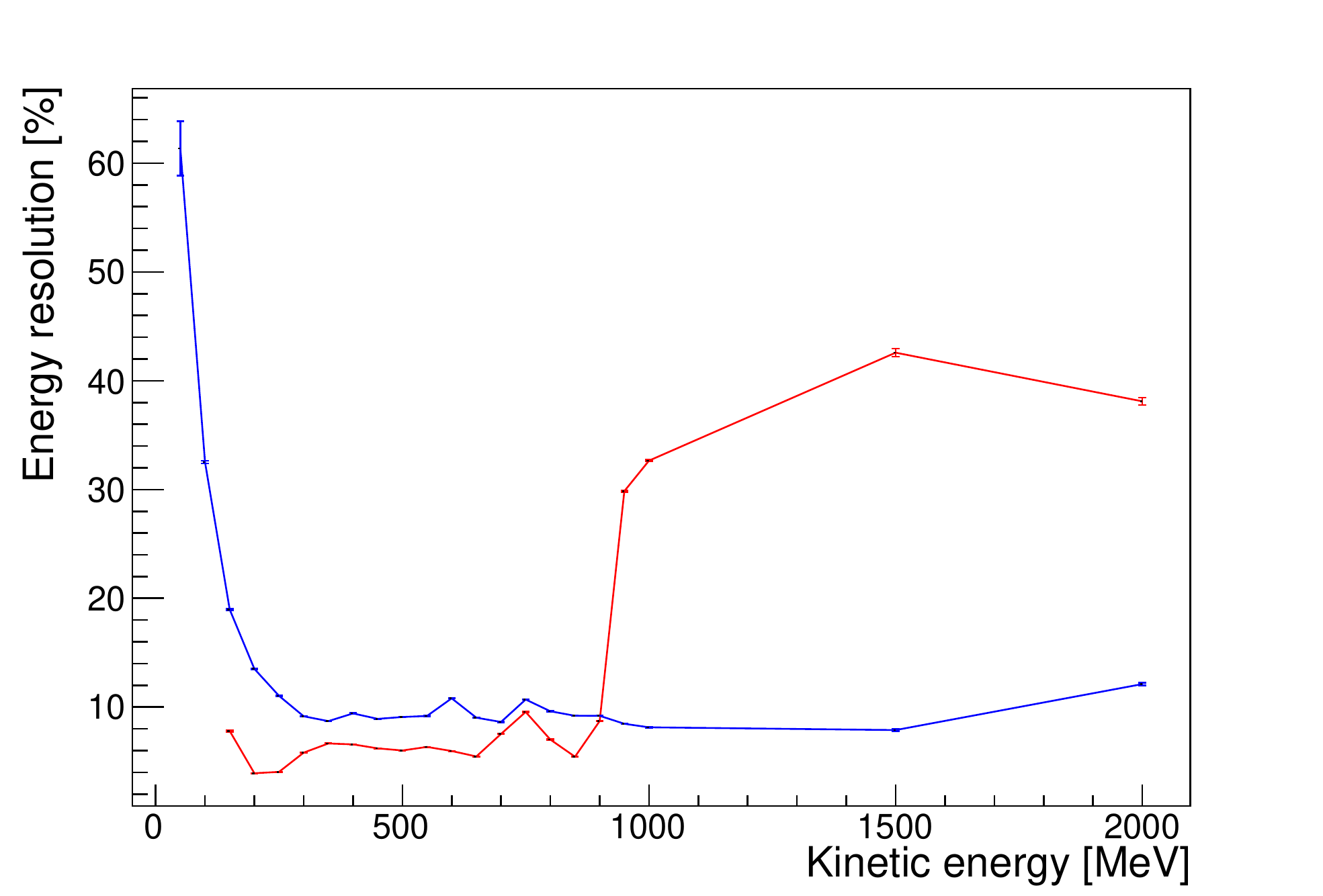}&
\includegraphics[width=0.4\textwidth]{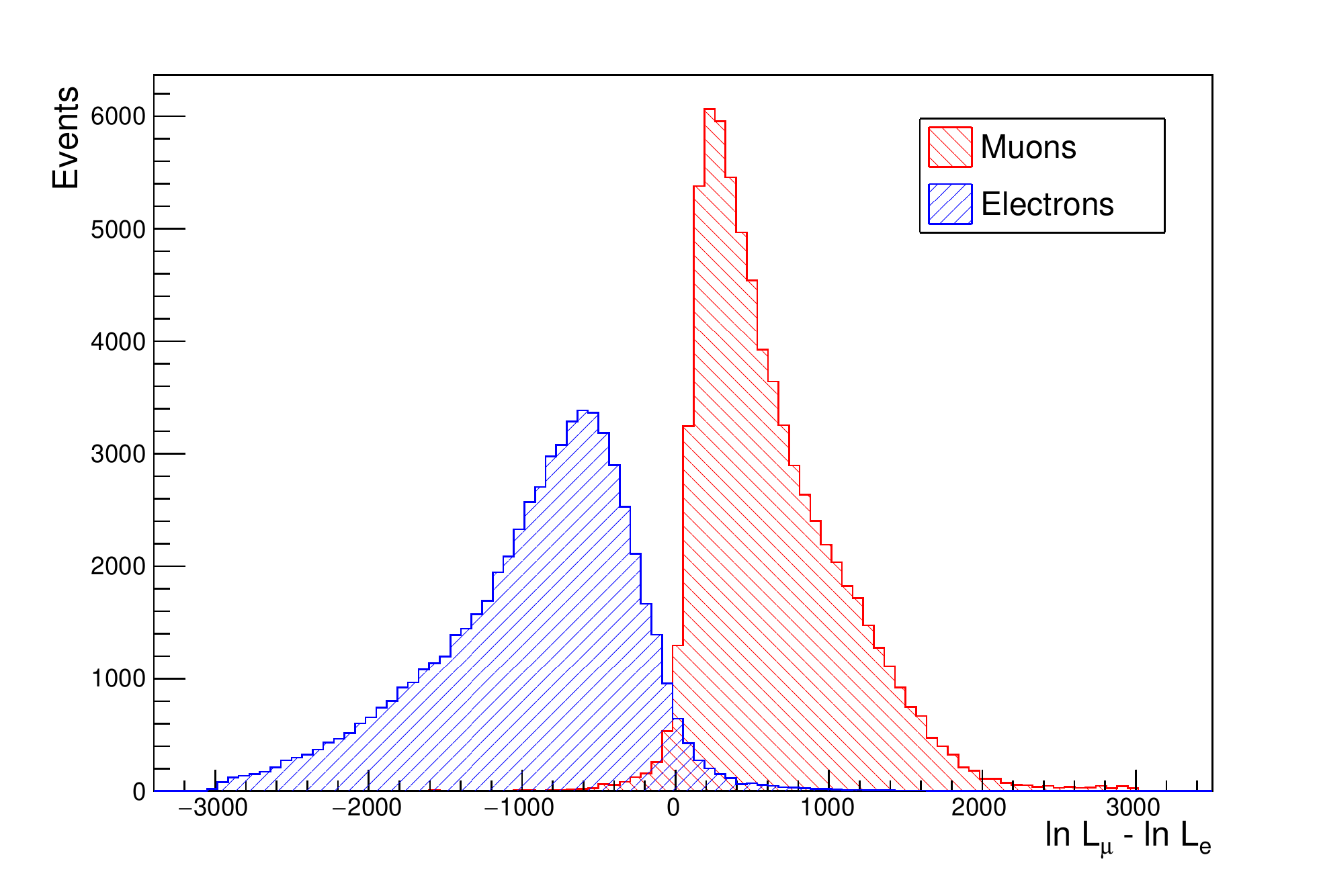}\\
\end{tabular}
\caption{
Resolutions of the reconstructed vertex position (top, left), lepton
track direction (top, right) and kinetic energy (bottom, left) for
muons (red) and electrons (blue) generated at kinetic energies of
50\,MeV to 2\,GeV for electrons and 150\,MeV to 2\,GeV for
muons. Particle identification for muon and electrons simulated in
TITUS (bottom, right).}
\label{fig:recoinfo}
\end{figure}

The total log-likelihood $\log L = \log L_\text{pattern} + \log
L_\text{time}$ is maximised, allowing the reconstructed interaction,
vertex position, time and direction to vary, for both electron and
muon hypotheses.  The difference between the log-likelihood value for
the electron hypothesis and that of the muon hypothesis is used to
discriminate between particle types. The resolutions obtained for the
vertex, direction and energy reconstruction as well as the particle
identification are shown in Figure~\ref{fig:recoinfo}, for muons and
electrons generated with uniformly distributed random positions within
the TITUS tank.

At low energies, the reconstruction performs poorly due to the smaller
number of Cherenkov photons produced. For muons, the vertex
reconstruction becomes poor below around 200\,MeV, and the direction
and energy reconstruction fails at around 100\,MeV. Above 200\,MeV,
the direction resolution is reasonable to high energies above 2\,GeV,
but the vertex resolution starts to become worse and energy
reconstruction fails above around 1\,GeV. This is due to the longer
range of higher energy muons resulting in their exiting the tank
before depositing the majority of their energy.  For those muons which
range out of the tank into the MRD it will be possible to improve the
reconstruction.  For electrons, the resolutions are generally worse,
but are stable for a wider range of energies; at low energies, the
direction and energy resolutions start to become significantly poorer
at around 100\,MeV, while all parts of the reconstruction remain
reasonable beyond 2\,GeV.

\subsection{MRD simulation}
\label{subsec-mrdsim}
A full simulation and reconstruction software infrastructure is being
developed in concert with work on the Baby-MIND
detector \cite{BabyMind}, a magnetised MRD with the same design as the
TITUS MRD, but a smaller transverse footprint.

The MRD detector, formed from magnetised iron plates interleaved with
scintillator, with air gaps which improve the charge reconstruction
efficiency of low-energy muons, has been simulated in Geant4.  Track
reconstruction by a Kalman Filter~\cite{RecPack} confirms the design
efficiencies reported earlier in section~\ref{sec:mrddesign}.  A
dedicated algorithm for the analysis of tracks which traverse a small
number of iron planes has been added to the TITUS software framework
to optimise the charge reconstruction efficiency at low muon
momenta. Although it is not used in the current work, it is planned to
be used in future improvements of the analysis.

\subsection{Full reconstruction selection}
\label{subsec-fullselection}
\label{sec:full selection}
For the current studies, samples have been produced for single
muon-like ring (1R$\mu$) and single electron-like ring (1R$e$)
events. A number of cuts are made which have been optimised based on
the current performance of the reconstruction. For the 1R$\mu$
selection, a single ring is required which is identified as muon-like,
with a reconstructed neutrino kinetic energy between 0 and
1.25\,GeV. A slightly expanded fiducial volume is used with respect to
that used in section~\ref{subsec-basicselection}, with cuts applied to
the reconstructed vertex requiring the vertex to be located at least
2\,m away from the tank wall, and at least 2\,m to the wall in the
reconstructed lepton track direction. For the 1R$e$ selection, a
single ring is required identified as electron-like, with a
reconstructed kinetic energy between 0 and 2.5\,GeV, and a
reconstructed vertex $>2$\,m from the tank wall and $>2$\,m to the
wall in the reconstructed track direction. Furthermore, for the
electron selection there is a requirement that no decay electrons are
identified, whereas in the muon sample up to one decay electron is
permitted. Decay electrons are identified as sub-events following the
single ring candidate within the time interval 135--8000\,ns and with
an energy $\ge$15\,MeV.

 With these selections, a 90\% $\nu_{\mu}$CC0$\pi$ purity\footnote{Purity is defined as the number of selected events of a given type divided by the total number of selected events.}  and 62\% efficiency\footnote{Efficiency is defined as number of selected events divided by the total number of events for specified sample.} within the selection fiducial volume (18\% efficiency overall) are achieved for the \FHC 1R$\mu$ sample.  A 12.1\% pure $\nu_e$CC0$\pi$ sample with 34\% efficiency within the selection fiducial volume (11.4\% efficiency overall) is achieved for the \FHC 1R$e$ sample.  The large contamination of the 1 ring electron sample is due in part to $\nu_\mu$NC$\pi^0$ events where the secondary ring associated with the $\pi^0$ is not positively identified, and part due to unidentified secondary pions. More sophisticated algorithms, e.g. fiTQun~\cite{Abe:2015awa,Patterson:2009ki} used for T2K oscillation analyses since Ref.~\cite{Abe:2013hdq}, have been shown to achieve significantly better $\pi^0$ rejection, reducing the remaining $\pi^0$ contribution by a factor of 9 in the Super-K detector.

\begin{table}[h] \centering
\begin{tabular}{c|c|c|c}
\hline\hline
\multicolumn{4}{c}{\FHC}\\
 Sample &  Purity & FV efficiency & Total Efficiency \\ \hline
1R$\mu$  & 90 .0& 61.9 & 18.1 \\ \hline
1R$\mu$ no n followers  & 95.0 & 42.2 & 12.3 \\ \hline
1Re  & 12.1 & 33.6 & 11.4 \\ \hline
1Re NC$\pi^0$ reduction & 32.6 & 33.6 & 11.4 \\ \hline
1Re no n followers  & 47.0 & 24.2 & 8.2 \\ \hline
NC$\pi^0$ & 44.0 & 87.5 & 29.0 \\ \hline \hline
\end{tabular}
\caption{
Purity and efficiency for the different CC0$\pi$ event samples considered from the full reconstruction chain
for the \FHC. For the NC$\pi^0$ the purity and efficiency refer to an NC$\pi^0$ sample.}
\label{tab:selpureff}
\end{table}

\begin{table}[h] \centering
\begin{tabular}{c|c|c|c}
\hline\hline
\multicolumn{4}{c}{\FHC}\\
Sample &  Purity & FV efficiency & Total Efficiency \\ \hline
% Sample &  CC0pi Purity & FV efficiency & Total Efficiency \\ \hline 
1R$\mu$ & 73.1 & 54.9 & 16.9 \\ \hline
1R$\mu$ no n followers  & 46.9 & 7.0 & 2.2 \\ \hline
1R$\mu$ n followers  & 79.6 & 47.9 & 14.8 \\ \hline
1Re  & 7.0 & 29.8 & 10.4 \\ \hline
1Re NC$\pi^0$ reduction  & 19.8 & 29.8 & 10.4 \\ \hline
1Re no n followers  & 6.9 & 3.4 & 1.19 \\ \hline
1Re n followers  & 26.2 & 26.4 & 9.2\\ \hline \hline
\end{tabular}
\caption{
Purity and efficiency for the different CC0$\pi$ event samples considered from the full reconstruction chain
for the  \RHC.}
\label{tab:selpureffrhc}
\end{table}

 Applying this reduction to events with a $\pi^0$ produced improves the 1 ring electron CC0$\pi$ sample purity to 33\% but further improvements with more sophisticated ring-finding algorithms can be expected for the \FHC 1R$e$ TITUS selection in the future. The CC0$\pi$ purity of the single ring samples is further increased, with an accompanying ~30-40\% drop in efficiency, if a neutron veto is imposed, which gives CC0$\pi$ electron purity of 47\% and CC0$\pi$ muon purity of 95\%.  The neutron reconstruction efficiency, for all neutron captures, is found to be 86\% in the selected fiducial volume, with around 90\% of captures on Gd.  The efficiency for reconstructing Gd neutron captures is 95\%.  

A summary of the selection efficiencies and
purities for the \FHC is given in Table~\ref{tab:selpureff} and
Table~\ref{tab:selpureffrhc} presents the efficiencies and purities
for the \RHC, where antineutrino CC events are expected to be
accompanied by a final state neutron.  A summary of the mean
resolutions for the correctly identified events over the selected
position and energy range is given in Table~\ref{tab:selres}. These
are fairly similar to the values assumed by Super-K, used in the basic
sensitivity studies, presented in Table~\ref{tab:res}.

\begin{table}[htb] 
\centering
\begin{tabular}{l r r}
\hline\hline
         & 1R$\mu$ & 1R$e$\\
\hline
Interaction vertex position [cm]& 17 & 17 \\
Outgoing lepton track direction [degrees] & 1.9 & 2.6 \\
Outgoing lepton kinetic energy [MeV] & 59 & 186 \\
Outgoing lepton kinetic energy [\%] & 13 & 21 \\
\hline\hline
\end{tabular}
\caption{
Resolutions for $\nu_\mu$CC0$\pi$ events in the single muon-like ring
(1R$\mu$) sample and for $\nu_e$CC0$\pi$ events in the single
electron-like ring (1R$e$) sample for a \FHC beam. }\label{tab:selres}
\end{table}

\subsection{Sensitivity studies with full reconstruction events}
\label{subsec-fullsensitivity}
Sensitivity studies produced using results from the full TITUS
reconstruction package detailed in the previous sections are presented
in the following sections.  The Hyper-K detector is simulated using
the Super-K Monte-Carlo 14a with exposure and detector errors scaled
appropriately. This study aims to be as close as possible to a full
Hyper-K oscillation analysis using currently available inputs and
tools.

The study is performed using the VaLOR analysis framework which is
used for official T2K analyses~\cite{valor}. Both the near detector,
ND, and the far detector, FD, samples are fitted simultaneously by
minimising the global Poisson log likelihood constructed from the
expected ($n^{\textrm{exp}}_{ij}$) and observed
($n^{\textrm{obs}}_{ij}$) number of events in each reconstructed
energy bin defined in Equation~\ref{eq:likelihood}.
\begin{equation}
-2\ln(\mathcal{L}) = \sum\limits_{i}^{Samples}\sum\limits_{j}^{Bins} 2 \cdot (n^{exp}_{ij} - n^{obs}_{ij} + n^{obs}_{ij} \cdot \ln(\frac{n^{obs}_{ij}}{n^{exp}_{ij} })) + \bold{(f - f_0)^TC^{-1}(f - f_0)}.
%\caption{Poisson likelihood function}
\label{eq:likelihood}
\end{equation}
With the predicted number of events in each $j^{th}$ reconstructed
energy bin is given by
\begin{equation}
n^{exp}_{ij} = \sum\limits_m^{modes}\sum\limits_t^{Etrue Bins}\sum\limits_{r'}^{ErecoBins} \textbf{P}_{i;m;t} \cdot \textbf{T}_{i;j;r'; \bold{f}} \cdot \textbf{S}_{i;m;t;r';\bold{f}} \cdot N^{MC}_{i;m;r';t}.
%\caption{Events in each reconstructed energy bin}
\label{eq:nev}
\end{equation}
where $N^{MC}_{i;m;r';t}$ denotes the number of events in the input MC
template for sample $i$ in the reconstructed energy bin $r'$, true
energy bin $t$ and of interaction mode $m$. \textbf{T} and \textbf{S}
are the systematic parameter weights applied to re-weight the MC for
any given set of systematic parameters \textbf{f} detailed in
Section~\ref{sec:full_cp_systematics} and \textbf{P} is the
oscillation probability applied to true energy bins. The oscillation
probability is calculated in a three flavour framework including
matter effects in constant density matter. The second term in Equation
\ref{eq:nev} is a penalty
with \textbf{C}, the error matrix, used to incorporate prior
constraints on systematic errors in the fit detailed below. Each
sample consists of 9 interaction modes (CCQE (charged current
quasi-elastic), CCnpnh (charged current with n-particle n-hole
interactions), CC1$\pi^{\pm}$ (charged current one charged pion),
CCcoh (charged current coherent), CCoth (any other charged current
interaction), NC1$\pi^{\pm}$ (neutral current with a charged pion),
NC1$\pi^0$ (neutral current with a neutral pion), NCcoh (neutral
current coherent)) and 8 neutrino types (beam $\nu_{e,\mu}$(bar) and
oscillated $\nu_{e,\mu}$(bar)).  The studies presented here are
performed fitting 4 Hyper-K samples, FHC$1R\mu$, FHC$1Re$, RHC$1R\mu$,
RHC$1Re$ currently available from the Super-K MC. For TITUS we include
neutron information from Gd doping, using binary tagged samples where
each sample is separated into events with $>$0 neutrons and 0 neutrons
for a total of 8 samples.

The following binning is used for the MC templates used in the analysis: 
\begin{itemize}
\item Hyper-K $e$-like : 25 reconstructed energy and 84 true energy bins;
\item Hyper-K $\mu$-like : 73 reconstructed energy and 84 true energy bins;
\item TITUS $e$-like : 6 reconstructed energy and 13 true energy bins.
\item TITUS $\mu$-like : 26 reconstructed energy and 13 true energy bins;
\end{itemize}
All sensitivity studies shown in this section are performed using an
Asimov dataset~\cite{Asimov}, where the nominal MC is taken to
represent the data. The Asimov dataset was chosen because it
represents nature, to the best of our knowledge, and is free from any
statistical or systematic fluctuations.

Significance and confidence regions are constructed with the constant
$\Delta\chi^2$ method, with nuisance parameters profiled
out. Minimisation is performed with Minuit, using the MIGRAD
method. Due to a possible degeneracy in the oscillation space; fits in
the parameters $\sin^2(\theta_{23})$ and $\delta_{CP}$ are performed
multiple times. After the initial fit a subsequent fit is performed
seeded at the mirror point, in $\sin^2(\theta_{23})$ and $\delta_{CP}$,
of the best fit point from the initial fit to ensure the global
minimum is found.

The studies presented here were performed for the Hyper-K baseline
design assuming a running ratio of 1:3 neutrino:antineutrino mode with
a nominal exposure of 1.3 MW$\times$10\,y assuming 320\,kA horn
current. Based on high intensity studies of the current accelerator
performance, it is expected that 1.3\,MW beam power can be achieved
after these
upgrades~\cite{jparc-status-upgrade} \cite{jparc-plan-2026}.

\subsection{\label{sec:full_cp_systematics}Treatment of systematic uncertainty}

The analysis considers 189 sources of systematic error from the
Hyper-K detector response uncertainty, TITUS detector response
uncertainty, flux prediction, and cross section model.

\subsubsection{Detector, pion FSI, SI and PN}
Detector, PFSI (Pion Final state interaction), SI (Secondary
interaction) and PN (Photo nuclear) uncertainties are found using the
MC predictions obtained by varying the underlying model parameters based on
prior uncertainty from external constraints. A covariance matrix,
binned in reconstructed energy bins for the interaction modes of
interest, is then produced from the variance in the MC
prediction. This allows us to use a detector+PFSI+SI+PN matrix
produced by the linear sum of the 4 matrices. Finally, there is also
an additional detection energy scale uncertainty on all events.

There are a total of 37 Hyper-K Detector+PFSI+SI+PN parameters which
affect reconstructed energy bins, 6 $\mu$-like and 12 $e$-like for
neutrino mode samples, as in Ref.~\cite{Abe:2015awa}, and the same
antineutrino mode samples. The extra parameter is the energy scale
which is used for both modes. Correlations are included between the
neutrino and antineutrino mode uncertainties. The detector error
matrix is scaled down by a factor of $\sqrt{20}$ to account for the
additional constraint from larger statistics from the control samples
in the Hyper-K era.

There are a total of 31 TITUS parameters which affect reconstructed
energy bins, 15 for neutrino mode and 15 for antineutrino mode, as for
Hyper-K with uncertainty on appearance events removed. Again, the last
systematic parameter is the energy scale parameter.  The detector
error matrix is taken unscaled from Ref.~\cite{Abe:2015awa}. Detector,
PFSI, SI and PN uncertainties are treated as fully correlated between
the neutron tagged and neutron untagged samples.

The PFSI uncertainties are included as described above and
correlations included between all Hyper-K and TITUS samples. SI and PN
uncertainties were calculated using only the SK MC and then applied to
both TITUS and Hyper-K samples with a 90\% correlation included
between the two.

\subsubsection{Flux}
A total of 100 parameters to account for flux uncertainties are
applied as normalisations on the true energy bins, 25 for each
detector and running mode combination. An error matrix for the flux
parameters is used assuming full replica target data from NA61/SHINE
which is expected during Hyper-K operation. As a result the pre-fit
uncertainty on the flux is greatly reduced compared to what is
currently used in T2K oscillation analyses~\cite{Abe:2015awa}.

\subsubsection{Cross section}

A total 15 cross section model parameter uncertainties applied and
implemented using response functions, as in the 2015 T2K oscillation
analyses~\cite{Abe:2015awa}.  These include the CCQE axial mass (MaQE
6\%), Fermi momentum (7\%), binding energy (33\%), nucleon to $\Delta$
axial form factor ($C^A_5$ 12\%), resonance production axial mass
scaling factor (MaRes 16\%), scale of isospin $\frac{1}{2}$
non-resonant background (15\%), CCoth shape (40\%), npnh normalisation
(102\%), npnh$\bar{\nu}$ normalisation (102\%), CCcoh normalisation
(100\%), NCcoh normalisation (30\%), NCoth normalisation (30\%), NC 
1$\gamma$ normalisation (100\%), CC $\nu_e$ normalisation (2\%) 
and CC $\bar{\nu}_e$ normalisation (2\%)

\subsubsection{Nucleon FSI}

Nucleon FSI uncertainties are included based on studies performed with
the GENIE event generator as described in
section~\ref{subsec-basicselection}. The result is a set of 6
parameters which parameterise the uncertainty due to nucleon FSI in
the studies.

\begin{table}
\centering
 \begin{tabular}{ c | c }
 \hline\hline
 Parameter(s)                                               		   		&      Value 	  						   \\
\hline
$\sin^{2}(\theta_{23})$                                       				&      0.528								  \\
$\sin^{2}(\theta_{13})$								&      0.0217							 	\\
$\sin^{2}(2\theta_{12})$  								&      0.846              							 		 \\
$|\Delta m^{2}_{32}|$ (NH) / $|\Delta m^{2}_{31}|$ (IH)      		&     	$2.509 \times 10^{-3}~\mbox{eV}^2/\mbox{c}^4$		 \\
$\Delta m^{2}_{21}$      								&       $7.53 \times 10^{-5}~\mbox{eV}^2/\mbox{c}^4$ 				\\
$\delta_{CP}$                                               				&       $-1.601$								    \\
Mass Hierarchy										&        Normal 	         							 \\  
\hline\hline
\end{tabular}
\caption{
Values of oscillation parameters used to compute the event rates,
systematic effects and sensitivity studies.  Each set of oscillation
parameters correspond to a different Asimov data set, which is the MC
expected distribution in a certain oscillation hypothesis.  }
\label{tab:osc_params_joint_asimov}
\end{table}

%\clearpage
\subsection{Predicted Hyper-K spectra}

\begin{figure}[!tb]
    \subfloat[FHC$1Re$]{
	\includegraphics[width=0.45\textwidth]{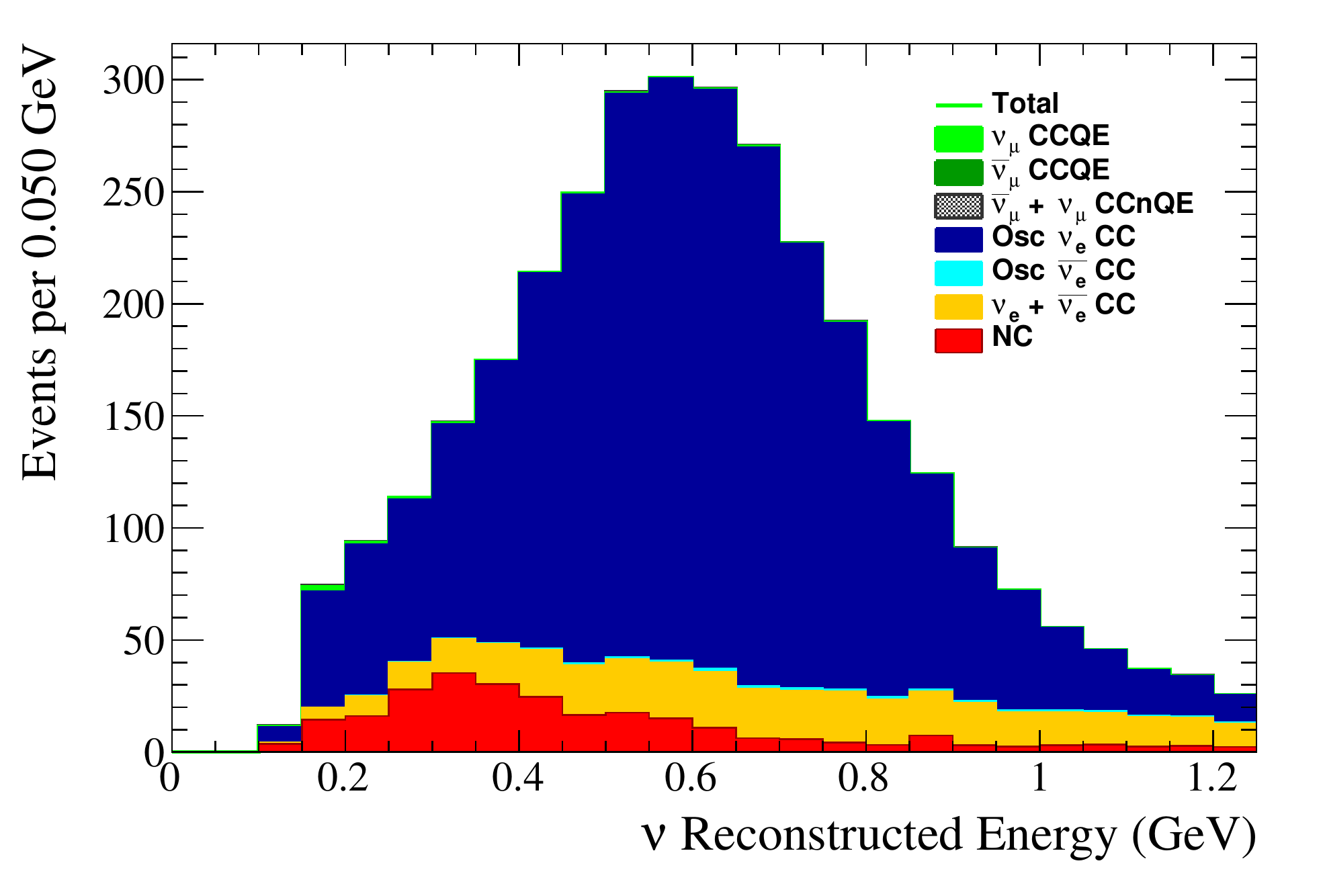} 
        \label{fig:predicted-hk-spectra_elike}
    }
    \subfloat[FHC$1R\mu$]{
\includegraphics[width=0.45\textwidth]{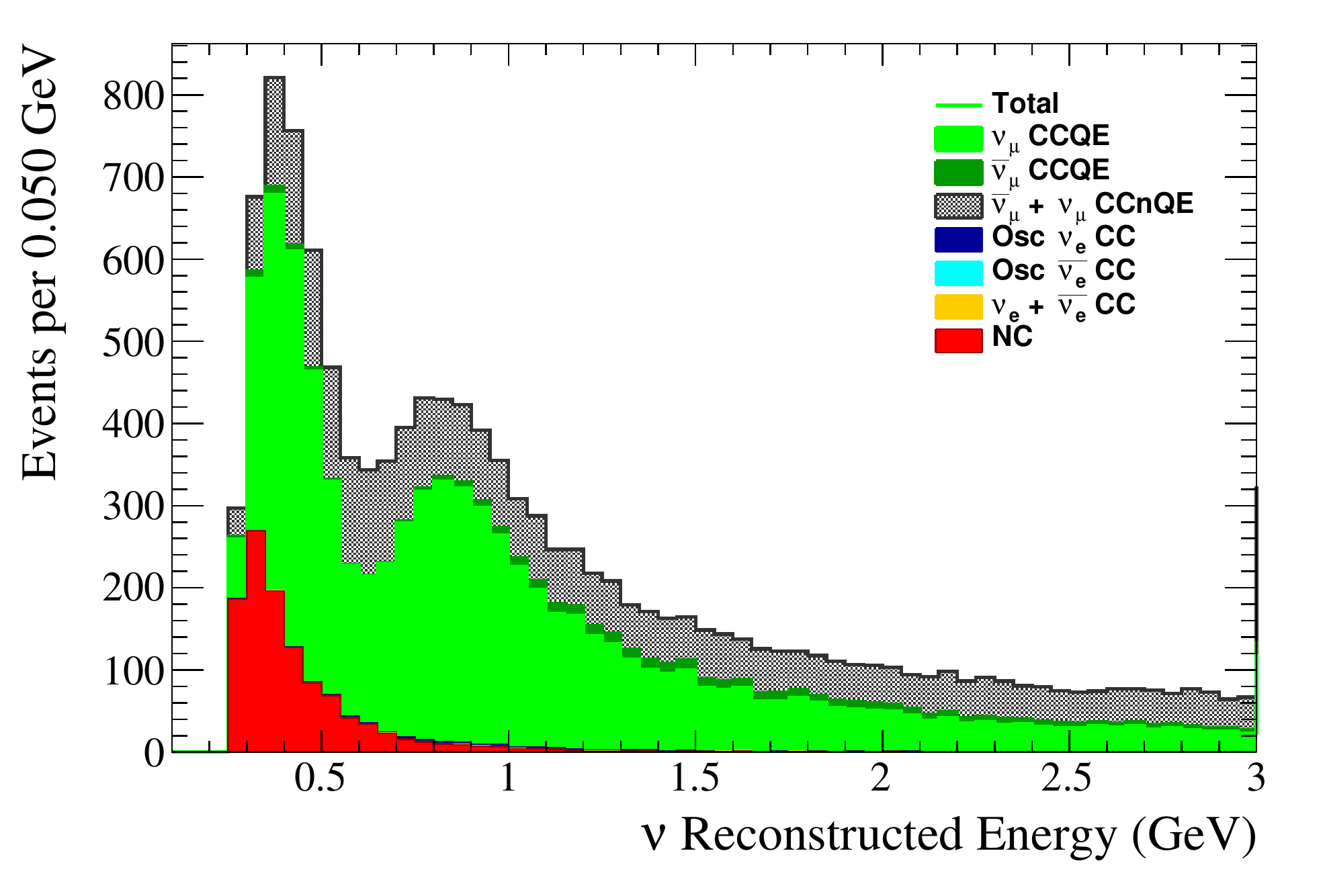} 
        \label{fig:predicted-hk-spectra_mulike}
    }
    
    \subfloat[RHC$1Re$]{
\includegraphics[width=0.45\textwidth]{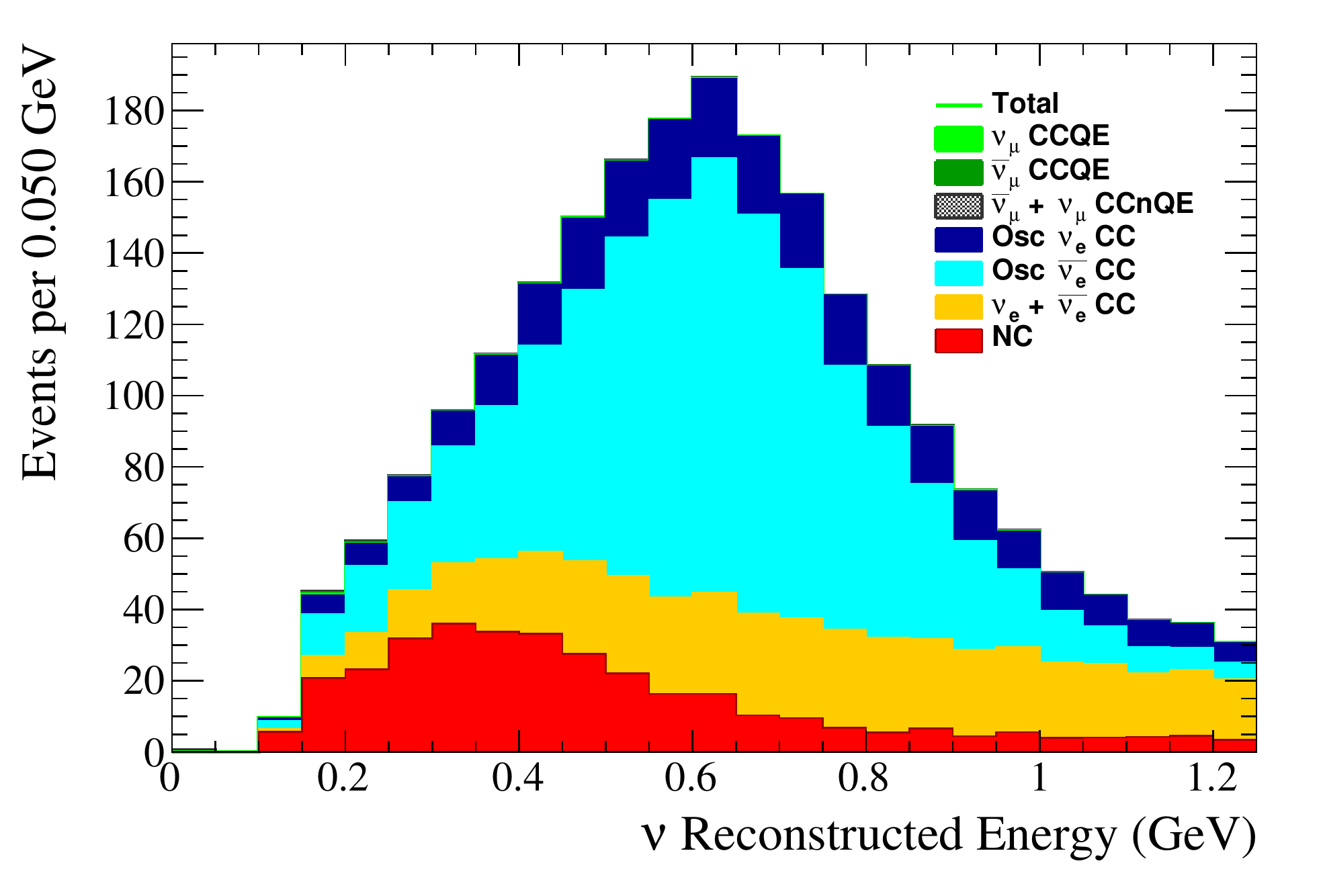} 
        \label{fig:predicted-hk-spectra_rhcelike}
    }
    \subfloat[RHC$1R\mu$]{
\includegraphics[width=0.45\textwidth]{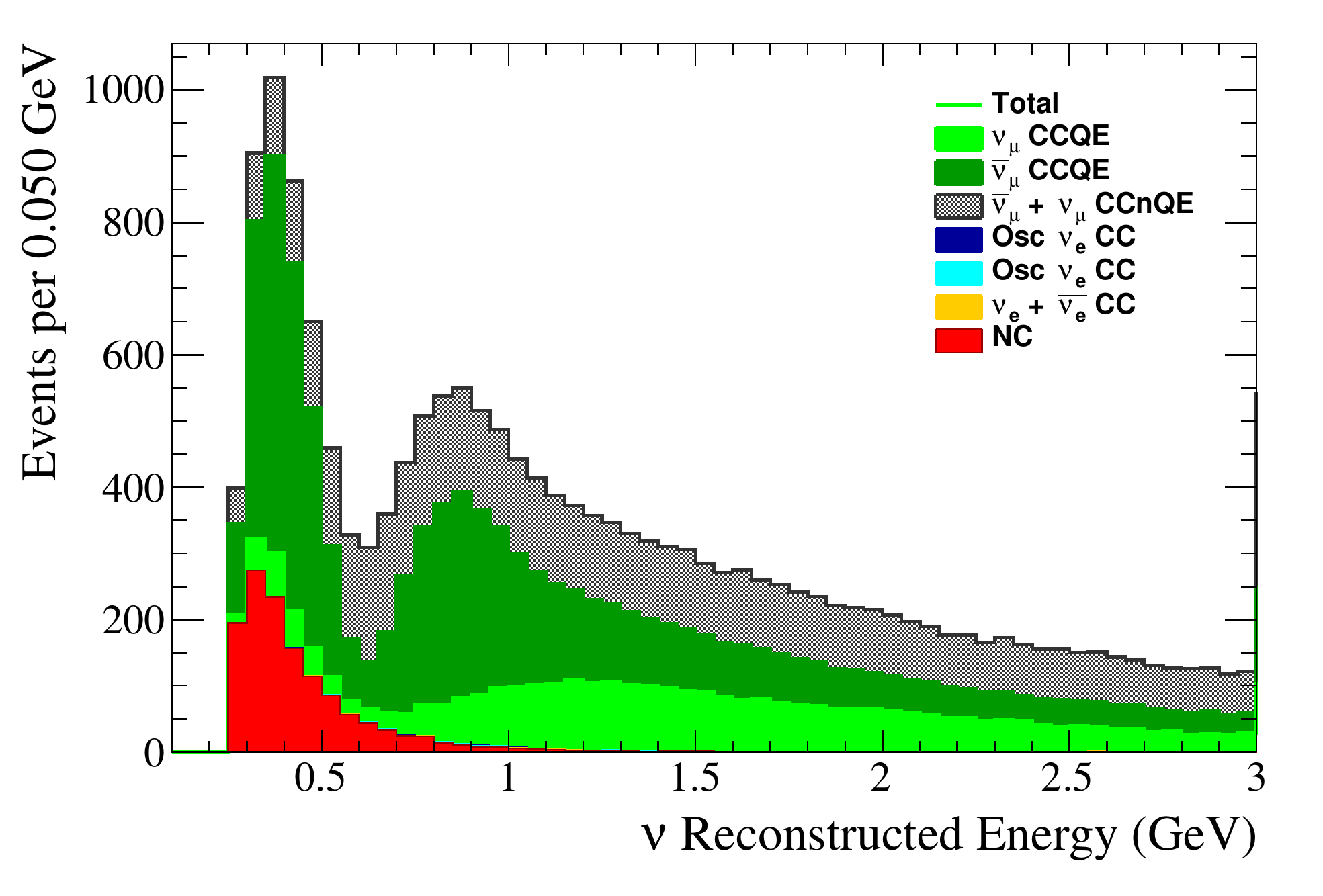} 
        \label{fig:predicted-hk-spectra_rhcmulike}
    }
    
\caption{Predicted Hyper-K spectra for the oscillation parameters in Table~\ref{tab:osc_params_joint_asimov}.}
\label{fig:predicted-hk-spectra}
\end{figure}

\begin{figure}[!tb]
    \subfloat[FHC$1Re$]{
	\includegraphics[width=0.45\textwidth]{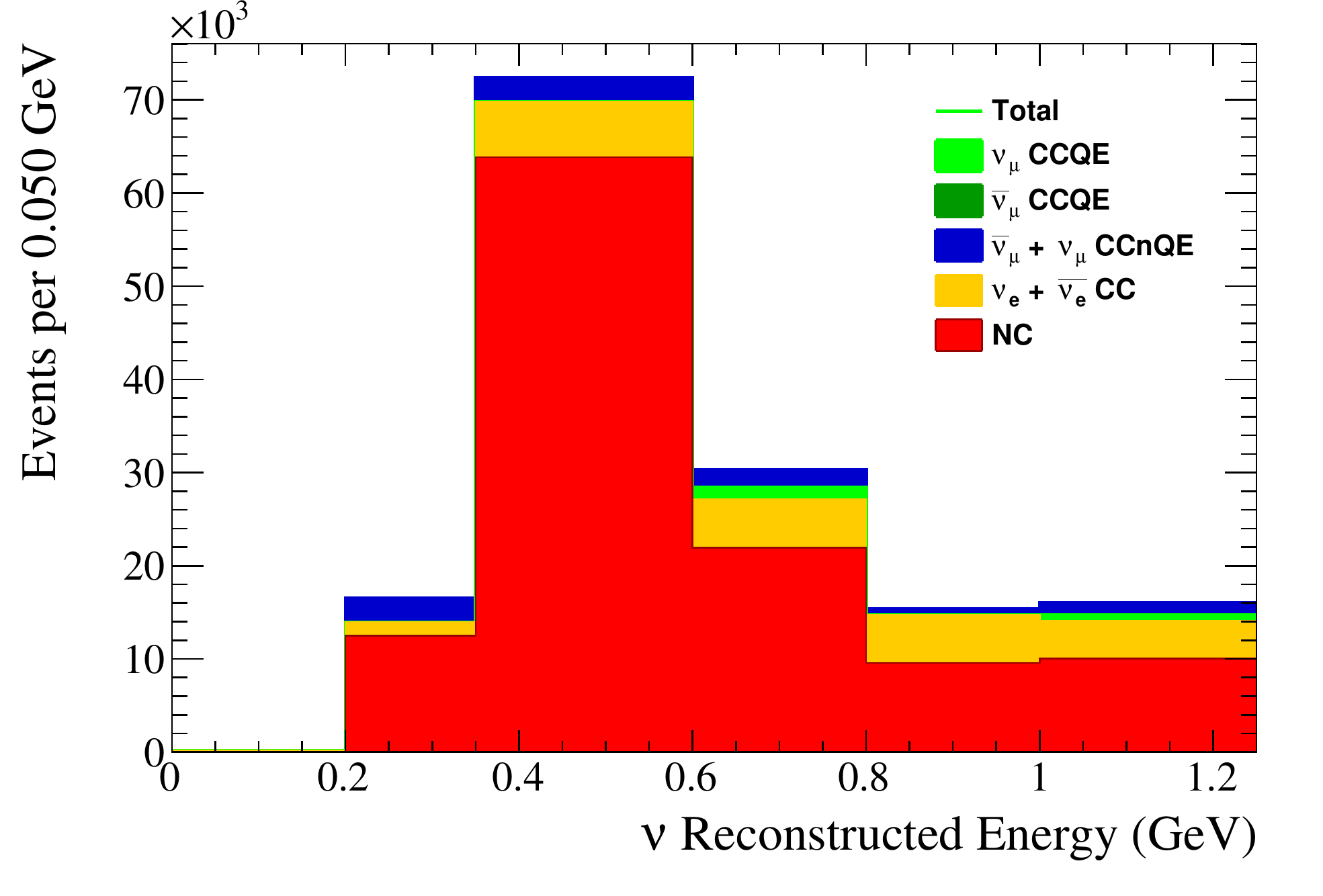} 
        \label{fig:predicted-titus-spectra_non_elike}
    }
    \subfloat[FHC$1R\mu$]{
\includegraphics[width=0.45\textwidth]{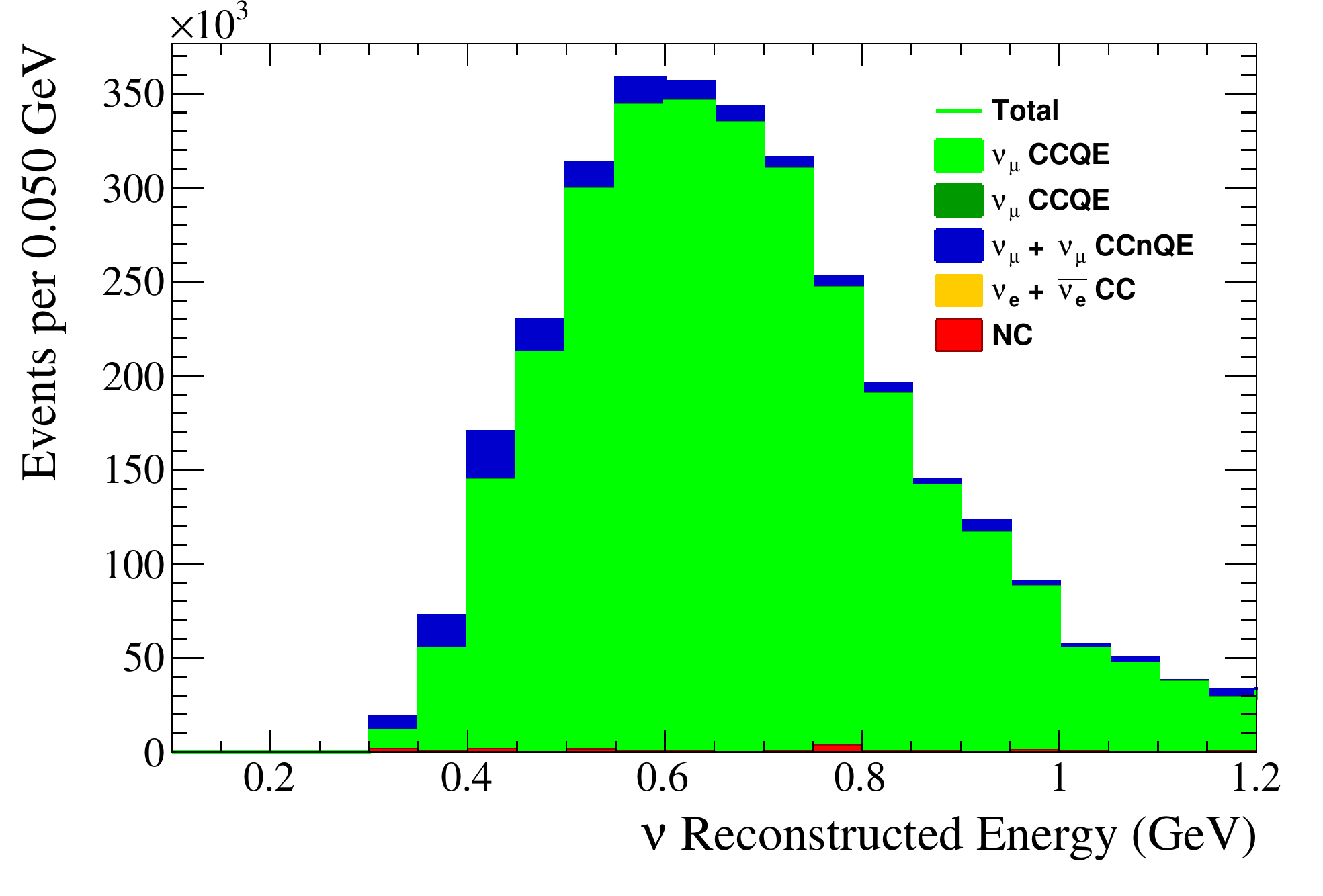} 
        \label{fig:predicted-titus-spectra_non_mulike}
    }
    
    \subfloat[RHC$1Re$]{
\includegraphics[width=0.45\textwidth]{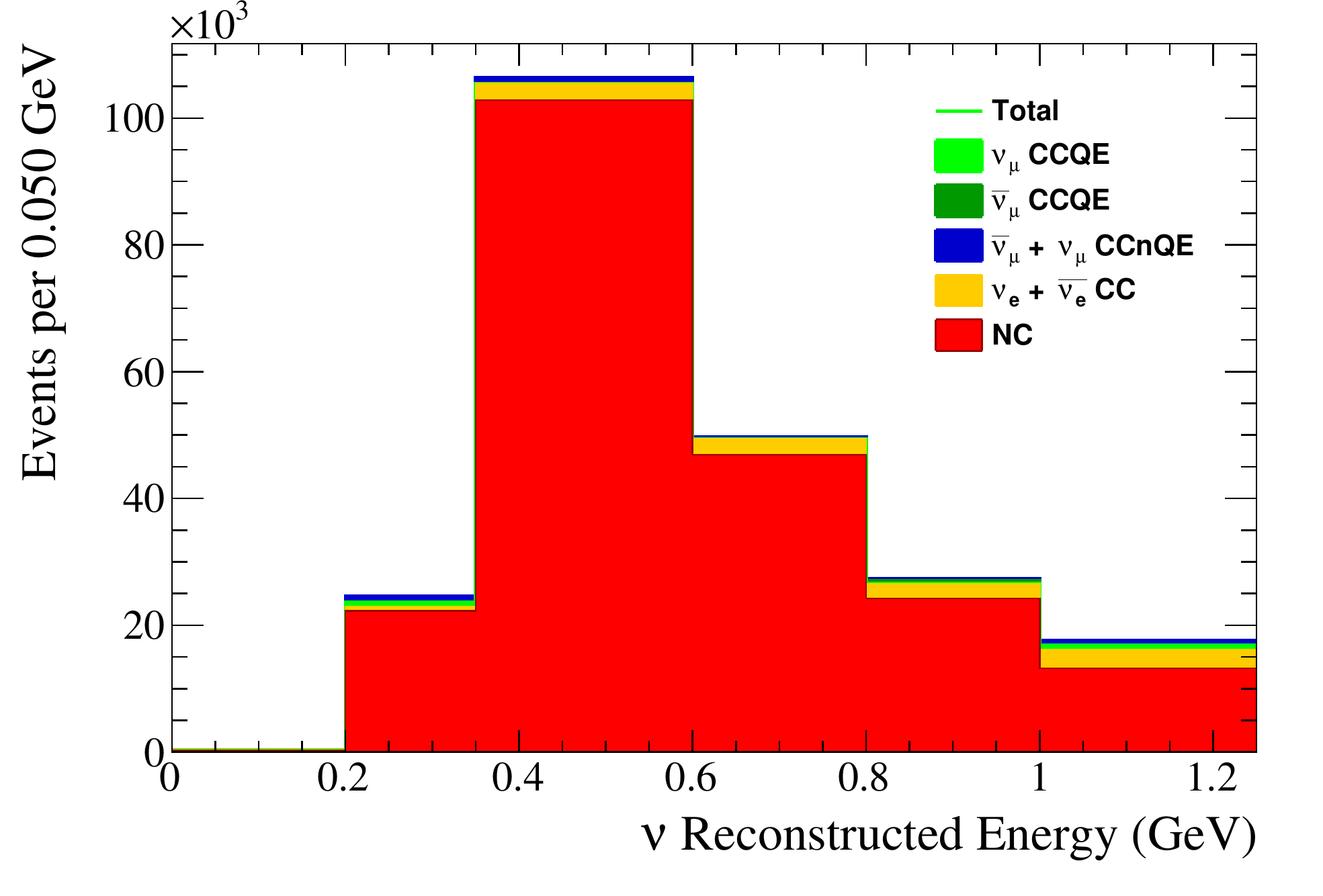} 
        \label{fig:predicted-titus-spectra_non_rhcelike}
    }
    \subfloat[RHC$1R\mu$]{
\includegraphics[width=0.45\textwidth]{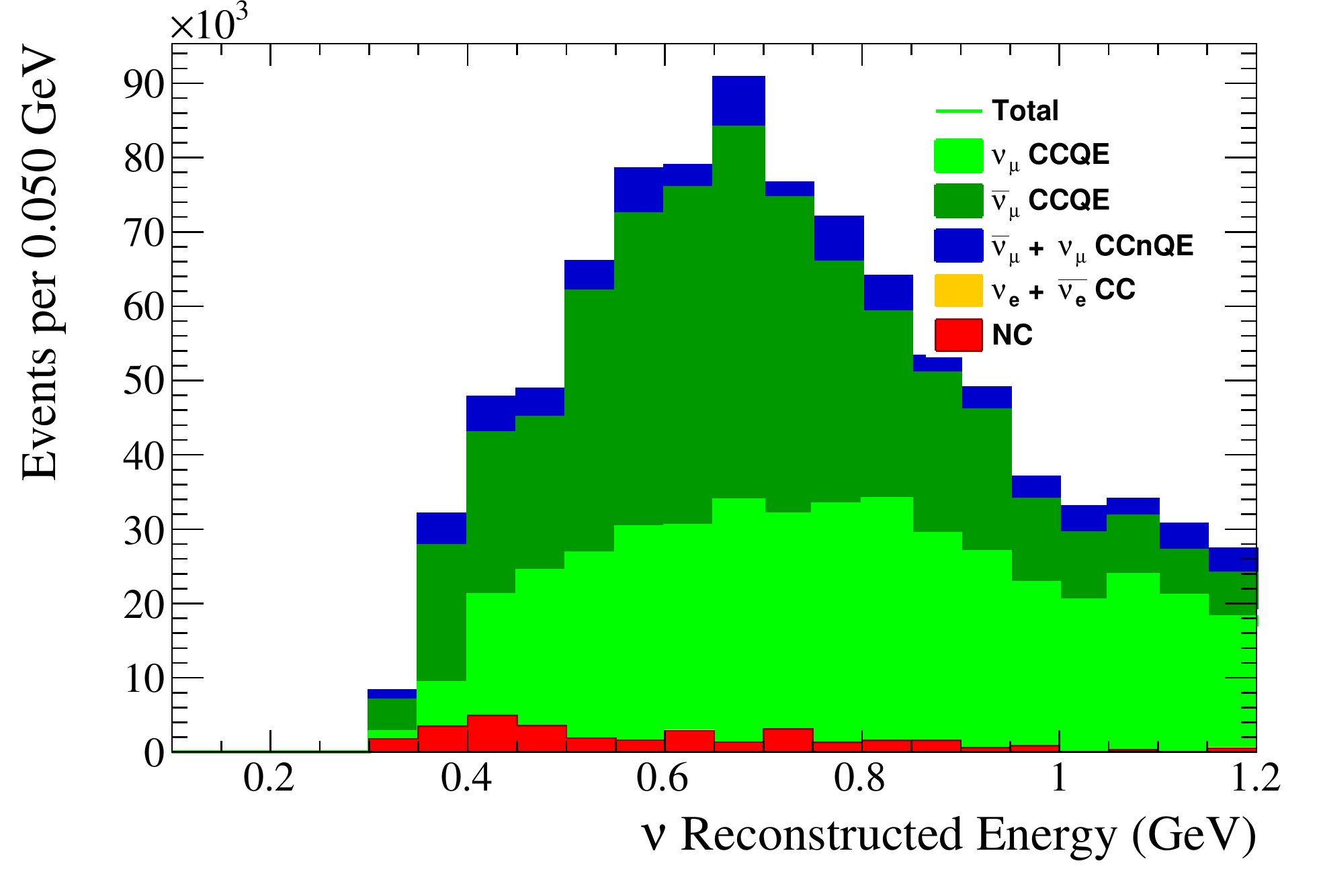} 
        \label{fig:predicted-titus-spectra_non_rhcmulike}
    }
    
\caption{Predicted no neutron tagged TITUS spectra.}
\label{fig:predicted-titus-spectra_non}
\end{figure}

\begin{figure}[!tb]
    \subfloat[FHC$1Re$]{
	\includegraphics[width=0.45\textwidth]{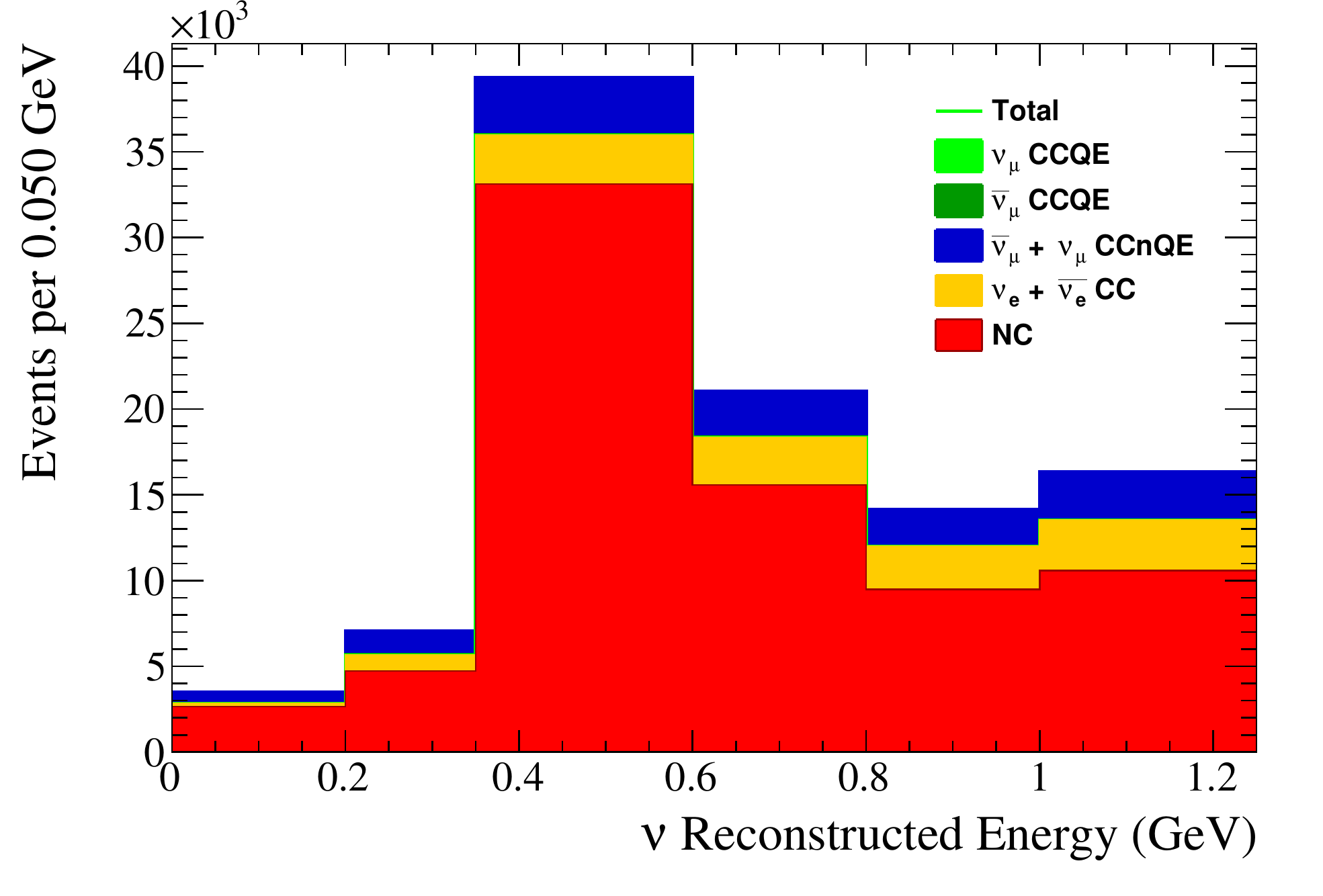} 
        \label{fig:predicted-titus-spectra_n_elike}
    }
    \subfloat[FHC$1R\mu$]{
\includegraphics[width=0.45\textwidth]{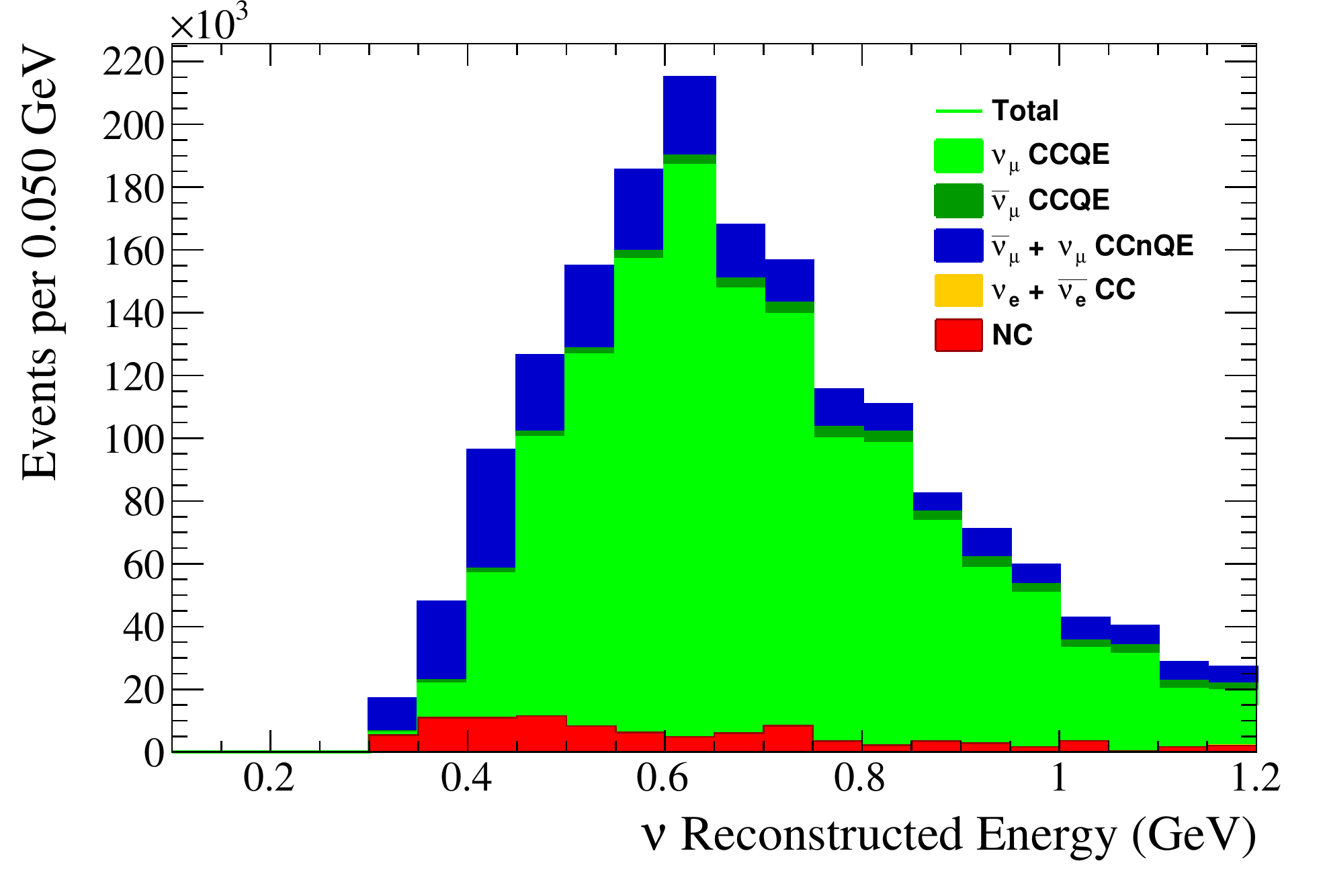} 
        \label{fig:predicted-titus-spectra_n_mulike}
    }
    
    \subfloat[RHC$1Re$]{
\includegraphics[width=0.45\textwidth]{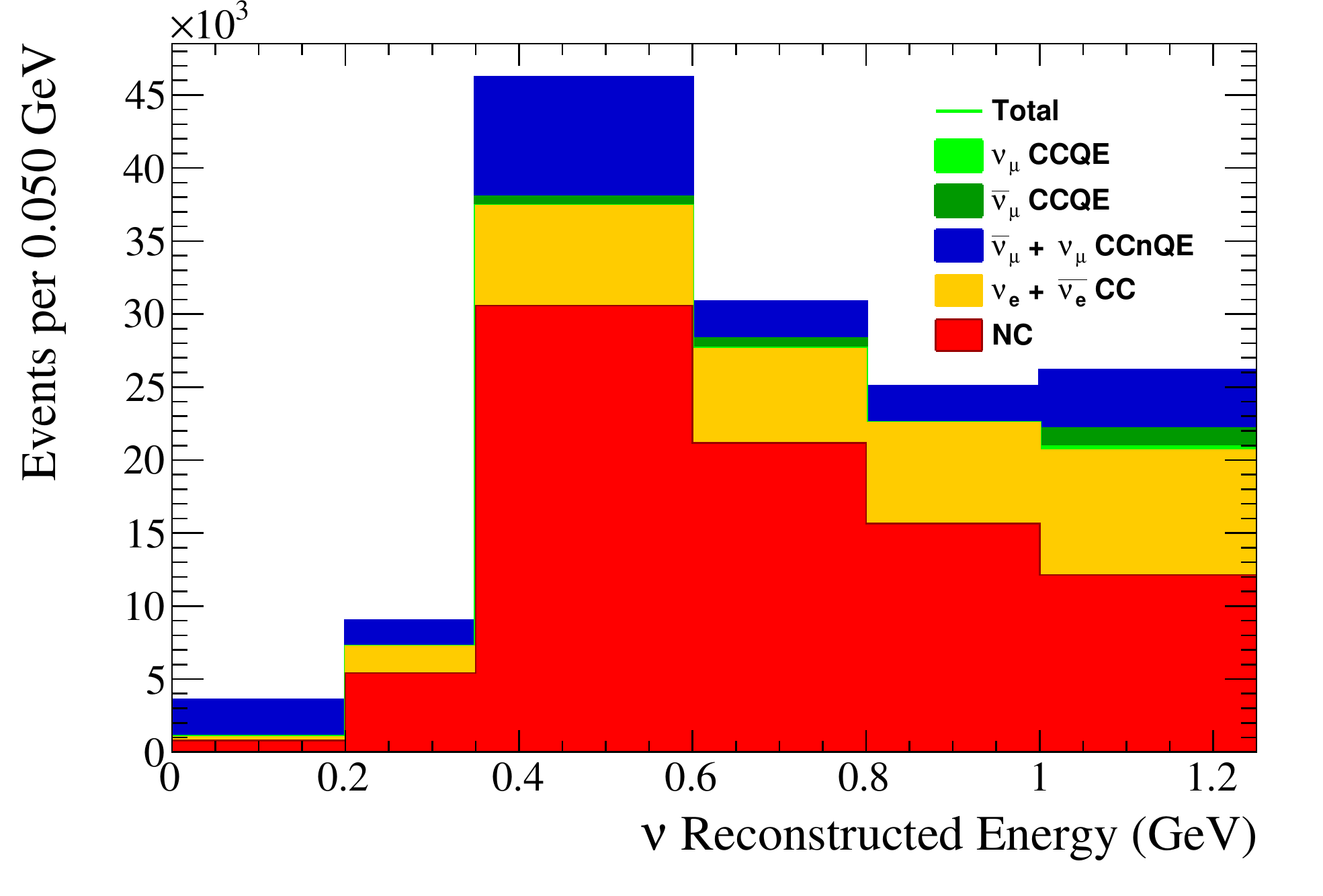} 
        \label{fig:predicted-titus-spectra_n_rhcelike}
    }
    \subfloat[RHC$1R\mu$]{
\includegraphics[width=0.45\textwidth]{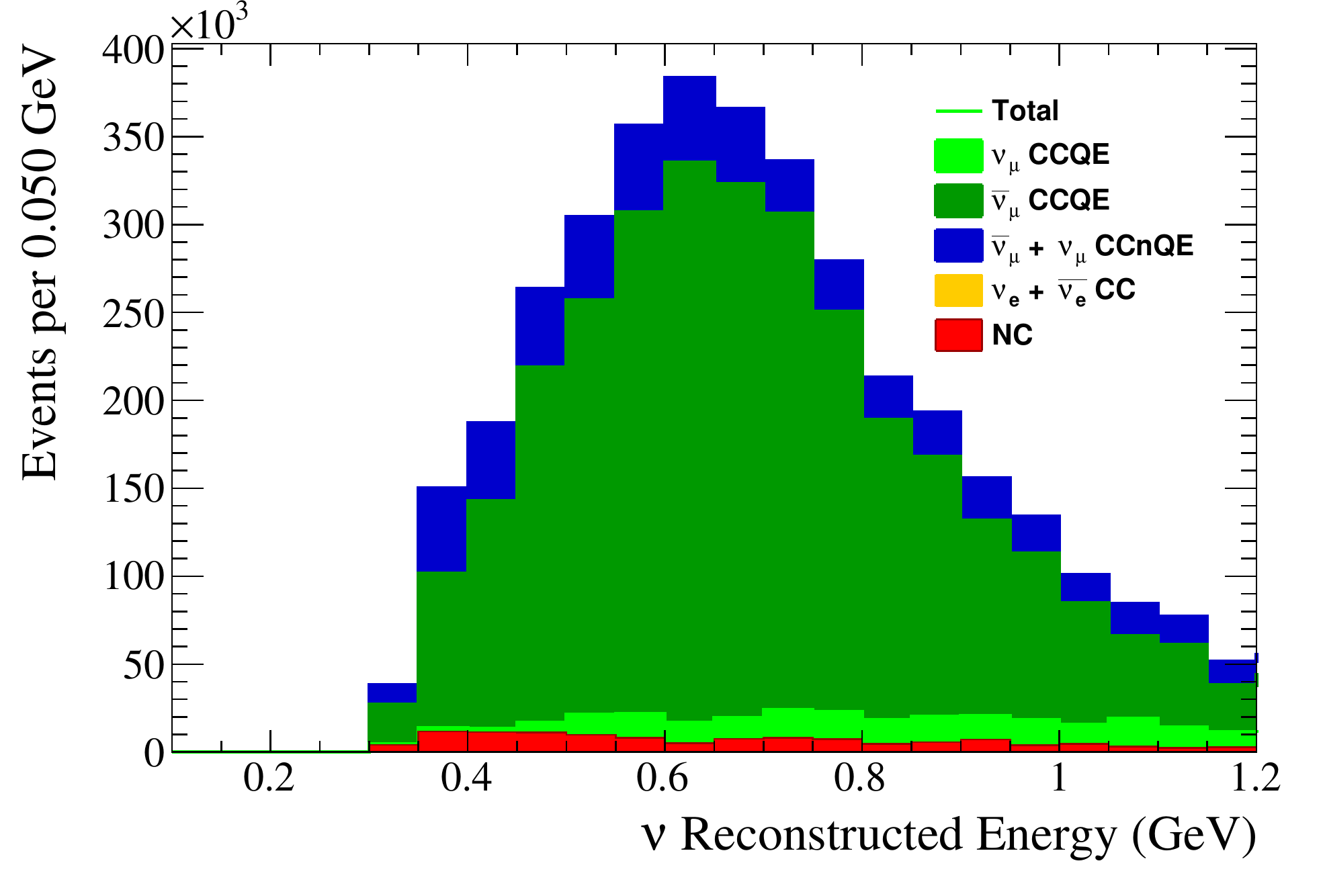} 
        \label{fig:predicted-titus-spectra_n_rhcmulike}
    }
    
\caption{Predicted neutron tagged TITUS spectra.}
\label{fig:predicted-titus-spectra_n}
\end{figure}

The numbers of predicted events rates at Hyper-K are shown in
Table~\ref{1re_event_table}, Table~\ref{1rmu_event_table} and
Figure~\ref{fig:predicted-hk-spectra} according to our tank and beam
assumptions, where FHC corresponds to the \FHC, and RHC corresponds to
the \RHC.  Notable features which are important for the measurement of
$\delta_{CP}$ and $\sin^2(\theta_{23})$ include the wrong sign
background component of the RHC spectra and the intrinsic beam
$\nu_{e}$ component.  The TITUS spectra are in
Figure~\ref{fig:predicted-titus-spectra_non} and
Figure~\ref{fig:predicted-titus-spectra_n} for no neutron tagged and
neutron tagged samples, respectively.

\begin{table}[htpb]
\begin{center}
\begin{tabular}{c c c c c c c c} \hline \hline
 & $\nu_\mu$ CC &  $\bar{\nu}_\mu$ CC & $\nu_e + \bar{\nu}_e$ CC  & NC & $\nu_\mu \rightarrow \nu_e$ CC & $\bar{\nu}_\mu \rightarrow \bar{\nu}_e$ CC & Total \\
 \hline
 $\nu$ mode 		&  10   &  0    &  383    &    204  &   2622  &   17       &   3236   \\
  $\bar{\nu}$ mode 	&  4     &  3    &  479   &   266   &  304     &  1076    &   2132   \\
  \hline\hline
\end{tabular}
\caption{1$R_e$ event rate breakdown.}
\label{1re_event_table}
\end{center}
\end{table}
\vspace{1cm}

\begin{table}[htpb]
\begin{center}
\begin{tabular}{c c c c c c c c}\hline \hline
 & $\nu_\mu$ CC &  $\bar{\nu}_\mu$ CC & $\nu_e + \bar{\nu}_e$ CC  & NC & $\nu_\mu \rightarrow \nu_e$ CC & $\bar{\nu}_\mu \rightarrow \bar{\nu}_e$ CC & Total \\
 \hline
 $\nu$ mode 		&  12315   &  899      & 9    &   1122   &   44  &   0     &   14389   \\
  $\bar{\nu}$ mode 	&  7139     &  12425    &  10   &   1321   &  6     &  14    &   20915   \\
\hline\hline
\end{tabular}
\caption{1$R\mu$ event rate breakdown.}
\label{1rmu_event_table}
\end{center}
\end{table}

\subsection{Systematic variations}
\begin{figure}[htpb]\centering
    \subfloat[FHC$1Re$]{
        
\includegraphics[width=0.45\textwidth]{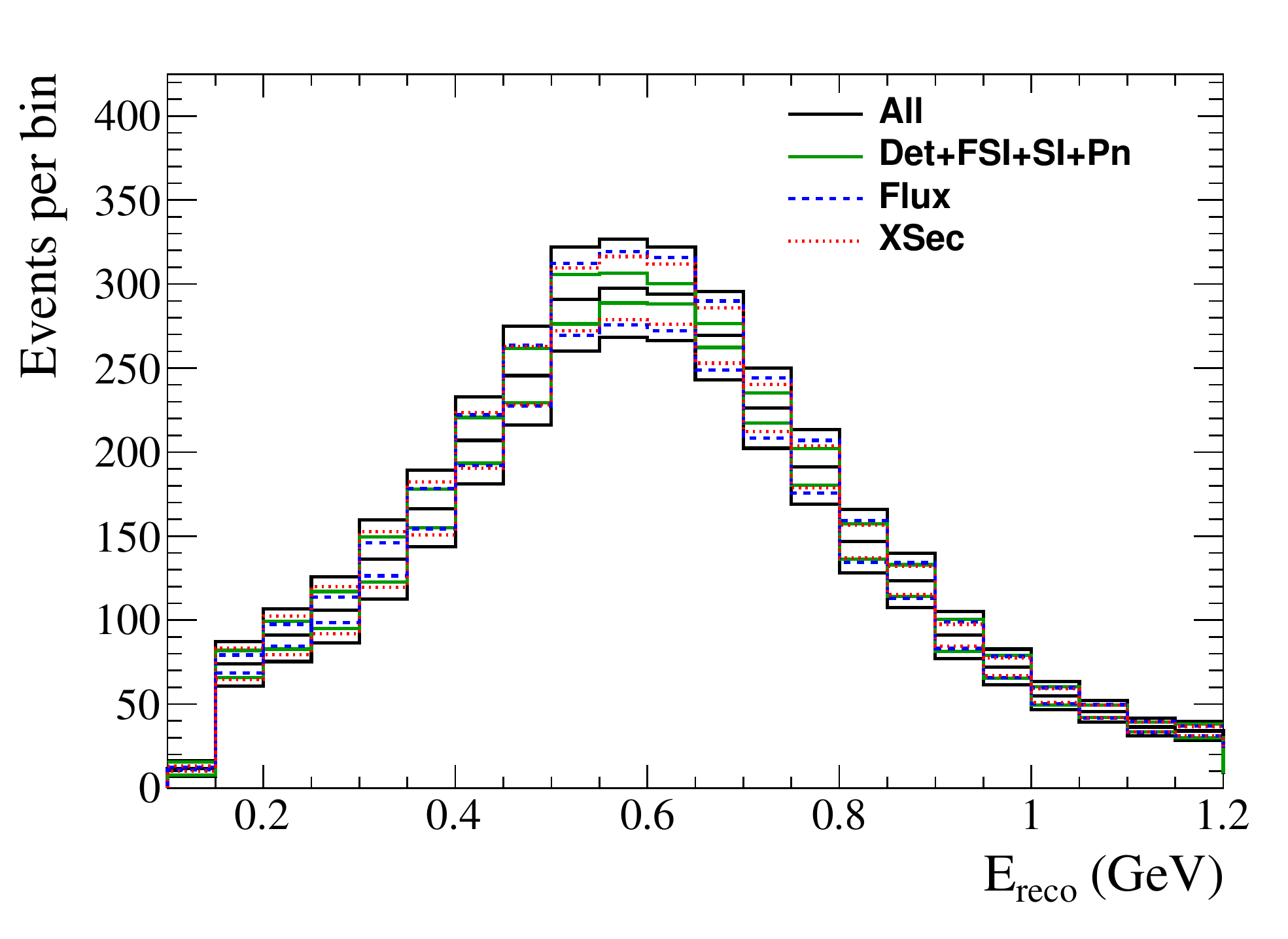} 
        \label{fig:systematic-variations_elike}
    }
    \subfloat[RHC$1Re$]{
        
\includegraphics[width=0.45\textwidth]{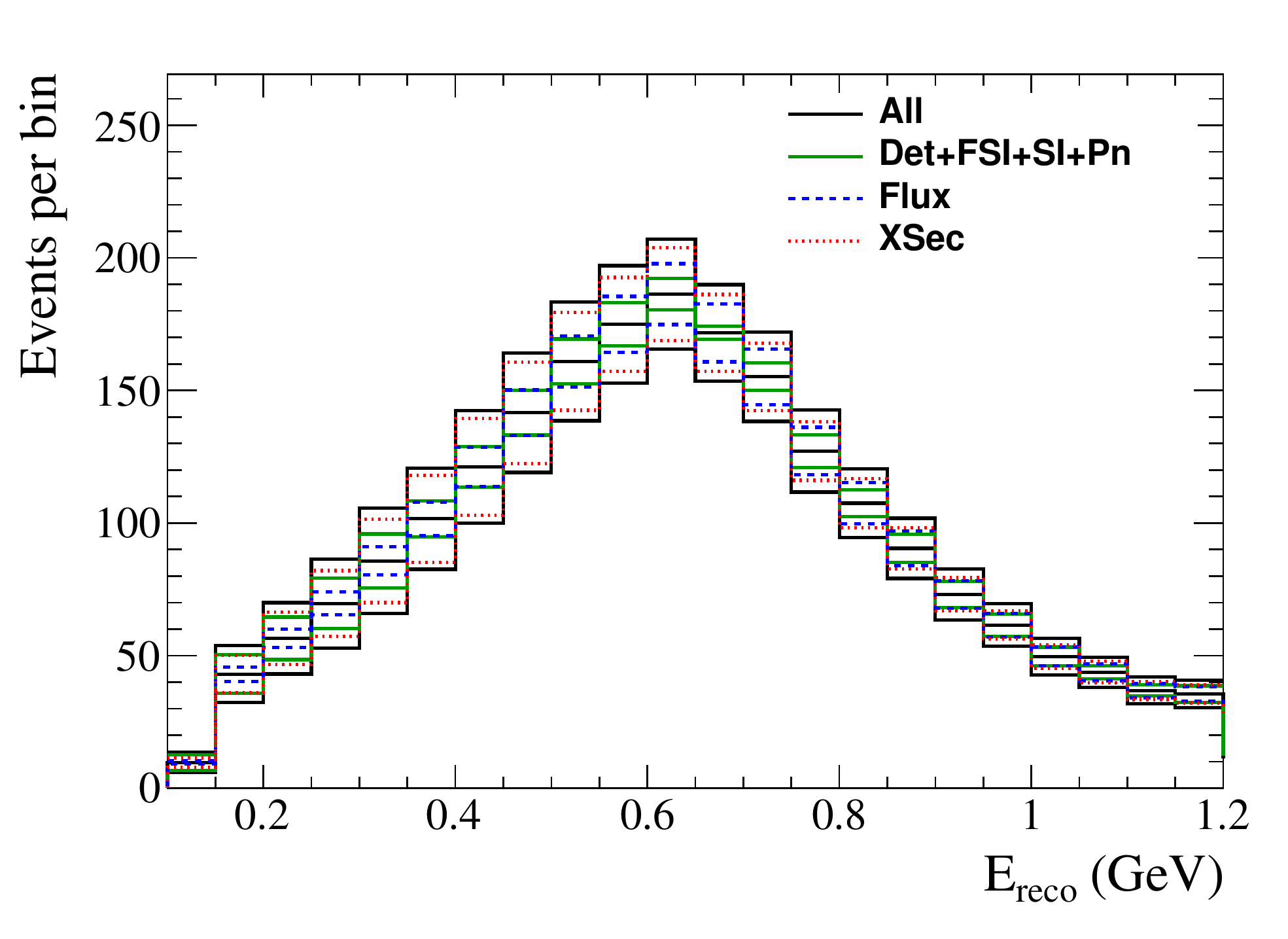} 
        
        \label{fig:systematic-variations_rhcelike}
    }
    
    \subfloat[FHC$1R\mu$]{
        
\includegraphics[width=0.45\textwidth]{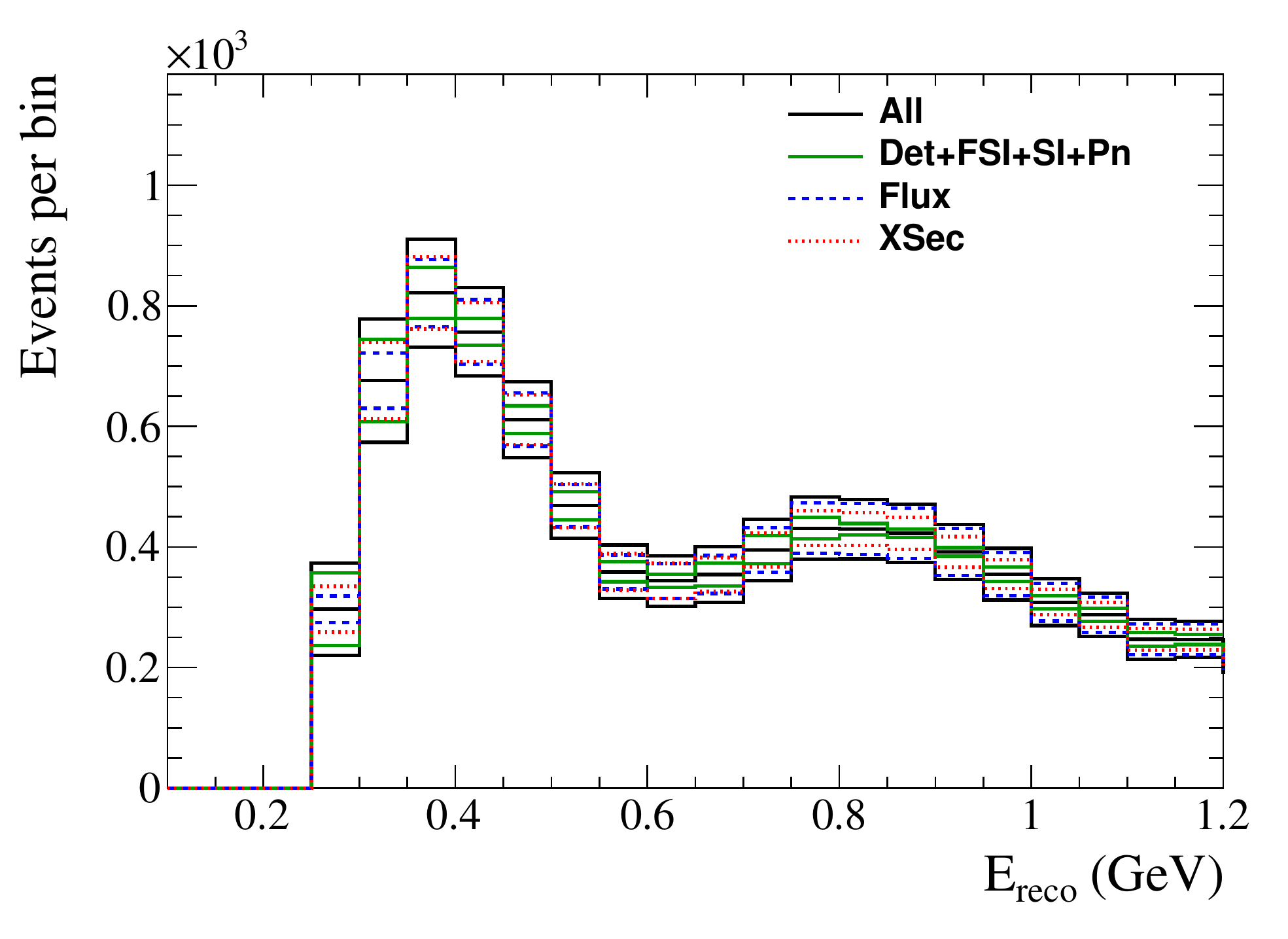} 
        
        \label{fig:systematic-variations_mulike}
    }
    \subfloat[RHC$1R\mu$]{
        
\includegraphics[width=0.45\textwidth]{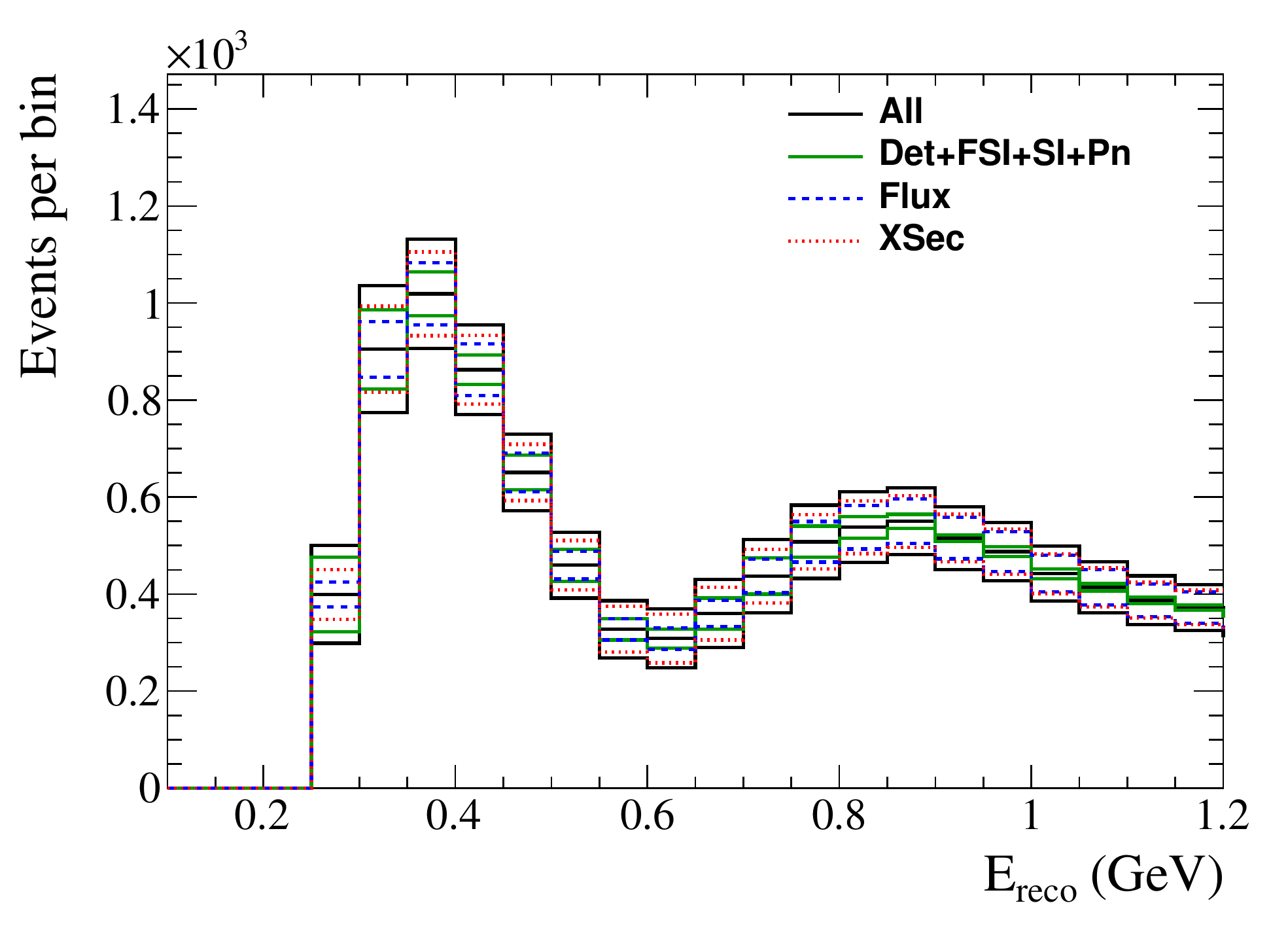} 
        
        \label{fig:systematic-variations_rhcmulike}
    }
    
        \subfloat[Ratio FHC$1Re$/RHC$1Re$]{
        
\includegraphics[width=0.45\textwidth]{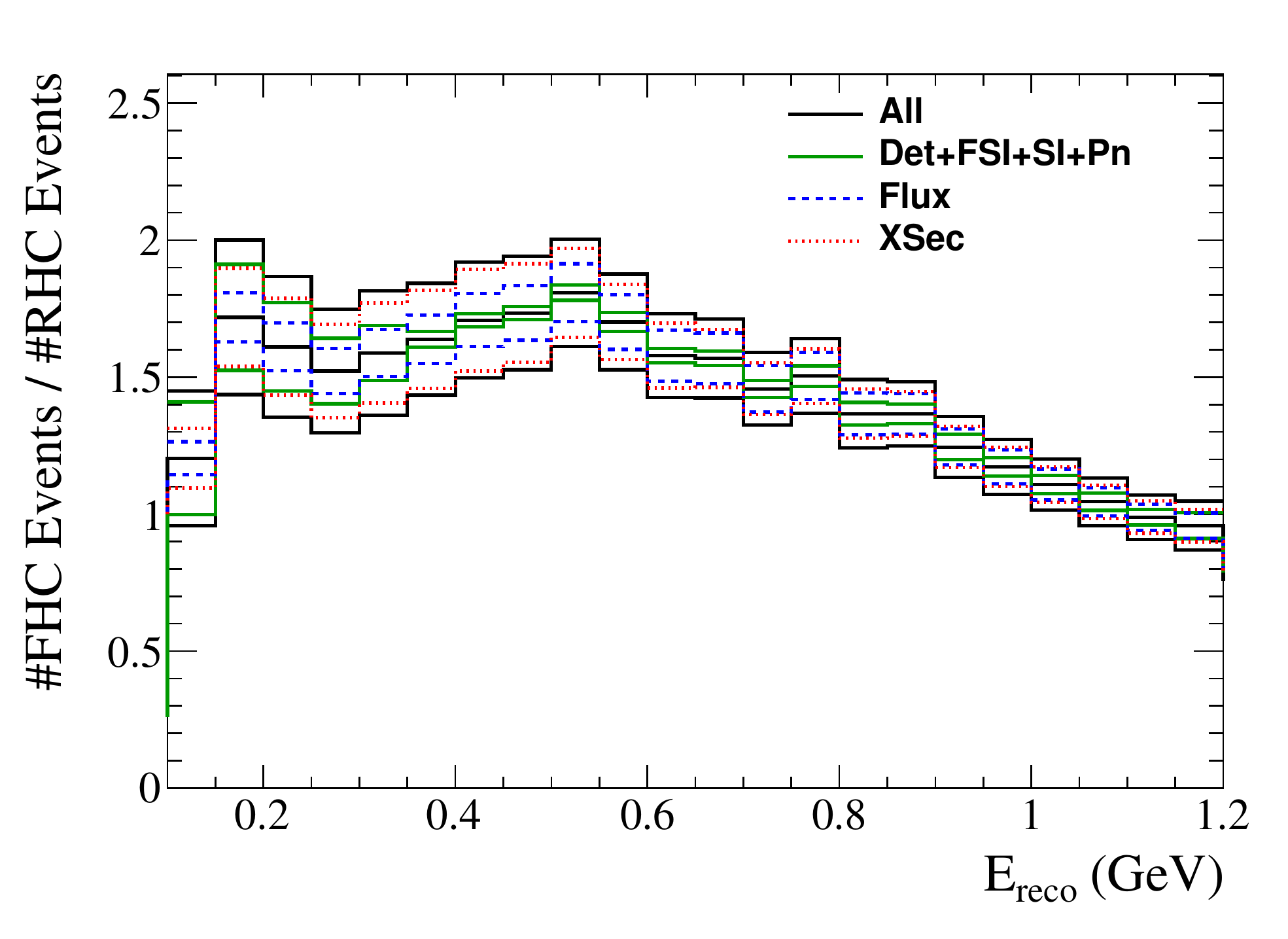} 
        
        \label{fig:systematic-variations_elike_ratio}
    }

\caption{
Systematic variations of Hyper-K energy spectra due to fluctuations of
systematic parameters produced by randomising the systematic
parameters 5,000 times and calculating the root mean squares of
each bin. For oscillation parameters given in Table~\ref{tab:osc_params_joint_asimov}.}
\label{fig:systematic-variations}
\end{figure}

%Oscillation parameters $\sin^2(\theta_{23})=0.528$, $\sin^2(\theta_{12})=0.306$, $\sin^2(\theta_{13})=0.025$, $\delta_{cp}=-1.6$, $\Delta m^2_{32} =2.5\cdot 10^{-3} eV^2$, $\Delta m^2_{12} =7.5\cdot 10^{-5} eV^2$, hierarchy=normal

Figure~\ref{fig:systematic-variations} shows that the uncertainty on
each bin in the Hyper-K samples. We see that the cross section and
flux errors are the dominant sources of uncertainty on all samples and
the detector+PFSI+SI+PN errors have a limited effect. In total the
uncertainty on each bin is approximately
10\%. Figure~\ref{fig:systematic-variations_elike_ratio} shows the
uncertainty on the ratio of events in each neutrino mode $1Re$ and
antineutrino mode $1Re$ bin, an important quantity to constrain
$\delta_{cp}$, is dominated by flux and cross section uncertainties
beyond 0.4 MeV where most events occur. This is important as the flux
and cross section parameters are what we expect the TITUS samples to
constrain most, although they also have some power to constrain
PFSI+SI+PN errors.

%For the TITUS samples the uncertainty is slightly larger at $\sim$15\%, the majority of which is from the detector+PFSI+SI+PN errors. Figure~\ref{fig:systematic-variations_elike_ratio} shows the uncertainty on the ratio of neutrino mode and antineutrino mode $e$-like spectra. 
\begin{center}
\begin{figure}[htpb]\centering
    \subfloat[FHC$1Re$]{
        
\includegraphics[width=0.45\textwidth]{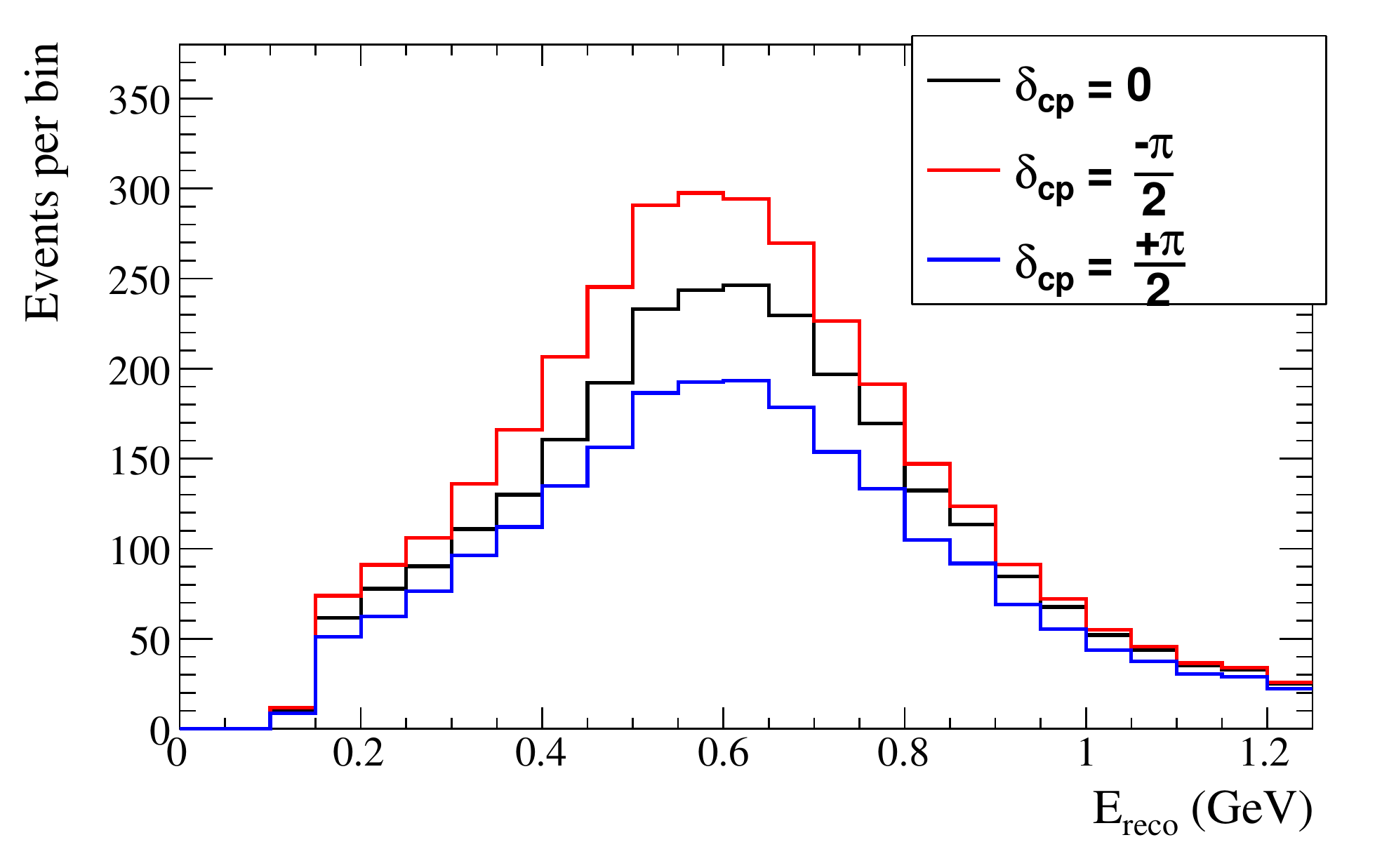} 
        
        \label{cp-variations_fhce}
    }
    \subfloat[RHC$1Re$]{
     
\includegraphics[width=0.45\textwidth]{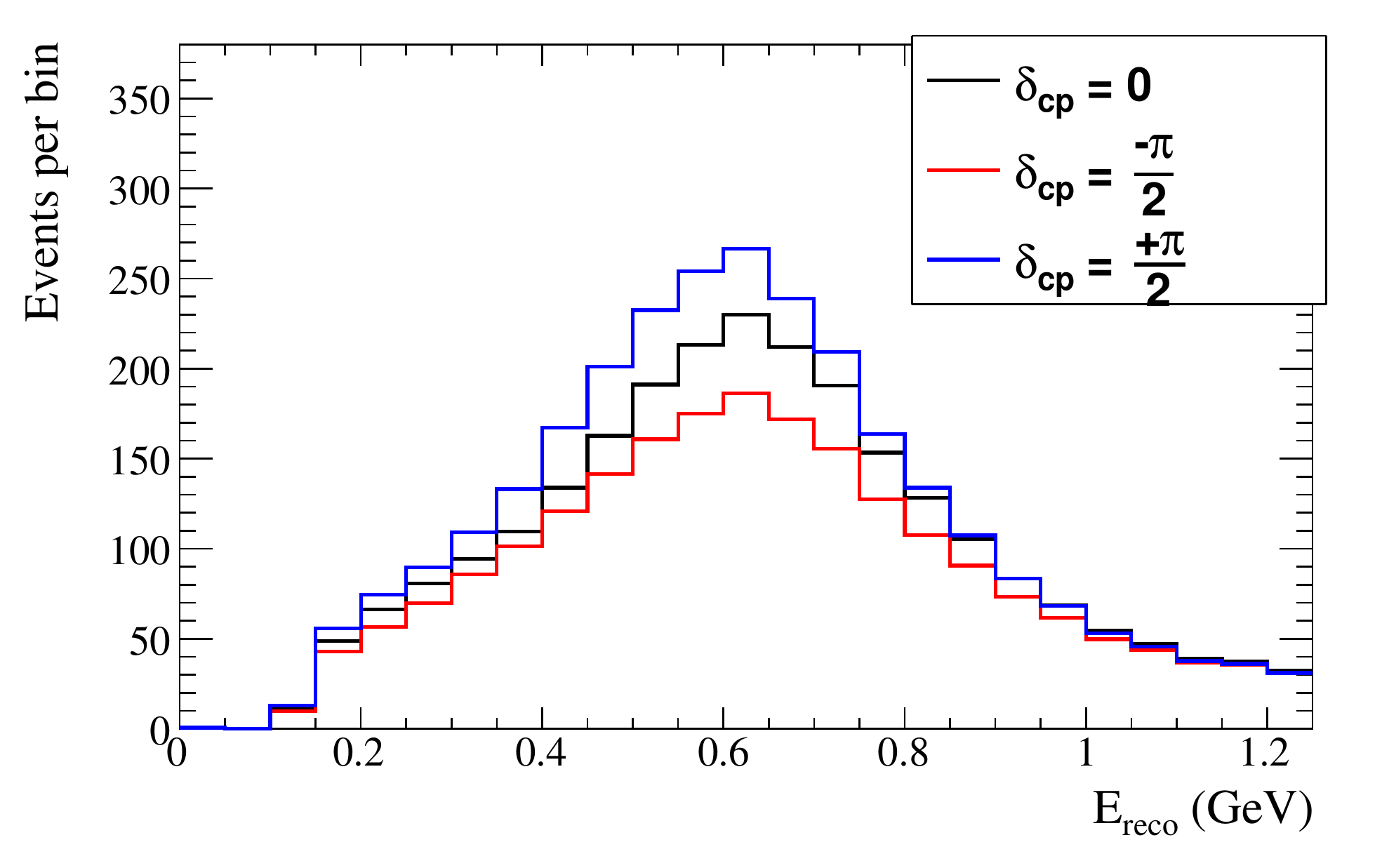} 
        
        \label{cp-variations_rhce}
    }
    
    \subfloat[FHC$1Re$ fractional change]{
        
\includegraphics[width=0.45\textwidth]{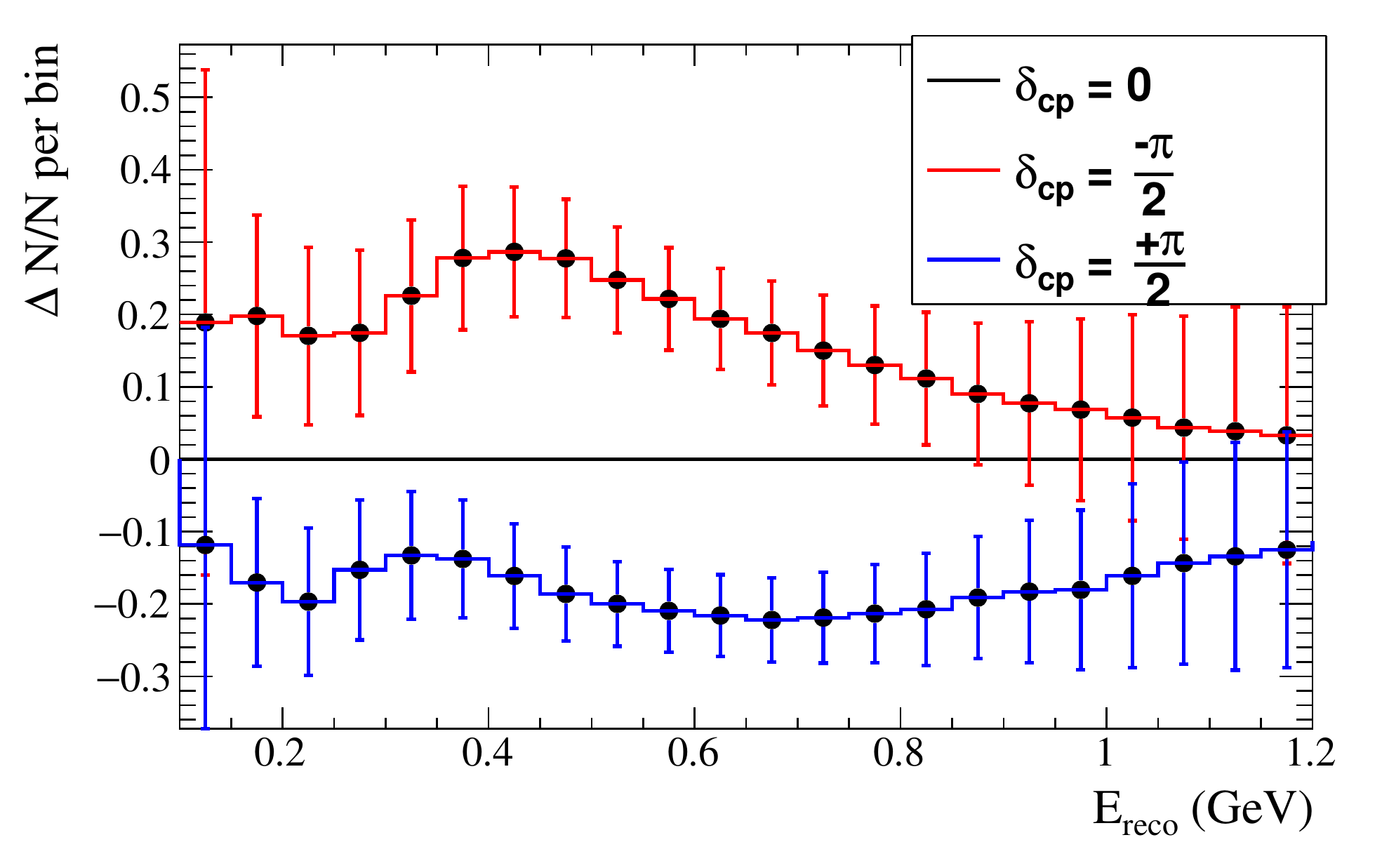} 
        
        \label{cp-variations_fhce_ratio}
    }
    \subfloat[RHC$1Re$ fractional change]{
        
\includegraphics[width=0.45\textwidth]{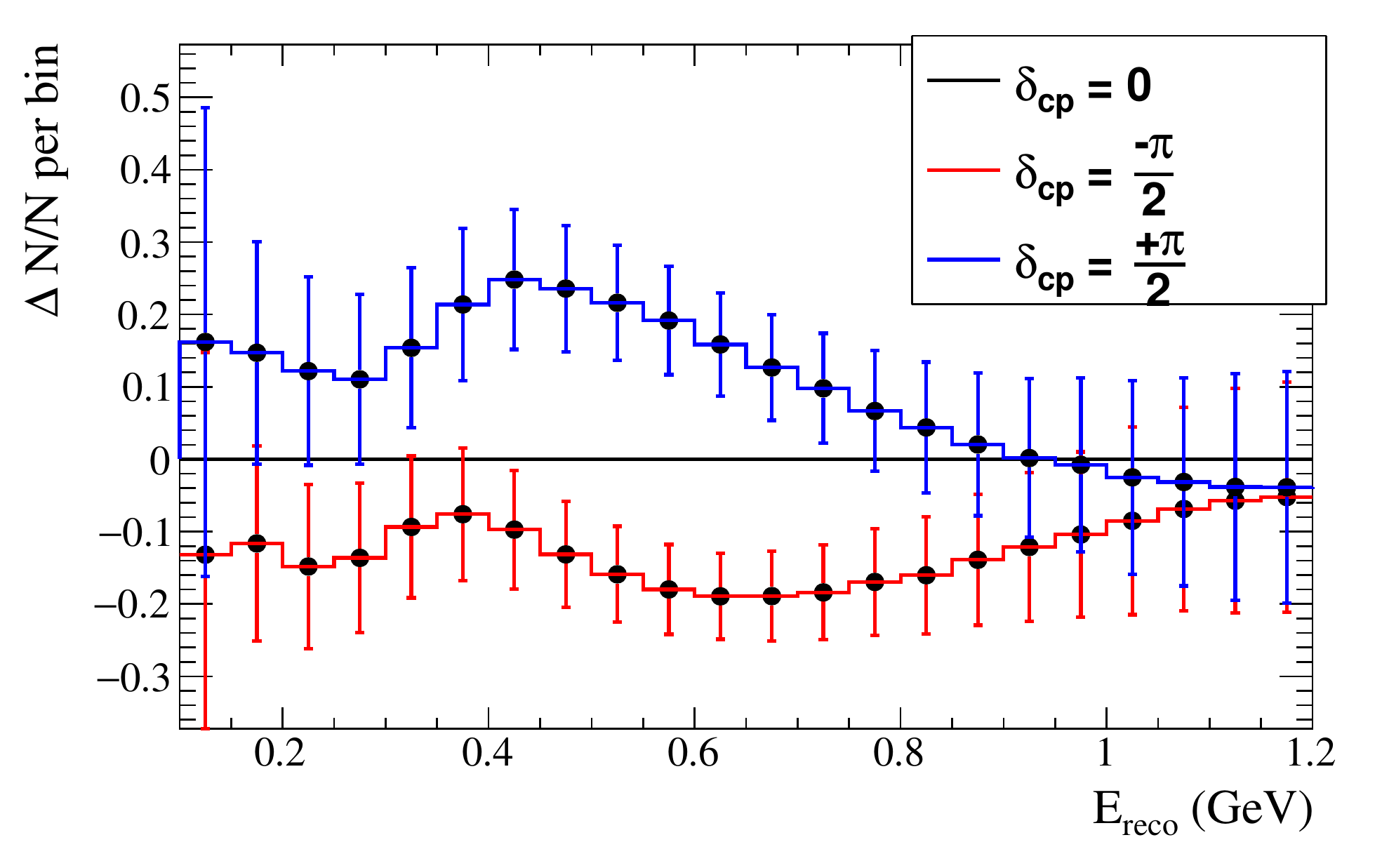} 
        
        \label{cp-variations_rhce_ratio}
    }
    
    \subfloat[FHC$1R\mu$]{
        
\includegraphics[width=0.45\textwidth]{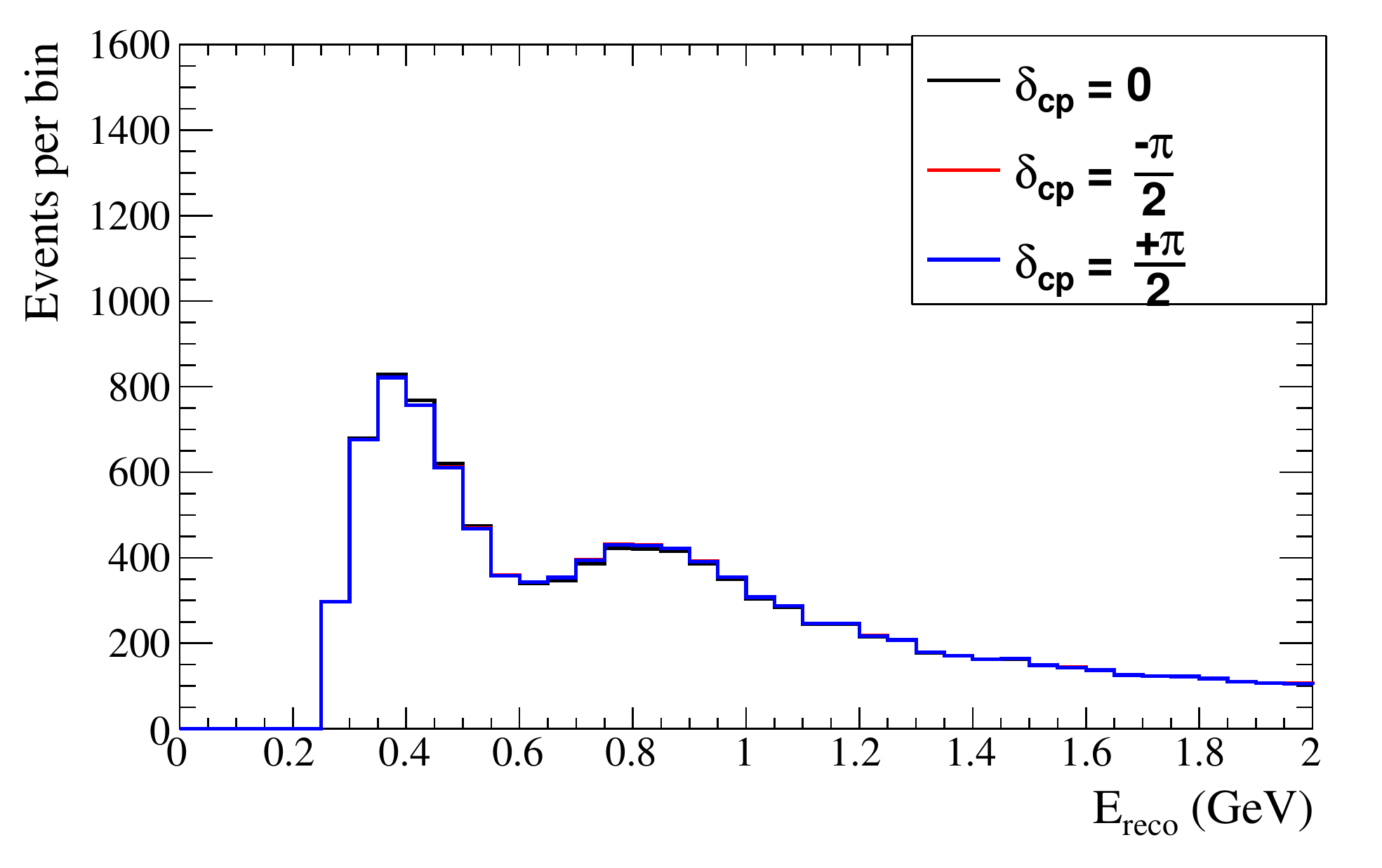} 
        
        \label{cp-variations_fhcmu}
    }
    \subfloat[RHC$1R\mu$]{
        
\includegraphics[width=0.45\textwidth]{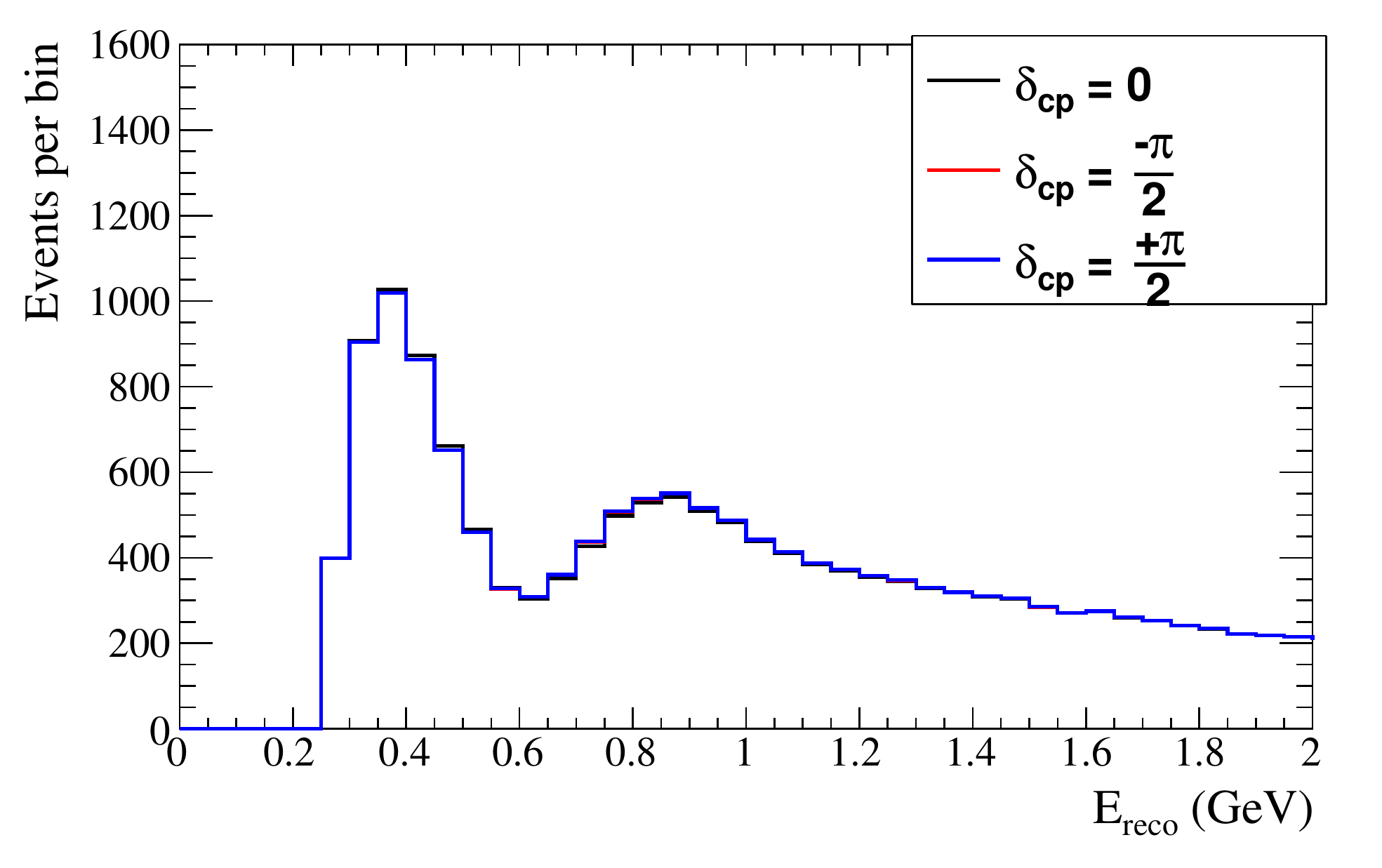} 
        
        \label{cp-variations_rhcmu}
    }
    
    \subfloat[FHC$1R\mu$ fractional change]{
        
\includegraphics[width=0.45\textwidth]{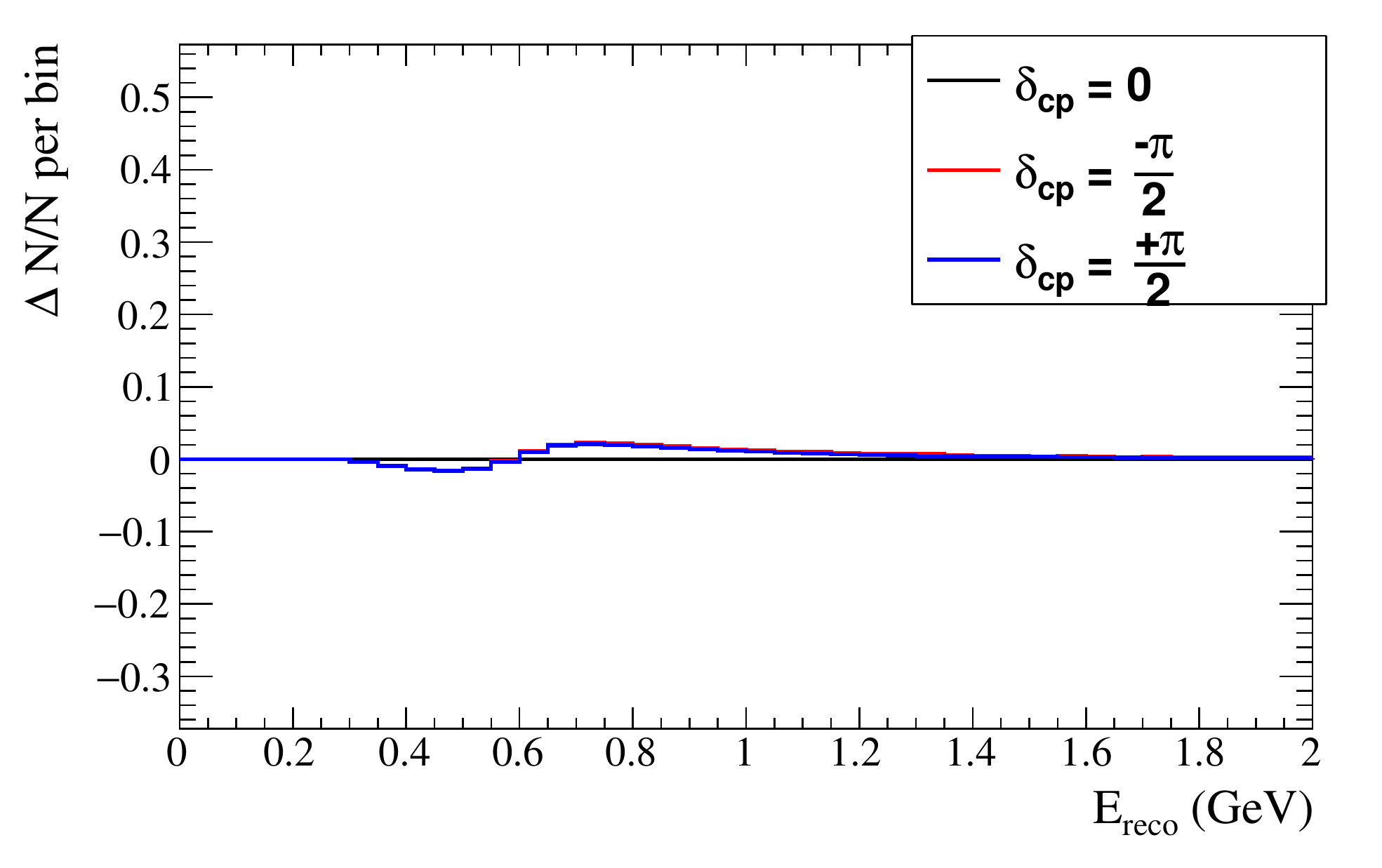}      
  
        \label{cp-variations_fhcmu_ratio}
    }
    \subfloat[RHC$1R\mu$ fractional change]{
        
\includegraphics[width=0.45\textwidth]{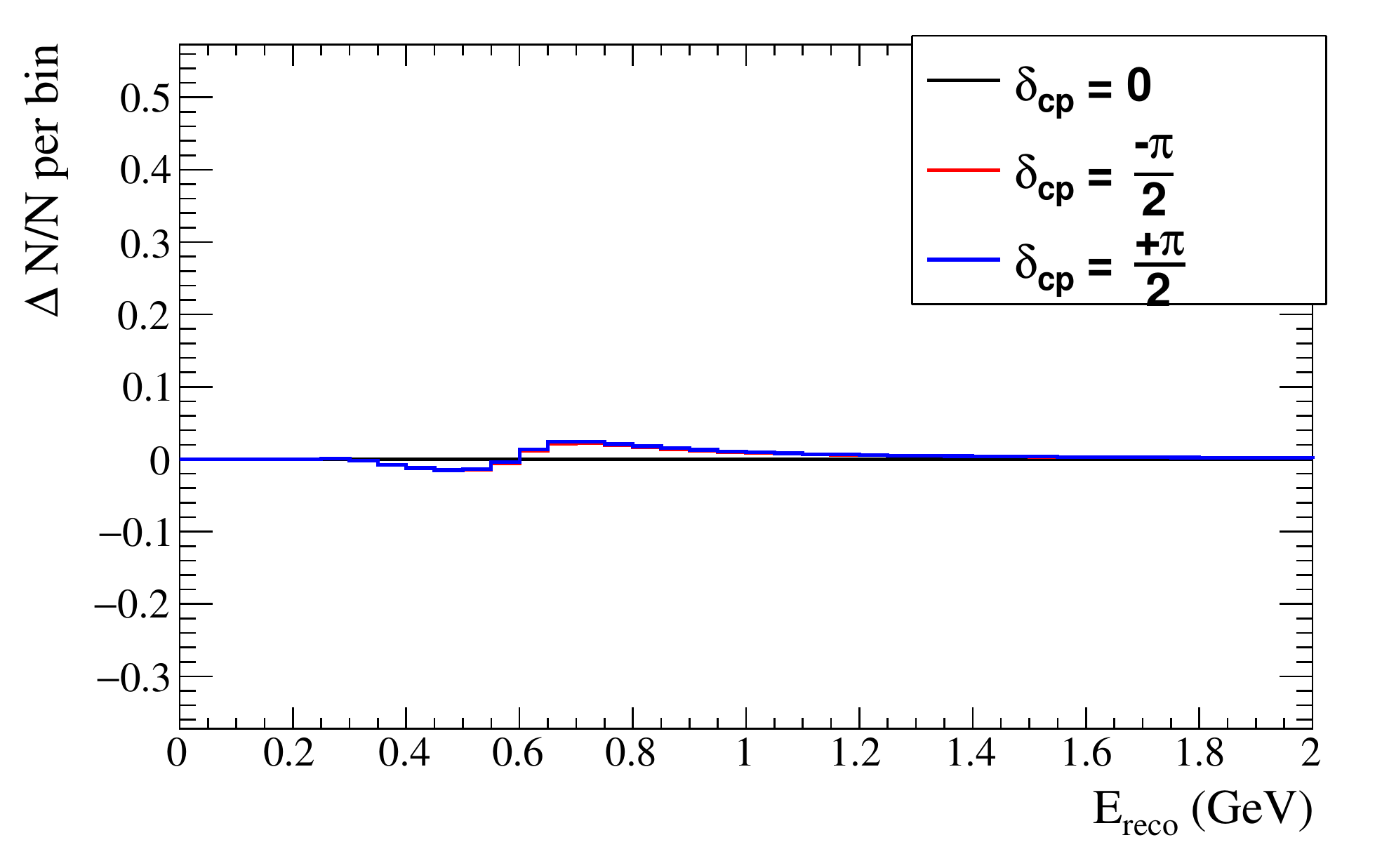} 
  
        \label{cp-variations_rhcmu_ratio}
    }

\caption{Effect of $\delta_{CP}$ on Hyper-K spectra.}
\label{cp-variations}
\end{figure}
\end{center}

%\subsection{Correlation matricies}
%\begin{figure}[htpb]\centering
%\begin{center} 
%Flux \\
%\includegraphics[width=450px]{./figs/full_sensitivity/matrix/Matricies_Flux.pdf} \\
%\begin{tabular}{cc}
%TITUS Det+FSI & Hyper-K Det+FSI \\
%\includegraphics[width=200px]{./figs/full_sensitivity/matrix/Matricies_NDDet.pdf}  &
%\includegraphics[width=200px]{./figs/full_sensitivity/matrix/Matricies_hkdet.pdf} \\
%\end{tabular}
%\caption{Correlation matrices for the 3 groups of systematic uncertainty considered in the fit. 0 correlation is assumed between the 3 groups}
%\end{center}
%\label{correlation-matrix}
%\end{figure}
%
%The error matrix \textbf{C} is constructed from a block diagonalization of the 3 independent error matrices. Figure~\ref{correlation-matrix} shows the strength of correlations between the $\nu$ and $\bar{\nu}$ reconstructed energy bins for the Hyper-K and TITUS Det+FSI errors which is particularly important as any asymmetry between them would have a big effect on the CP violation search. The flux matrix also shows the importance of constraining the TITUS flux which is very strongly correlated with the Hyper-K flux in both $\nu$ and $\bar{\nu}$ mode.

\subsection{$\delta_{CP}$ sensitivity}
In Figure~\ref{cp-variations} we see the effect of $\delta_{CP}$ on
the Hyper-K samples. It is clear from this how the effect of
$\delta_{CP}$ is confined to the appearance samples and causes a shift
in the total event rate by up to 20\%. We also see the effect of
$\delta_{CP}$ on the shape of the Hyper-K spectra which highlights the
importance of the energy resolution for both appearance and
disappearance spectra.

\subsubsection{CP violation sensitivity}
\begin{figure}[htb]
\centering
\includegraphics[width=0.6\textwidth]{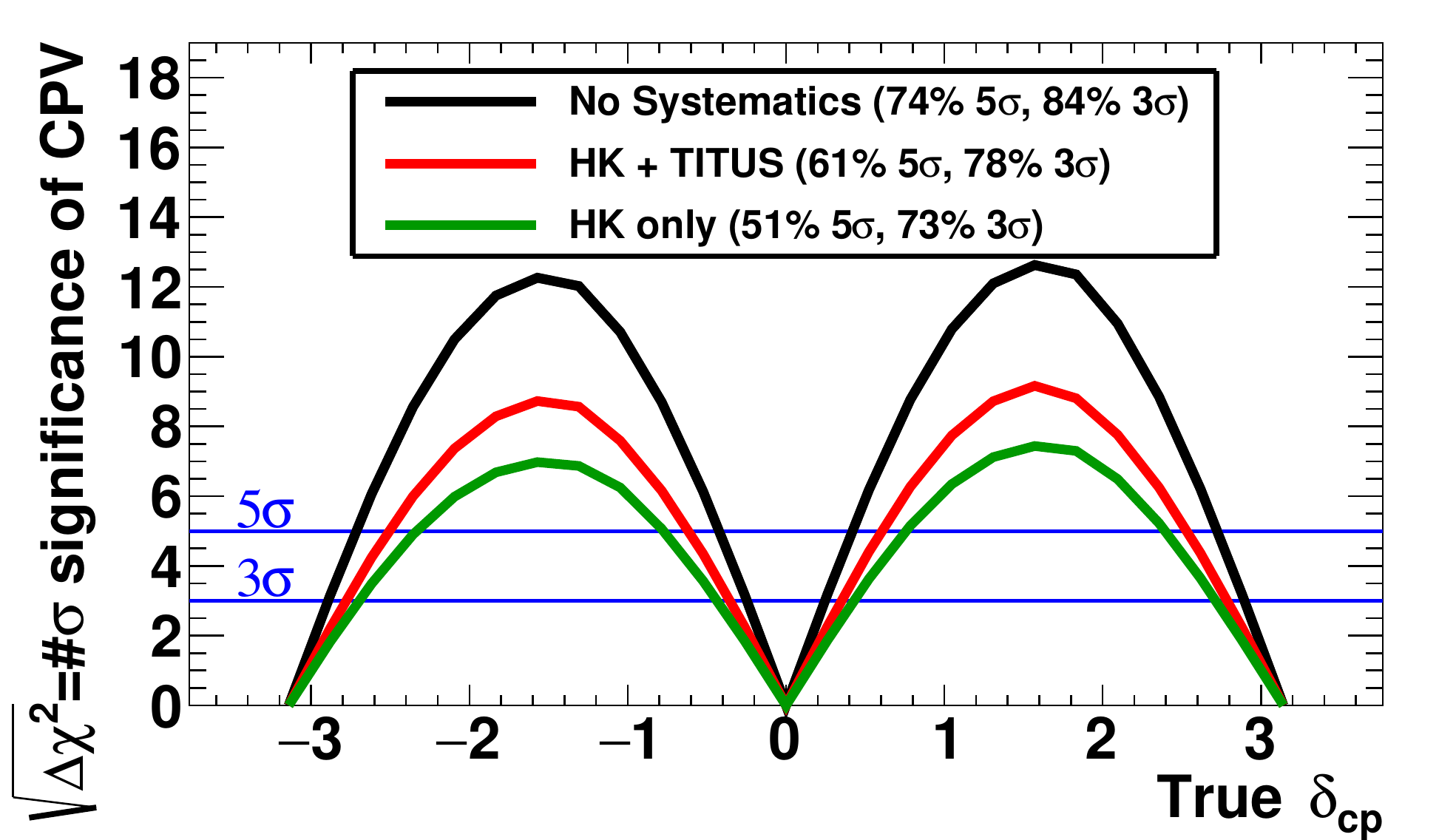} 
\caption {
Significance to measure CP violation as a function of $\delta_{CP}$
with Hyper-K only, TITUS (Gd-doped) and Hyper-K (not Gd-doped), and Hyper-K without considering
sources of systematic uncertainty.
For the studies, the true values of
oscillation parameters to generate the data sets were taken from 
Table~\ref{tab:osc_params_joint_asimov} but with $\sin^2(\theta_{23})=0.5$.
The fraction of $\delta_{CP}$ for
which CP violation will be measured at 5$\sigma$ and 3$\sigma$ is
given in the legend. The sensitivity is for the case of normal
hierarchy and assuming it is known.}
\label{fig:cp-violation}
\end{figure}

The sensitivity to CP violation is shown in
Figure~\ref{fig:cp-violation}. A significant improvement in the
sensitivity can be seen with the addition of TITUS samples in the
fit. Hyper-K alone, without any near detector constraint, can
determine CP violation at the 5$\sigma$ level for 50\% of
$\delta_{cp}$ space and can provide a 3$\sigma$ measurement for
72\%. With the TITUS constraint, Hyper-K will be able to provide a
5$\sigma$ measurement for 62\% of $\delta_{cp}$ space, close to the
74\% achieved without considering systematic uncertainties.

\subsubsection{1D $\delta_{cp}$ fits }
\begin{figure}[htb]
\centering
\includegraphics[width=0.6\textwidth]{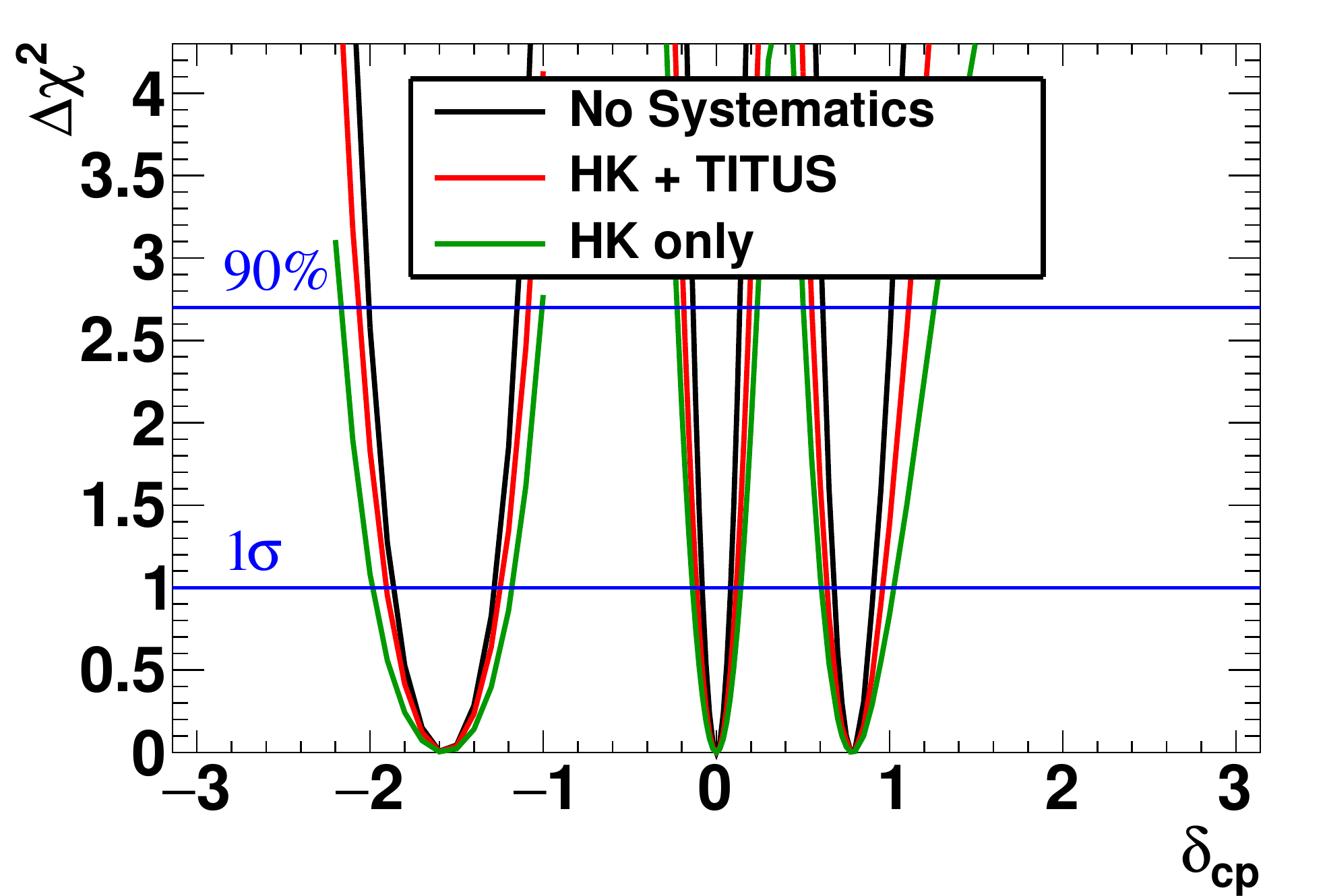} 
\caption{
Significance of measuring CP violation as a function of $\delta_{CP}$
with Hyper-K only, TITUS and Hyper-K, and Hyper-K without considering
sources of systematic uncertainty for 3 Asimov fits to $\delta_{CP}$.
The sensitivity is for the case of normal hierarchy and assuming it is
known. The uncertainty on $\delta_{CP}$ is given in
Table~\ref{tab:dcp_error}.}
\label{fig:cp-fit}
\end{figure}

The results of 3 Asimov fits to $\delta_{CP}$ are shown in
Figure~\ref{fig:cp-fit} and the constraints achieved given in
Table~\ref{tab:dcp_error}. We find that Hyper-K alone provides a weak
measurement with the uncertainty in $\delta_{CP}$ up to seven times
greater in comparison to the case where no systematic uncertainty is
considered. Figure~\ref{fig:cp-fit} also shows how the likelihood can
become asymmetric due to a slight symmetry in the appearance spectra
around $\pm \frac{\pi}{2}$, which also act as boundaries, and how the
constraint on $\delta_{CP}$ can depend on the true value.

\begin{table}[htpb]
\begin{center}
\begin{tabular}{c| c |c |c } 
\hline
 \hline & \multicolumn{3}{c}{Error (radians/\textit{degrees}) } \\ \hline True
 					$\delta_{cp}$ & No Systematics
 					& Hyper-K only & TITUS +
 					Hyper-K \\

 \hline
 $0$ 	   			&   0.082 {\it(4.7)} 	&	0.139 {\it(8.0)}	& 	 0.115 {\it(6.6)} 	  \\
 \hline
 $\frac{\pi}{4}$ 	   	&   0.115 {\it(6.6)} 	&		0.211 {\it(12.1)} 	& 	 0.161 {\it(9.2)}   \\
 \hline
 $-\frac{\pi}{2}$ 	   	&  0.296 {\it(16.9)}	&		0.405 {\it(23.2)} 	& 	 0.332 {\it(19.0)}   \\
  \hline\hline
\end{tabular}
\end{center}
\caption{1 $\sigma$ error (radians/\textit{degrees}) of $\delta_{cp}$ for each fit shown in Figure~\ref{fig:cp-fit}.}
\label{tab:dcp_error}
\end{table}

\subsection{$23$ sector sensitivity}
\subsubsection{$\sin^2(\theta_{23})\neq0.5$ sensitivity}
\begin{figure}[htb]
\centering
\includegraphics[width=0.6\textwidth]{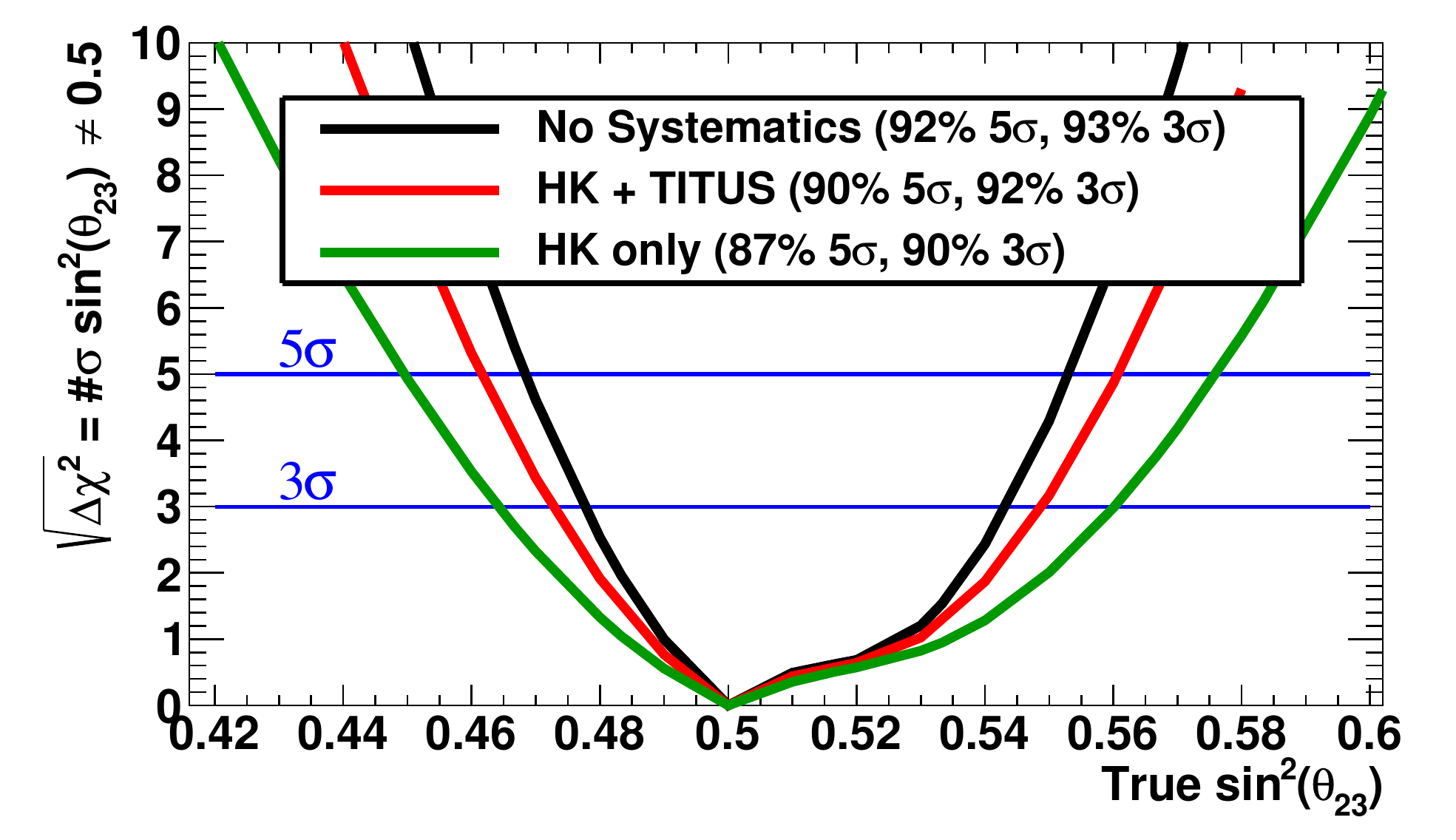} 
\caption{
Significance to exclude $\sin^2(\theta_{23})=0.5$ as a function of
$\sin^2(\theta_{23})$ with the combinations of samples and constraints
given in the legend and the fraction of $\sin^2(\theta_{23})$ Hyper-K
will be able to exclude $\sin^2(\theta_{23})=0.5$ (at 5$\sigma$ and
3$\sigma$). For the studies, the true values of oscillation parameters
to generate the data sets were taken from
Table~\ref{tab:osc_params_joint_asimov}. Sensitivity is for the case
of normal hierarchy and assuming it is known.  Due to a non zero
$\sin^2(\theta_{13})$, the disappearance spectra have a symmetry
around $\sin^2(\theta_{23})=0.511$, which is the cause of the flat
feature in the sensitivity between 0.5 and 0.511.  }
\label{fig:ss23-neq0.5}
\end{figure}

The sensitivity to $\sin^2(\theta_{23})\neq0.5$ is shown in
Figure~\ref{fig:ss23-neq0.5}. The sensitivity increases as true
$\sin^2(\theta_{23})$ is away from 0.5 however the exclusion is not
symmetric around 0.5.  Due to the non zero value of
$\sin^2(\theta_{13})$, the disappearance spectra have a symmetry
around $\sin^2(\theta_{23}) = 0.513$, which is the cause of the flat
feature in the sensitivity between 0.5 and 0.53.  As expected, there
is an improvement in the sensitivity to $\sin^2(\theta_{23})\neq0.5$
when TITUS is included in the fit increasing the amount of
$\sin^2(\theta_{23})$ space by 2\%, halfway to the possible 91\% in
the case of no systematic uncertainty. The improvement is however not
as drastic as in the case of the CP violation measurement, as
constraints on the cross section model parameters and flux prior to
the fit lead to a small uncertainty on the shape of the oscillation
dip in the disappearance spectra.
\begin{figure}[htb]
\centering
\includegraphics[width=0.6\textwidth]{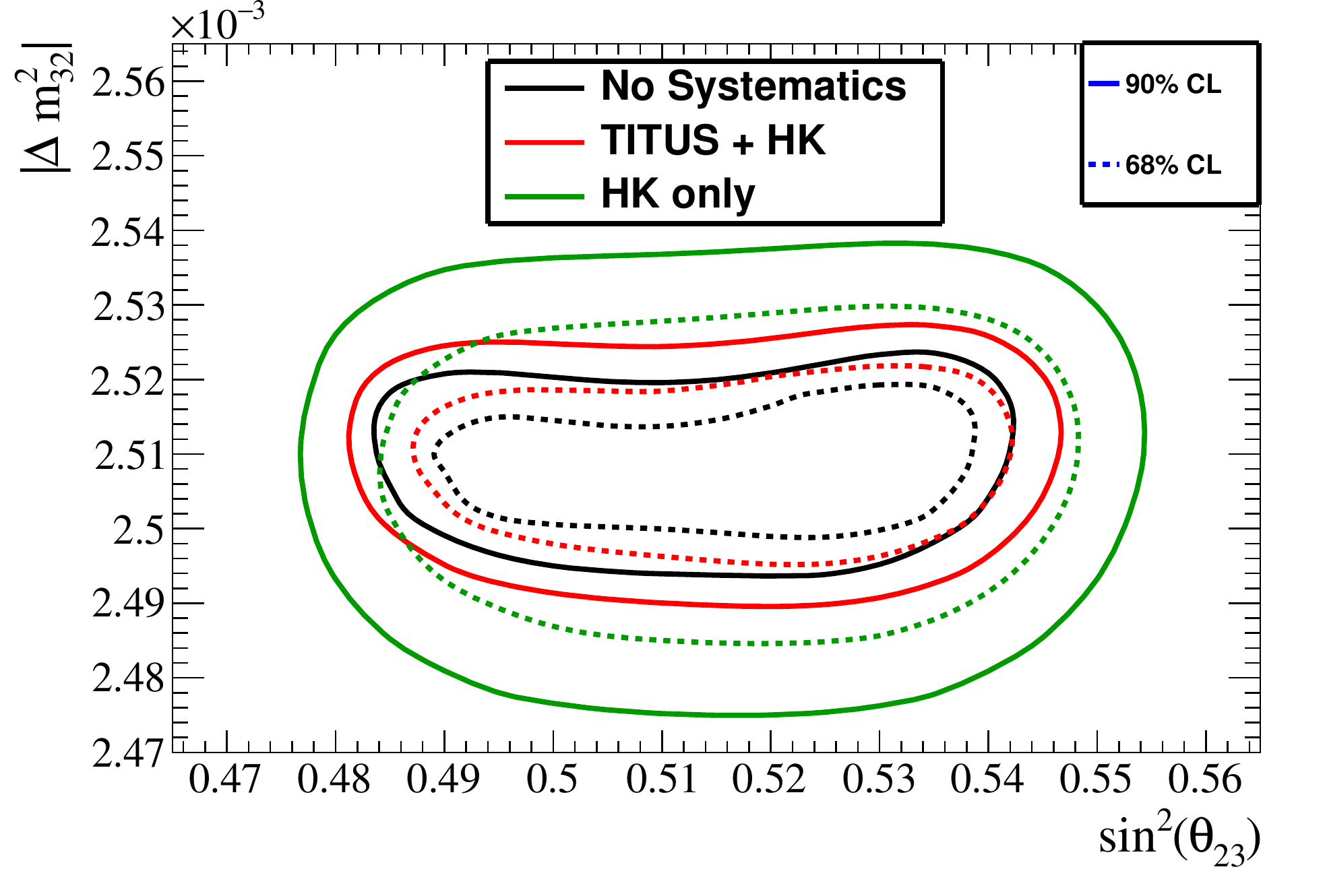} 
\caption{
The 68\% and 90\% confidence level contours for an Asimov fit of
$\sin^2(\theta_{23})$ and $\Delta m^2_{32}$ for the cases of no
systematic uncertainty, only HK constraint and both TITUS and HK
constraints, are shown. Confidence contours were made with the
constant $\Delta \chi^2$ method. For the studies, the true values of
oscillation parameters to generate the data sets were taken from
Table~\ref{tab:osc_params_joint_asimov}.  All oscillation and
systematic parameters were profiled in the production of the contours
except the hierarchy which was fixed to normal. }
\label{fig:ss23-dmsq}
\end{figure}

\subsubsection{$\sin^2(\theta_{23})$ vs $\Delta m^2_{32}$}
The sensitivity of an Asimov fit for the parameters
$\sin^2(\theta_{23})$ vs $\Delta m^2_{32}$ is shown in
Figure~\ref{fig:ss23-dmsq}.  In all fits we see the symmetry around
the point of maximal disappearance at $\sin^2(\theta_{23}) = 0.511$
however, as the true value is close to the point of maximal
disappearance, there is almost no sensitivity to the octant of
$\sin^2(\theta_{23})$ in any of the fits and the contours are
symmetric.  The TITUS samples greatly improve the sensitivity to the
23 sector parameters; note that the 90\% CL contour from the
TITUS+Hyper-K fit falling within the Hyper-K only 68\% CL
contour. With the TITUS constraint the expected 90\% CL ranges are
approximately 0.480 - 0.548 in $\sin^2(\theta_{23})$ and 0.00249 -
0.00253 (eV/c$^2$)$^2$ in $\Delta m^2_{32}$.

A summary of the sensitivities are given in Table~\ref{tab:titus_cpv_sensitivity_summary}
and Table~\ref{tab:titus_ss23_sensitivity_summary}.

\begin{table}[htb]
\centering
 \begin{tabular}{ c | c | c } \hline\hline Parameter(s) & 3 $\sigma$ &
 5 $\sigma$ \\
\hline
TITUS + HK 				                                  	&       78\%				&		61\%				  \\
HK  only										&       73\%				&		51\%			 	\\
No systematics	 								&       84\%       		 	  	&		74\%				 		 \\
\hline\hline
\end{tabular}
\caption{
Summary of the percentage of $\delta_{CP}$ space for which there is expected to be a measurement of 
CP violation at 3 $\sigma$ and 5 $\sigma$ significance.
}
\label{tab:titus_cpv_sensitivity_summary}
\end{table}

\begin{table}[hbt]
\centering
 \begin{tabular}{ c | c | c }
 \hline\hline
 Parameter(s)                                               		   	&      3 $\sigma$ 			& 5 $\sigma$ 	  						   \\
\hline
TITUS + HK 				                                  	&       92\%				&		90\%				  \\
HK  only										&       90\%				&		87\%			 	\\
No systematics	 								&       93\%       		   		&		92\%				 		 \\
\hline\hline
\end{tabular}
\caption{
Summary of the percentage of $\sin^2(\theta_{23})$ space for which there is expected to be a measurement which
excludes $\sin^2(\theta_{23})=0.5$ at 3 $\sigma$ and 5 $\sigma$ significance.
}
\label{tab:titus_ss23_sensitivity_summary}
\end{table}

\subsection{T2K phase-2 sensitivity}
\begin{figure}[htb]
\centering
\includegraphics[width=0.8\textwidth]{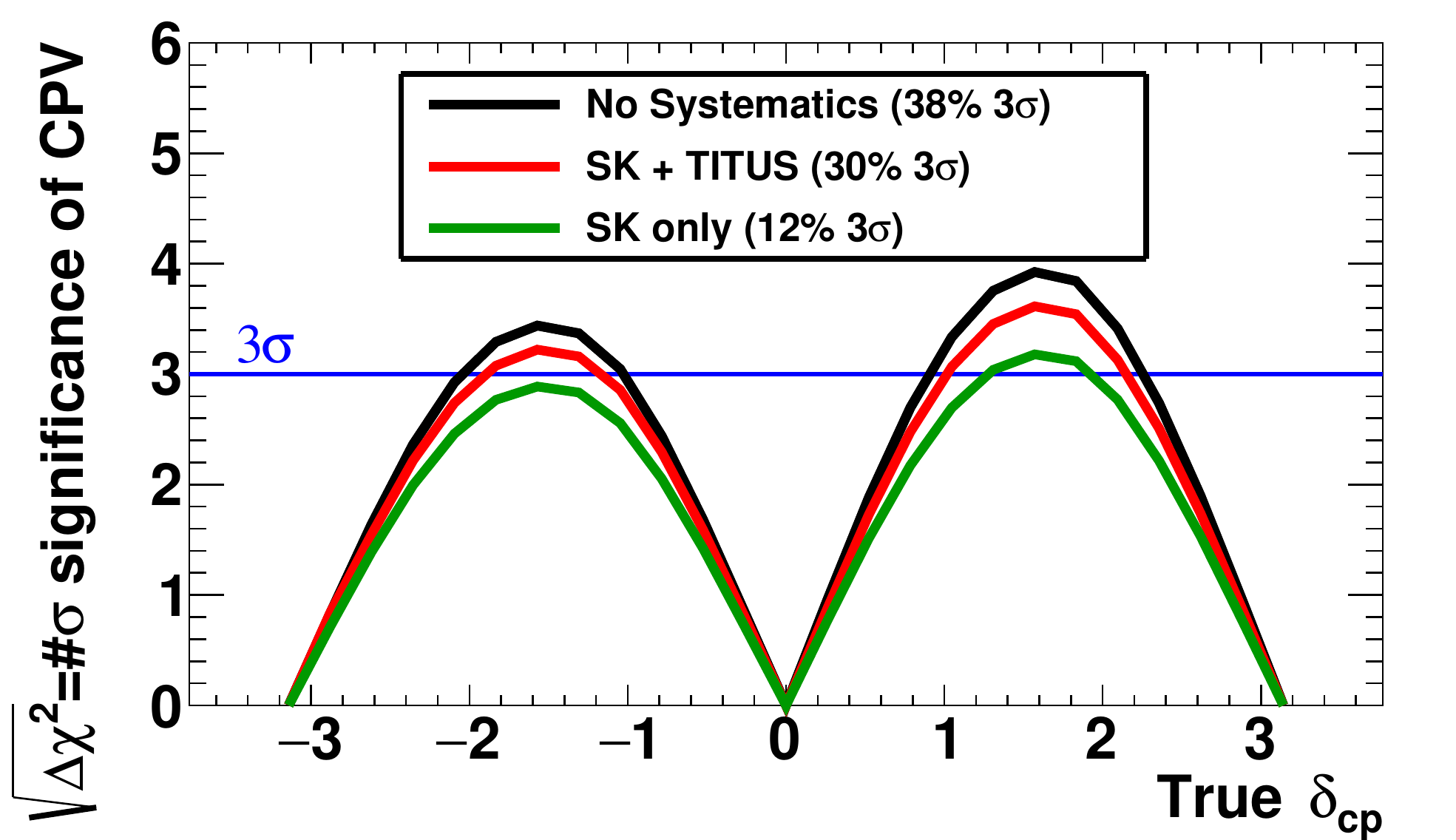} 
\caption {
Significance to measure CP violation as a function of $\delta_{cp}$
with T2K phase-2 only, TITUS and T2K phase-2, and T2K phase-2 without
considering sources of systematic uncertainty assuming T2K phase-2
statistics.  For the studies, the true values of oscillation
parameters to generate the data sets were taken from
Table~\ref{tab:osc_params_joint_asimov} but with
$\sin^2(\theta_{23})=0.5$.  The fraction of $\delta_{cp}$ for which CP
violation will be measured at 5$\sigma$ and 3$\sigma$ is given in the
legend. The sensitivity is for the case of normal hierarchy and
assuming it is known.}
\label{fig:cp-violation-t2k2}
\end{figure}

\begin{figure}[htb]
\centering
\includegraphics[width=0.8\textwidth]{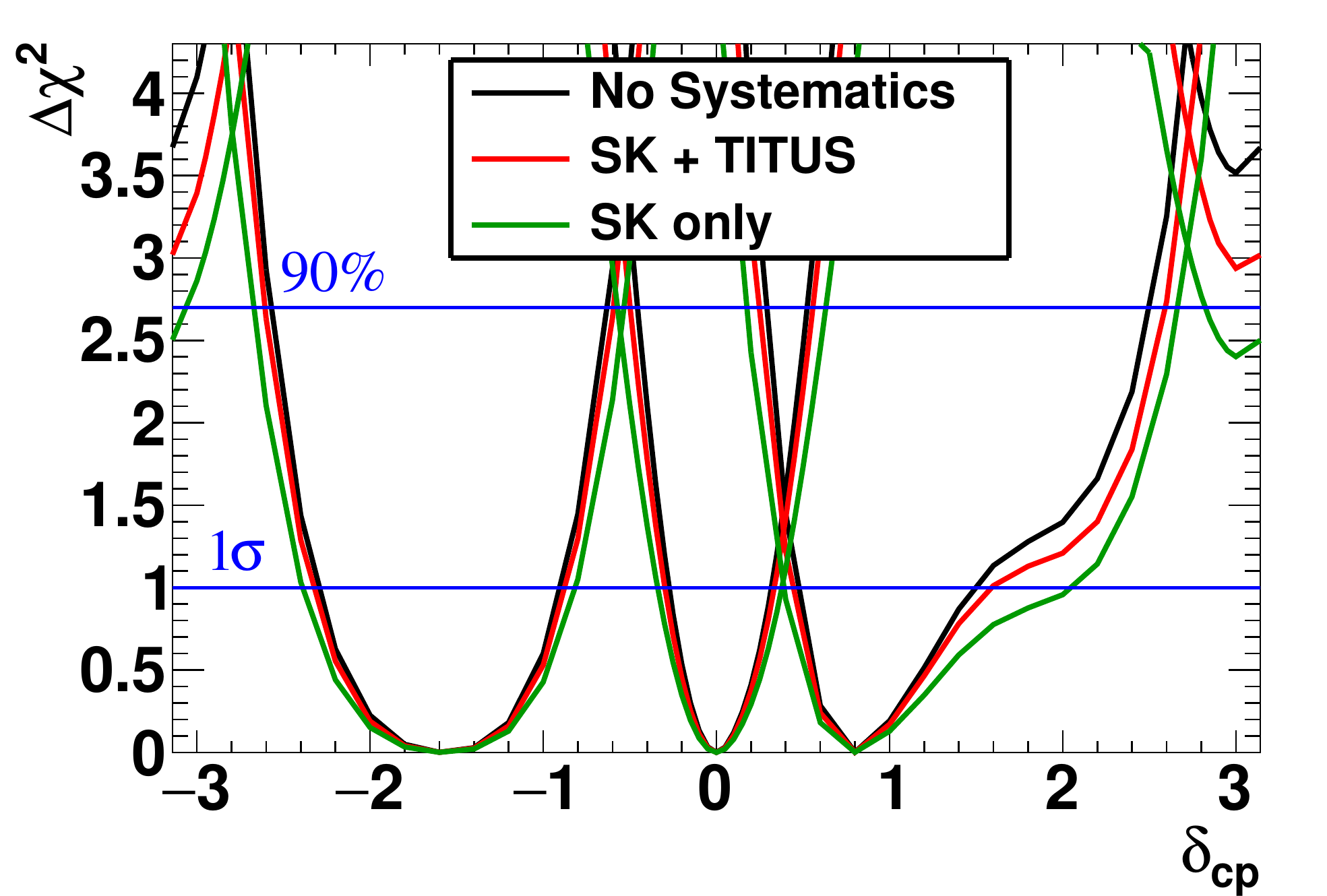} 
\caption{
Significance to measure CP violation as a function of $\delta_{cp}$
with T2K phase-2 only, TITUS and T2K phase-2, and T2K phase-2 without
considering sources of systematic uncertainty for 3 Asimov fits to
$\delta_{cp}$ assuming T2K phase-2 statistics.  The sensitivity is for
the case of normal hierarchy and assuming it is known. The uncertainty
on $\delta_{cp}$ is given in Table~\ref{tab:dcp_error-t2k2} and the
other oscillation parameters are given in
Table~\ref{tab:osc_params_joint_asimov}.}
\label{fig:cp-fit-t2k2}
\end{figure}

The $\delta_{cp}$ fit and CP violation sensitivity studies were also
performed for the T2K phase-2 setup with the addition of a TITUS
constraint. In these studies the systematic model was maintained from
the Hyper-K fits but exposures of each detector were scaled to the
size and beam power $\times$ time. Additionally, the exposure was
scaled to account for the expected increase in fiducial volume expected for 
Super Kamiokande~\cite{t2k2}. The sensitivity to CP violation is
shown in Figure~\ref{fig:cp-violation-t2k2} and the $\delta_{cp}$
sensitivity is shown in Figure~\ref{fig:cp-fit-t2k2} and
Table~\ref{tab:dcp_error-t2k2}.

\begin{figure}[htb]
\centering
\includegraphics[width=0.6\textwidth]{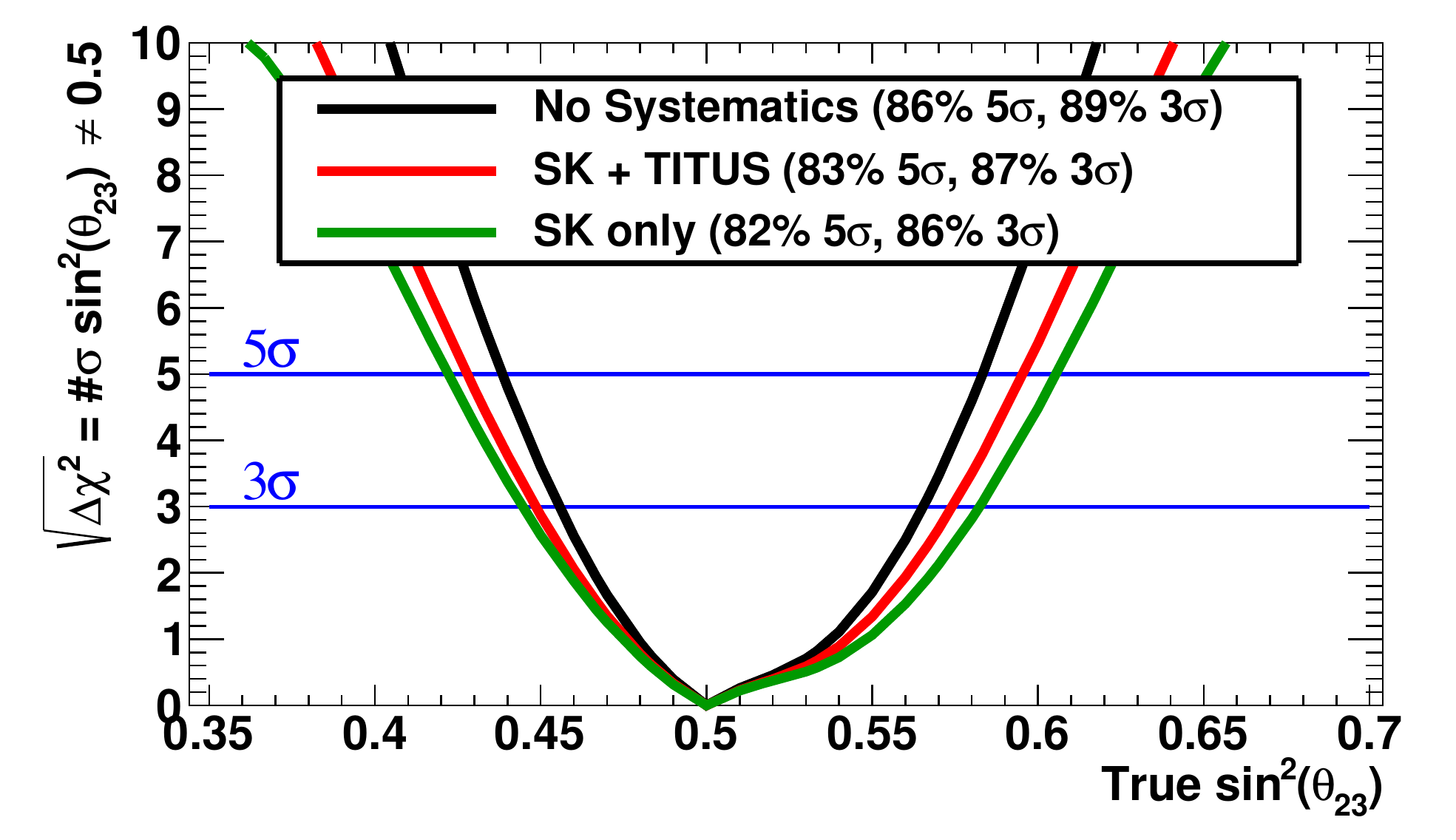} 
\caption{
Significance to exclude $\sin^2(\theta_{23})=0.5$ as a function of
$\sin^2(\theta_{23})$ with the combinations of samples and constraints
given in the legend and the fraction of $\sin^2(\theta_{23})$ Super-K
will be able to exclude $\sin^2(\theta_{23})=0.5$ (at 5$\sigma$ and
3$\sigma$). For the studies, the true values of oscillation parameters
to generate the data sets were taken from
Table~\ref{tab:osc_params_joint_asimov}. Sensitivity is for the case
of normal hierarchy and assuming it is known.  Due to a non zero
$\sin^2(\theta_{13})$, the disappearance spectra have a symmetry
around $\sin^2(\theta_{23})=0.511$, which is the cause of the flat
feature in the sensitivity between 0.5 and 0.511.}
\label{fig:ss23-neq0.5-t2k2}
\end{figure}

The sensitivity to non maximal mixing is shown in Figure~\ref{fig:ss23-neq0.5-t2k2}.

From these studies we see that TITUS is able to limit the dominant systematics effectively for 
the case of a $\delta_{cp}$ and CPV measurement as the Super Kamiokande + TITUS combination
has very similar sensitivity to the case with no systematic uncertainty. In the case of excluding maximal 
mixing, TITUS offers some improvement in the sensitivity but could also do better with optimisation. 

\begin{table}[htb]
\begin{center}
\begin{tabular}{c| c |c |c } 
\hline
 \hline & \multicolumn{3}{c}{Error (radians/\textit{degrees}) } \\ \hline True
 					$\delta_{cp}$ & No Systematics
 					& T2K phase-2 only & TITUS +
 					T2K phase-2 \\

 \hline
 $0$ 	   			&   0.296 {\it(17.0)}		& 	0.359 {\it(20.6)} 	&		0.316 {\it(18.1)}	  \\
 \hline
 $\frac{\pi}{4}$ 	   	&   0.514 {\it(29.5)} 		& 	 0.839 {\it(48.1)} 	&		0.577 {\it(33.1)} 	 \\
 \hline
 $-\frac{\pi}{2}$ 	   	&    0.709 {\it(40.6)}		& 	0.790 {\it(45.3)} 	&		0.737 {\it(42.2)} 	  \\
  \hline\hline
\end{tabular}
\end{center}
\caption{1$\sigma$ error (radians/\textit{degrees}) of $\delta_{CP}$ for each fit for T2K phase-2 shown in Figure~\ref{fig:cp-fit-t2k2}.}
\label{tab:dcp_error-t2k2}
\end{table}

\section{Other physics}
\label{sec-other}
The physics programme of the TITUS detector goes beyond oscillation
analysis. The detector will be able to provide new important
measurements on neutrino cross sections; it will have the unique
ability to measure the neutron content of the cross sections.  It will
also be able to address possible precision measurements like the weak
mixing angle, the strangeness content of the nucleus and isospin
physics.  Moreover, the experiment maybe able to generate a supernova
alarm and perform interesting physics with the detected events.
Finally, it can look at WIMP signatures and non-standard neutrino
interactions.  In the following sections, a more detailed look at
these processes will be given.

\subsection{Neutrino cross sections}
\label{subsec-xsection}
Cherenkov detectors exhibit excellent kinematic coverage due to their
4$\pi$ angular acceptance~\cite{MB1}. Located near to the beam
production point, a sizeable Cherenkov detector will offer significant
statistics for cross-section studies.  In Table~\ref{tab-ndcross} we
give estimated statistics for various event samples in a 2\,kton water
Cherenkov detector, with a 1\,m from wall fiducial volume cut, at
2\,km (2.5$^\circ$ off-axis) and a 1.3\,MW beam, corresponding to a
total of 27.05$\times10^{21}$ POT for 10 years.  The event samples for
this analysis are defined by topology, i.e. the number of mesons in
the final state, such that 1$\pi$ is one charged pion in the final
state with no other mesons.
%Since the full TITUS reconstruction is still under
%development, reconstruction efficiencies comparable to those achieved
%for Super-K (with 40\% photocoverage) were assumed to obtain these
%numbers. The event numbers obtained for Super-K are scaled by baseline
%and detector fiducial volume but the efficiencies and purities
%obtained from Super-K simulations are used directly and shown in the
%final column of the table.
We use a similar approach to that adopted for the basic selection in
section~\ref{sec-basicstrategy}.
\begin{table}[htbp]
\begin{center}
\begin{tabular}{c|p{3.5cm}|c|c} \hline\hline
Selection & Cuts & Nevents & Selection Characteristics \\ \hline\hline
$\nu_e$CC0$\pi$ enhanced & 1Re FCFV, 0DE, $E_{vis}>100$\,MeV & 350k      & $\epsilon\approx$ 41\%, $P\approx $ 61\% \\ \hline
$\nu_\mu$CC0$\pi$ enhanced & 1R$\mu$, $<2$DE, $p_{\mu}>$200\,MeV& 25M   &  $\epsilon\approx$ 58\%, $P\approx $ 90\% \\ \hline
$\nu_e$CC1$\pi$ enhanced &1Re FCFV, 1DE, $E_{vis}>$100\,MeV& 180k       &  $\epsilon\approx$ 12\%, $P\approx $ 21\%   \\ \hline
$\nu_\mu$CC1$\pi$ enhanced &1R$\mu$, $2$DE, $p_{\mu}>$200\,MeV& 2.3M    &  $\epsilon\approx$ 17\% , $P\approx $ 90\% \\ \hline
NC$\pi^0$ enhanced &2Re FCFV, 0DE, $85<m_{\rm invariant}<185$\,MeV/c$^2$ & 1.5M    &  $\epsilon\approx$ 40\%, $P\approx $ 88\%   \\ \hline
\hline
\end{tabular}
\caption{
Some of the primary cross-section measurements accessible with a
2\,kton water Cherenkov detector at 2\,km from the beamline.  The
predicted numbers of events have been evaluated for 27.05$\times10^{21}$POT.  The
efficiency, $\epsilon$ is the number of selected events divided by the
total number of events for the given topology, the purity $P$ is the
number of events of a given topology divided by the total events
selected. Key to Cuts: FCFV = fully contained in the fiducial volume
of a 2.09\,kton inner detector. 1Re/$\mu$ = 1 electron/muon type ring,
DE = decay electrons identified, $m_{\rm invariant}$ = reconstructed
invariant mass. }

\label{tab-ndcross}
\end{center}
\end{table}%

%In order to predict the accuracy of cross-section measurements with
%these samples, the existing T2K flux and cross-section uncertainty
%covariance matrices were used to establish the baseline
%accuracy\cite{Abe:2015awa}. 
Improvements in our understanding of the underlying interaction
mechanism are expected in the near future with ongoing T2K analyses
and hence these estimates are considered conservative.
%Individual and cross-section ratio uncertainties were
%established by random gaussian selection from the covariance matrices. 
Expressing cross-sections as a ratio takes advantage of the
cancellation of correlated systematic sources, in particular flux
uncertainties, to maximise the accuracy of extracted
information. These studies indicate that the
$\nu_e$CC0$\pi$/$\nu_\mu$CC0$\pi$ cross section ratio can be measured
to 3.2\% accuracy, dominated by a 3.1\% contribution from cross
section uncertainties. The NC$\pi^0$/$\nu_\mu$CC0$\pi$ cross section
ratio can be measured to an accuracy of 14.8\% again dominated by
cross section uncertainties.

To put these measurement predictions into context, the $\nu_e$
cross-section ratio would give significant improvement over existing
results: the T2K experiment has published a $\nu_e$ charged current
cross-section measurement on carbon to 18\% precision\cite{nue-t2k}
and compared the interaction rate on water to the predicted rate with
45\% precision\cite{pod-t2k}. TITUS can improve upon these
measurements due to the large active water target volume. Furthermore,
with further reduction of other cross-section uncertainties from
proceeding T2K analyses, the 3.2\% accuracy predicted here can be
considered a conservative bound. For the NC$\pi^0$ ratio, the
MiniBooNE experiment have published a measurement of the single
$\pi^0$ production cross-section on mineral oil with 16\%
accuracy\cite{miniboone-pi0} and the ArgoNeuT collaboration have
measured the NC$\pi^0$ to CC cross-section ratio on an argon target
with 28\% precision\cite{argoneut-pi0}. The only existing measurement
on a water target is from the K2K collaboration, who give a
non-differential NC$\pi^0$ to total charged current cross section
ratio with 11\% precision\cite{k2k-ncpi0}, but with a number of
model-dependent assumptions that yield significantly lower
uncertainties than current thinking on neutrino interaction model
uncertainty. Despite the lower beam energy, the larger exposure and
target volume should yield nearly 24 times as many selected NC$\pi^0$
events in TITUS as K2K.

%The challenge of such measurements is to achieve the necessary event
%purity with sufficient efficiency to maintain useful sample
%statistics. In particular, to measure electron neutrino
%cross-sections, % statement about how interesting these are.... % a
%high level of muon interaction rejection is required as electrons
%make up only a small percentage of the initial beam flux.  Further
%optimisations for the ring PID are envisaged, but currently from the
%full simulation chain we achieve ...  assumptions -> % accuracy.
%Compare to T2K measurements - different target so meaningful numbers
%cannot be extracted from reconstructed TITUS Monte Carlo for electron
%neutrino cross-section studies at this stage. However, to understand
%how well the broad Super-K reconstruction assumptions apply to a
%smaller detector, with higher granularity of photo-detector coverage,
%the neutral current with a final state neutral pion selection has
%been studied in more detail as described in
%section~\ref{subsubsec-ncxsection}.
%
%Despite their superior kinematic coverage, Cherenkov detectors tend to
%be poorer at event topology classification but there is a movement
%amongst next generation neutrino cross-section measurements, such as
%MicroBooNE~\cite{Teppei_uB}, to further classify event topologies to
%measure more exclusive differential cross-sections to understand
%genuine neutrino interactions as well as nuclear effects.  
The selections in Table~\ref{tab-ndcross} do not include any neutron
tagging information.  However, gadolinium doped water Cherenkov
detectors, such as TITUS, provide a natural path to future neutrino
cross-section measurements with neutron tagging of the final state.

\begin{figure}[htb]
\centering
\begin{tabular}{cc}
\includegraphics[width=0.45\textwidth]{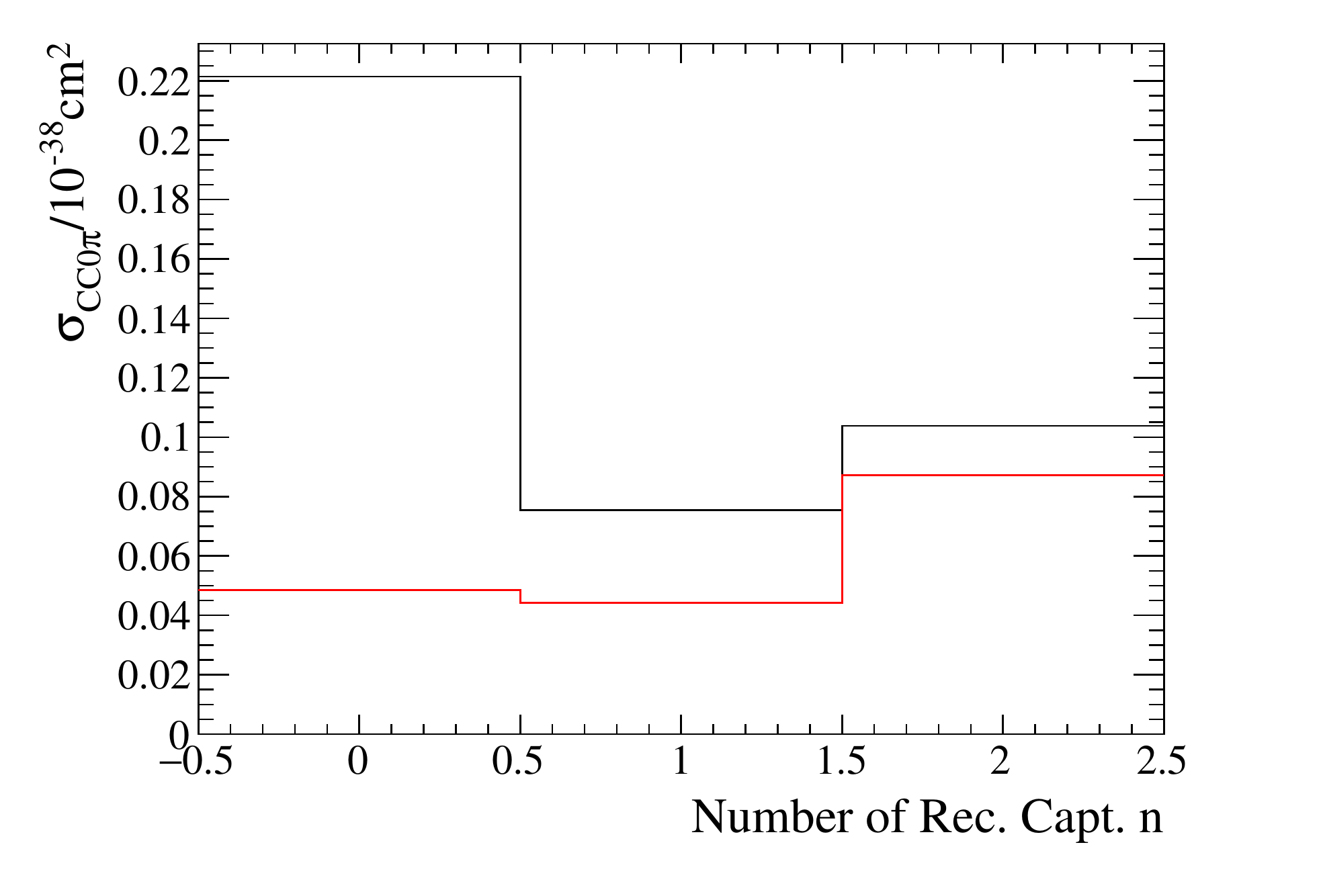}&
\includegraphics[width=0.45\textwidth]{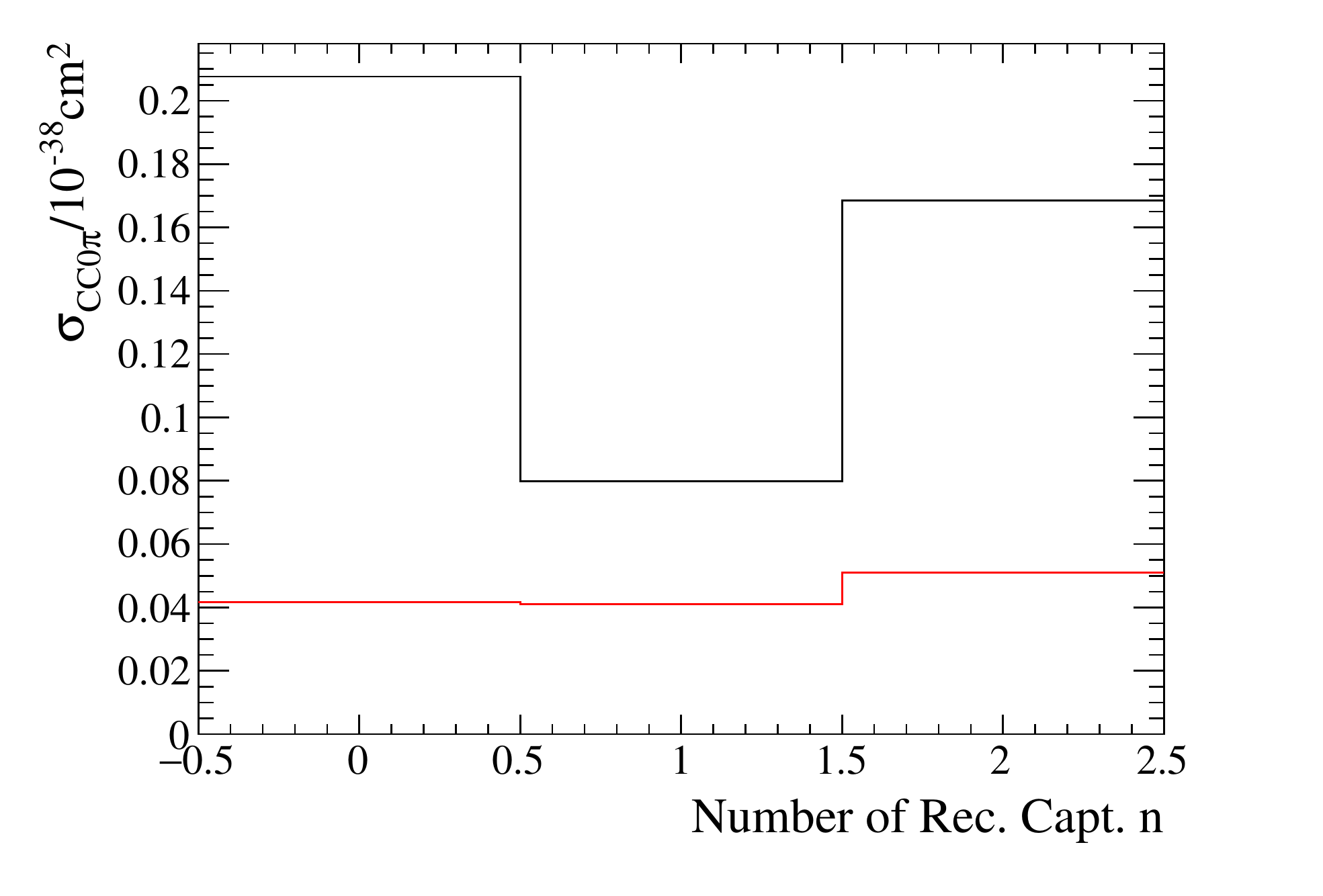}\\
\end{tabular}
\caption{
True flux-integrated cross section as a function of neutron
multiplicity for $\nu_\mu$ (black) and $\overline{\nu}_{\mu}$ (red)
CC0$\pi$ using the NEUT generator.  The \FHC(\RHC) measurement is on
the left(right).  Statistical and systematic uncertainties have not
been considered.  The last bin is integrated for all events with
neutron captures greater than 1.}
\label{fig:antinuCC0pi}
\end{figure}

A statistical separation of interaction types based on neutron
multiplicity described in section~\ref{subsec-neutron} can offer a new
way to measure exclusive differential cross-sections by Cherenkov
detectors, and could provide the first measurement of a genuine CCQE
cross-section by a Cherenkov detector. An example of which would be
the $\overline{\nu}_{\mu}$CC0$\pi$ cross section as a function of
neutron multiplicity, as seen in Figure~\ref{fig:antinuCC0pi}.  It is
also expected that the inelastic channels are accompanied by nucleon
emissions~\cite{ArgoNeuT_Hammer}.  Therefore, detailed measurements of
neutron multiplicity also opens a way to study inelastic channels,
mainly the $\Delta$ resonance.

Furthermore, TITUS can help to reduce the main background to proton
decays in e$^+\pi^0$, which comes from atmospheric neutrinos. Neutron
tagging is used to reject background events by requiring signal
without neutron candidate events in the final state. A measurement of
the neutron content in the candidate events will provide an important
input to the background determination for proton decays.
%In the following, we will focus on two cross sections in particular:
%the neutral current cross section with a neutral pion in the final
%state (NC$\pi^0$) and the electron (anti)neutrino cross section.

\subsection{Standard Model measurements}
\label{subsec-SM}
Neutrinos are natural probes of electroweak physics and TITUS can
perform several important measurements, a few of which are discussed
here.  The $4\pi$ angular coverage of the detector will allow to look
at both low Q$^2$ and Deep Inelastic Scattering (DIS) regions that are
needed for the measurements below.

Perhaps the cleanest and most direct method of determining the proton
d/u quark ratio at large Bjorken-$x$ is from neutrino and
antineutrino DIS on hydrogen. Existing neutrino data on hydrogen have
relatively large errors and do not extend beyond
$x\, \sim \,$0.5~\cite{Yang:2000ju}. A new measurement of neutrino and
antineutrino DIS from hydrogen with significantly improved
uncertainties would therefore make an important discovery about the
d/u behaviour as $x\rightarrow 1$.

\begin{figure}[htb]
\centering
\begin{tabular}{cc}
\includegraphics[width=0.45\textwidth]{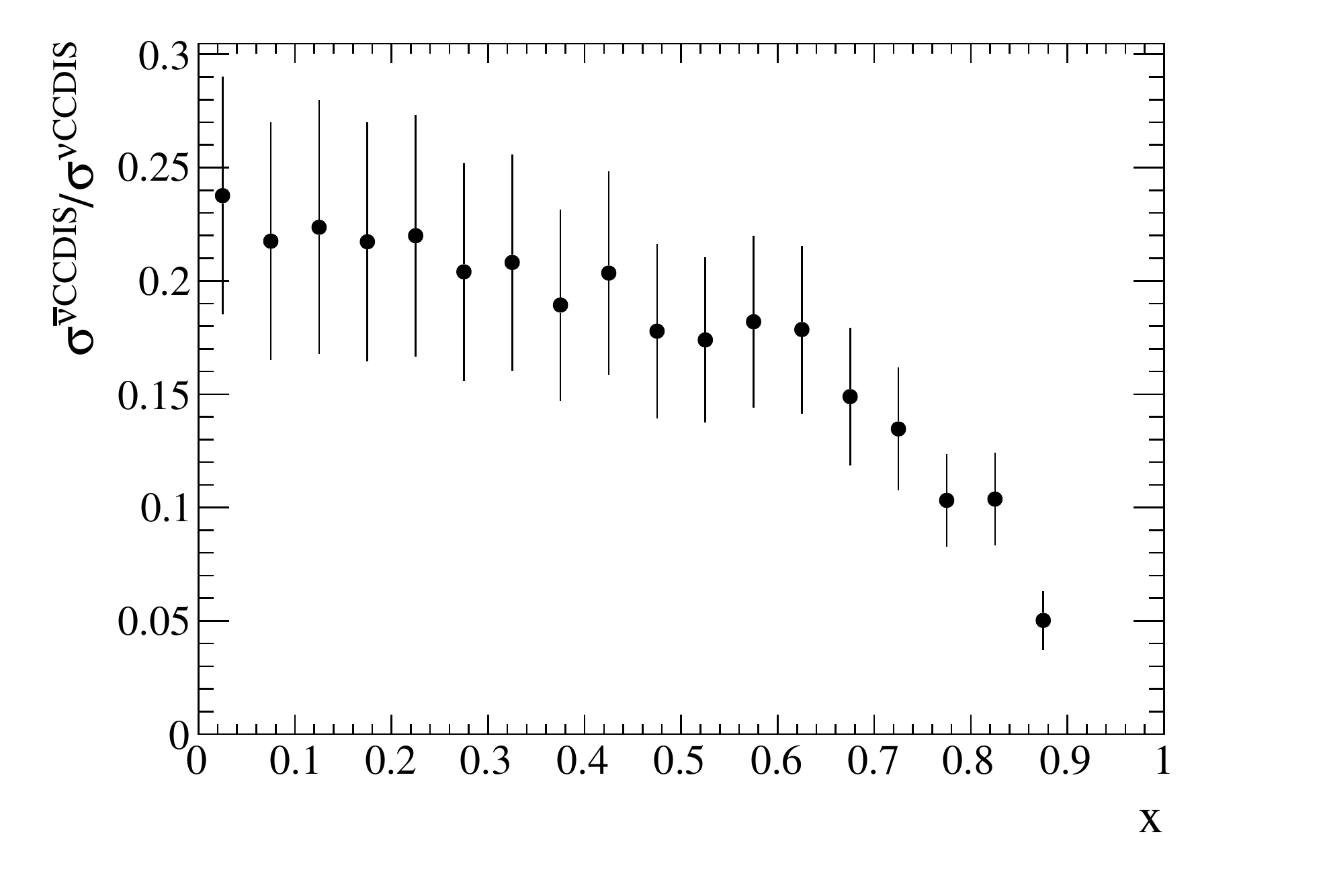}&
\includegraphics[width=0.45\textwidth]{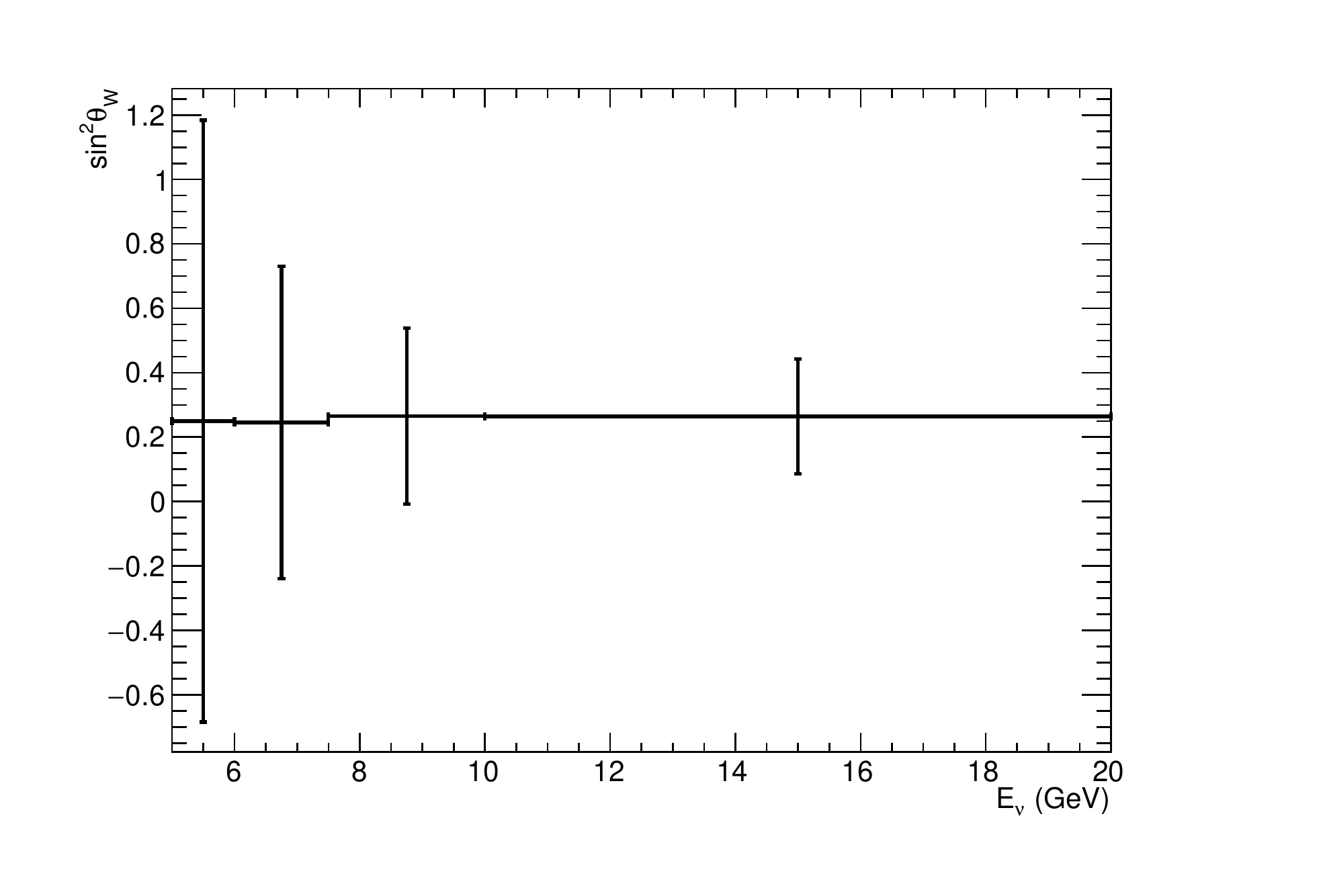}\\
\end{tabular}
\caption{
Left: Ratio of the $\overline{\nu}_{\mu}$CCDIS to $\nu_{\mu}$CCDIS
cross sections as a function of $x$ with statistical and systematic
errors given in the text. Right: Measurement of $\sin^{2}\theta_{W}$
as a function of neutrino energy starting from 5\,GeV.}
\label{fig:du_ratio}
\end{figure}

TITUS will have neutrino DIS interactions with $x>0.5$, as shown in
Figure~\ref{fig:du_ratio}.  NEUT was used to simulate $\nu_{\mu}$
interactions using a \FHC flux of $6.8\times 10^{21}$\,POT and
$\overline{\nu}_{\mu}$ interactions with the \RHC flux with
$20.3\times10^{21}$\,POT.  Gaussian systematic errors were thrown with
three assumptions for its source: an uncertainty on the cross section
of 0.4/E$_{\nu}$, where the neutrino energy is in GeV; a flat flux
uncertainty of 9\%, independently thrown for the \RHC and \FHC
configurations; and a flat detector uncertainty of 5\%.  Uncertainties
for $x>0.5$ may be on the order of 20\%.  A more precise estimation
would require a full event reconstruction that would likely need to
select multi-ring interactions, more precise treatment of background
subtraction mainly of wrong sign neutrino interactions and propagation
of hadrons through the nucleus; this would provide new data where
there are currently no measurements. Moreover, the above error
estimates are conservative, so it is possible to refine them with
further study.

In addition, the strange quark content of the proton and its
contribution to the proton-spin can be inferred from TITUS data using
ratio of NCEL and CCQE cross sections, a procedure which reduces
systematic uncertainties.  A large observed value of the strange quark
contribution to the nucleon spin (axial current), $\Delta$s, would
change our understanding of the proton structure. The spin structure
of the nucleon also affects the couplings of axions and supersymmetric
particles to dark matter.

A measurement of the ratios $\frac{\sigma(\nu_\mu
p \rightarrow \nu_\mu p )}{\sigma(\nu_\mu n \rightarrow \mu^- p )}$
and $\frac{\sigma(\overline{\nu}_\mu p \rightarrow \overline{\nu}_\mu
p )}{\sigma(\overline{\nu}_\mu n \rightarrow \mu^+ p )}$ to 1\%
precision would enable the extraction of $\Delta$s with an uncertainty
of 0.8\% and 0.5\% for neutrino and antineutrino ratios,
respectively. Current measurements have $\sim$50\%~\cite{MBNCEL}. The
$\Delta$s analysis requires that TITUS is able to reconstruct protons
at just above the proton's Cherenkov threshold.

A measurement of $\frac{\sigma(\nu_\mu p \rightarrow \nu_\mu p
)}{\sigma(\nu_\mu n \rightarrow \nu_\mu n )}$ is also possible due to
the fact that TITUS is Gd-doped.  Assuming $\Delta$s has a value of
$-0.1$, an uncertainty of 10\% on $\Delta$s requires that the ratio
has an uncertainty of 30\%.  While many systematic uncertainties will
cancel in this ratio, the kinematic threshold of the proton,
background neutrino interaction subtraction, external neutron
backgrounds, and the choice of nuclear model need to be
well-understood to attain a measurement of this precision, which would
be the first of its kind and would provide the best sensitivity to
$\Delta$s if it were non-zero.

Finally, by measuring the ratio of the DIS neutrino NC and CC
interactions, TITUS can perform a measurement of the weak mixing angle
$\theta_W$:
\begin{equation}
R_{\nu} \equiv \frac{\sigma^{\nu}_{NC} - \sigma^{\overline{\nu}}_{NC}}{\sigma^{\nu}_{CC} - \sigma^{\overline{\nu}}_{CC}} \sim \rho^{2}(1-\sin^{2}\theta_{W})
%R_\nu \equiv \frac{\sigma^\nu_{NC}}{\sigma^\nu_{CC}}\sim \rho^2 (\frac{1}{2}-sin^2\theta_{W}+\frac{5}{9}(1+r)\sin^4\theta_W)
\end{equation}
where $\rho$ is from the Paschos-Wolfenstein relations and is taken to
be one \cite{Paschos:1972kj}. 

A single bin Monte Carlo truth study was performed to get an idea of
the precision of such a measurement on neutrino interactions on oxygen
using the same generated sample as the $d/u$ measurement discussed
above. It was found that an uncertainty of 2.6\% on the measurement of
$\sin^{2}\theta_{W}$ can be reached. This would be the first
measurement of its kind by a water Cherenkov detector and will
potentially shed light on the NuTeV anomaly \cite{ZellerNuTeV}.

%Here $\rho$ is the relative coupling strength of the
%neutral-to-charged current interactions ($\rho$ = 1 at tree-level in
%the Standard Model) and $r$ is the ratio of antineutrino to neutrino
%cross section. 

\subsection{Supernova burst}
\label{subsec-sn}
Core collapse supernova explosions are the final evolutionary stage of
massive stars, $\gtrsim8M_\odot$. Such explosions release $99\%$ of
their energy, estimated to be
$\mathcal{O}\left(10^{53}\right)\text{ergs}$, in neutrinos. To date,
the only observation made is that of 25 neutrinos from supernova 1987A
by the Kamiokande~\cite{kamiokande1987a}, IMB~\cite{imb1987a} and
Baksan~\cite{baksan1987a} detectors. Supernova 1987A occurred in the
Large Magellanic Cloud, at a distance of
$(50.0\pm1.1)\,$kpc~\cite{pietrzynski13}; if a supernova were to occur
within the Milky Way Galaxy, at a likely distance of
$\mathcal{O}\left(10\right)\text{kpc}$, then the combination of a
5$\times$ shorter distance and the much larger target masses of
current and future detectors leads to an expected event rate several
orders of magnitude greater than SN 1987A. These large event rates
precede by a few hours the observation of light from supernova
explosions, and so can be used as an early warning for astronomers to
prepare visible observations through the SuperNova Early Warning
System (SNEWS)~\cite{SNEWS}, as well as offering information about the
formation of the neutron star and the first few seconds of the
subsequent explosion. Additionally, such observations probe
interesting neutrino physics such as the neutrino mass ordering and
neutrino oscillations.

The TITUS detector, with its reasonably large mass $\left( \sim
2 \, \text{kton} \right)$ and potential use of gadolinium doping,
should be a useful contributor to the SNEWS network and probe some
supernova neutrino interactions in ways unavailable to traditional
(not doped) water Cherenkov detectors.

Prediction of the expected neutrino flux from core-collapse supernovae
remains a challenging problem in astrophysics. Although the
``neutralisation pulse'' of $\nu_e$ from the formation of the neutron
star is easily calculated, it does not dominate the neutrino flux:
most neutrinos are produced thermally in the course of the
explosion. Modelling the propagation of neutrinos through the
explosion is difficult, as the extremely high densities of both
electrons and neutrinos produce not only matter-enhanced MSW
oscillations, but also collective effects caused by neutrino-neutrino
interactions. The latter can be very dramatic, resulting in wholesale
exchange of flavours above a particular energy
threshold~\cite{scholberg12}.  In addition, 3D simulations of
supernova explosions indicate that they are quite asymmetric, and the
flux of neutrinos, particularly its detailed time structure, may
depend on direction~\cite{tamborra14}.

\begin{figure}[htbp]
\centering
\includegraphics[width=0.45\linewidth]{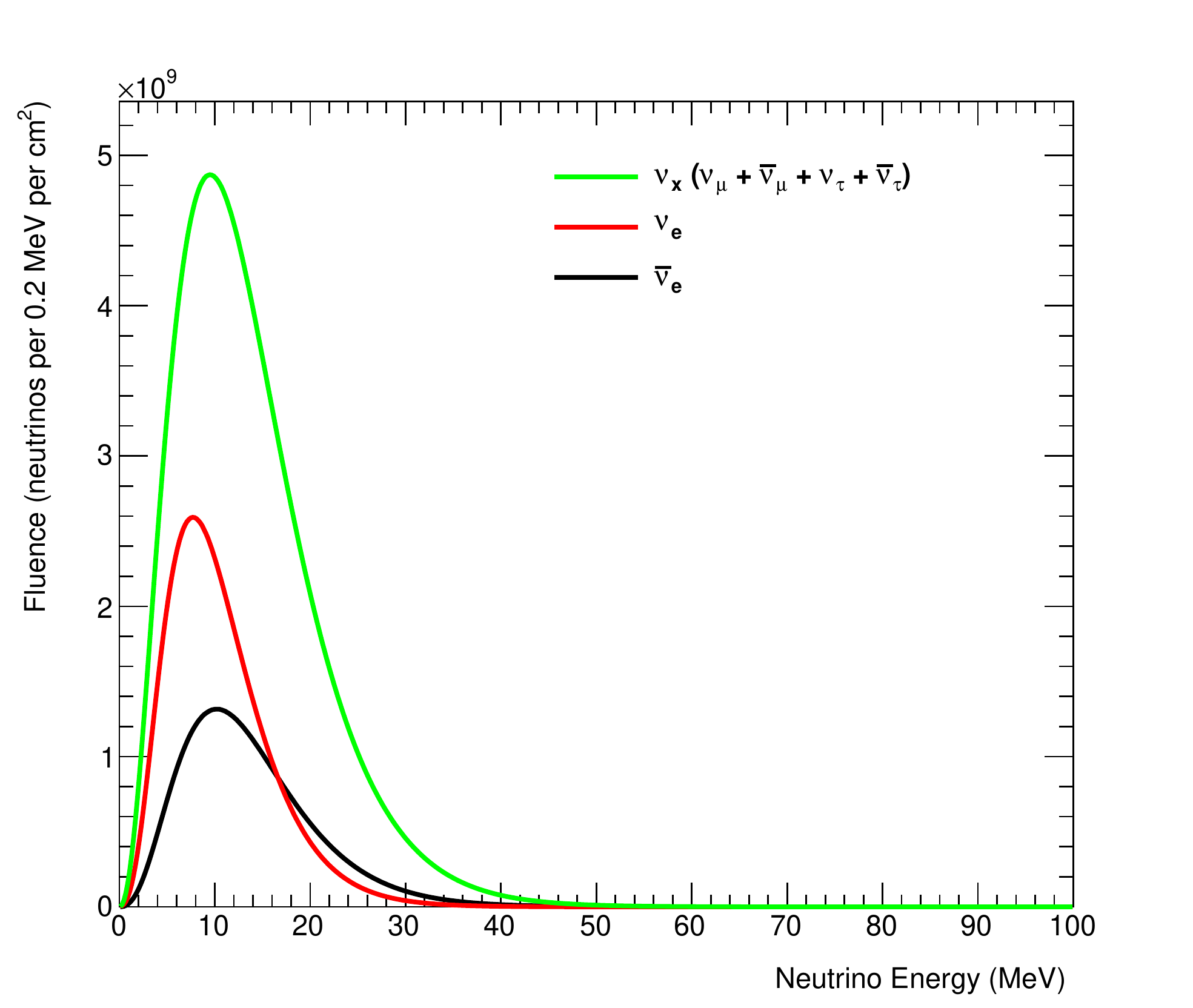}
\includegraphics[width=0.45\linewidth]{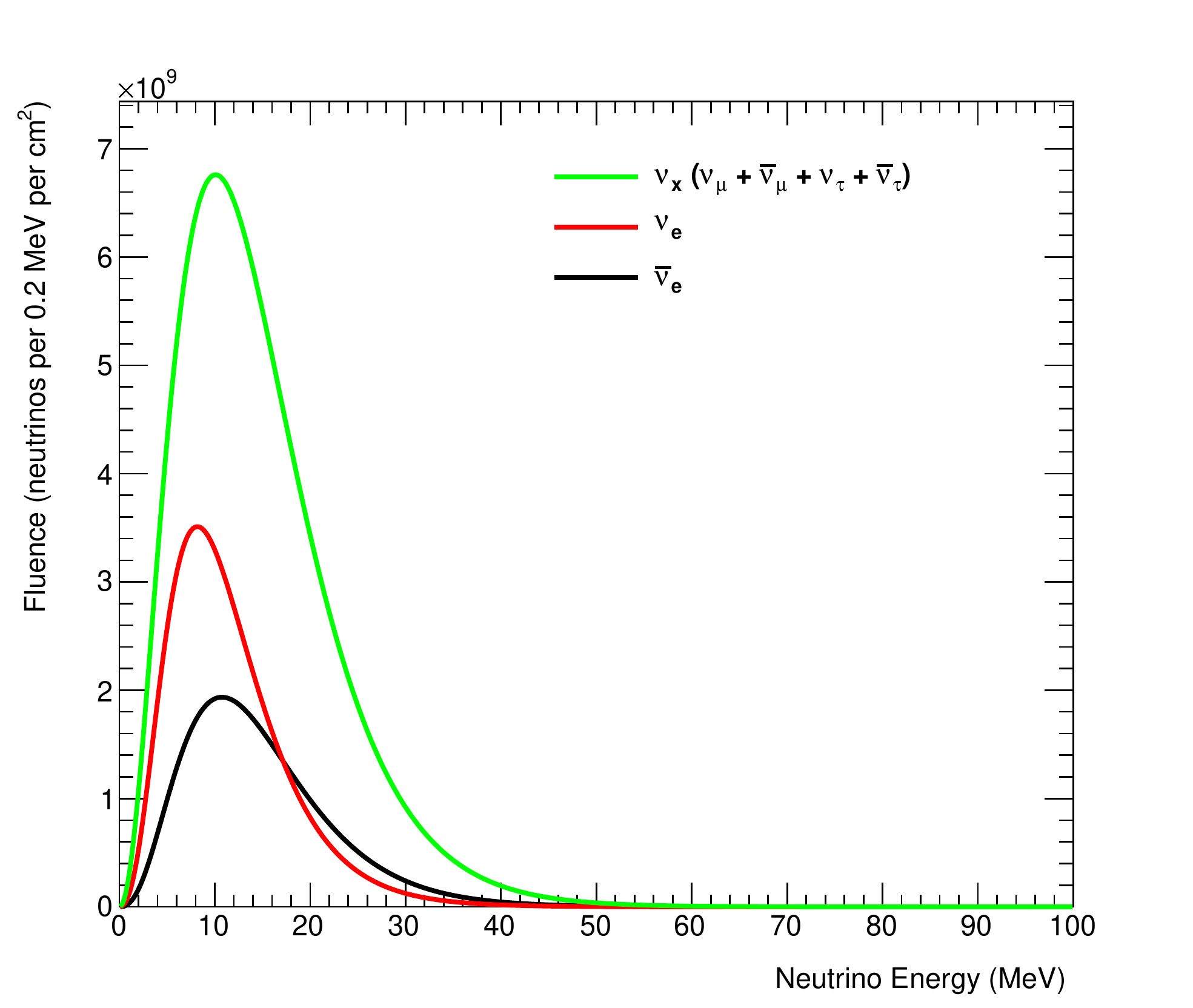}
\caption{\label{fig:sn_flux}Total predicted neutrino flux for $11.2M_\odot$ (left) and $27.0M_\odot$ (right) progenitor mass stars at a distance of 10\,kpc.}
\end{figure}

\begin{figure}[htbp]
\centering
\includegraphics[width=0.45\linewidth]{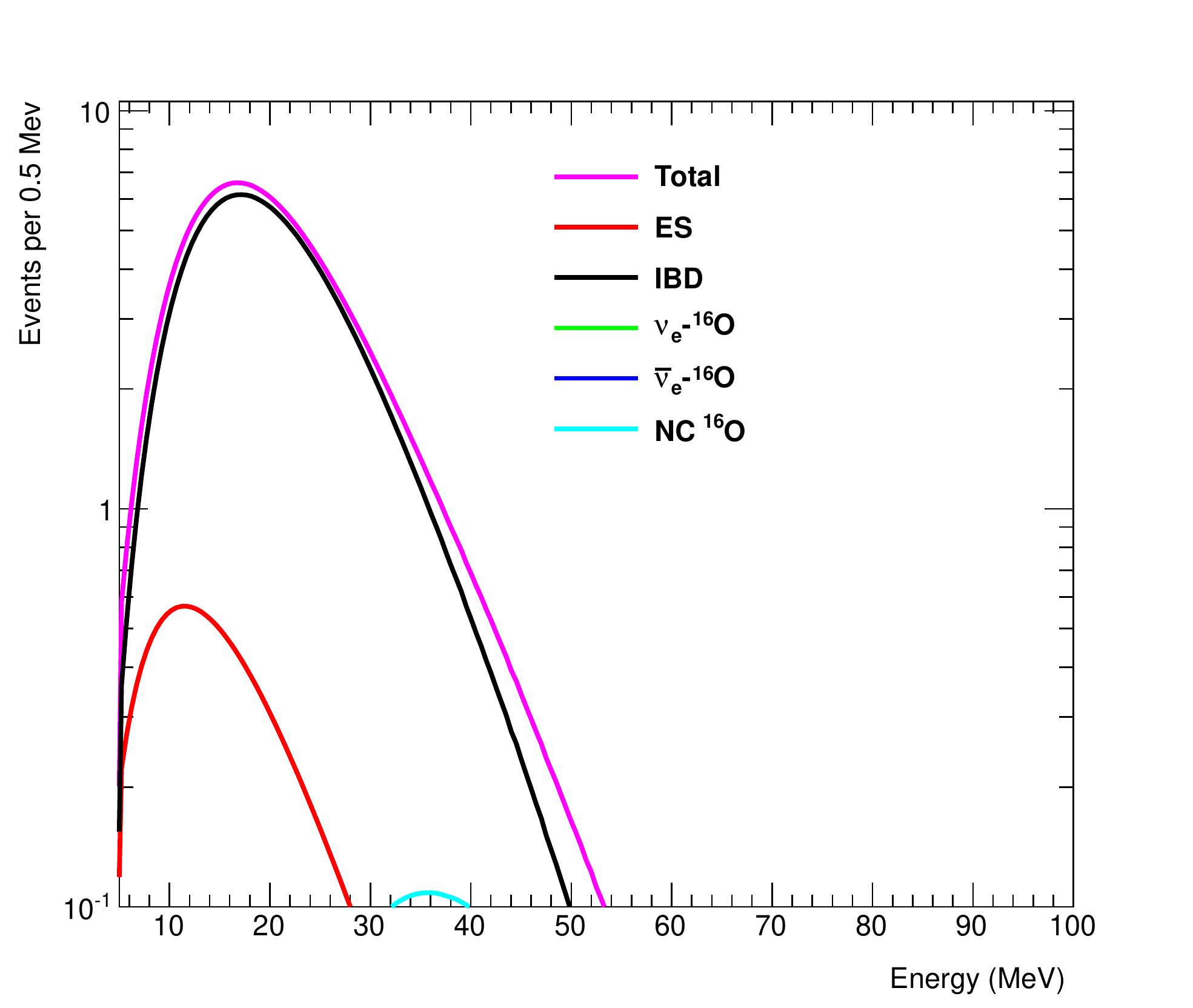}
\includegraphics[width=0.45\linewidth]{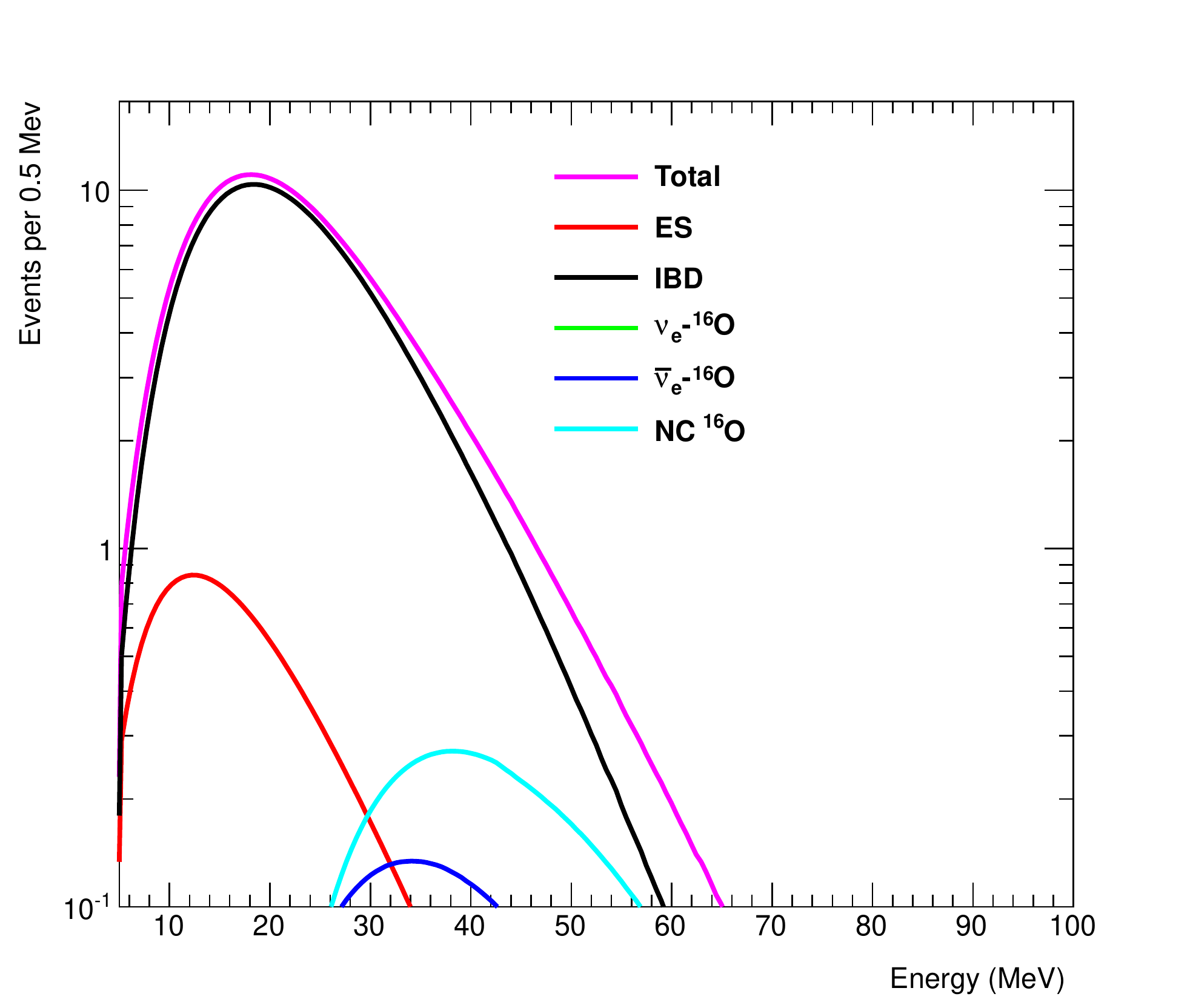}
\caption{\label{fig:sn_spec}Predicted neutrino interaction spectrum, broken down by interaction type, for $11.2M_\odot$ (left) and $27.0M_\odot$ (right) progenitor mass stars at a distance of 10\,kpc in the TITUS detector. Legend defined in Table~\ref{table:sn_rate}.}
\end{figure}

Initial studies of the expected neutrino interaction rate from a
supernova explosion within the TITUS detector have been undertaken
using the SNOwGLoBES software package~\cite{snowglobes}, in
conjunction with numerical flux predictions developed by Hans-Thomas
Janka and colleagues at the Max Planck Institute for Astrophysics, and
detailed in the thesis of Lorenz H\"{u}depohl~\cite{hudepohl}. The
available models include a range of different supernova progenitor
masses, two different equations of state for the nuclear environment,
and potential inclusion of corrections to the treatment of convection
and neutrino-nucleon opacities. In the future the full range of these
models will be considered in detail, but for the moment, results are
shown for progenitor masses of $11.2M_\odot$ and $27.0M_\odot$, with
the Lattimer and Swesty equation of state~\cite{lattimer} with a
nuclear compressibility factor $K=220$~MeV and the inclusion of
convection and opacity corrections. The predicted integrated fluxes
from these two models are shown in Figure~\ref{fig:sn_flux} for
progenitors at a distance of 10\,kpc. Note that these results do not
currently include the effects of neutrino oscillations.
The integrated neutrino flux is then convolved with the relevant
neutrino interaction cross-sections for a water target to produce an
expected event rate and interaction spectrum within the TITUS
detector, as shown in Figure~\ref{fig:sn_spec} and
Table~\ref{table:sn_rate}.
\begin{table}[hbtp]
\begin{center}
\begin{tabular}{ c || c | c }
\hline\hline
Interaction & \multicolumn{2}{c}{Event Rate} \\ Type & $11.2M_\odot$ &
$27.0M_\odot$ \\
\hline
Elastic Scatter & 17.9 & 29.9 \\
Inverse Beta Decay & 233.8 & 440.2 \\
$\nu_e-^{16}$O & 0.4 & 2.1 \\
$\bar{\nu}_e-^{16}$O & 2.5 & 7.1 \\
NC $^{16}$O & 5.3 & 15.0 \\
\hline
Total & 260.0 & 494.2 \\
\hline\hline
\end{tabular}
\caption{\label{table:sn_rate}Predicted interaction rates broken down by interaction type, for $11.2M_\odot$ and $27.0M_\odot$ progenitor mass stars at a distance of 10\,kpc in the TITUS detector.}
\end{center}
\end{table}
Work is being undertaken to calculate and apply the TITUS detector
efficiency and energy reconstruction smearing to the predicted event
rates. However, the significant event rate predicted at the TITUS
detector should allow it to usefully contribute to the SNEWS network
and potentially contribute to interesting physics measurements,
particularly through unambiguous identification of inverse beta decay
events by neutron capture on gadolinium, offering both background
reduction and flavour tagging of $\bar{\nu}_e$.

\subsection{Dark matter}
\label{subsec-DM}
Astronomical and cosmological observations indicate that a large
amount ($\sim 85\%$) of the mass content of the Universe is made of
dark matter~\cite{planck2015}. Particle candidates under the generic
name of weakly interacting massive particles (WIMPs) arise naturally
in many theories beyond the Standard Model of particle physics and can
be experimentally detected~\cite{goodman1985}. However, the existing
evidence for dark matter provides limited information about its
non-gravitational interactions, and many candidates are sufficiently
non-relativistic and weakly interacting. This situation may open a
more complex hidden sector containing additional light
states.~\cite{bohem2004, pospelov2008} Direct detection experiments,
such as LUX or XENON100, impose stringent constraints on dark matter
with a weak-scale mass, excluding a significant region of the
spin-independent cross section versus mass phase
space~\cite{xenon100_si, lux_2015}. Along with this, astrophysical and
cosmological observations pose constraints on the dark matter relic
abundance and self-interaction cross sections, while collider
experiments can set bounds on the $U(1)'$ dark sector gauge coupling
and on the kinetic mixing coefficient~\cite{babar}. Near detector
neutrino beam experiments can probe each region, but it is
particularly important in the sub-GeV dark matter mass range, where
direct-detection experiments tend to have poor sensitivity.

This ``dark force'' phenomenology has seen increased interest in
recent years, and the presence of light mediators coupled to the
Standard Model opens up the possibility to probe it experimentally via
the production of dark matter beams directly in fixed target
facilities.~\cite{battaglieri2014} The scattering of the final
``dark'' light states in a detector spatially separated from the
production point may represent the most efficient search
strategy. Moreover, owing to the potentially large production rate,
and the existence of large volume near detectors, proton fixed-target
facilities focusing on neutrino physics appear to be an optimal means
for exploring these scenarios~\cite{deNiverville}.

Light dark matter particle  ($\chi$) beams can be generated via two processes:
{\it i)} direct production, $pp(n) \rightarrow
V^* \rightarrow \bar{\chi}\chi$, where $V$ is the vector mediator; and
{\it ii)} indirect production, $pp(n) \rightarrow \phi +
... \rightarrow V + ... \rightarrow \bar{\chi}\chi$ + ..., where
$\phi$ is a generic hadron state and $V$ is the vector mediator. Once
produced, the dark matter beam propagates along with the
neutrinos. The beam, weakly scattering with normal matter, is
detectable through neutral current-like elastic scattering processes
with electrons or, of most relevance here, with nucleons within the
neutrino near detector. 

TITUS will be particularly suitable for this measurement. In direct
$\chi$-production forward direction emission of dark matter is
suppressed, therefore the TITUS off-axis alignment is ideal for
capturing a large flux of dark matter, as compared to an on-axis
detector. As for signal identification, the dominant background from
neutrino elastic scattering can be rejected by exploiting some
distinctive characteristics. The dark matter beam has a higher average
energy than the neutrino beam ($\sim12$ GeV for a WIMP of 100 MeV, and
$\sim10$ GeV for a WIMP of 300 MeV, for direct production process and
off-axis detector angle of $\theta = 2^{\circ}$). This would permit a
relatively high cut in momentum transfer in scattering. In addition to
that, a much higher cutoff can be observed when the scattering energy
approaches the energy of the primary proton beam. In particular, a
cusp at the kinematic limits for larger $m_{\chi}$ can be observed, as
a result of a degeneracy in the angle between $\chi$ and the beam
direction in the lab frame, $\theta$, as a function of its value in
the $V$ rest frame. This feature, which can be optimally exploited by
off-axis experiments such as TITUS, is an additional tool for signal
identification.  Another feature lies in the fact that the dark matter
beam will be relatively unaffected by turning off or switching the
polarity of the magnetic focusing horns, which would alter the
neutrino beam significantly. If TITUS is equipped with LAPPDs
(section~\ref{sec:lappds}) this will further enhance the time
resolution and will provide useful information on the timing
structure, as the production mechanism for vector-portal-coupled dark
matter is very distinct from the neutrino beam.
% The addition of a water-soluble gadolinium compound, already foreseen for the TITUS detector, will further enhance the background rejection capability for dark matter searches, as also shown in~\cite{Rott}. 
The expected number of elastic nucleon dark matter scattering events
can be expressed as
\begin{equation}
N_{N\chi \rightarrow N\chi} \propto n_{N} \cdot \sum_{prod.} \left(
N_{\chi} \cdot \sum_{traj. i} R_i \sigma_{N\chi}(E_i)
f(\theta_i,p_i) \right),
\end{equation}
where $n_N$ is the nucleon density in the detector, $N_{\chi}$ is the
number of dark matter particle produced (in turn proportional to
$N_{\textrm{POT}}$ ), $R_i$ is a parameter corresponding to the
trajectory of the dark matter path within the detector and
$f(\theta_i,p_i)$ describes the production via direct and indirect
processes, with the former including the $(1-\cos^2\theta)$ factor
suppressing the $\chi$-production in the forward direction. Based on
studies previously performed on the ND280 expected
sensitivity~\cite{deNiverville}, TITUS could reach a WIMP-nucleus
cross section of the better than $\sim 10^{-40}$ cm$^{2}$ for 300 MeV
WIMP from direct production and $5 \times 10^{-39}$ cm$^{2}$ for 100
MeV WIMP from indirect production (under the assumptions of the
mediator mass, $m_V$, between 1 GeV and 400 GeV, and the coupling
$\alpha'$ to be equal to $\alpha$, so that the coupling with the dark
sector is equal to the square of the kinetic mixing coefficient).

\section{Conclusions}
\label{sec-conclusions}
A description of the TITUS detector, a $\sim$2\,ton Gd-doped WC
detector with an MRD downstream, as proposed near detector for the
Hyper-Kamiokande experiment at about 2\,km from the beam target is
presented, along with an overview of its physics potential. The
physics potential described in this document is only based on the
analysis of the WC data and does not include the MRD.\\ The main goal
of the TITUS detector is the improvement of the CP potential of the
Hyper-Kamiokande experiment. Two sensitivity studies for CP
violation that give consistent results are presented in the text.
Using the VaLOR fitting method~\cite{valor}, assuming a 1.3\,MW beam,
1:3 POT ratio for \FHC and \RHC beams, the cross section uncertainties
as in the T2K oscillation analysis presented in
Ref.~\cite{Abe:2015awa} Hyper-K and TITUS can determine CP violation
at the 5$\sigma$ level for 62\% of $\delta_{cp}$ space and can provide
a 3$\sigma$ measurement for 79\%, close to the 84\% achieved without
considering systematic uncertainties.\\ Improvements are also observed
in the 23 sector using the TITUS detector.  \\ Finally, the detector
is also able to address non-oscillation physics, such as cross section
measurements, SM measurements, supernova neutrino and dark matter.

\end{document}